\documentclass[apj,numberedappendix]{emulateapj}
\usepackage{amssymb,amsmath}
\usepackage{epsfig}
\usepackage{longtable}
\usepackage{multirow}
\usepackage{natbib}

\newcommand{\etal}{et~al.~}

\begin{document}


%\title{The Redshift Evolution of Cool Cores in a Sample of \\83 SZ-Selected Galaxy Clusters from $0.3 < \lowercase{z} < 1.2$}
\title{The Growth of Cool Cores and Evolution of Cooling Properties in a \\Sample of 83 Galaxy Clusters at $0.3<\lowercase{z}<1.2$ Selected from the SPT-SZ Survey}

%\title{The Cool Core Properties of a Mass-Limited Sample of 80 Galaxy Clusters at $0.3<\lowercase{z}<1.2$ Discovered by the South Pole Telescope}


%\author{The SPT Team}

\def\hubble{$\dagger$}
\def\MIT{1}
\def\KICPChicago{2}
\def\EFIChicago{3}
\def\CfA{4}
\def\PhysicsUChicago{5}
\def\McGill{6}
\def\Caddo{7}
\def\UChicago{8}
\def\Harvard{9}
\def\Munich{10}
\def\ExcellenceCluster{11}
\def\Miss{12}
\def\AAUChicago{13}
\def\ANL{14}
\def\NIST{15}
\def\PUC{16}
\def\Berkeley{17}
\def\UFlorida{18}
\def\Colorado{19}
\def\NASA{20}
\def\Davis{21}
\def\LBNL{22}
\def\Caltech{23}
\def\Arizona{24}
\def\Michigan{25}
\def\MPE{26}
\def\CaseWestern{27}
\def\Minnesota{28}
\def\STScI{29}
\def\SAIC{30}
\def\LLNL{31}
\def\Dunlap{32}
\def\UofT{33}
\def\BCCP{34}

\altaffiltext{\hubble}{Hubble Fellow}
\altaffiltext{\MIT}{Kavli Institute for Astrophysics and Space Research, Massachusetts Institute of Technology, 77 Massachusetts Avenue, Cambridge, MA 02139}
\altaffiltext{\KICPChicago}{Kavli Institute for Cosmological Physics, University of Chicago, 5640 South Ellis Avenue, Chicago, IL 60637}
\altaffiltext{\EFIChicago}{Enrico Fermi Institute, University of Chicago, 5640 South Ellis Avenue, Chicago, IL 60637}
\altaffiltext{\CfA}{Harvard-Smithsonian Center for Astrophysics, 60 Garden Street, Cambridge, MA 02138}
\altaffiltext{\PhysicsUChicago}{Department of Physics, University of Chicago, 5640 South Ellis Avenue, Chicago, IL 60637}
\altaffiltext{\McGill}{Department of Physics, McGill University, 3600 Rue University, Montreal, Quebec H3A 2T8, Canada}
\altaffiltext{\Caddo}{Caddo Parish Magnet High School, Shrevport, LA 71101}
\altaffiltext{\UChicago}{University of Chicago, 5640 South Ellis Avenue, Chicago, IL 60637}
\altaffiltext{\Harvard}{Department of Physics, Harvard University, 17 Oxford Street, Cambridge, MA 02138}
\altaffiltext{\Munich}{Department of Physics, Ludwig-Maximilians-Universit\"{a}t, Scheinerstr.\ 1, 81679 M\"{u}nchen, Germany}
\altaffiltext{\ExcellenceCluster}{Excellence Cluster Universe, Boltzmannstr.\ 2, 85748 Garching, Germany}
\altaffiltext{\Miss}{Department of Physics and Astronomy, University of Missouri, 5110 Rockhill Road, Kansas City, MO 64110}
\altaffiltext{\AAUChicago}{Department of Astronomy and Astrophysics, University of Chicago, 5640 South Ellis Avenue, Chicago, IL 60637}
\altaffiltext{\ANL}{Argonne National Laboratory, 9700 S. Cass Avenue, Argonne, IL, USA 60439}
\altaffiltext{\NIST}{NIST Quantum Devices Group, 325 Broadway Mailcode 817.03, Boulder, CO, USA 80305}
\altaffiltext{\PUC}{Departamento de Astronomia y Astrosifica, Pontificia Universidad Catolica, Chile}
\altaffiltext{\Berkeley}{Department of Physics, University of California, Berkeley, CA 94720}
\altaffiltext{\UFlorida}{Department of Astronomy, University of Florida, Gainesville, FL 32611}
\altaffiltext{\Colorado}{Department of Astrophysical and Planetary Sciences and Department of Physics, University of Colorado, Boulder, CO 80309}
\altaffiltext{\NASA}{Department of Space Science, VP62, NASA Marshall Space Flight Center, Huntsville, AL 35812}
\altaffiltext{\Davis}{Department of Physics, University of California, One Shields Avenue, Davis, CA 95616}
\altaffiltext{\LBNL}{Physics Division, Lawrence Berkeley National Laboratory, Berkeley, CA 94720}
\altaffiltext{\Caltech}{California Institute of Technology, 1200 E. California Blvd., Pasadena, CA 91125}
\altaffiltext{\Arizona}{Steward Observatory, University of Arizona, 933 North Cherry Avenue, Tucson, AZ 85721}
\altaffiltext{\Michigan}{Department of Physics, University of Michigan, 450 Church Street, Ann  Arbor, MI, 48109}
\altaffiltext{\MPE}{Max-Planck-Institut f\"{u}r extraterrestrische Physik, Giessenbachstr.\ 85748 Garching, Germany}
\altaffiltext{\CaseWestern}{Physics Department, Center for Education and Research in Cosmology and Astrophysics, Case Western Reserve University, Cleveland, OH 44106}
\altaffiltext{\Minnesota}{Physics Department, University of Minnesota, 116 Church Street S.E., Minneapolis, MN 55455}
\altaffiltext{\STScI}{Space Telescope Science Institute, 3700 San Martin Dr., Baltimore, MD 21218}
\altaffiltext{\SAIC}{Liberal Arts Department, School of the Art Institute of Chicago, 112 S Michigan Ave, Chicago, IL 60603}
\altaffiltext{\LLNL}{Institute of Geophysics and Planetary Physics, Lawrence Livermore National Laboratory, Livermore, CA 94551}
\altaffiltext{\Dunlap}{Dunlap Institute for Astronomy \& Astrophysics, University of Toronto, 50 St George St, Toronto, ON, M5S 3H4, Canada}
\altaffiltext{\UofT}{Department of Astronomy \& Astrophysics, University of Toronto, 50 St George St, Toronto, ON, M5S 3H4, Canada}
\altaffiltext{\BCCP}{Berkeley Center for Cosmological Physics, Department of Physics, University of California, and Lawrence Berkeley National Labs, Berkeley, CA 94720}

\author{
M.~McDonald\altaffilmark{\hubble,\MIT},
B.~A.~Benson\altaffilmark{\KICPChicago,\EFIChicago},
A. Vikhlinin\altaffilmark{\CfA},
B.~Stalder\altaffilmark{\CfA},
L.~E.~Bleem\altaffilmark{\KICPChicago,\PhysicsUChicago},
T.~de~Haan\altaffilmark{\McGill},
H.~W.~Lin\altaffilmark{\Caddo},
K.~A.~Aird\altaffilmark{\UChicago},
M.~L.~N.~Ashby\altaffilmark{\CfA},
M.~W.~Bautz\altaffilmark{\MIT},
M.~Bayliss\altaffilmark{\CfA,\Harvard}, 
S.~Bocquet\altaffilmark{\Munich,\ExcellenceCluster},
M.~Brodwin\altaffilmark{\Miss},
J.~E.~Carlstrom\altaffilmark{\KICPChicago,\PhysicsUChicago,\EFIChicago,\AAUChicago,\ANL}, 
C.~L.~Chang\altaffilmark{\KICPChicago,\EFIChicago,\ANL}, 
H.~M. Cho\altaffilmark{\NIST}, 
A.~Clocchiatti\altaffilmark{\PUC},
T.~M.~Crawford\altaffilmark{\KICPChicago,\AAUChicago},
A.~T.~Crites\altaffilmark{\KICPChicago,\AAUChicago},
S.~Desai\altaffilmark{\Munich,\ExcellenceCluster},
M.~A.~Dobbs\altaffilmark{\McGill},
J.~P.~Dudley\altaffilmark{\McGill},
R.~J.~Foley\altaffilmark{\CfA}, 
W.~R.~Forman\altaffilmark{\CfA},
E.~M.~George\altaffilmark{\Berkeley},
D.~Gettings\altaffilmark{\UFlorida},
M.~D.~Gladders\altaffilmark{\KICPChicago,\AAUChicago},
A.~H.~Gonzalez\altaffilmark{\UFlorida},
N.~W.~Halverson\altaffilmark{\Colorado},
F.~W.~High\altaffilmark{\KICPChicago,\AAUChicago}, 
G.~P.~Holder\altaffilmark{\McGill},
W.~L.~Holzapfel\altaffilmark{\Berkeley},
S.~Hoover\altaffilmark{\KICPChicago,\EFIChicago},
J.~D.~Hrubes\altaffilmark{\UChicago},
C.~Jones\altaffilmark{\CfA},
M.~Joy\altaffilmark{\NASA},
R.~Keisler\altaffilmark{\KICPChicago,\PhysicsUChicago},
L.~Knox\altaffilmark{\Davis},
A.~T.~Lee\altaffilmark{\Berkeley,\LBNL},
E.~M.~Leitch\altaffilmark{\KICPChicago,\AAUChicago},
J.~Liu\altaffilmark{\Munich,\ExcellenceCluster},
M.~Lueker\altaffilmark{\Berkeley,\Caltech},
D.~Luong-Van\altaffilmark{\UChicago},
A.~Mantz\altaffilmark{\KICPChicago},
D.~P.~Marrone\altaffilmark{\Arizona},
J.~J.~McMahon\altaffilmark{\Michigan},
J.~Mehl\altaffilmark{\KICPChicago,\AAUChicago,\ANL},
S.~S.~Meyer\altaffilmark{\KICPChicago,\PhysicsUChicago,\EFIChicago,\AAUChicago},
E.~D.~Miller\altaffilmark{\MIT},
L.~Mocanu\altaffilmark{\KICPChicago,\AAUChicago},
J.~J.~Mohr\altaffilmark{\Munich,\ExcellenceCluster,\MPE},
T.~E.~Montroy\altaffilmark{\CaseWestern},
S.~S.~Murray\altaffilmark{\CfA},
D.~Nurgaliev\altaffilmark{\Harvard}, 
S.~Padin\altaffilmark{\KICPChicago,\AAUChicago,\Caltech},
T.~Plagge\altaffilmark{\KICPChicago,\AAUChicago},
C.~Pryke\altaffilmark{\Minnesota}, 
C.~L.~Reichardt\altaffilmark{\Berkeley},
A.~Rest\altaffilmark{\STScI},
J.~Ruel\altaffilmark{\Harvard},
J.~E.~Ruhl\altaffilmark{\CaseWestern}, 
B.~R.~Saliwanchik\altaffilmark{\CaseWestern}, 
A.~Saro\altaffilmark{\Munich},
J.~T.~Sayre\altaffilmark{\CaseWestern}, 
K.~K.~Schaffer\altaffilmark{\KICPChicago,\EFIChicago,\SAIC}, 
E.~Shirokoff\altaffilmark{\Berkeley,\Caltech}, 
J.~Song\altaffilmark{\Michigan},
R.~\v{S}uhada\altaffilmark{\Munich},
H.~G.~Spieler\altaffilmark{\LBNL},
S.~A.~Stanford\altaffilmark{\Davis,\LLNL},
Z.~Staniszewski\altaffilmark{\CaseWestern},
A.~A.~Stark\altaffilmark{\CfA}, 
K.~Story\altaffilmark{\KICPChicago,\PhysicsUChicago},
%C.~W.~Stubbs\altaffilmark{\CfA,\Harvard}, width]{plots/
A.~van~Engelen\altaffilmark{\McGill},
K.~Vanderlinde\altaffilmark{\Dunlap,\UofT},
J.~D.~Vieira\altaffilmark{\KICPChicago,\PhysicsUChicago,\Caltech},
R.~Williamson\altaffilmark{\KICPChicago,\AAUChicago}, 
O.~Zahn\altaffilmark{\Berkeley,\BCCP},
and
A.~Zenteno\altaffilmark{\Munich,\ExcellenceCluster},
}

%\affiliation{MIT Kavli Institute for Astrophysics and Space Research, Cambridge, MA 02139, USA}
\email{Email: mcdonald@space.mit.edu}   %optional

%\altaffiltext{1}{Kavli Institute for Astrophysics and Space Research, MIT, Cambridge, MA 02139, USA}
%\altaffiltext{2}{Email: mcdonald@space.mit.edu}


\begin{abstract}

We present first results on the cooling properties derived from Chandra X-ray observations of 83 high-redshift ($0.3 < z < 1.2$) massive galaxy clusters selected by their Sunyaev-Zel'dovich signature in the South Pole Telescope data. We measure each cluster's central cooling time, central entropy, and mass deposition rate, and compare these properties to those for local cluster samples. 
We find no significant evolution from $z\sim0$ to $z\sim1$ in the distribution of these properties, suggesting that cooling in cluster cores is stable over long periods of time. 
We also find that the average cool core entropy profile in the inner $\sim$100 kpc has not changed dramatically since z $\sim$ 1, implying that feedback must be providing nearly constant energy injection to maintain the observed ``entropy floor'' at $\sim$10 keV cm$^2$.
While the cooling properties appear roughly constant over long periods of time, we observe strong evolution in the gas density profile, with the normalized central density ($\rho_{g,0}/\rho_{\textrm{crit}}$) increasing by an order of magnitude from $z\sim1$ to $z\sim0$. 
When using metrics defined by the inner surface brightness profile of clusters, we find an apparent lack of classical, cuspy, cool-core clusters at $z > 0.75$, consistent with earlier reports for clusters at $z > 0.5$ using similar definitions.
Our measurements indicate that cool cores have been steadily growing over the 8 Gyr spanned by our sample, consistent with a constant, $\sim$150 M$_{\odot}$ yr$^{-1}$ cooling flow that is unable to cool below entropies of 10 keV cm$^2$ and, instead, accumulates in the cluster center. 
We estimate that cool cores began to assemble in these massive systems at $z_{\textrm{cool}}=1.0^{+1.0}_{-0.2}$, which represents the first constraints on the onset of cooling in galaxy cluster cores.
At high redshift ($z\gtrsim0.75$), galaxy clusters may be classified as ``cooling flows'' (low central entropy, cooling time) but not ``cool cores'' (cuspy surface brightness profile), meaning that care must be taken when classifying these high-$z$ systems.
We investigate several potential biases that could conspire to mimic this cool core evolution and are unable to find a bias that has a similar redshift dependence and a substantial amplitude. 
\end{abstract}



\setcounter{footnote}{1}
%================================================================%
%============== INTRODUCTION ====================================%
%================================================================%
\section{Introduction}
%- Basic introduction

In the inner $\sim$100~kpc of a galaxy cluster, the hot (10$^7-10^8$~K) intracluster medium is often sufficiently dense that the cooling time, which is roughly $t_{cool} \propto T_{\rm{x}}^{1/2} n_e^{-1}$, is shorter than a Hubble time. In these so-called ``cool core'' clusters, cooling gas should sink toward the center of the cluster, establishing a cooling flow which could deposit as much as $\sim$1000 M$_{\odot}$ yr$^{-1}$ of cold gas onto the central brightest cluster galaxy \citep[for a review, see][]{fabian94}. The fact that brightest cluster galaxies (hereafter BCGs) are rarely forming stars at such prodigious rates (with the exception of the newly-discovered Phoenix cluster; McDonald \etal 2012, 2013) is prime evidence that some form of feedback offsets this cooling. The most likely culprit is mechanical feedback from the central active galactic nucleus \citep[AGN; see][]{churazov01, mcnamara07,mcnamara12, fabian12}, although other heat sources such as particle heating \citep{mathews09}, blazars \citep{pfrommer12}, and mergers \citep{gomez02} are also viable.

If the balance between energy input from feedback and energy loss due to cooling is not exact, one would expect a residual cooling flow to develop. There is substantial evidence for such ``reduced cooling flows''. Clumps and filaments of cooling intracluster gas have been detected at 10$^6$--10$^7$~K in the cores of clusters via high resolution X-ray spectroscopy \citep[e.g.,][]{peterson06,sanders10} and OVI emission in the far ultraviolet \citep[e.g.,][]{bregman01,oegerle01,bregman06}. \cite{sparks12} recently reported evidence for 10$^5$~K gas (as traced by the C IV $\lambda$1549\AA\ emission line) in the core of the Virgo cluster. Warm (10$^4$~K) gas is nearly ubiquitous in cool core clusters \citep[e.g.,][]{Hu85,johnstone87, heckman89, crawford99, edwards07, hatch07, mcdonald10, mcdonald11a}, as are both warm \citep[e.g.,][]{jaffe05, edge10, donahue11,lim12} and cold \citep[e.g.,][]{edge01,edge03,salome03,salome08,lim08,mcdonald12b} molecular gas components. Finally, perhaps the most convincing evidence that a fraction of the cooling intracluster medium (ICM) is reaching low temperatures is the fact that nearly all cool core clusters have star-forming BCGs, with star formation rates that correlate with the ICM cooling rate \citep[e.g.,][]{mcnamara89,odea08,rafferty08,donahue10,hicks10,mcdonald11b}. Thus, while there is significant evidence that some form of feedback is offsetting a large fraction of energy loss due to cooling in the ICM, it is also clear that this balance is imperfect and likely to vary on both short (periodic outbursts) and long (evolution) timescales.

%- Properties of z $\sim$ 0 cool cores (White+97, Hudson+10, etc)
While the physical processes that conspire to prevent or allow the formation of a dense, cool core in the ICM are not fully understood, there has been significant effort towards understanding the overall properties of these systems. Early, large surveys, including those by \cite{white97}, \cite{peres98}, and \cite{allen00}, have formed the basis of our understanding of cooling flows (or lack thereof). These studies established the distribution of cooling properties, including the mass deposition rate, the cooling radius, and the central cooling time, for large X-ray flux-limited samples of nearby galaxy clusters. These studies showed that, among other things, clusters with strong cooling signatures tend to have multi-phase (i.e., H$\alpha$-emitting) gas, radio-loud BCGs, and cooling rates that correlate with the total X-ray luminosity. Studies mentioned in the previous paragraph have largely built upon these early, pioneering works to classify the cooling properties of the intracluster medium.

%- Evolution of cool cores (Vikhlinin+07, Santos+07, Santos+10, Samuele+11, McDonald 11)
While the properties of nearby ($z\lesssim0.3$) cool core clusters are well documented, very little is presently known about how cooling flows have evolved. Early work by \cite{donahue92} reported that, while the general properties of cooling flows appear to be unchanged since $z\sim0.3$, cool cores were more common by roughly a factor of two at this epoch. More recently, utilizing higher quality data from the \emph{Chandra X-ray Observatory}, as well as ground-based optical data, on much larger, more complete samples, various studies have found evidence that there may be a decline in the fraction of clusters harboring strong (cuspy) cool cores with increasing redshift \citep{vikhlinin07, santos10, samuele11,mcdonald11c}. These studies all report cool core fractions $\lesssim$10\% at $z\gtrsim0.5$, indicating that cool core clusters are a recent phenomenon. Indeed, only a small number of clusters with strong cool cores are known at $z>0.5$ \citep[e.g.,][]{siemiginowska10, russell12, santos12, mcdonald12c}.
It has been suggested that, since most of these samples were drawn from early surveys with the ROSAT X-ray telescope, they may be biased against cool cores at high redshifts due to their point-like appearance compared to the ROSAT resolution. The fact that optically-selected samples \citep{mcdonald11c} show the same evolution suggests that such a bias may not be a serious issue.

One significant issue affecting our understanding of the evolution of ICM cooling is the lack of large samples of high-redshift clusters with a well understood selection.  The South Pole Telescope \citep[SPT;][]{carlstrom11} recently completed a 2500 square degree survey that has discovered hundreds of massive, high-redshift clusters using the Sunyaev-Zel'dovich \citep[SZ;][]{sunyaev72} effect.  
Unlike X-ray and optical surveys, which have strong surface brightness biases, the SPT selection is nearly redshift-independent \citep[at $z>0.3$, see][]{song12,reichardt13} and, based on simulations, is not expected to be significantly biased by the presence of cool cores \citep[][Lin \etal in prep]{motl05,pipino10}.
In principle, such a survey should be able to trace the evolution of cool cores in the most massive clusters out to $z>1$. Indeed, \cite{semler12} showed, in a pilot study of 13 SPT-selected clusters, that there is a significant population of cool core clusters at $z>0.5$, contrary to the majority of the results reported in the literature at the time. Furthermore, the most extreme cool core cluster known is at $z=0.597$, the Phoenix cluster \citep[SPT-CLJ2344-4243;][]{mcdonald12c}, and was discovered by the SPT. Taken together, these results suggest that ICM cooling has not changed drastically in the past $\sim$8 Gyr.

In this work, we expand significantly on \cite{semler12}, presenting \emph{Chandra} X-ray observations of 83 massive, SPT-selected clusters. The majority of these observations were completed as part of a recent \emph{Chandra X-ray Visionary Project} (PI B. Benson). With these data we are able to address two outstanding questions about the evolution of the cooling intracluster medium: i) Were cool cores less common at $z>0.5$? and ii) How have the properties of cooling flows evolved in the most massive galaxy clusters over the past $\sim$8 Gyr? In \S2 we present the sample, describing first the selection and observations, followed by the analysis. In \S3 we present the major results of this work, following in spirit the early works of \cite{white97} and \cite{peres98} which identified the cooling flow properties of low-redshift, X-ray selected clusters. The implications of these results are discussed in \S4. Throughout this work, we assume H$_0$=70 km s$^{-1}$ Mpc$^{-1}$, $\Omega_M$ = 0.27, and $\Omega_{\Lambda}$ = 0.73.

%================================================================%
% ======= DATA ANALYSIS ==========================================%
%================================================================%

\section{Data and Analysis}
\subsection{Sample Definition}

The clusters used in this work were selected based on their Sunyaev-Zel'dovich (SZ) signature in the 2500 deg$^2$ SPT-SZ 
survey.  The SPT-SZ survey was completed in November 2011, producing maps in three frequency bands (95, 150, and 220 GHz), 
with a key science goal of discovering clusters via the SZ effect \citep{staniszewski09, vanderlinde10, williamson11, benson13,reichardt13}.

The clusters considered in this work have additionally been observed with the \emph{Chandra X-ray Observatory},
%The present cluster sample has additionally been observed with the Chandra X-ray Observatory, 
with exposures typically sufficient to obtain $\sim$2000 X-ray source counts.  
The majority of the clusters have been observed through a Chandra X-ray Visionary Project to obtain X-ray imaging 
of the 80 clusters detected with the highest SZ significance ($\xi$) in the first 2000 deg$^2$ of the 2500 deg$^2$ SPT-SZ survey at $z > 0.4$ (hereafter B13; Benson \etal in prep.)  While B13 analyze the full XVP sample, we exclude six of the 80 clusters only observed with XMM-\emph{Newton}, which does not have sufficient angular resolution to resolve the cool cores in typical high-redshift clusters.  
In addition, we include nine clusters at $z > 0.3$ also detected in the SPT-SZ survey that were observed by 
Chandra, a sub-sample that primarily consists of clusters observed either in previous Chandra GO and GTO proposals from the 
SPT-SZ collaboration, or in other proposals to observe SZ-selected clusters from the Atacama Cosmology Telescope \citep[ACT;][]{marriage11} and Planck \citep{planck11} collaborations. We note that every SPT-selected cluster that was targeted with \emph{Chandra} yielded an X-ray detection -- perhaps unsurprising due to the dependence of both techniques on a rich ICM.

The final sample used in this work, referred to hereafter as SPT-XVP, is summarized in Table 1.  The sample consists of 83 clusters, spanning a redshift range of $0.3 < z < 1.2$ and a mass range of $\sim 2 \times 10^{14} < $~M$_{500} < 20 \times 10^{14}$~M$_{\odot} / h_{70}$.    The clusters were all identified in the SPT-SZ survey maps with a SPT detection significance, $\xi$, spanning a range from $5.7 < \xi < 43$. 
As was done in \cite{vanderlinde10}, we predict the SPT survey completeness using cosmological and scaling relation constraints of the $\xi$-mass relation. We assume the $\Lambda$CDM cosmological constraints from \cite{reichardt13} when using a CMB data set and the SPT cluster catalog.
At our median redshift of $z\sim 0.7$, the SPT-XVP sample is expected to be $\sim$50\% complete at $M_{500} = 4 \times 10^{14} M_{\odot} / h_{70}$ and nearly 100\% complete at  $6 \times 10^{14} M_{\odot} / h_{70}$.  These completeness thresholds are nearly redshift independent, varying by $\lesssim$15\% over the redshift range of the sample.

%\begin{table*}[t]
%\centering

%\begin{longtable}{p{0.15\textwidth} p{0.055\textwidth} p{0.055\textwidth} p{0.14\textwidth}}
\begin{longtable}{@{\extracolsep{\fill}}l c c c c}
\hline\hline
Name & $\alpha$ [$^{\circ}$] & $\delta$ [$^{\circ}$]& OBSIDs \\
\hline
SPT-CLJ0000-5748 & 0.250 & -57.809 & 9335\\
SPT-CLJ0013-4906 & 3.331 & -49.116 & 13462\\
SPT-CLJ0014-4952 & 3.690 & -49.881 & 13471\\
SPT-CLJ0033-6326 & 8.469 & -63.444 & 13483\\
SPT-CLJ0037-5047 & 9.447 & -50.788 & 13493\\
SPT-CLJ0040-4407 & 10.208 & -44.132 & 13395\\
SPT-CLJ0058-6145 & 14.586 & -61.768 & 13479\\
SPT-CLJ0102-4603 & 15.677 & -46.072 & 13485\\
SPT-CLJ0102-4915 & 15.734 & -49.266 & 12258\\
SPT-CLJ0123-4821 & 20.796 & -48.358 & 13491\\
SPT-CLJ0142-5032 & 25.546 & -50.540 & 13467\\
SPT-CLJ0151-5954 & 27.857 & -59.908 & 13480\\
SPT-CLJ0156-5541 & 29.042 & -55.698 & 13489\\
SPT-CLJ0200-4852 & 30.141 & -48.872 & 13487\\
SPT-CLJ0212-4657 & 33.108 & -46.950 & 13464\\
SPT-CLJ0217-5245 & 34.304 & -52.763 & 12269\\
SPT-CLJ0232-5257 & 38.202 & -52.953 & 12263\\
SPT-CLJ0234-5831 & 38.677 & -58.523 & 13403\\
SPT-CLJ0236-4938 & 39.258 & -49.637 & 12266\\
SPT-CLJ0243-5930 & 40.865 & -59.515 & 13484,15573\\
SPT-CLJ0252-4824 & 43.212 & -48.415 & 13494\\
SPT-CLJ0256-5617 & 44.106 & -56.298 & 13481,14448\\
SPT-CLJ0304-4401 & 46.067 & -44.033 & 13402\\
SPT-CLJ0304-4921 & 46.067 & -49.357 & 12265\\
SPT-CLJ0307-5042 & 46.961 & -50.705 & 13476\\
SPT-CLJ0307-6225 & 46.830 & -62.436 & 12191\\
SPT-CLJ0310-4647 & 47.634 & -46.785 & 13492\\
SPT-CLJ0324-6236 & 51.053 & -62.598 & 12181,13137,13213\\
SPT-CLJ0330-5228 & 52.728 & -52.473 & 0893\\
SPT-CLJ0334-4659 & 53.547 & -46.996 & 13470\\
SPT-CLJ0346-5439 & 56.733 & -54.649 & 12270\\
SPT-CLJ0348-4515 & 57.075 & -45.247 & 13465\\
SPT-CLJ0352-5647 & 58.241 & -56.798 & 13490,15571\\
SPT-CLJ0406-4805 & 61.731 & -48.082 & 13477\\
SPT-CLJ0411-4819 & 62.814 & -48.320 & 13396\\
SPT-CLJ0417-4748 & 64.347 & -47.813 & 13397\\
SPT-CLJ0426-5455 & 66.520 & -54.918 & 13472\\
SPT-CLJ0438-5419 & 69.575 & -54.322 & 12259\\
SPT-CLJ0441-4855 & 70.451 & -48.924 & 13475,14371,14372\\
SPT-CLJ0446-5849 & 71.514 & -58.830 & 13482,15560\\
SPT-CLJ0449-4901 & 72.275 & -49.025 & 13473\\
SPT-CLJ0456-5116 & 74.118 & -51.278 & 13474\\
SPT-CLJ0509-5342 & 77.339 & -53.704 & 9432\\
SPT-CLJ0528-5300 & 82.023 & -52.998 & 11747,11874,12092,13126\\
SPT-CLJ0533-5005 & 83.406 & -50.096 & 11748,12001,12002\\
SPT-CLJ0542-4100 & 85.709 & -41.000 & 0914\\
SPT-CLJ0546-5345$^a$ & 86.655 & -53.759 & 9332,9336\\
SPT-CLJ0551-5709 & 87.896 & -57.147 & 11743,11871\\
SPT-CLJ0555-6406 & 88.864 & -64.105 & 13404\\
SPT-CLJ0559-5249 & 89.933 & -52.827 & 12264,13116,13117\\
SPT-CLJ0616-5227 & 94.144 & -52.453 & 12261,13127\\
SPT-CLJ0655-5234 & 103.974 & -52.568 & 13486\\
SPT-CLJ2031-4037 & 307.966 & -40.623 & 13517\\
SPT-CLJ2034-5936 & 308.537 & -59.605 & 12182\\
SPT-CLJ2035-5251 & 308.793 & -52.855 & 13466\\
SPT-CLJ2043-5035 & 310.823 & -50.592 & 13478\\
SPT-CLJ2106-5844$^b$ & 316.518 & -58.743 & 12180\\
SPT-CLJ2135-5726 & 323.912 & -57.439 & 13463\\
SPT-CLJ2145-5644 & 326.468 & -56.749 & 13398\\
SPT-CLJ2146-4632 & 326.645 & -46.549 & 13469\\
SPT-CLJ2148-6116 & 327.181 & -61.279 & 13488\\
SPT-CLJ2218-4519 & 334.746 & -45.316 & 13501\\
SPT-CLJ2222-4834 & 335.712 & -48.577 & 13497\\
SPT-CLJ2232-5959 & 338.141 & -59.998 & 13502\\
SPT-CLJ2233-5339 & 338.319 & -53.654 & 13504\\
SPT-CLJ2236-4555 & 339.219 & -45.930 & 13507,15266\\
SPT-CLJ2245-6206 & 341.260 & -62.116 & 13499\\
SPT-CLJ2248-4431 & 342.183 & -44.530 & 4966\\
SPT-CLJ2258-4044 & 344.706 & -40.740 & 13495\\
SPT-CLJ2259-6057 & 344.752 & -60.960 & 13498\\
SPT-CLJ2301-4023 & 345.471 & -40.389 & 13505\\
SPT-CLJ2306-6505 & 346.734 & -65.090 & 13503\\
SPT-CLJ2325-4111 & 351.302 & -41.196 & 13405\\
SPT-CLJ2331-5051 & 352.963 & -50.865 & 9333\\
SPT-CLJ2335-4544 & 353.785 & -45.739 & 13496\\
SPT-CLJ2337-5942 & 354.352 & -59.706 & 11859\\
SPT-CLJ2341-5119 & 355.300 & -51.329 & 11799\\
SPT-CLJ2342-5411 & 355.692 & -54.185 & 11741,11870,12014,12091\\
SPT-CLJ2344-4243$^c$ & 356.183 & -42.720 & 13401\\
SPT-CLJ2345-6405 & 356.250 & -64.099 & 13500\\
SPT-CLJ2352-4657 & 358.068 & -46.960 & 13506\\
SPT-CLJ2355-5055 & 358.948 & -50.928 & 11746\\
SPT-CLJ2359-5009 & 359.933 & -50.170 & 9334,11742,11864,11997\\
\hline
\\
\multicolumn{4}{l}{Table 1. Summary of Chandra X-ray observations. Positions listed} \\ 
\multicolumn{4}{l}{here are of the X-ray centroid (\S2.2). The fourth column provides}\\
\multicolumn{4}{l}{the observational IDs from the Chandra X-ray Observatory.} \\
\multicolumn{4}{l}{$^a$: \cite{brodwin10}}\\
\multicolumn{4}{l}{$^b$: \cite{foley11}}\\
\multicolumn{4}{l}{$^c$:  \cite{mcdonald12c}}
%\end{tabular}
\label{table:sample}
%}
%\end{table*}
\end{longtable}

\subsection{Data Reduction and Analysis}
Our basic data reduction and analysis follows closely that outlined in \cite{vikhlinin05} and \cite{andersson11}. Briefly, this procedure includes filtering for background flares, applying the latest calibration corrections, and determining the appropriate blank sky background. In addition to using blank-sky backgrounds, we simultaneously model additional background components from Galactic sources as well as unresolved cosmic X-ray background (CXB) sources in off-source regions. Point sources were identified using an automated routine following a wavelet decomposition technique \citep{vikhlinin98}, and then visually inspected. Clumpy, asymmetric substructure was masked by hand, and excluded in calculations of the global temperature. The center of the cluster was chosen by iteratively measuring the centroid in a 250--500~kpc annulus. This choice, rather than the peak of emission, can play a significant role in whether or not the cluster is ultimately classified as a cool core or not -- a subject we will return to in \S4.

Global cluster properties (L$_{X,500}$, M$_{500}$, T$_{X,500}$, M$_{g,500}$) used in this work are derived in B13,
following closely the procedures described in \cite{andersson11}. For each of these quantities, the subscript refers to the quantity measured within R$_{500}$ -- the radius within which the average enclosed density is 500 times the critical density. We estimate R$_{500}$ by requiring the measured quantities (T$_X$, M$_g$, Y$_X$) to satisfy a set of scaling relations between T$_{X,500}$, M$_{g,500}$, and Y$_{X,500}$ and M$_{500}$ \citep{vikhlinin09}.
Each of these three scaling relations are individually satisfied by iteratively adjusting R$_{500}$. In this paper, we use R$_{500}$ from the Y$_X$--M relation only. Further details on both the data reduction and the derivation of global X-ray properties can be found in \cite{vikhlinin05} and \cite{andersson11}, respectively.

\subsection{Surface Brightness Profiles and Concentration Measurements}
\label{sec:sb}
%-- start with ne(r)

The surface brightness profile for each cluster, extracted in the energy range 0.7--2.0~keV, is measured in a series of 20 annuli, with the outer radii for each annulus defined as:
\begin{equation}
%r_i = \left(\frac{i}{n}\right)^c\times r_{max}, ~~~i=1...n
r_i=1.5\textrm{R}_{500}\left(\frac{i}{20}\right)^{1.5} ~~~i=1...n~.
\end{equation}
%
%where $n=20$, $r_{max}=1.5r_{500}$, and $c=1.5$. 
Following the techniques described in \cite{vikhlinin06}, we correct these surface brightness profiles for spatial variations in temperature, metallicity, and the telescope effective area. Calibrated surface brightness profiles (see Appendix A) are expressed as a projected emission measure integral, $\int n_en_p dl$, where $n_e$ and $n_p$ are the electron and proton densities, respectively. We model the calibrated surface brightness profile with a modified beta model:

\begin{equation}
n_en_p = n_0^2\frac{(r/r_c)^{-\alpha}}{(1+r^2/r_c^2)^{3\beta-\alpha/2}}\frac{1}{(1+r^3/r_s^3)^{\epsilon/3}} ,
\label{eq:ne}
\end{equation}
%\\
%\\
where $n_0$ is the core density, and $r_c$ and $r_s$ are scaling radii of the core and extended components, following \cite{vikhlinin06}. This 3-dimensional model is numerically projected along the line of sight, yielding a model emission measure profile that is fit to the data. We estimate the 3-dimensional gas density assuming $n_e=Zn_p$ and $\rho_g=m_pn_eA/Z$, where $A=1.397$ and $Z=1.199$ are the average nuclear charge and mass, respectively, for a plasma with metal abundance 30\% of solar (0.3$Z_{\odot}$). The calibrated surface brightness profiles and best-fit projected gas density models for the full sample are shown in Appendix A.

In recent studies of high-redshift cool core clusters \citep[e.g.,][]{vikhlinin07,santos08,santos10,semler12}, the presence of a cool core has been quantified solely by the central cuspiness of the surface brightness profile. While measuring the central deprojected temperature and cooling time typically requires $>$10,000 X-ray counts, the surface brightness profile can often be constrained in the central region with as few as $\sim$500 counts, making this an inexpensive method of classifying cool cores. To classify a sample of high-redshift clusters as cool core or non-cool core, \cite{vikhlinin07} defined a ``cuspiness'' parameter, 
\begin{equation}
\alpha \equiv \left. \frac{d\log\rho_g}{d\log r}\right|_{r=0.04\textrm{R}_{500}}
\end{equation}
This parameter has been shown to correlate well with the central cooling time for galaxy clusters at $z\sim0$ \citep{vikhlinin07,hudson10}. \cite{vikhlinin07} showed that $\alpha$ was typically higher for a low-redshift sample of clusters ($z<0.5$), suggesting a rapid evolution in cool core strength \citep{vikhlinin07}. While easily measurable, this parameter has the drawback that it assumes that the cool core radius evolves at the same rate as the cluster radius (R$_{500}$). 

An alternative measure of the surface brightness cuspiness is the ``concentration'' parameter, as defined by \cite{santos08}: 
\begin{equation}
c_{SB} \equiv \frac{F_{0.5-5.0\rm{keV}}(r <40~\rm{kpc})}{F_{0.5-5.0\rm{keV}}(r<400~\rm{kpc})} ,
\end{equation}
where $F_{0.5-5.0\rm{keV}}$ is the X-ray flux in the energy bandpass 0.5--5.0 keV. This value can range from $\sim$0 (no flux peak), to 1 (all flux in central 40~kpc). This choice of parameter is relatively insensitive to redshift effects, such as worsening spatial resolution, reduced counts, and $k$-corrections \citep{santos08}, but has the potential drawback that it assumes no evolution in the cooling radius. 

We will use both the full 3-dimensional density profile ($n_e(r)$), as well as the commonly-used single-parameter estimates of profile peakedness ($\alpha$, $c_{SB}$) to trace the evolution of cool cores in this unique sample.

\subsection{Deprojecting Radial X-ray Profiles}
\label{sec:deprojection}
\subsubsection{$\rho_g(r)$, $\Phi(r)$, T$_X(r)$}
Many recent works have verified the presence, or lack, of high-redshift cool cores via surface brightness quantities, as discussed in the previous section. We wish to extend this analysis further and quantify the cooling properties of the ICM. With only $\sim$2000 X-ray counts per cluster, we cannot perform a full temperature and density deprojection analysis, as is typically done at low redshift \citep[e.g.,][]{vikhlinin06, sun09a}. Instead, motivated by earlier studies with the Einstein X-ray observatory \citep{white97, peres98}, we combine our knowledge of the X-ray surface brightness and a coarse temperature profile with assumptions about the underlying dark matter distribution to produce best-guess 3-dimensional temperature profiles. This procedure, which will be described in complete detail in an upcoming paper (McDonald \etal in prep), is summarized below.

First, a 3-bin temperature profile is derived by extracting X-ray spectra in logarithmically-spaced annuli over the range $0 < r < \rm{R}_{500}$. These spectra were fit in {\sc xspec} \citep{arnaud96} with a combined {\sc phabs(mekal)} model\footnote{\url{http://heasarc.gsfc.nasa.gov/docs/xanadu/xspec/manual/\\XspecModels.html}}. In cases requiring additional background components, as determined from off-source regions of the field (see B13 for more details), an additional {\sc mekal} (Galactic) or {\sc bremss} (CXB) component was used, with temperatures fixed to 0.18~keV and 40~keV, respectively. For the source model, we fix the abundance to $0.3Z_{\odot}$ and the hydrogen absorbing column, n$_H$, to the average from the Leiden-Argentine-Bonn survey \citep{kalberla05}. The resulting temperature profiles for the full sample are shown in Appendix A.

In order to model the underlying dark matter potential, we use a generalized Navarro-Frenk-White \citep[GNFW;][]{zhao96,wyithe01} profile:

\begin{equation}
\rho_{D}=\frac{\rho_{D,0}}{(r/r_s)^{\beta_{D}}(1+r/r_s)^{3-\beta_{D}}} ,
\label{eq:nfw}
\end{equation}

\noindent{}where $\rho_{D,0}$ is the central dark matter density, $r_s$ is a scale radius related to the halo concentration by $C=\rm{R}_{200}/r_s$, and $\beta_{D}$ is the inner slope of the dark matter profile. This model is similar to the NFW \citep{nfw} profile at large radii, but has a free ``cuspiness'' parameter at small radii. We estimate the initial values of $\rho_{D,0}$ and $r_s$ using the measured M$_{500}$ and the mass-concentration relation \citep{duffy08}. 

Given an assumed 3-dimensional functional form of both the dark matter (Eq.\ \ref{eq:nfw}) and gas density profiles (\S\ref{sec:sb}, Eq.\ \ref{eq:ne}), and further assuming a negligible contribution from stars to the total mass, we can derive the 3-dimensional temperature profile by combining hydrostatic equilibrium,
\begin{equation}
\frac{d\textrm{P}}{dr} = -\frac{\textrm{GM}(r)\rho(r)}{r^2} ,
\end{equation}
with the ideal gas law (P$=n_Tk$T, where $n_T = n_e + n_p$). This temperature profile is projected along the line of sight (weighted by $n_e^2$T$^{1/2}$), producing a 2-dimensional temperature profile which is compared to the data, allowing a calculation of $\chi^2$. We repeat this process, varying both the normalization of the GNFW halo ($\rho_{D,0}$), and thus the total dark matter mass, as well as the inner slope ($\beta_{D}$), while requiring that the mass--concentration \citep{duffy08} and M$_{500}$--P$_{500}$ \citep{nagai07} relations are always satisfied (removing $r_s$ and P$_{500}$ as free parameters), until a stable minimum in the $\chi^2$($\rho_{D,0}$, $\beta_{D}$) plane is found. The net result of this process is a 3-dimensional model of the gas density, gas temperature, and gravitational potential for each cluster (see Appendix A).

\subsubsection{$t_{\textrm{cool}}(r)$, K$(r)$, \.{M}$(r)$}
While a centrally-concentrated surface brightness profile is an excellent indicator that the ICM is cooling rapidly \citep[e.g.,][]{hudson10}, we ultimately would like to quantify, in an absolute sense, the strength of this cooling. Classically, clusters have been identified as ``cooling flows'' if the cooling time in the central region is less than the age of the Universe, with the cooling time defined as:

\begin{equation}
t_{\textrm{cool}} = \frac{3}{2}\frac{n_Tk\rm{T}}{n_en_H\Lambda(\rm{T,Z})} ,
\end{equation}

\noindent{}where $\Lambda(\rm{T,Z})$ is the cooling function for an optically-thin plasma \citep{sutherland93}. Similarly, the specific entropy of the gas is defined as:

\begin{equation}
\textrm{K} = \frac{k\textrm{T}}{n_e^{2/3}} .
\end{equation}

Both the central cooling time and central entropy are smallest in the centers of cool core clusters, and are distributed bimodally over the full cluster population \citep[e.g.,][]{cavagnolo09,hudson10}. In nearby, well-studied clusters, the central cooling time and entropy are well-defined, as both of these functions tend to flatten at radii less than $\sim$10 kpc. However, for lower signal-to-noise data, the measurement of central cooling time is strongly dependent on the choice of bin size \citep[see e.g.,][]{white97,peres98}, and inwards extrapolation can be risky if a flattening of the surface brightness profile is not observed due to poor spatial sampling. For this work, we choose as our central bin $r<0.012\textrm{R}_{500}$ ($r\lesssim10$ kpc), which roughly corresponds to the first data point in our surface brightness profiles. While these quantities are not truly ``central'', this choice allows us to avoid the increasingly large uncertainty associated with extrapolating our temperature and density fits as $r\rightarrow0$.

Following \cite{white97}, we estimate the classical mass deposition rate, \.M(r), using the following formula:
\begin{equation}
\frac{d\textrm{M}}{dt}(r_i)=\frac{\textrm{L}_X(r_i)-(\Delta\phi(r_i)+\Delta h(r_i))\dot{\textrm{M}}(<r_{i-1})}{h(r_i)+f(r_i)\Delta\phi(r_i)} ,
\end{equation}
where L$_X(r_i)$ is the X-ray luminosity in shell $i$, $\Delta\phi(r_i)$ is the change in the gravitational potential across shell $i$, $h(r_i)$ is the temperature in units of energy per particle mass, $h(r_i)=\frac{5}{2}(k$T$(i)/\mu m_p)$, and $f(r_i)$ is the fraction of the shell that the gas crosses before dropping out of the flow. 
This equation calculates the cooling rate due to X-ray radiation ($\frac{dM}{dt} \propto \frac{L_X}{kT}$) corrected for the gravitational work done on the gas as it falls inwards towards the center of the cluster's gravitational potential.
There are currently no constraints on what $f(r_i)$ should be, so we choose the mid-point ($f(r_i)=0.5$). We note that varying $f(r_i)$ from 0 to 1 typically alters the estimate of $d$M$/dt$ by only $\sim$5\%. We integrate the mass deposition rate out to the radius at which the cooling time equals the age of the Universe \emph{at the epoch of the cluster}. 
 The resulting $\frac{dM}{dt}(r<r_{cool}$) represents the time-averaged cooling rate if the cluster as we currently observe it has been in equilibrium for all time. We note that, by this definition, our sample ought to have overall smaller cooling radii due to the fact that these high-redshift clusters have had less time to cool than their low-redshift counterparts.

\subsection{Comparing Aperture and 3-D Model Temperatures}
In previous studies \citep[e.g.,][]{hudson10}, the central entropy and cooling time are calculated from a combination of 3-dimensional, central electron density, $n_{e,0}$, and a 2-dimensional, aperture temperature measured in some small aperture (e.g., $r\lesssim0.05$R$_{500}$). For clusters with only $\sim$2000 X-ray counts, this central aperture may only contain $\sim$100 counts, making the estimate of a central temperature complicated. However, in cool core clusters, where a significant fraction of the flux originates from this small aperture, we can measure a reliable temperature and compare to our 3-dimensional models described above.

\begin{figure}[htb]
\centering
\includegraphics[width=0.49\textwidth]{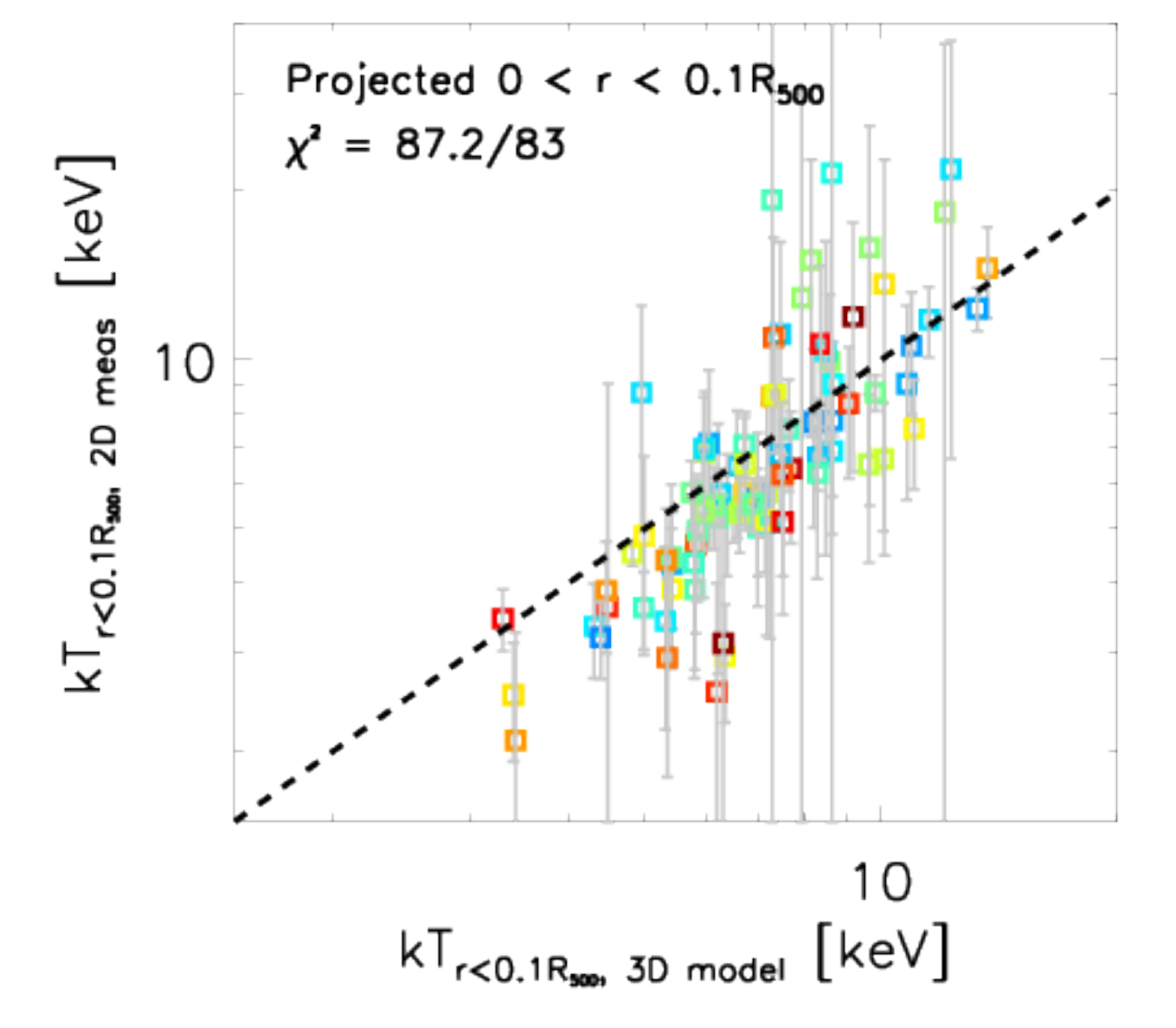}
\caption{Comparison of 2-dimensional spectroscopic temperatures measured in an aperture of $r<0.1$R$_{500}$ (with AGN masked) to the 3-dimensional model temperature projected onto the same annulus. The one-to-one correspondence (dashed line) between the data and models suggest that our mass-modeling approach (\S2.4) yields reliable estimates of the central temperature for clusters that are relaxed. Point color corresponds to redshift, from $z=0.3$ (blue) to $z=1.2$ (red), indicating that the scatter in this plot is largely independent of redshift.
}
\label{fig:compare_kT0}
\end{figure}

In Figure \ref{fig:compare_kT0}, we show the measured spectroscopic temperature (kT$_{0,2D}$) in an aperture of $r<0.1$R$_{500}$ (with AGN masked), where the outer radius was chosen to maximize the number of X-ray counts, while still capturing the central temperature drop in cool core clusters, following the universal profile shown in \cite{vikhlinin06}. The modeled 3-dimensional temperature profile was projected onto this same aperture (kT$_{0,3D}$) to enable a fair comparison of the two quantities. While the uncertainty in the 2-D temperature for these small X-ray apertures is high, there appears to be good agreement between the models and data (reduced $\chi^2$ = 87.2/83), suggesting that this technique is able to recover the ``true'' central temperature of the cluster.

Based on this agreement, we feel confident extrapolating inwards into a regime without sufficient X-ray counts to measure the spectroscopic temperature and proceed throughout \S3 utilizing the central ($r<0.012$R$_{500}$), deprojected model temperatures to calculate $t_{\textrm{cool},0}$ and K$_0$. In \S4 we will return to the comparison between 2-D and 3-D quantities to determine the dependence of our results on this extrapolation and our choice of models.

%================================================================%
% ======= RESULTS ================================================%
%================================================================%
\section{Results}

This sample represents the largest and most complete sample of galaxy clusters with X-ray observations at $z\gtrsim0.4$. Given the depth of our X-ray exposures, combined with the high angular resolution of the \emph{Chandra X-ray Observatory}, we are in a unique position to study the evolution of cooling in the ICM for similar-mass clusters over timescales of $\sim$8~Gyr. In this section we will present the broad results of this study, drawing comparisons to samples of nearby clusters \citep[e.g.,][]{white97,vikhlinin06, vikhlinin09,cavagnolo09,hudson10}.  The interpretation of these results, as well as systematic errors that may affect them, are discussed in \S4.

\subsection{Cooling Time and Entropy Profiles}
% Show like Cavagnolo entropy profiles

\begin{figure*}[h!]
\centering
\includegraphics[width=0.95\textwidth]{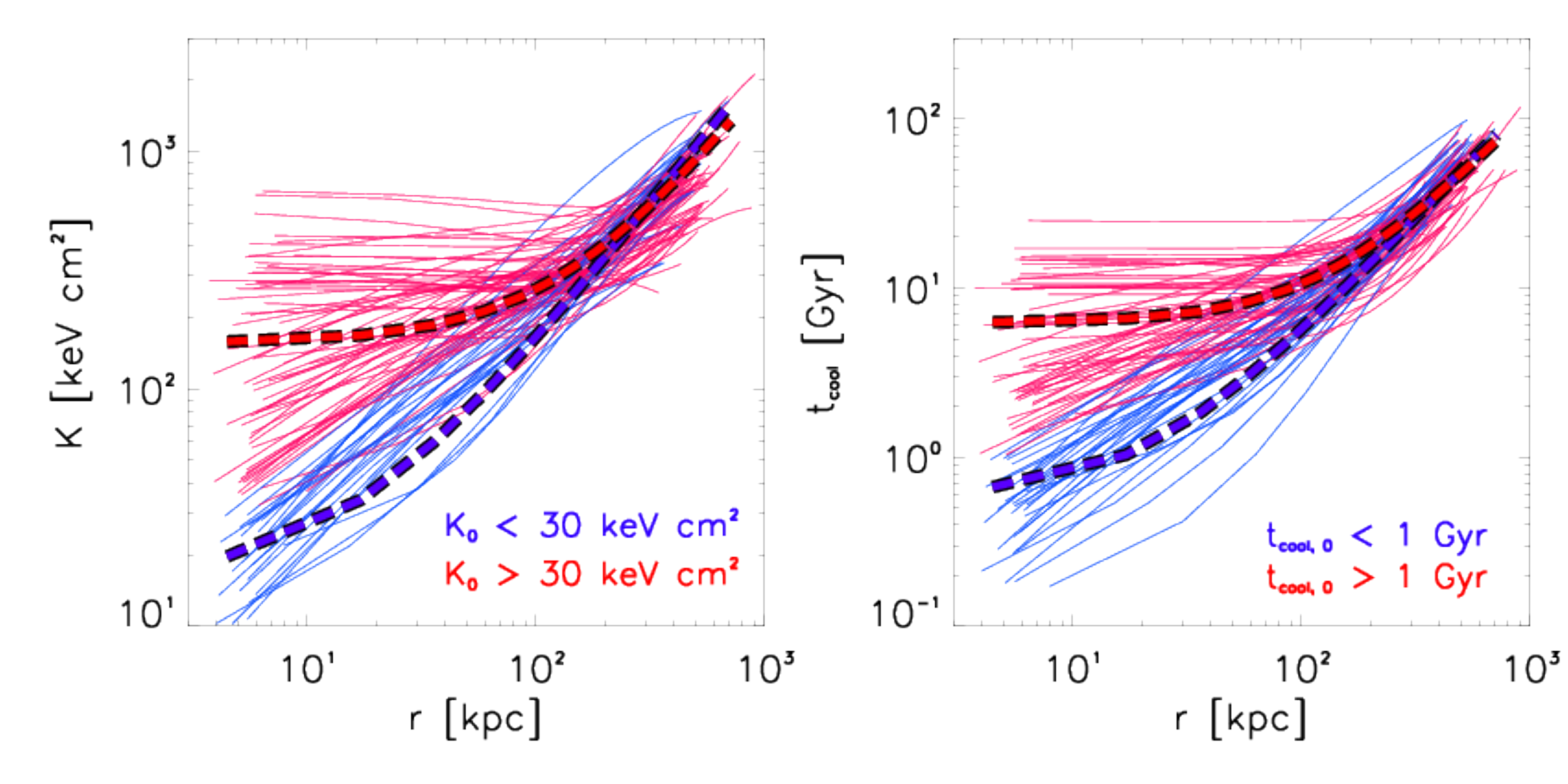}
\caption{Radial entropy (K) and cooling time (t$_{cool}$) profiles for our full sample of SPT-selected clusters. The average entropy and cooling time curves from \cite{cavagnolo09} for both cool core (blue) and non-cool core (red) clusters are shown in thick dashed lines, which are found to be in good qualitative agreement with our high-$z$, more massive clusters. Overall, these profiles have similar shapes and normalization to low-redshift clusters, suggesting little evolution in the cooling properties of massive clusters over the past $\sim$8 Gyr.}
\label{fig:kprof}
\end{figure*}

In Figure \ref{fig:kprof}, we present the radial entropy and cooling time profiles for our full sample, based on the deprojection procedures described in \S\ref{sec:deprojection}. For comparison, we show the average entropy profiles for low-redshift cool core and non-cool core clusters from \cite{cavagnolo09}. Perhaps surprisingly, there is no qualitative difference in the entropy profiles between this sample of high-redshift, massive, SZ-selected clusters and the low-redshift sample of groups and clusters presented in \cite{cavagnolo09}. 

Naively, one might expect the mean central entropy to decrease with time, as clusters have had more time to cool. 
However, the similarity between low-redshift clusters and this sample, which has a median redshift of 0.63 and a median age nearly half of the $z\sim0$ sample, indicates that the entropy and cooling time profiles are unchanging.
This suggests that the characteristic entropy and cooling time profiles, having minimal core entropies of $\sim$10 keV cm$^2$ and cooling radii of $\sim$100 kpc, were established at earlier times than we are probing with this sample ($z\gtrsim1$).

%
%In the rightmost panel of Figure \ref{fig:kprof}, we show the radial profile of the ICM cooling time, normalized to the age of the Universe at the redshift of this cluster. This choice of scaling is more meaningful if we are most interested in whether the cooling ICM has had time to cool over the age of the cluster. The overall shape of the profiles after this scaling has been applied is qualitatively similar to the case without scaling.

\begin{figure*}[h!]
\centering
\begin{tabular}{c c}
\includegraphics[width=0.48\textwidth]{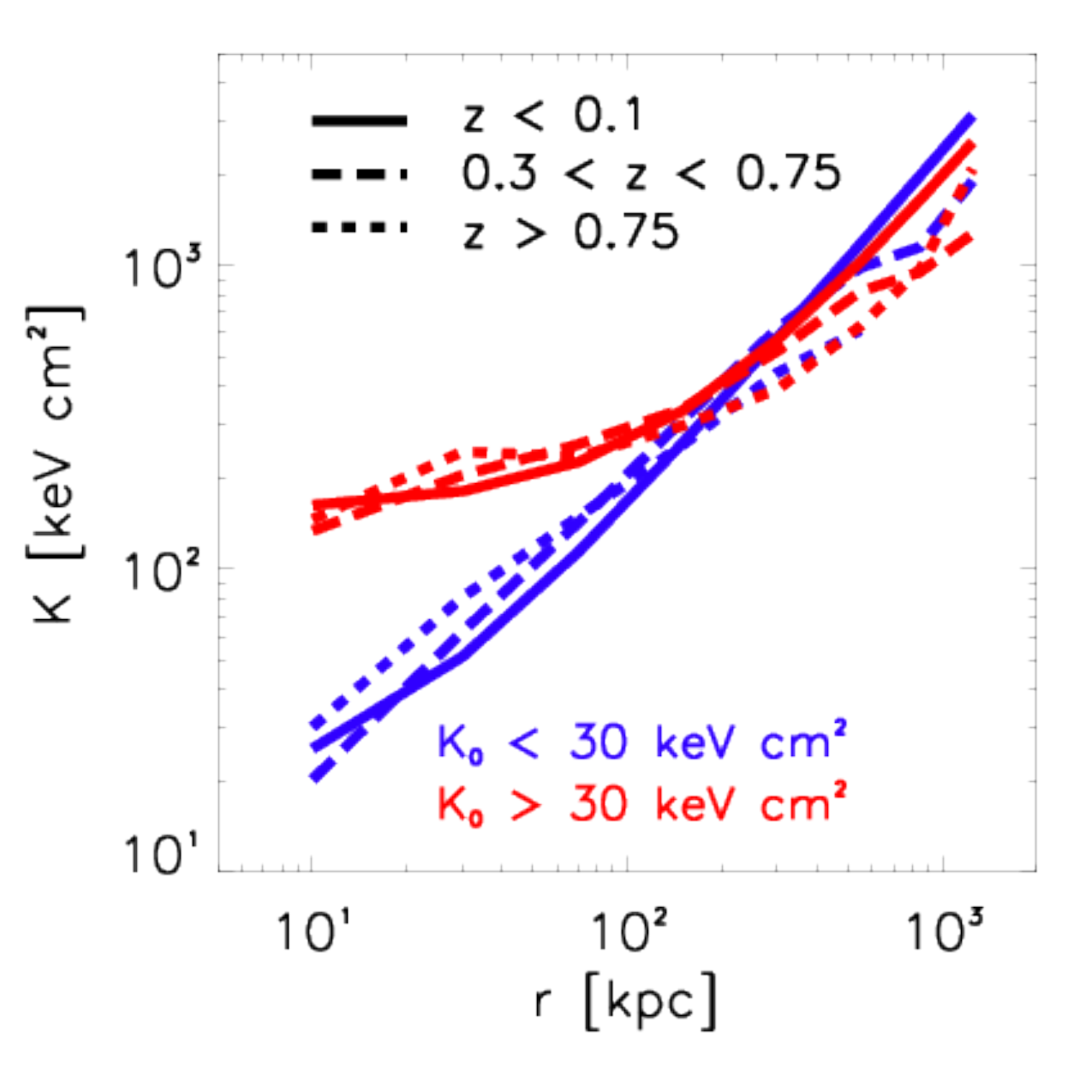}&
\includegraphics[width=0.48\textwidth]{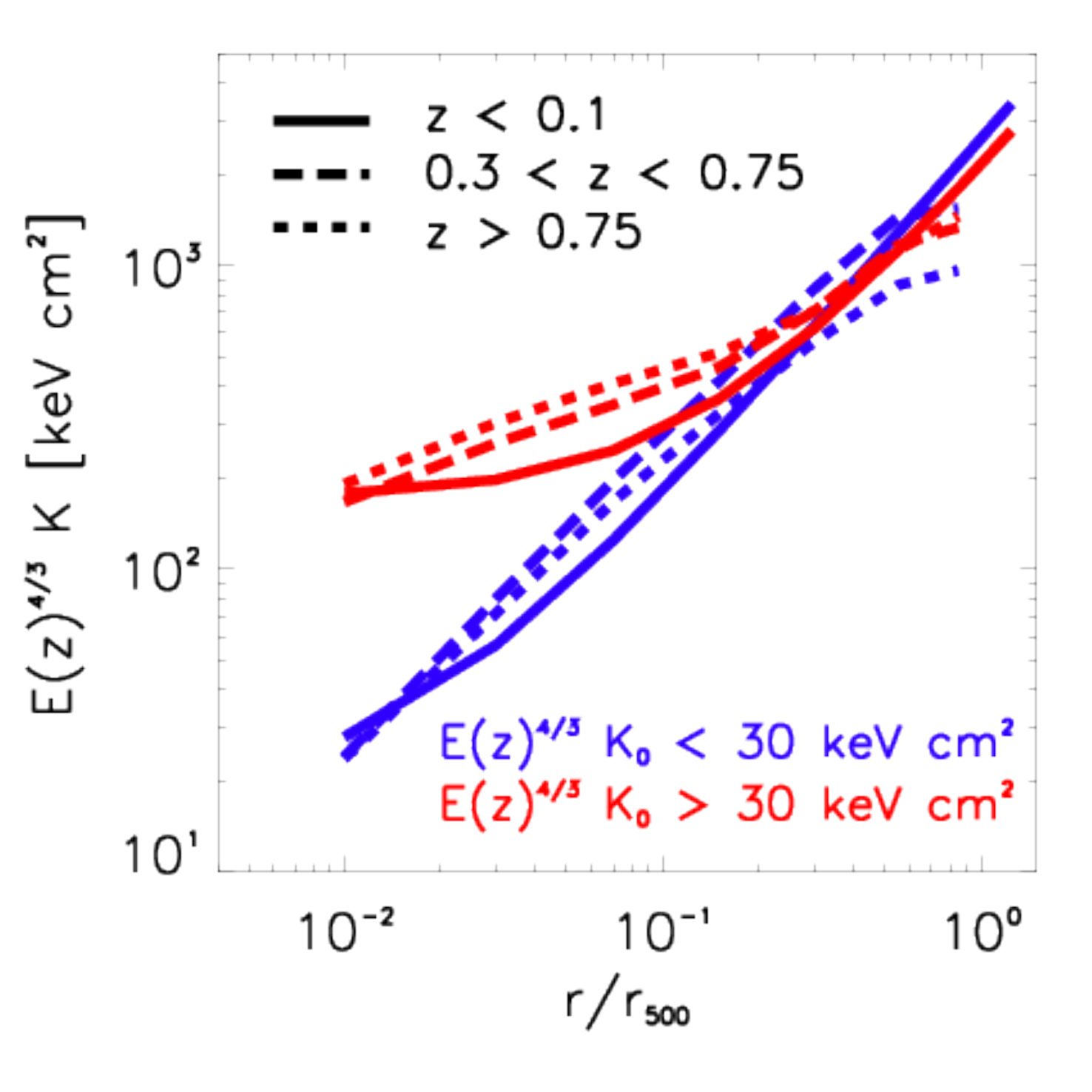}\\
\end{tabular}
\caption{Average entropy profiles for low-redshift ($z<0.1$) clusters \citep{cavagnolo09}, as well as intermediate ($0.3<z<0.75$) and high ($0.75<z<1.1$) redshift clusters from this work, divided into cool core and non-cool core bins based on their central entropy. This plot demonstrates that the inner $\sim$200~kpc of the cluster experiences very little evolution in both the shape and normalization of the entropy profile over $\sim$8~Gyr. At large radii, the entropy appears to decrease with increasing redshift, leading to overall shallower entropy profiles at early times. In the right panel, we apply the self-similar scaling from \cite{pratt10}, $E(z)^{4/3}$, which shows that the central entropy is becoming slightly higher \emph{relative to the outer entropy profile} as a function of redshift. The central entropy evolution in the right panel could be thought of as the effect of ``forcing'' an evolutionary scaling in a regime where the profile is unevolving. Combined, these two plots suggest that the outer profile is evolving as expected based on cosmological models, while the inner $\sim$100kpc has no measurable evolution.}
\label{fig:kprof_stack}
\end{figure*}

In order to look for evolution in the entropy profile, we plot in Figure \ref{fig:kprof_stack} the average entropy profile for cool core (K$_0<30$ keV cm$^2$) and non-cool core  (K$_0>30$ keV cm$^2$) clusters in the SPT-XVP sample, divided into two redshift bins corresponding to $z<0.75$ and $z>0.75$. These are compared to the average profiles from \cite{cavagnolo09}, for clusters at $z\lesssim0.1$. In general, the average profiles are indistinguishable in the inner few hundred kiloparsecs, with high-redshift clusters having slightly higher entropy at small radii ($r\lesssim200$~kpc) than their low-redshift counterparts. At larger radii, the profiles vary according to the self-similar E$(z)^{4/3}$ scaling \citep{pratt10}. These results suggest that the outer entropy profile is following the gravitational collapse of the cluster, while the inner profile has some additional physics governing its evolution, most likely baryonic cooling. The mild central entropy evolution in the right panel of Figure \ref{fig:kprof_stack} could be thought of as the effect of ``forcing'' an evolutionary scaling in a regime where the profile is unevolving

The combination of Figures \ref{fig:kprof} and \ref{fig:kprof_stack} suggests that the cooling properties of the intracluster medium in the inner $\sim$200 kpc have remained relatively constant over timescales of $\sim$8 Gyr. The short cooling times at these radii imply that the core entropy profile should change on short timescales. The fact that this is not observed suggests that some form of feedback has offset cooling on these exceptionally long timescales, keeping the central entropy at a constant value. There is evidence that mechanical feedback from AGN is stable over such long periods of time \citep{hlavacek-larrondo12}, perhaps maintaining the observed entropy floor of 10 keV cm$^2$ since $z\sim1.2$.

\subsection{Distribution of Central Entropy, Cooling Time, and Cooling Rate}
\begin{figure}[htb]
\centering
\includegraphics[width=0.48\textwidth]{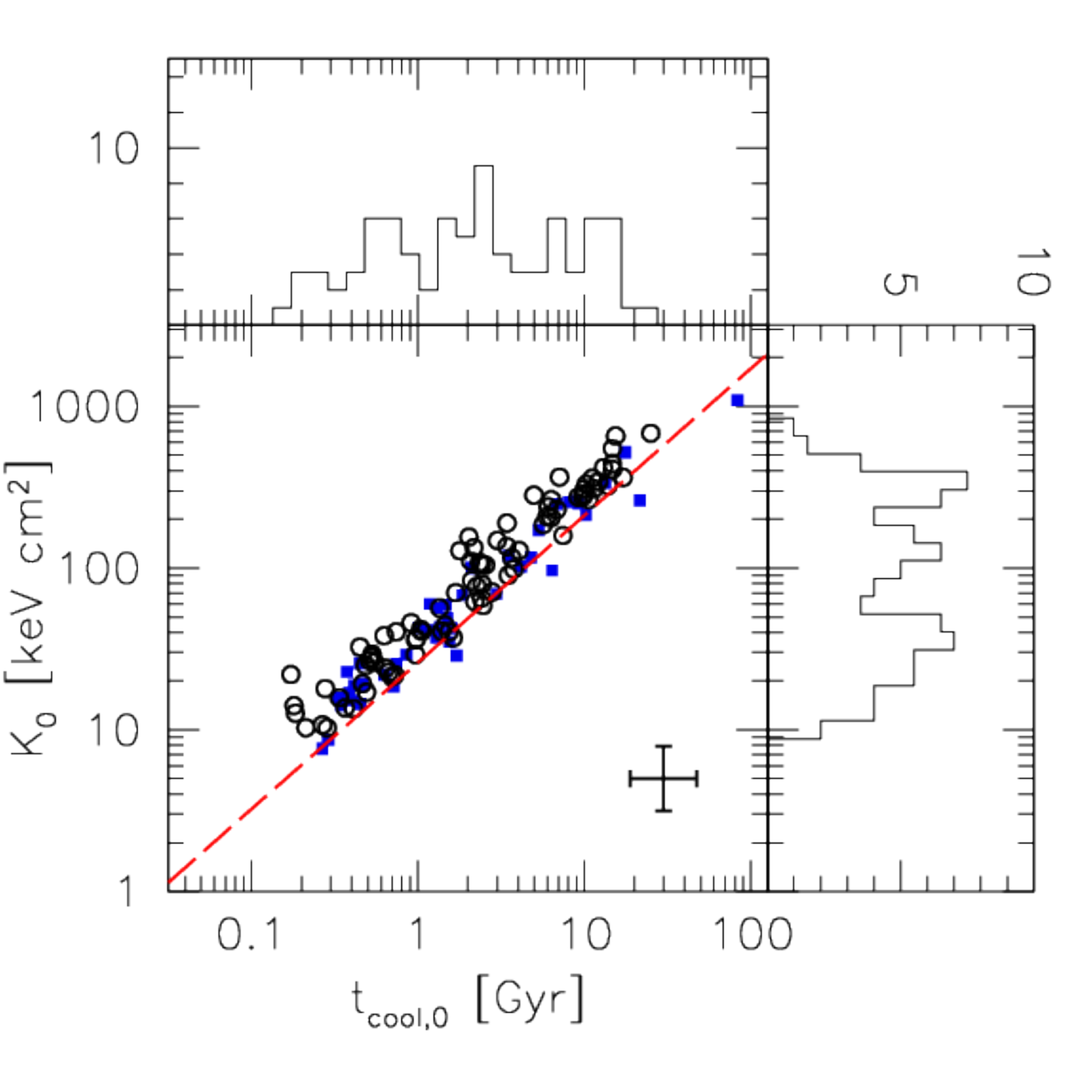}
\caption{Central entropy ($K_0$) versus central cooling time ($t_{\textrm{cool},0}$) for our full sample of SPT-selected clusters, with the typical uncertainty shown in the bottom right. Low-redshift clusters from the CCCP \citep{vikhlinin09} are shown as blue squares, and the best fit for a low-$z$ sample from \cite{hudson10} is shown in red. Both the SPT-XVP and CCCP data lie slightly above this line, as is expected for higher-mass samples. \cite{cavagnolo09} found a bimodal distribution of both $t_{cool,0}$ and K$_0$ around $\sim$1 Gyr and 30 keV cm$^2$, respectively. We find similar, though less significant, minima in our cluster distributions around these same values.}
\label{fig:tc_ent}
\end{figure}

%The presence of a cooling flow, or a cool core, is typically quantified in nearby clusters via the central cooling time ($t_{\textrm{cool},0}$), entropy (K$_0$), or mass deposition rate ($d$M$/dt$). As discussed in \S\ref{sec:deprojection}, we do not have sufficient sampling to reliably interpolate our 3-dimensional temperature and density profiles to $r\sim0$. Instead, we calculate ``central'' quantities interior to 0.012R$_{500}$, or $\lesssim$10~kpc. 
In Figure \ref{fig:tc_ent}, we compare the derived central entropy and cooling time (see \S\ref{sec:deprojection} for details on deriving central quantities) for the SPT-XVP sample to those for the low-redshift clusters in the Chandra Cluster Cosmology Project \cite[hereafter CCCP;][]{vikhlinin06,vikhlinin09}. Overall, we find excellent agreement between the two samples. While it is unsurprising that these two quantities are correlated, due to their similar dependence on both T$_X$ and n$_e$, the normalization and distribution of points along the sequence is reassuring. Both the SPT-XVP and the CCCP clusters have a slightly higher normalization than found by \cite{hudson10}, which can be accounted for by the fact that both of these samples target more massive clusters. Indeed, \cite{hudson10} showed that the scatter about the $t_{\textrm{cool},0}$--K$_0$ correlates with the cluster temperature, with high-T$_X$ clusters lying above the relation and low-T$_X$ groups lying below the relation. Similar to previous low-redshift studies \citep[e.g.,][]{cavagnolo09,hudson10}, we see hints of multiple peaks in both $t_{\textrm{cool},0}$ and K$_0$, with minima at $\sim$1 Gyr  and 50 keV cm$^2$, respectively \citep[e.g.,][]{cavagnolo09,hudson10}. This threshold, separating cool core from non-cool core clusters, appears to be unchanged between the low-redshift and high-redshift samples. We will return to this point in \S3.3.

\begin{figure}[htb]
\centering
\begin{tabular}{c}
\includegraphics[width=0.4\textwidth]{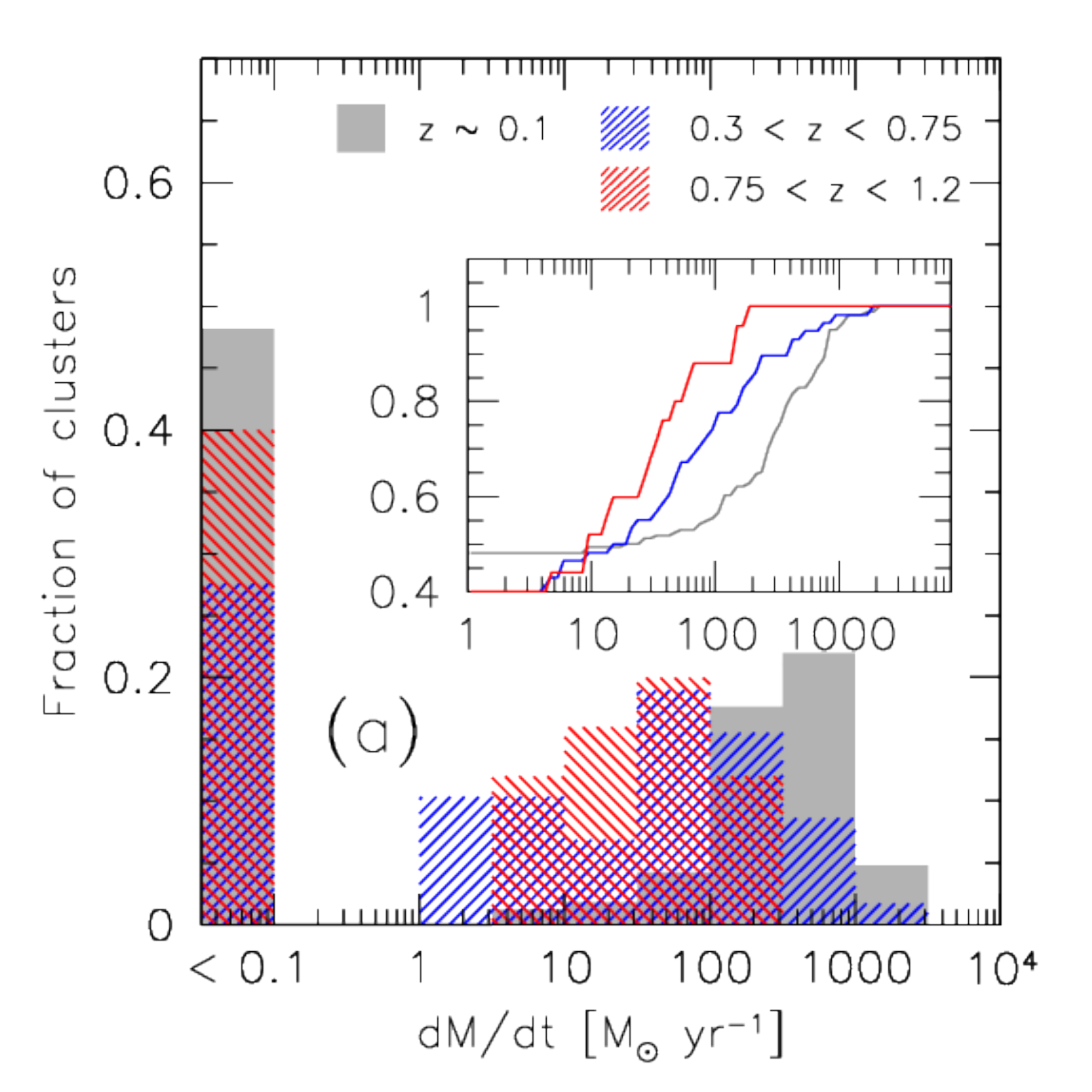} \\
\includegraphics[width=0.4\textwidth]{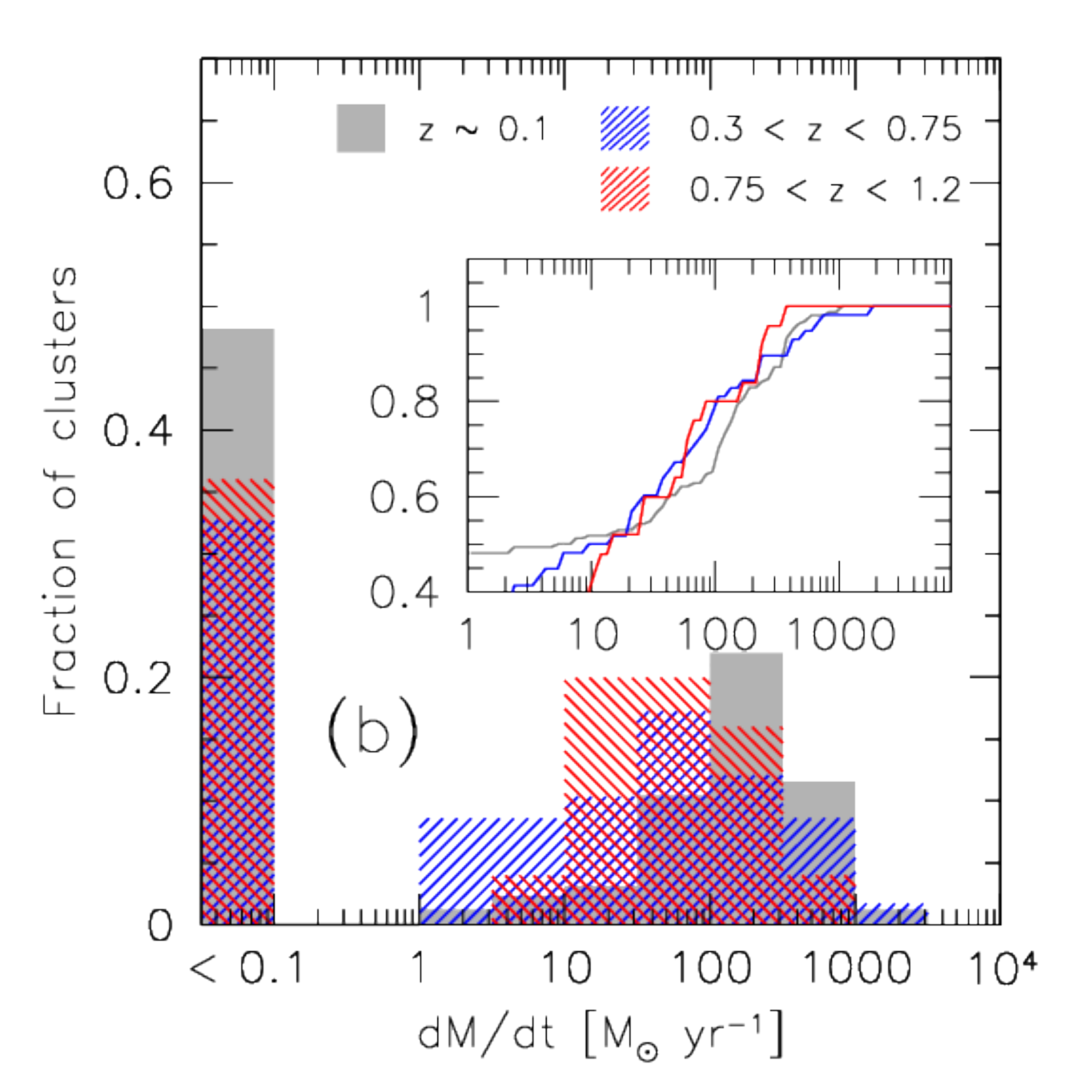} \\
\end{tabular}
\caption{Distribution of classical mass deposition rates (dM/dt) for the SPT-XVP sample, divided into intermediate-redshift (blue) and high-redshift (red) bins. For comparison, we also show a sample of nearby clusters from \citep[gray histogram;][L$_X \ge 1.5\times10^{44}$ erg s$^{-1}$]{white97}. 
In the top (a) and bottom (b) panels, we consider two different definitions of the cooling radius: (a) based on the age of the Universe at the epoch of the cluster (\S2.4.2), and (b) a constant value of 7.7 G-manyr \citep[e.g.,][]{odea08}. 
We find that the evolution observed in the mass deposition rate in the top panel (a) is due to our definition of the cooling radius, which is based on the cluster age -- if we assume that all clusters have been cooling for the same amount of time, the three samples are statistically identical.
In the insets we show the cumulative distribution, which further highlights the similarities between the three samples.}
\label{fig:dmdt}
\end{figure}

In Figure \ref{fig:dmdt}, we plot the distribution of mass deposition rates, $d$M$/dt$, for our SPT-selected sample. We show the integrated cooling rate within two radii: $r$($t_{\textrm{cool}} = t_{\textrm{Univ}})$ and $r(t_{\textrm{cool}} = 7.7$ Gyr). The former is more physically motivated, representing the amount of gas that has had time to cool since the cluster formed. The latter is motivated by the desire to have the definition of the cooling radius be independent of redshift -- the choice of a 7.7 Gyr timescale is arbitrary and was chosen simply to conform with the literature \citep[e.g.,][]{odea08}.

We find that clusters at high-$z$ have overall smaller time-averaged cooling rates, which is unsurprising given that they have had less time to cool. If we remove this factor by instead computing the cooling rate within a non-evolving aperture ($r[t_{\textrm{cool}} = 7.7$ Gyr]), we find no significant difference between the mass deposition rates measured in intermediate redshift ($0.4<z<0.75$) and high redshift ($0.75<z<1.1$) clusters. These sub-samples have median mass deposition rates of 49 M$_{\odot}$ yr$^{-1}$ and 57 M$_{\odot}$ yr$^{-1}$ (excluding non-cooling systems), respectively. For comparison, we also show the distribution of cooling rates for nearby clusters from \cite{white97}, for which the distribution is cut off at $d$M$/dt$ $\lesssim$ 50 M$_{\odot}$ yr$^{-1}$ due to poorer sampling and, thus, reduced sensitivity to modest cooling rates. However, as evidenced by the cumulative distribution, at $d$M$/dt$ $>$ 50 M$_{\odot}$ yr$^{-1}$ the three samples are nearly identical, suggesting very little evolution in the rate of cooling in the ICM over timescales of $\sim$8 Gyr.

%\begin{figure}[htb]
%\includegraphics[width=0.45\textwidth]{plots/tc0_hist.pdf}
%\caption{}
%\label{fig:alphahist}
%\end{figure}

% Show histograms of tcool, K, dM/dt

\subsection{Evolution of Cooling Flow Properties}

\begin{figure}[htb]
\centering
\includegraphics[width=0.49\textwidth]{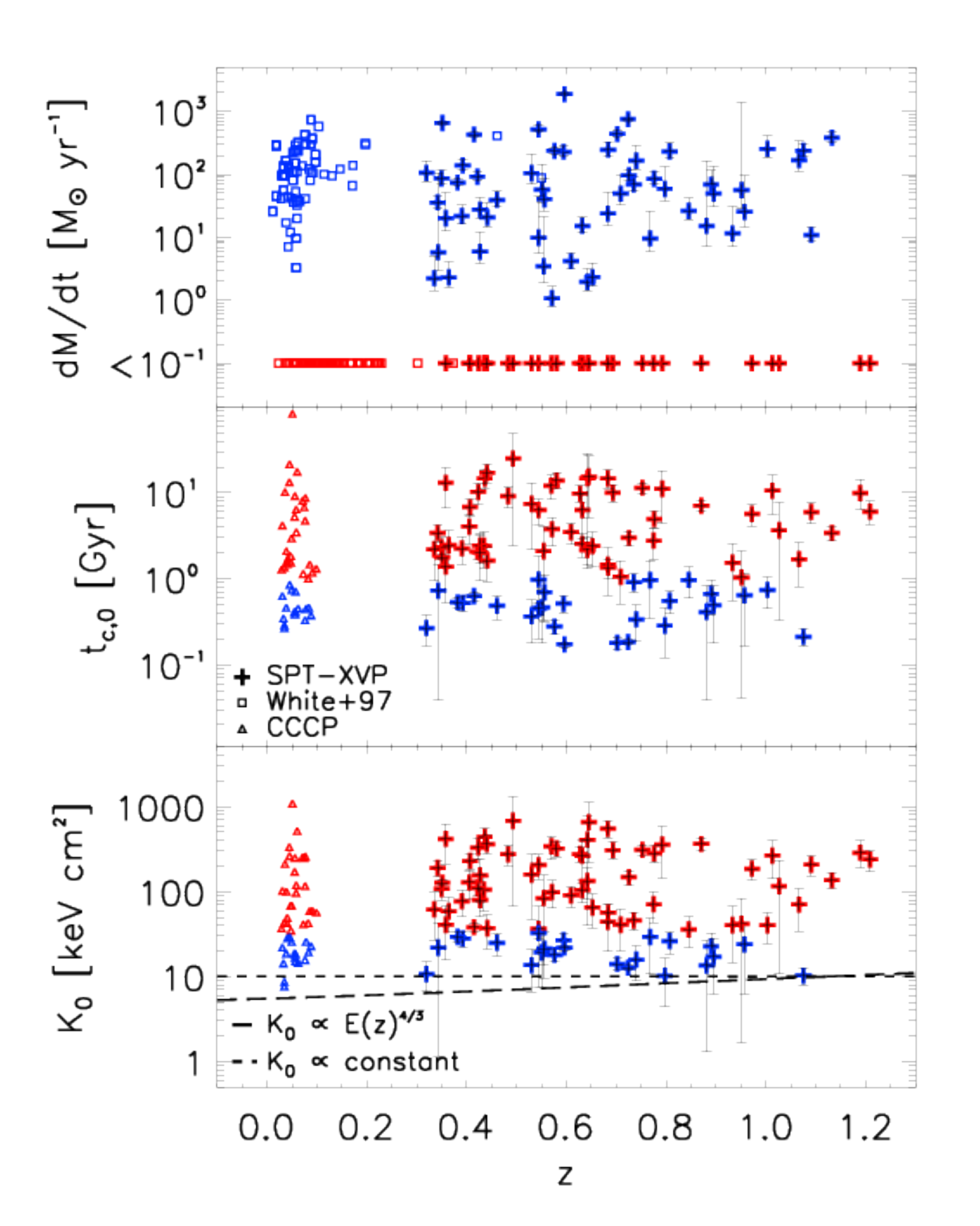}
\caption{Redshift evolution of the mass deposition rate ($d$M$/dt$), central cooling time ($t_{c,0}$), and central entropy (K$_0$) for the sample presented in this paper. Blue and red points represent cooling and non-cooling clusters, respectively, with divisions following Figure \ref{fig:kprof}. For comparison, we show nearby X-ray selected samples \citep{white97, vikhlinin09}, with cuts made to mimic the SPT selection (L$_X>1.5\times10^{44}$ ergs s$^{-1}$, M$_{500}>3\times10^{14}$ M$_{\odot}$). This plot suggest that there has been little evolution in the cooling properties of cluster cores over the range $0<z<1.2$. The lower panel shows the expectation for self-similar evolution of the central entropy \citep[$E(z)^{4/3}$;][]{pratt10}, which is consistent with the observations. The fact that the gas in central $\sim$100~kpc appears not to be cooling suggests that the balance between cooling and feedback has been stable for several Gyr. }
\label{fig:dmdt_evolution}
\end{figure}

% Show that tcool, K0, dM/dt isn't evolving
Figures \ref{fig:kprof_stack} and \ref{fig:dmdt} suggest that the cooling properties of the ICM vary little from $z\sim0$ to $z\sim1$. In order to directly quantify this, we show in Figure \ref{fig:dmdt_evolution} the evolution of $d$M$/dt$, K$_0$, and $t_{\textrm{cool},0}$ with redshift.  For each quantity, we separate cool core and non-cool core clusters using the following thresholds: dM/dt $> 0$ M$_{\odot}$ yr$^{-1}$, K$_0 < 30$ keV cm$^2$, and $t_{\textrm{cool},0} < 1$ Gyr. This figure more clearly shows that there is very little, if any, evolution in the cooling properties of SPT-selected clusters over the range $0.3<z<1.2$. We compare the range of $d$M$/dt$, K$_0$, and t$_{c,0}$ observed for these clusters to samples of nearby clusters from \cite{white97} and the CCCP and find no appreciable change. The entropy floor, at K$_{0} \sim 10$ keV cm$^2$, is constant over the full redshift range of our sample, consistent with earlier work by \cite{cavagnolo09} which covered clusters at $0 < z\lesssim0.5$. The data are also consistent with the self-similar expectation \citep[$E(z)^{4/3}$;][]{pratt10}, which predicts only a factor of $\sim$1.6 change in central entropy from $z=1$ to $z=0$. This self-similar evolution is based on gravity alone -- the fact that it is an adequate representation of the data in the central $\sim$100~kpc of clusters, where cooling processes should be responsible for shaping the entropy profile, suggests that cooling is offset exceptionally well over very long over the past $\sim$7 Gyr. In the absence of feedback, cool core clusters at $z\sim$1 should have $K_0\rightarrow0$ in $<$1 Gyr.

\subsection{Evolution of Cluster Surface Brightness Profiles}
% Show that SB profile evolution
Figures \ref{fig:kprof}--\ref{fig:dmdt_evolution} suggest that there is little change in the ICM cooling properties in the cores of X-ray- and SZ-selected clusters since $z\sim1.2$, in agreement with earlier studies at lower redshift \citep[e.g.,][]{cavagnolo09}. However, Several recent studies have argued that there are fewer cool core clusters at high redshift \citep{vikhlinin07, santos08, santos10} based on measurements of surface brightness concentration, suggesting that ``cool cores'' and ``cooling flows''  may not have the same evolution and, thus, are not necessarily coupled at high redshift as they are now. 

\begin{figure}[htb]
\centering
\includegraphics[width=0.49\textwidth]{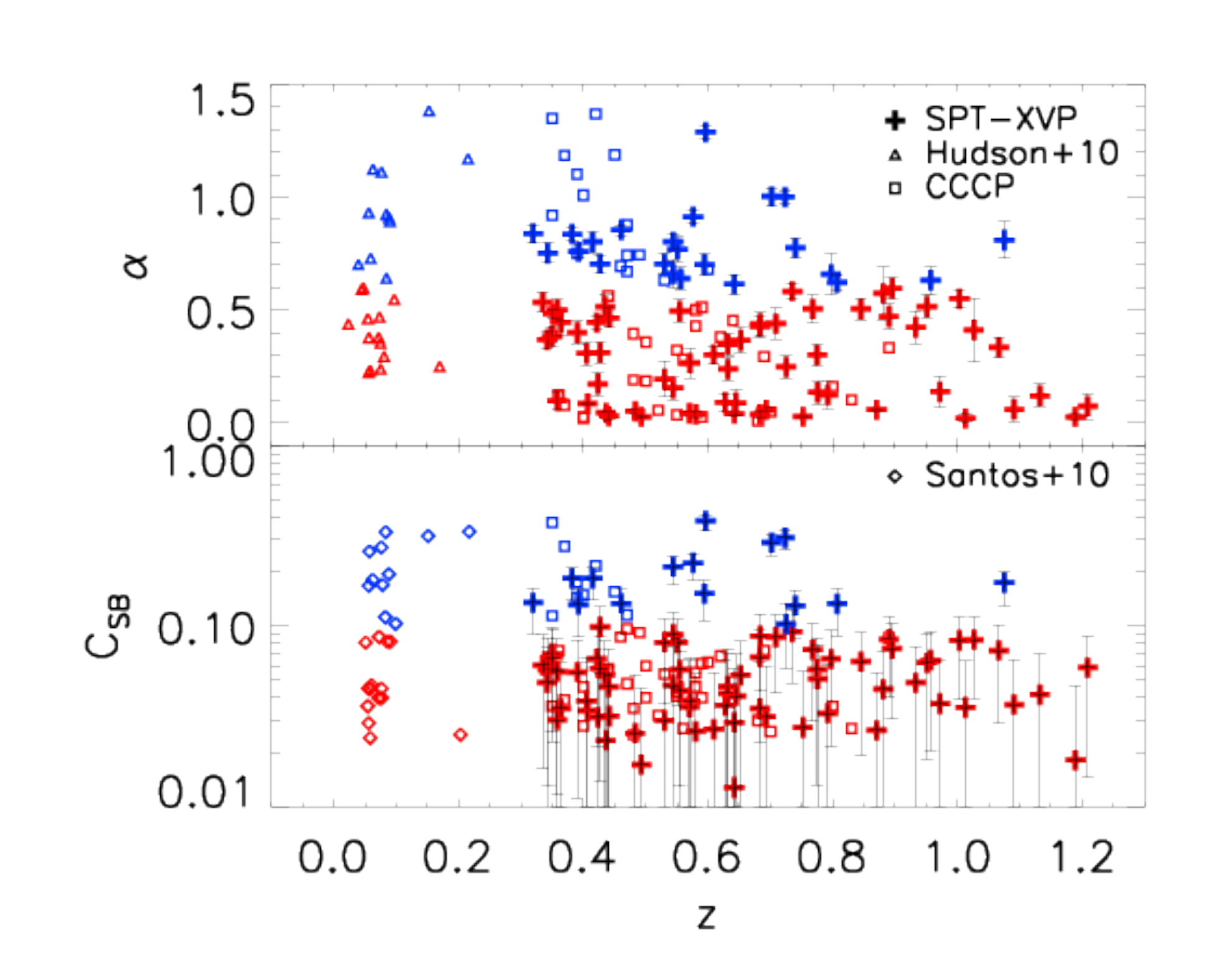}
\caption{Evolution of surface brightness quantities from \cite{vikhlinin07}, $\alpha$, and \cite{santos08}, $c_{SB}$. For comparison we show measurements for low redshift clusters \citep{vikhlinin09, hudson10, santos10}. This figure confirms the strong evolution in cool core strength reported by both \cite{vikhlinin07} and \cite{santos08}, suggesting that the cuspy surface brightness profiles associated with nearby cooling flows were not present at $z\gtrsim0.7$.
}
\label{fig:csb_evol}
\end{figure}

% C_SB, alpha
In Figure \ref{fig:csb_evol}, we duplicate the analyses of \cite{vikhlinin07}, \cite{santos08}, and \cite{semler12} in order to look for evolution in the cool core properties.  We find that the number of galaxy clusters classified as ``cool core'' by both $\alpha$ and $c_{SB}$  (see \S2.3) decreases with redshift, from $\sim$40\% at $z\sim0$ to $\sim$10\% at $z\gtrsim0.75$. These results confirm the evolution in cool core strength reported by \cite{vikhlinin07} and \cite{santos08} for X-ray selected samples, however the evolution appears to be a bit slower for the SZ-selected sample, with several strong cool cores in the range $0.5<z<0.75$ \citep{semler12}. The higher fraction of ``moderate'' cool cores in Figure \ref{fig:csb_evol} at higher redshift in consistent with recent work by \cite{santos10}. All samples agree that there is a lack of strong, classical cool cores at $z>0.75$, which seems to be in opposition to the results presented in Figures \ref{fig:kprof}--\ref{fig:dmdt_evolution} which suggest no evolution in the cooling properties. 
One possible explanation for this discrepancy is that both the concentration parameter, $c_{SB}$, and the cuspiness parameter, $\alpha$, (see eqs. 3 \& 4) assume no evolution in the scale radius of the cool core: $c_{SB}$ assumes a radius of 40~kpc, while $\alpha$ uses $0.04$R$_{500}$.

\begin{figure*}[htb]
\centering
\includegraphics[width=0.99\textwidth]{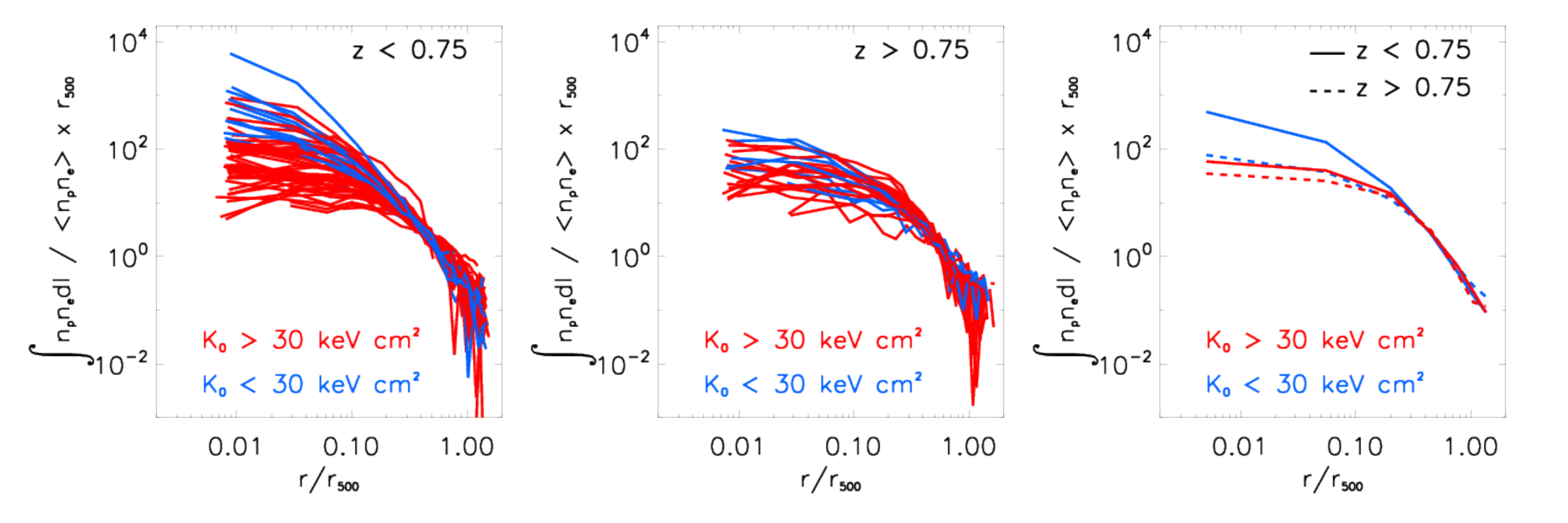}
\caption{
X-ray surface brightness profiles in the 0.5--2.0 keV energy band for the sample presented in this paper, normalized such that they are self-similar at large radii. Profiles are separated into cool core (blue; K$_0 < 30$ keV cm$^2$) and non-cool core (red; K$_0 >30$ keV cm$^2$). The centrally-peaked surface brightness profile, characteristic of a cooling-flow cluster, is present only in the low-redshift ($z<0.75$) sample. At high redshift ($z>0.75$) both cool-core and non-cool core clusters, as defined by their central entropy, are indistinguishable from their surface brightness profiles alone. In the right panel, average surface brightness profiles are shown, which demonstrate the similarity between low-redshift non-cool cores, high-redshift non-cool cores, and high-redshift cool cores.
}
\label{fig:sb_evol}
\end{figure*}

% profile evolution
To further investigate the surface brightness evolution, we move away from single-parameter measures of surface brightness concentration and, instead, directly compare the X-ray surface brightness profiles for low- and high-redshift clusters in Figure \ref{fig:sb_evol}. At $z<0.75$, we confirm that, overall, clusters with low central entropy (K$_0<30$ keV cm$^2$) have more centrally-concentrated surface brightness profiles than those with high central entropy (K$_0>30$ keV cm$^2$). However, at high redshift ($z>0.75$) we find a lack of strongly-concentrated clusters, with only a weak increase in concentration for the clusters with low central entropy, consistent with earlier studies of distant X-ray-selected clusters \citep[e.g.,][]{vikhlinin07, santos08, santos10}. This difference in surface brightness concentration is not a result of increased spatial resolution for low-redshift clusters -- the difference in spatial resolution between the centers of these redshift bins is only $\sim$30\%.

\begin{figure*}[htb]
\centering
\begin{tabular}{c}
\includegraphics[width=0.99\textwidth]{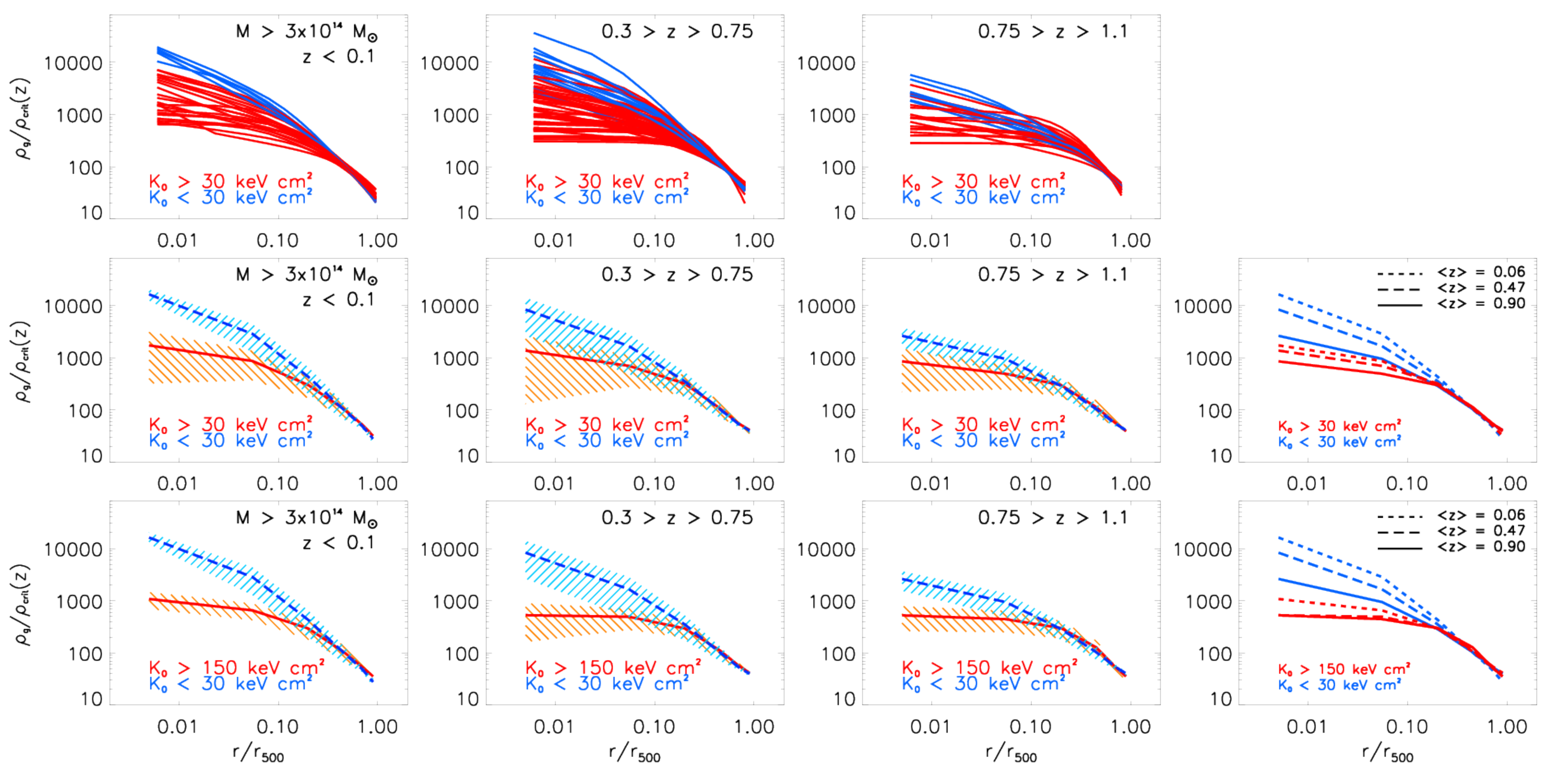}\\
\end{tabular}
\caption{
Gas density profiles for an X-ray-selected sample of nearby clusters \citep[CCCP;][]{vikhlinin09}, as well as the sample of SPT-selected clusters presented in this work. In the upper panels we show all of the profiles, scaled in terms of the critical density ($\rho_{crit}$) and R$_{500}$.  In the middle row, we show the median profiles for clusters with K$_0< 30$ keV cm$^2$ (blue) and K$_0>30$ keV cm$^2$ (red). In the bottom row we classify ``non-cool cores'' as having K$_0>150$ keV cm$^2$ \citep{hudson10}, which further highlights the difference between cool cores and non-cool cores. In the right-most column, we show all of the median profiles together, demonstrating the substantial evolution in the median gas density profile as a function of redshift.  This figure shows clearly that the 3-dimensional gas density is becoming more centrally concentrated over time in cool core clusters, while remaining nearly constant in non-cool core clusters.
}
\label{fig:rho_evol}
\end{figure*}

This result becomes even more dramatic when the data are deprojected into gas density, rather than surface brightness, and including $z\sim0$ clusters for comparison. In Figure \ref{fig:rho_evol}, we compare the ICM gas density profiles for the sample presented in this work to a low-redshift sample \citep[Chandra Cluster Cosmology Project;][]{vikhlinin09}. We restrict the low-redshift sample to clusters with M$_{500}>3\times10^{14}$ M$_{\odot}$, in order to approximate the SPT-selection cut. This comparison is particularly appropriate since our reduction and analysis pipeline is identical to that used by \cite{vikhlinin09}. Figure \ref{fig:rho_evol} shows that the 3-dimensional gas density profiles become more centrally concentrated with decreasing redshift, with nearly an order of magnitude difference in central gas density between cool core clusters at $z\sim0.1$ and $z\sim0.9$. On the contrary, non-cooling clusters (K$_0>30$ keV cm$^2$) experience no appreciable evolution in the central physical density over the same timescale. The combination of Figures \ref{fig:sb_evol} and \ref{fig:rho_evol} seem to suggest that the dense cores which are associated with cooling flows have built up slowly over the past $\sim$8 Gyr. This scenario would explain the lack of centrally-concentrated, or ``cuspy'', clusters at high redshift.

%\begin{figure}[htb]
%\includegraphics[width=0.45\textwidth]{plots/alpha_hist.pdf}
%\caption{}
%\label{fig:alphahist}
%\end{figure}

%================================================================%
% ======= DISCUSSION=============================================%
%================================================================%

%\clearpage

\section{Discussion}

Figures \ref{fig:kprof}--\ref{fig:rho_evol} present an interesting story. The cooling properties of the intracluster medium in the most massive galaxy clusters appear to be relatively constant -- that is, classical cooling rates are not getting any higher and central entropies are not getting any lower -- since $z\sim1.2$. Over the same timescale, the central gas density has increased by roughly an order of magnitude in clusters exhibiting cooling signatures, leading to considerably more concentrated surface brightness profiles in low-redshift cool core clusters. Below, we discuss potential explanations for these results, along with systematics that may be confusing the issue.

\begin{figure}[htb]
\centering
\includegraphics[width=0.49\textwidth]{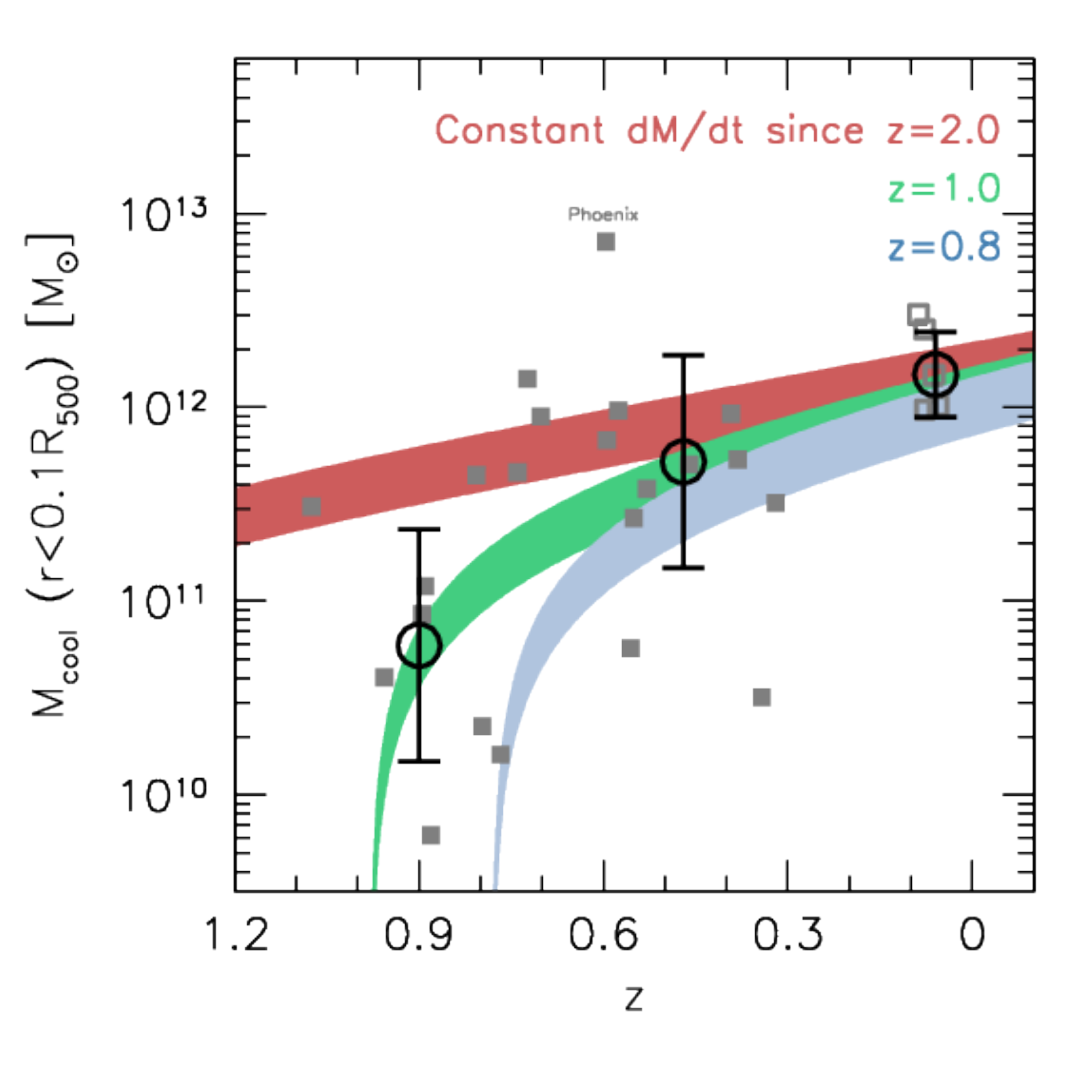}\\
\caption{Cool core gas mass within 0.1R$_{500}$ as a function of redshift for clusters with K$_0<30$~keV cm$^2$ from the CCCP (open squares) and SPT-XVP (filled squares). Here, the cool core mass represents the volume-integrated difference between the median ``non-cool core'' (K$_0>150$ keV cm$^2$) profile and the cluster density profile. Individual clusters are shown as filled grey squares, while the averages in three redshift bins are shown as black circles. 
Black circles represent the median values in three bins: $z<0.1$, $0.3<z<0.75$, and $0.75<z<1.2$. 
The median growth of cool cores is well-modeled by a constant cooling flow ($d$M$/dt$ = 150 M$_{\odot}$ yr$^{-1}$) that began at $z=1$, with the full range of points being consistent with cooling flows beginning at $0.8<z<2$. This plot suggests that cooling flows do bring low-entropy material into the core of the cluster, but that some form of feedback prevents this gas from cooling completely, leading to the build-up of low-entropy gas in the cluster core. This scenario is in agreement with the lack of evolution of dM/dt and its peak value reported in Figure \ref{fig:dmdt}, coupled with the constant entropy floor of 10 keV cm$^2$ shown in Figure \ref{fig:dmdt_evolution}.
}
\label{fig:ccev}
\end{figure}

\begin{figure}[htb]
\centering
\includegraphics[width=0.49\textwidth]{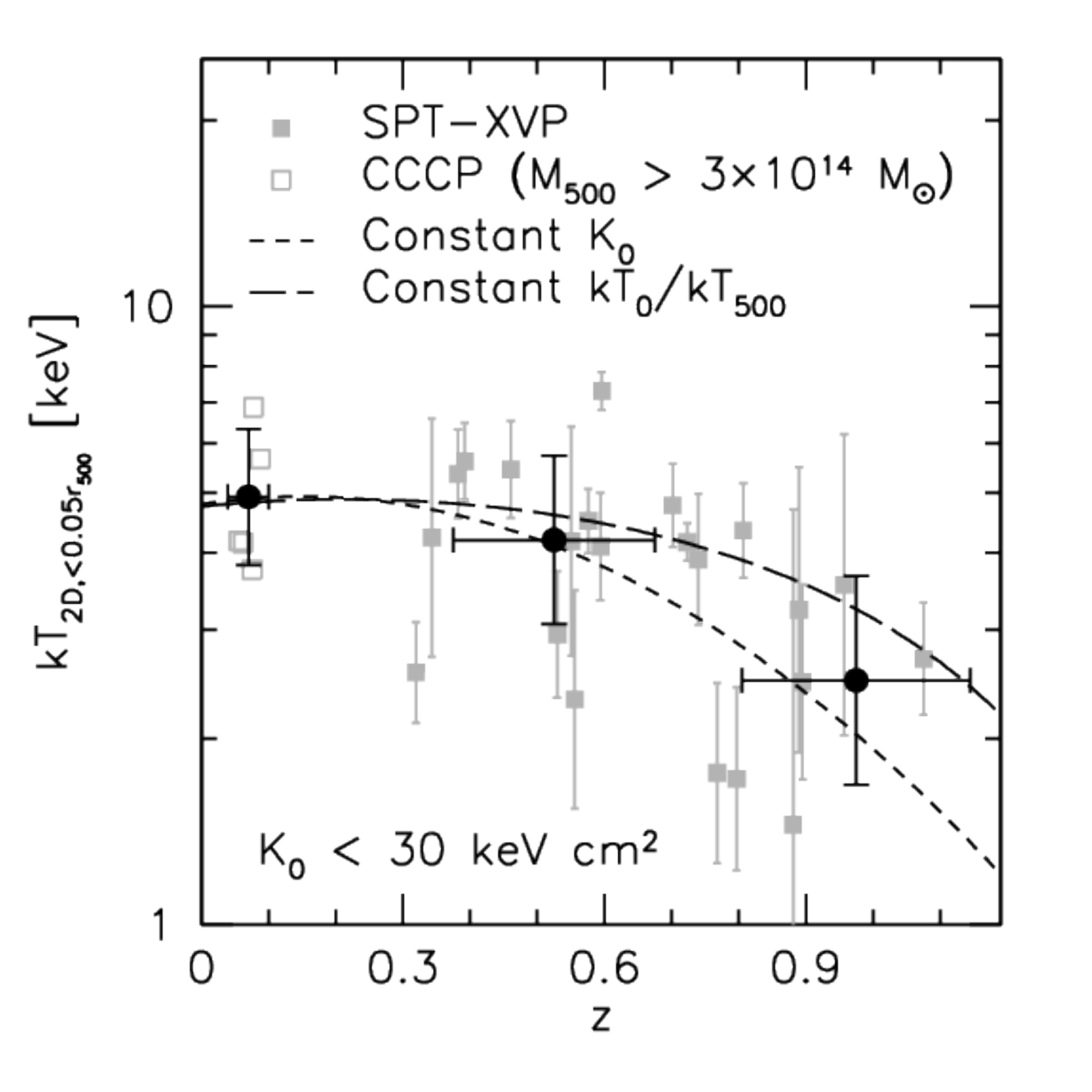}\\
\caption{Central temperature, measured within $r<0.05$R$_{500}$, for cool core clusters as a function of redshift. Black circles represent the mean redshift and central temperature in three redshift bins. The binned points are inconsistent with the hypothesis of no evolution ($dk$T$/dz=0$) at the $\gtrsim1\sigma$ level ($\gtrsim$67\% confidence). The long-dashed line here represents the self-similar expectation (i.e., factoring in that high-$z$ clusters are lower mass and, thus, cooler in general), while the short-dashed line represents no evolution in the central entropy, K$_0$ (see Figure \ref{fig:dmdt_evolution}). This figure suggests that present day cool cores are warmer than their high-$z$ counterparts, although we are unable to distinguish if this is purely due to self-similar evolution, or if there may be some contribution from feedback in order to prevent the central entropy from reaching $<$10 keV cm$^2$.
}
\label{fig:ktcore}
\end{figure}

\subsection{The Origin of Cool Cores}
The results presented thus far suggest that the dense cores associated with cooling flow clusters were not as pronounced $\sim$8 Gyr ago, despite the fact that clusters at these early times had similar cooling rates and central entropy (Figure \ref{fig:dmdt_evolution}). We propose that these cores have grown over time as a direct result of cooling flows being halted by feedback.  In this scenario, gas from larger radii cools and flows inwards, but it hits a ``cooling floor'' at $\sim$10 keV cm$^2$, below which cooling is less efficient. This floor is likely a result of some form of feedback, with the most promising explanation currently being mechanical energy injection from the central AGN \citep[e.g.,][]{fabian12, mcnamara12}. This concept of a cooling floor is supported by observations. \cite{peterson06} show, using high resolution X-ray spectroscopy of nearby galaxy clusters, that gas at temperatures less than $\sim$1/3 of the ambient ICM temperature is cooling orders of magnitude less effectively than predicted. Further, in agreement with \cite{cavagnolo09}, we show in Figure \ref{fig:dmdt_evolution} that there is a lower entropy limit in the cores of galaxy clusters of $\sim$10 keV cm$^2$ which is roughly constant over the range $0<z<1.2$. The fact that the ICM appears unable to cool efficiently below $\sim$1/3 of the ambient temperature, or $\sim$10 keV cm$^2$, implies that, if material is indeed flowing into the cluster core, then we should observe a build up of low-entropy gas in the core.

To test this hypothesis, we measure the ``cool core mass'' for all clusters in our sample with K$_0<30$ keV cm$^2$. The cool core mass is defined as:

\begin{equation}
\textrm{M}_{cool} = 4\pi\int_0^{0.1\textrm{R}_{500}} (\rho_{g}-\left<\rho_{g,\textrm{NCC}}\right>)r^2dr ,
\end{equation}

\noindent{}where $\left<\rho_{g,\textrm{NCC}}\right>$ represents the median non-cool core (K$_0>150$ keV cm$^2$) density profile (Figure \ref{fig:rho_evol}) and the outer radius of 0.1R$_{500}$ is roughly where the uncertainty in $\rho_g$ is similar in scale to the difference between the median cool core and non-cool core profiles. In Figure \ref{fig:ccev}, we plot the cool core mass, M$_{cool}$, versus redshift for the full sample of SPT-XVP clusters with K$_0<30$ keV cm$^2$, including 5 nearby ($z<0.1$) clusters from \cite{vikhlinin06}. We find a rapid evolution in the total cool core mass, with an order of magnitude increase in M$_{cool}$  between $z\sim1$ and $z\sim0.5$. As we show in Figure \ref{fig:ccev}, the median growth is fully consistent with a constant cooling flow since $z=1$ with $d$M$/dt$ = 150 M$_{\odot}$ yr$^{-1}$. The range of cool core masses is consistent with cooing flowing initiating at $0.8\lesssim z \lesssim 2$, providing the first constraints on the onset of cooling in galaxy clusters.

Figure \ref{fig:ccev} suggests that cool cores at $z\sim0$ are a direct result of long-standing cooling flows (Figure \ref{fig:dmdt}) coupled with a constant entropy floor (Figure \ref{fig:dmdt_evolution}) -- most likely the result of AGN feedback. The long-standing balance between cooling and feedback prevents gas from cooling completely and, instead, leads to an accumulation of cool gas in the core of the cluster.

This simple evolutionary scenario, which we offer as an explanation for the increase in central gas density in clusters from $z\sim1$ to $z\sim0.5$, is based on a sample of massive (M$_{500} > 2\times10^{14}$ M$_{\odot}$), rich galaxy clusters. While the cooling rate (dM/dt) is proportional to cluster mass \citep{white97}, there is no evidence that the presence of a cool core is dependent on whether the host is a rich galaxy cluster or a poor group \citep{mcdonald11a}. Thus, if these trends hold at $z\gg0.1$, we would expect future surveys of low-mass, high-redshift clusters, via optical \citep[e.g., LSST;][]{lsst} or X-ray \citep[e.g., eRosita;][]{erosita} detection, to see a similar decline in the cuspiness of cool cores at high redshift. At this point, however, such an extrapolation to lower masses is purely speculative.
	
\subsection{The Evolution of ICM Cooling}
In Figure \ref{fig:dmdt_evolution}, we show that there is no measurable evolution in the minimum central entropy, K$_0$, over the range $0.3<z<1.2$. Coupled with the apparent increase in central density, this would imply that cool cores today are \emph{warmer} than their high-$z$ counterparts. Since a detailed spatial comparison, such as we did for density, is more challenging with a spectroscopically-measured quantity, we reduce the problem to a single measurement of ``central'' temperature. Here we define the central temperature as the spectroscopically-measured temperature within 0.05R$_{500}$, with central point sources masked. We consider here only systems with K$_0<30$ keV cm$^2$, which are the most centrally concentrated systems in our sample, allowing us to use a smaller aperture than in Figure 1.

In Figure \ref{fig:ktcore}, we see that, indeed, for clusters classified as cool core on the basis of their central entropy (K$_{0}<30$ keV cm$^2$), central temperature increases with decreasing redshift. The slope of this relation is equally consistent with the expected self-similar evolution, as well as what is required to have no evolution in K$_0$ over this redshift interval (dashed line, see also Figure \ref{fig:dmdt_evolution}). This figure seems to suggest that there may be some additional heating in low-redshift cluster cores above the self-similar expectation, perhaps resulting from AGN feedback. However, we stress that these data are insufficient to distinguish between these two scenarios, and so we defer any further speculation on the evolution of the central temperature to an upcoming paper which will perform a careful stacking analysis of these clusters.
 
% Show T_(r<0.15R500)/T_500 histogram for z>0.75 and z<0.75 -- Are cool cores cooler at high z?
% Show model of growing cool core with dM/dt = 100 Msun/yr

The combined evidence presented in \S4.1 and \S4.2 support the scenario that cooling material has been gradually building up in cluster cores over the last $\sim$8 Gyr, but has been prevented from completely cooling by an almost perfectly balanced heating source.
While at first glance it would appear that the amount of energy injected by feedback must increase rapidly to offset increased cooling (L$_{\textrm{cool}} \propto n_e^2$T$^{1/2}$), much of this is offset by the fact that the gravitational potential in the core is increasing, leading to more heating as cooling material falls into the potential well. So, while cool cores are becoming denser with decreasing redshift, the actual cooling rates, and by extension the energy needed for feedback to offset cooling, have remained nearly constant since $z\sim1$.

Recent observations of the ``Phoenix Cluster'' \citep{mcdonald12c, mcdonald13} suggest that some clusters may undergo episodes of runaway cooling, perhaps before the feedback responsible for establishing the cooling flow was fully established, or that the feedback mechanism is strongly episodic.

\subsection{The Cool Core / Cooling Flow Fraction}

\begin{figure}[htb]
\centering
\includegraphics[width=0.49\textwidth]{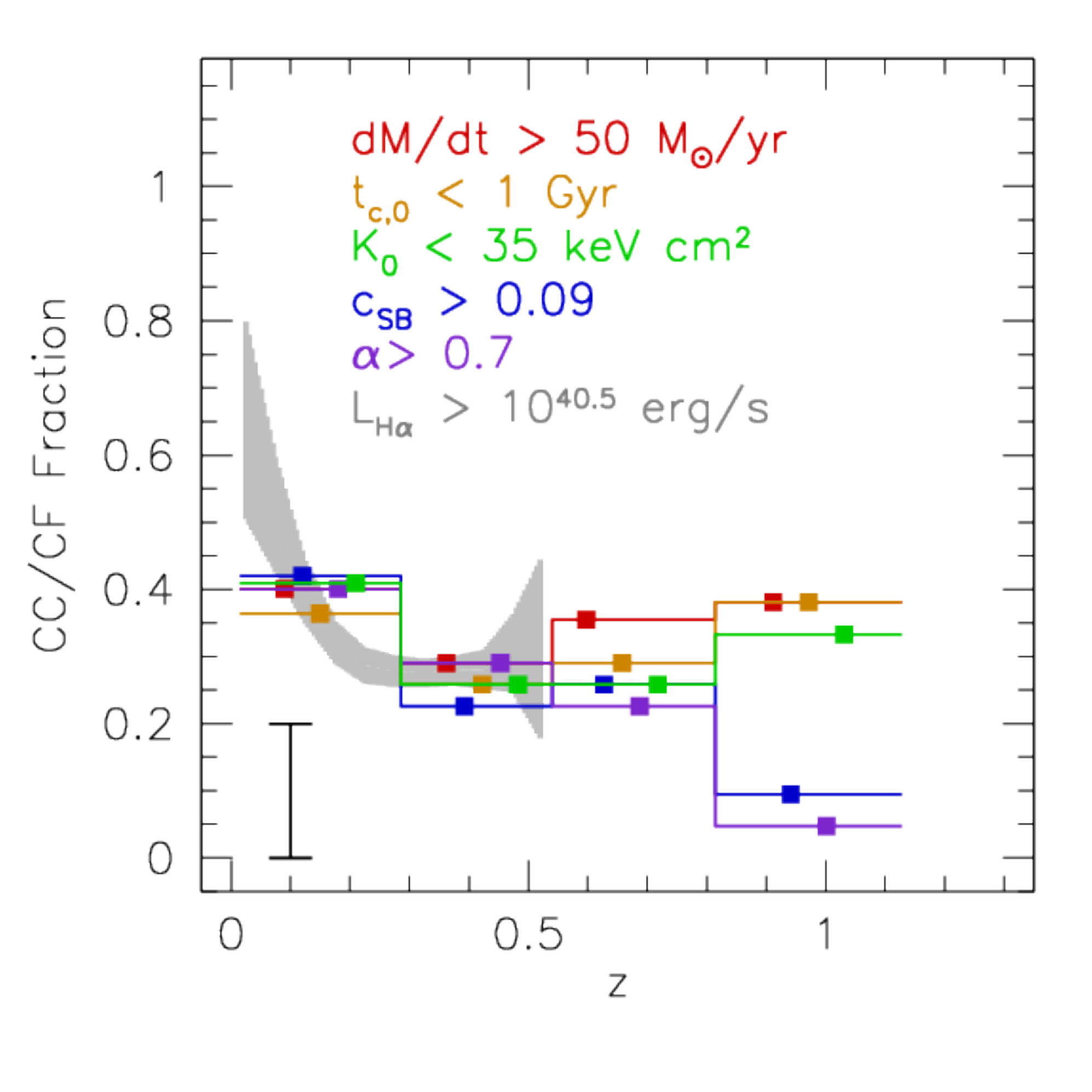}
\caption{The fraction of clusters harboring a cooling flow or cool core, as determined by a variety of indicators, versus redshift. This figure demonstrates the difficulty in classifying cool core clusters at $z>0.75$, where the cooling rate is high, but the density profile is not cuspy. Low redshift points here come from Figures \ref{fig:dmdt_evolution} and \ref{fig:csb_evol}. We have chosen slightly different thresholds for K$_0$ and c$_{\rm{SB}}$ than in previous plots, in order to reduce the scatter at $z<0.5$ where these parameters should all agree on the cool core fraction. We show in the lower left corner the typical uncertainty on the cool core fraction for each bin. For comparison, we also show the fraction of low-redshift clusters with strong emission line nebulae, which are generally indicative of a cool core \citep[gray shaded region;][]{mcdonald11c}.}
\label{fig:ccfrac}
\end{figure}

Much effort has recently focused on determining the fraction of high-redshift clusters which harbor a cool core \citep{vikhlinin07,santos08,santos10,samuele11,mcdonald11c,semler12}. However, based on the results of this paper, we now know that the inferred evolution in the cool core fraction depends strongly on the criteria that are used to classify cool cores. 
In Figure \ref{fig:ccfrac} we demonstrate this point, showing that the measured fraction of high-$z$ cool cores is drastically different if the classification of cool cores is based on the presence of \emph{cooling} (K$_0$, t$_{\textrm{cool},0}$, $d$M$/dt$; $\sim$35\%) or \emph{cooled} ($\alpha$, c$_{\rm{SB}}$; $\sim$5\%) gas.
This figure shows that, at high redshift, it is important to differentiate between ``cooling flows'' and ``cool cores'' when classifying galaxy clusters as cooling or not -- a distinction which is unnecessary in nearby clusters.

Figure \ref{fig:ccfrac} shows that the fraction of strongly-cooling clusters ($d$M$/dt>50$ M$_{\odot}$ yr$^{-1}$, K$_0<35$ keV cm$^2$, $t_{\textrm{cool},0}<1$ Gyr) undergoes little evolution over the range $0.3<z<1.2$. This is qualitatively apparent in Figure \ref{fig:dmdt_evolution}. The fraction of strongly-cooling clusters increases from 25$^{+11}_{-5}$\% to 35$^{+10}_{-10}$\%, consistent at the $1\sigma$ level with no change. This figure shows that, not only is the rate at which the ICM is cooling roughly constant since $z\sim1$ (Figure \ref{fig:dmdt} and \ref{fig:dmdt_evolution}), but the fraction of clusters which experience strong cooling is also nearly constant over similar timescales.

It is also worth noting here the overall agreement in Figure \ref{fig:ccfrac} between the evolution of ICM cooling inferred from X-ray properties (this work) and optical properties \citep{mcdonald11c} at $z\leq0.5$. The latter sample was drawn from optically-selected catalogues, and used emission-line nebulae as a probe of ICM cooling. This overall agreement suggests that the evolution of cooling properties is relatively independent of how the clusters are selected (optical vs SZ). The steep rise in the cooling fraction at $z<0.5$ was interpreted by \cite{mcdonald11c} as being due to timing -- we are seeing clusters transitioning from ``weak'' to ``strong'' cool cores as the central cooling time drops over time.

\subsection{Potential Biases}
While the observed cool core evolution presented here is interesting, it may suffer from a combination of several biases in both the sample selection and in the analysis. Below we address four biases and quantify their effects on the observed cool core evolution.

\subsubsection{3-Dimensional Mass Modeling}

\begin{figure}[htb]
\centering
\includegraphics[width=0.49\textwidth]{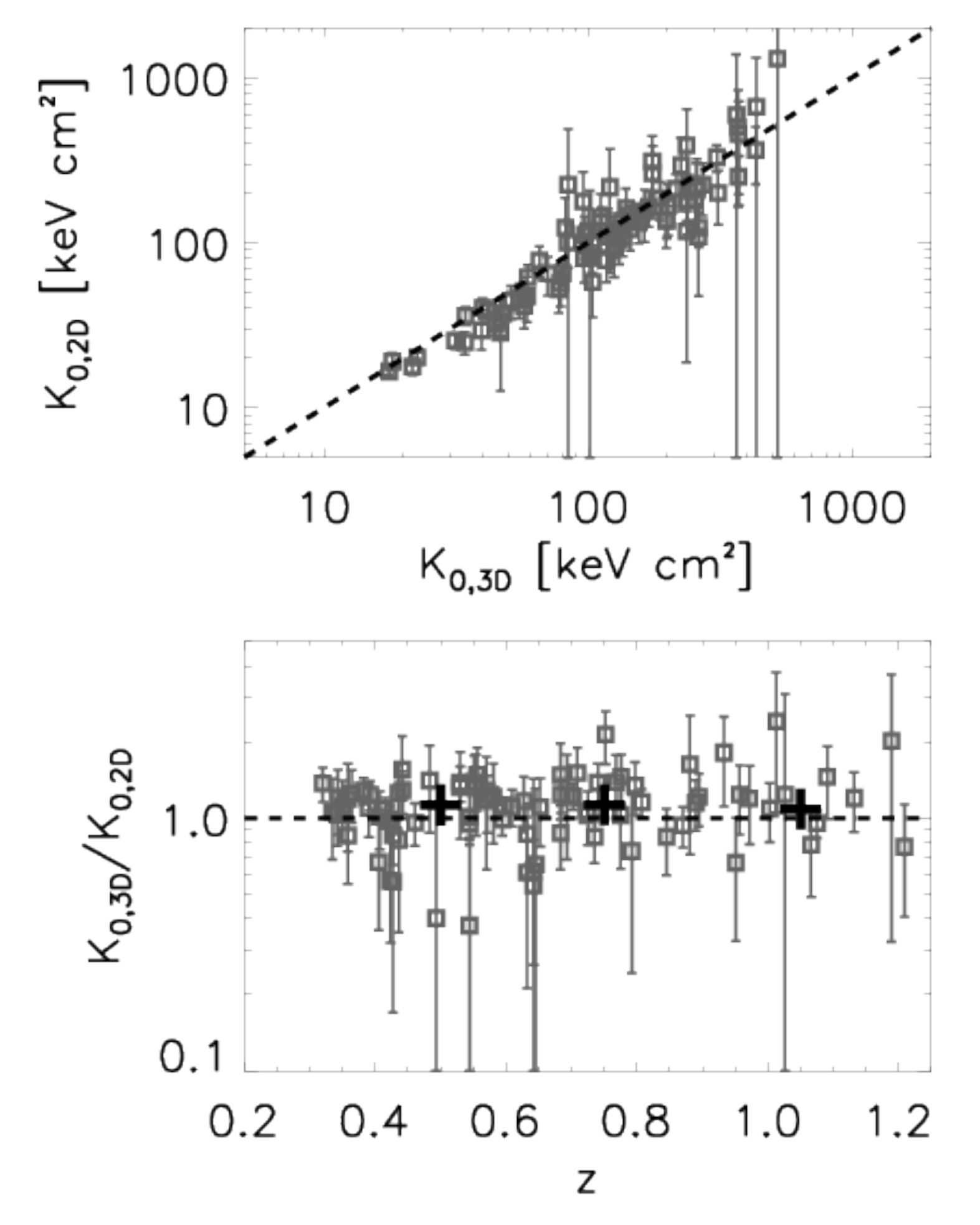}\\
\caption{Comparison of central entropy (K$_0$) measured in a 2-dimensional aperture with $r<0.1$R$_{500}$ to our 3-dimensional models projected onto the same annulus. In the lower panel the mean residuals in three different redshift bins are shown as black crosses. This figure demonstrates that the 2-dimensional and 3-dimensional measurements agree well with each other, and any differences are redshift independent.
}
\label{fig:compare_k0}
\end{figure}

\begin{figure}[htb]
\centering
\includegraphics[width=0.45\textwidth]{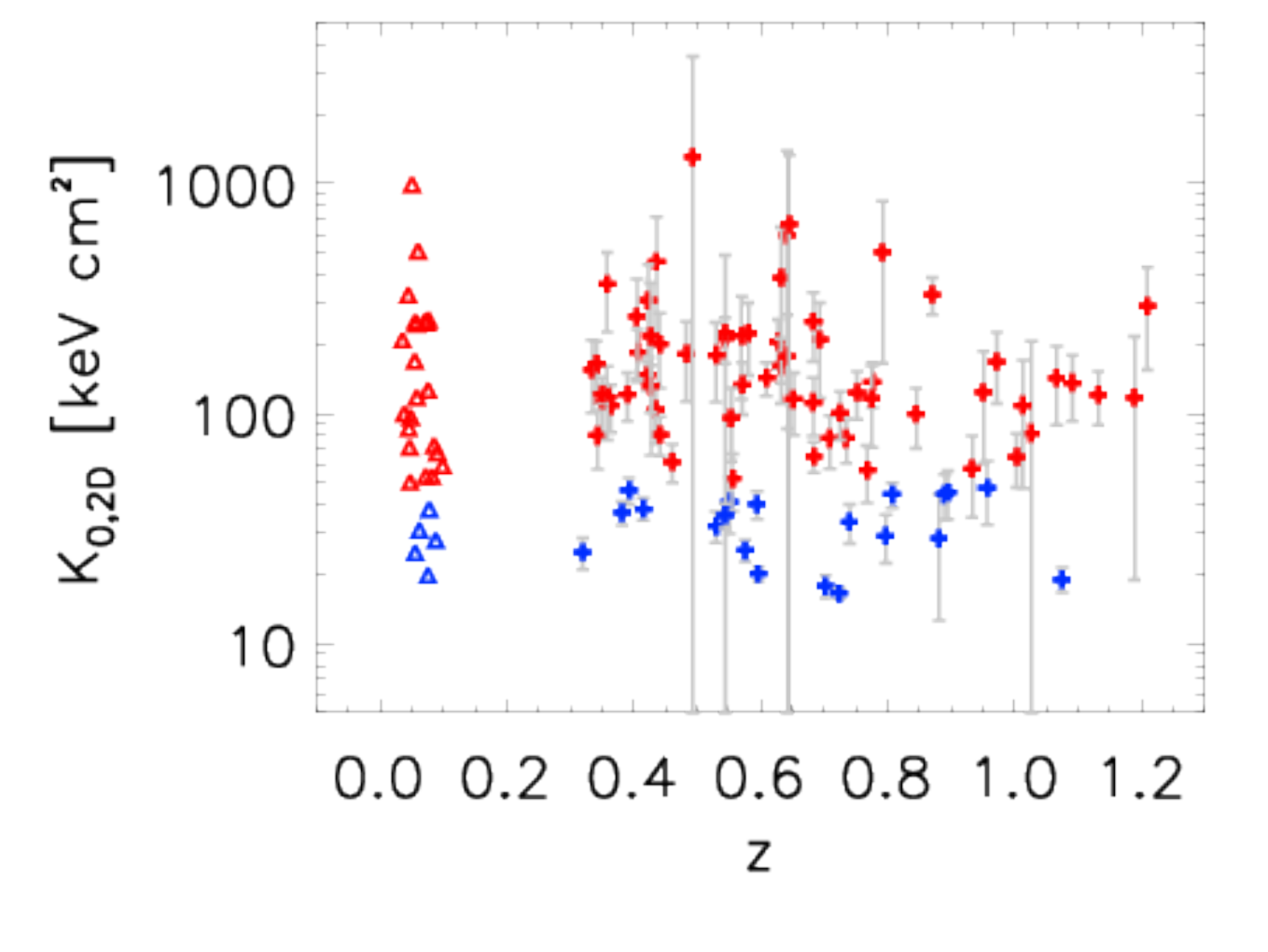}\\
\caption{Similar to Figure \ref{fig:dmdt_evolution}, but with the central entropy calculated using a 2-dimensional aperture. Open triangles show low-redshift clusters from the CCCP, while filled crosses represent data from this work (SPT-XVP). This figure demonstrates that, regardless of the method used to estimate the central entropy, there appears to be no evolution over the range $0<z<1.2$.
}
\label{fig:compare_k0evol}
\end{figure}

The results presented thus far rely on our ability to estimate the central temperature based on assumptions about the dark matter halo and hydrostatic equilibrium (see \S2.3). While we established in \S2.5 that these estimates are reliable, it is worthwhile to investigate whether any of the observed evolutionary trends are due to this approach.  In Figure \ref{fig:compare_k0}, we compare the central entropy calculated using a 2-dimensional, spectroscopically-measured temperature (see \S2.5) to our 3-dimensional model calculation. We find very good one-to-one agreement between these two quantities, with the scatter being uncorrelated with redshift. This suggests that the observed lack of evolution in K$_0$ is not a result of our modeling technique. We further demonstrate this in Figure \ref{fig:compare_k0evol}, showing the evolution of the central entropy based on the 2-dimensional central temperature. Comparing to low-redshift clusters from the CCCP, with K$_{0,2D}$ calculated in the same way, we confirm the lack of evolution in the central entropy from Figure \ref{fig:dmdt_evolution} over the range $0<z<1.2$. 

\subsubsection{Increased SZ Signal in Cool Cores}

In the central regions of cool core clusters, the increased density leads to a substantial increase in the inner pressure profile \citep{planck12}. This should, in turn, lead to an increase in the SZ detection significance, biasing SZ-selected samples towards detecting clusters with dense cores. However, given the small relative volume and mass of cool cores to the rest of the cluster, we expect this bias to be small. 
This bias was first quantified via detailed numerical simulations by \cite{motl05}. 
These authors found that integrated SZ quantities were relatively unbiased.  In simulations that allowed unrestricted radiative cooling, the logarithmic slope, $\alpha$, of the $M_{500}-y_{SZ}$ relation increased by $\sim$7.5\% over their non-cooling counterparts.
It is well understood that galaxy clusters which are simulated without feedback or star formation become too centrally concentrated \citep[the ``over-cooling problem'';][]{balogh01}, so this represents the upper limit of the SZ bias due to the presence of ICM cooling. When \cite{motl05} include star formation and stellar feedback -- which yields the most realistic-looking cool core clusters -- the difference in $\alpha$ between cooing and non-cooling clusters is reduced to $\sim$1\%. 
The relatively small bias of SZ integrated quantities was confirmed by \cite{pipino10}, who explicitly simulated the bias for a SPT-like survey and found that, at masses above $\sim2\times10^{14}$ M$_{\odot}$, the observed fraction of non-cool cores with the SPT should be nearly identical to the true fraction.
In an upcoming publication (Lin \etal in prep), we further show that this small bias is nearly redshift independent. Thus, while there is a small bias in the SZ signal due to the presence of a low-entropy core, we do not expect this bias to seriously alter our results, due both to its small magnitude and weak redshift dependence.

\subsubsection{X-ray Centroid Determination}
In \S2.2 we describe our method of determining the cluster center, which is based on the X-ray emission in an annulus of 250--500~kpc. This method, which reduces scatter in X-ray scaling relations by finding the large-scale center rather than what may be the displaced core, will yield less centrally-concentrated density profiles than if we chose the X-ray peak as the cluster center. Regardless of how the center is defined, we can investigate how our choice can affect the resulting density profile.

\begin{figure}[htb]
\centering
\includegraphics[width=0.49\textwidth]{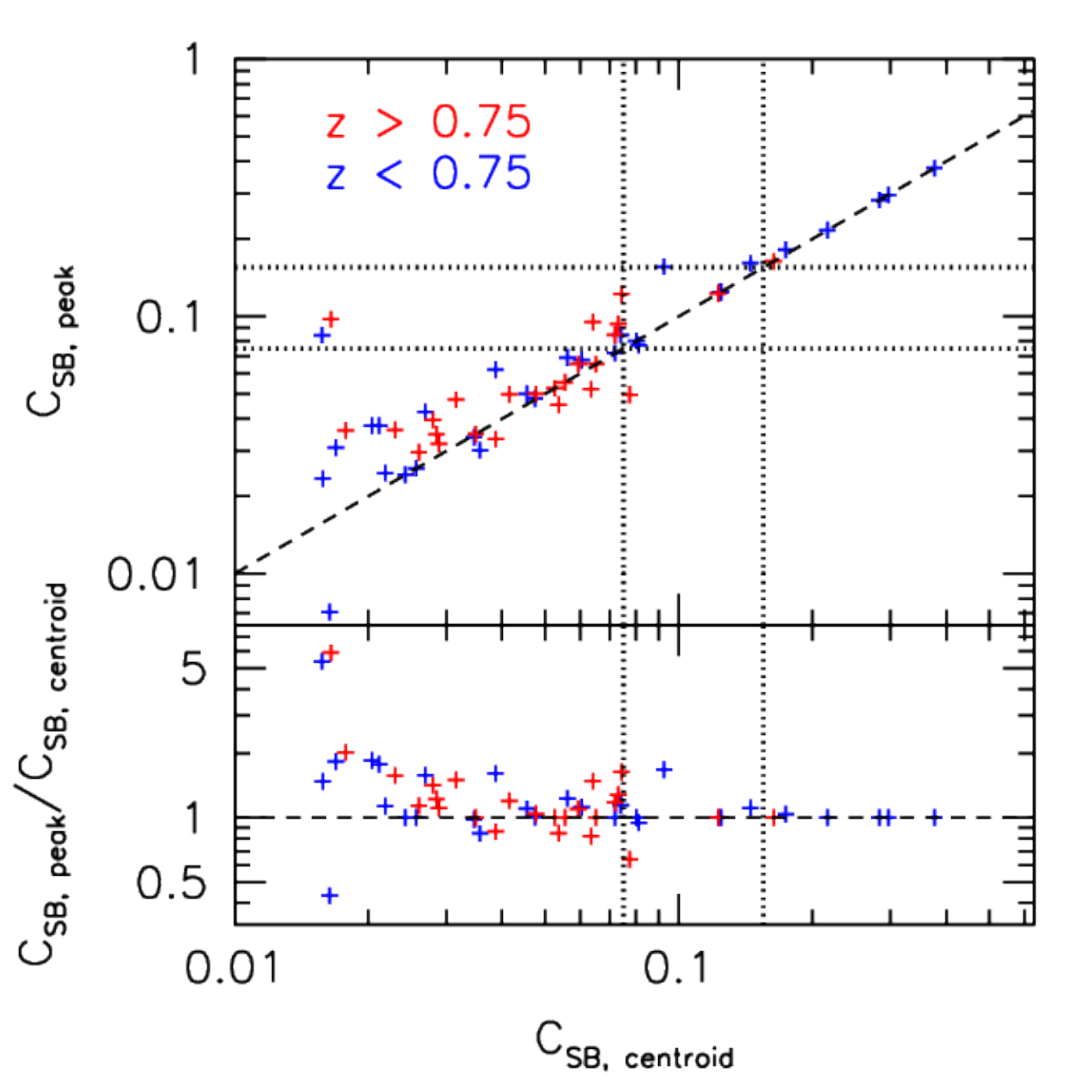}
\caption{Comparison of X-ray surface brightness concentration (c$_{SB}$) measured around the X-ray peak and the large-scale centroid. Dotted lines represent the thresholds for weak cool cores (c$_{SB}>0.075$) and strong cool cores (c$_{SB}>0.155$ from \cite{santos08}). This figure demonstrates that, while c$_{SB}$ is biased high when measuring around the X-ray peak, this bias appears to have no redshift dependence and is likely not responsible for the observed evolution in cool core density profiles. The three outliers in this plot are SPT-CLJ0102-4915, SPT-CLJ0411-4819, and SPT-CLJ0307-6226 -- all of which have cores displaced from the centroid of the large-scale emission.}
\label{fig:csb_compare}
\end{figure}

In Figure \ref{fig:csb_compare}, we compare the measured surface brightness concentration \citep[c$_{SB}$; ][]{santos08} based on our large-scale centroid and the X-ray peak. This figure confirms that, for relaxed strong cool cores, the X-ray peak and the large-scale centroid are nearly equivalent. By switching to c$_{SB,peak}$, the number of high-$z$ strong cool cores (c$_{SB} > 0.155$) would remain constant at 1, while the number of low-$z$ strong cool cores would increase from 5 to 7. Repeating this exercise for moderate cool cores ($0.075 < c_{SB} < 0.155$) we find increases of 5 (from 1 to 6) and 2 (from 3 to 5) for high and low redshift clusters, respectively. This large difference is primarily due to the arbitrary definition of moderate cool cores -- if we switched to a threshold of 0.07 rather than 0.075, the number of high-$z$ clusters which would be re-classified as moderate cool cores by re-defining the center would decrease from 5 to 2. Perhaps most importantly, the increase in c$_{SB}$ resulting from changing the center to the X-ray peak appears to have no dependence on redshift, as the lower panel of Figure \ref{fig:csb_compare} demonstrates. Thus, while the choice of center certainly affects the shape of the density profile, there is no evidence that this could result in low-$z$ clusters being measured to be more centrally concentrated than their high-$z$ counterparts.

\subsubsection{Radio-Loud and Star-forming BCGs}
% Radio loud BCGs washing out signal - Sayers et al. 2012 - radio-loud BCG contamination is small
There is a strong correlation in the local Universe between the presence of a cool core and radio emission from the BCG \citep{sun09b}, which may conspire to fill in the SZ signal for the strongest cool cores. Intuitively, this bias should tend to be strongest for nearby clusters (since the SZ signal is nearly redshift independent, but radio flux is not), which would lead to a bias \emph{against} detecting strong cool cores at low redshift with the SZ effect. This is exactly the opposite of what we observe -- the strongest cool cores in our sample are all at $z<0.75$, with a general lack of such systems at high redshift. We can further quantify this bias by appealing to Figure 3 from \cite{sayers13}, which presents a correlation between radio flux density and the SZ bias for Bolocam, which has a similar frequency coverage to SPT. This figure demonstrates that a 140 GHz flux density of $>$0.5 mJy is required to produce more than a 1\% change in the SZ S/N measurement. Assuming a typical radio luminosity for strong cool cores of 10$^{32}$ erg s$^{-1}$ Hz$^{-1}$ \citep{sun09b}, and a spectral index of $\alpha=-0.8$, we find a typical 140 GHz flux at $z=0.3$ of 0.9 mJy, corresponding to an SZ bias of $\sim3$\%. 
While there certainly may be systems with higher radio luminosity in this sample, we note that this bias becomes substantially weaker with increasing redshift. 
This conclusion qualitatively agrees with estimates from radio observations of clusters which found that correlated radio emission is negligible relative to the SZ signal at 150 GHz for typical clusters in the SPT mass and redshift range \citep{lin09,sehgal10}
Thus, we conclude that radio-loud BCGs should not substantially bias our sample against cool core clusters, and certainly can not drive the observed growth of cool cores that we observe.

The most star-forming BCG in this sample is in the Phoenix cluster \cite[SPT-CLJ2344-4243;][]{mcdonald12c,mcdonald13}, with a star formation rate of $\sim$800 M$_{\odot}$ yr$^{-1}$. In \cite{mcdonald12c} we demonstrated that, at 1.5mm and 2.0mm, the flux of this source would be $\sim$0.5 mJy and $\sim$0.1 mJy, respectively. This is significantly lower than the detection limit of the SPT ($\sim$20 mJy), suggesting that star formation has a negligible effect on the SZ signal. Since the other 82 clusters in this sample have \emph{significantly} lower star formation rates, we conclude that star-forming BCGs are not biasing our selection for or against cool cores.

\subsubsection{X-ray Underluminous Clusters}
If there is a population of galaxy clusters at high-z which meet our mass threshold (M$_{500}>2\times10^{14}$ M$_{\odot}$) but are gas-poor ($f_{gas} \ll 0.125$) and, as a result, go undetected in the SPT survey, then our estimate of the cool core fraction (e.g., Figure \ref{fig:ccfrac}) would be biased high. Such ``X-ray underluminous'' clusters have been identified in large optical surveys, and may be the result of either delayed assembly of the ICM or strong interactions that strip a substantial fraction of the hot gas. However, these systems are, in general, lower mass than the clusters which we consider here. Indeed, \cite{koester07} show, using a sample of 13,823 optically-selected galaxy clusters, that there is a near 1-to-1 correspondence between optically-selected and X-ray selected clusters at the high mass end, with the fraction of X-ray underluminous clusters increasing with decreasing cluster richness. Thus, assuming we can extrapolate this to high redshift, we do not expect that our results are seriously biased by the presence of a significant population of massive, high-redshift, gas-poor galaxy clusters.

\section{Summary}

%We present the first results from a large \emph{Chandra X-ray Observatory} program to observe 83 massive, SZ-selected clusters from the first 2000 deg$^2$ of the South Pole Telescope survey. 
%
We present X-ray observations of 83 massive SZ-selected clusters from the 2500 deg$^2$ South Pole Telescope SZ (SPT-SZ) survey, which includes the first results of a large \emph{Chandra X-ray Observatory} program to observe the 80 most-significant clusters detected at $z > 0.4$ from the first 2000 deg$^2$ of the SPT-SZ survey.
This uniformly-selected sample provides a unique opportunity to study the evolution of the cooling intracluster medium in clusters from $z=0.3$ to $z=1.2$. 

We find no evolution in the cooling properties of the intracluster medium over this large redshift range, with the average entropy and cooling time profiles remaining roughly constant in the inner $\sim$100 kpc despite the outer profile ($r>200$ kpc) following self-similar evolution. The distribution of the central entropy (K$_0$), central cooling time ($t_{cool,0}$), and mass deposition rate ($d$M$/dt$) in cool core clusters remains unchanged from $z=0$ to $z=1.2$. Further, the fraction of clusters experiencing strong cooling ($\sim$30\%) has not changed significantly over the 8~Gyr sampled here. The fact that the cooling properties of galaxy clusters are not evolving suggests that feedback is balancing cooling on very long ($\sim$8 Gyr) timescales.

We observe a strong evolution in the central density of galaxy clusters over this same timescale, with the average $\rho_{g,0}$/$\rho_{crit}$ increasing by a factor of $\sim$10  in this same redshift interval. We find a general lack of centrally concentrated cool cores at $z>0.75$, consistent with earlier reports of a lack of cool cores at high redshift from X-ray surveys. We show that this steady growth of cool cores from $z>1$ to $z=0$ is consistent with a cooling flow of $\sim$150 M$_{\odot}$ yr$^{-1}$ which is unable to reach entropies below 10 keV cm$^2$, leading to an accumulation of cool gas in the central $\sim$100~kpc. In order to build cool cores of the observed masses at $z\sim0$, we estimate that cooling flows would need to begin at $0.8<z<2.0$ in most massive galaxy clusters.
This work represents the first observations of galaxy clusters that span a broad enough redshift range and are sufficiently well-selected to track the growth of cool cores from their formation. These measurements give further evidence that stable, long-standing feedback is required to both halt cooling of the ICM to low temperatures and grow cool, dense cores.

This dataset, which contains dozens of new clusters, both cooling and non-cooling, at $0.3<z<1.2$ will prove invaluable for understanding the complex interplay between cooling and feedback in galaxy cluster cores, the formation and evolution of galaxy clusters, and the growth of massive central galaxies in cluster cores.

\section*{Acknowledgements} 

M. M. acknowledges support by NASA through a Hubble Fellowship grant HST-HF51308.01-A awarded by the Space Telescope Science Institute, which is operated by the Association of Universities for Research in Astronomy, Inc., for NASA, under contract NAS 5-26555. 
The South Pole Telescope program is supported by the National Science Foundation through grant ANT-0638937. Partial support is also provided by the NSF Physics Frontier Center grant PHY-0114422 to the Kavli Institute of Cosmological Physics at the University of Chicago, the Kavli Foundation, and the Gordon and Betty Moore Foundation. Support for X-ray analysis was provided by NASA through Chandra Award Numbers 12800071, 12800088, and 13800883 issued by the Chandra X-ray Observatory Center, which is operated by the Smithsonian Astrophysical Observatory for and on behalf of NASA. Galaxy cluster research at Harvard is supported by NSF grant AST-1009012 and at SAO in part by NSF grants AST-1009649 and MRI-0723073. Argonne National Laboratory's work was supported under U.S. Department of Energy contract DE-AC02-06CH11357.

%\bibliographystyle{apj}
%\bibliography{ref}

\begin{thebibliography}{}
\expandafter\ifx\csname natexlab\endcsname\relax\def\natexlab#1{#1}\fi

\bibitem[{{Allen}(2000)}]{allen00}
{Allen}, S.~W. 2000, \mnras, 315, 269

\bibitem[{{Andersson} {et~al.}(2011){Andersson}, {Benson}, {Ade}, {Aird},
  {Armstrong}, {Bautz}, {Bleem}, {Brodwin}, {Carlstrom}, {Chang}, {Crawford},
  {Crites}, {de Haan}, {Desai}, {Dobbs}, {Dudley}, {Foley}, {Forman},
  {Garmire}, {George}, {Gladders}, {Halverson}, {High}, {Holder}, {Holzapfel},
  {Hrubes}, {Jones}, {Joy}, {Keisler}, {Knox}, {Lee}, {Leitch}, {Lueker},
  {Marrone}, {McMahon}, {Mehl}, {Meyer}, {Mohr}, {Montroy}, {Murray}, {Padin},
  {Plagge}, {Pryke}, {Reichardt}, {Rest}, {Ruel}, {Ruhl}, {Schaffer}, {Shaw},
  {Shirokoff}, {Song}, {Spieler}, {Stalder}, {Staniszewski}, {Stark}, {Stubbs},
  {Vanderlinde}, {Vieira}, {Vikhlinin}, {Williamson}, {Yang}, {Zahn}, \&
  {Zenteno}}]{andersson11}
{Andersson}, K., {Benson}, B.~A., {Ade}, P.~A.~R., {et~al.} 2011, \apj, 738, 48

\bibitem[{{Arnaud}(1996)}]{arnaud96}
{Arnaud}, K.~A. 1996, in Astronomical Society of the Pacific Conference Series,
  Vol. 101, Astronomical Data Analysis Software and Systems V, ed.
  {G.~H.~Jacoby \& J.~Barnes}, 17--+

\bibitem[{{Balogh} {et~al.}(2001){Balogh}, {Pearce}, {Bower}, \&
  {Kay}}]{balogh01}
{Balogh}, M.~L., {Pearce}, F.~R., {Bower}, R.~G., \& {Kay}, S.~T. 2001, \mnras,
  326, 1228

\bibitem[{{Benson} {et~al.}(2013){Benson}, {de Haan}, {Dudley}, {Reichardt},
  {Aird}, {Andersson}, {Armstrong}, {Ashby}, {Bautz}, {Bayliss}, {Bazin},
  {Bleem}, {Brodwin}, {Carlstrom}, {Chang}, {Cho}, {Clocchiatti}, {Crawford},
  {Crites}, {Desai}, {Dobbs}, {Foley}, {Forman}, {George}, {Gladders},
  {Gonzalez}, {Halverson}, {Harrington}, {High}, {Holder}, {Holzapfel},
  {Hoover}, {Hrubes}, {Jones}, {Joy}, {Keisler}, {Knox}, {Lee}, {Leitch},
  {Liu}, {Lueker}, {Luong-Van}, {Mantz}, {Marrone}, {McDonald}, {McMahon},
  {Mehl}, {Meyer}, {Mocanu}, {Mohr}, {Montroy}, {Murray}, {Natoli}, {Padin},
  {Plagge}, {Pryke}, {Rest}, {Ruel}, {Ruhl}, {Saliwanchik}, {Saro}, {Sayre},
  {Schaffer}, {Shaw}, {Shirokoff}, {Song}, {Spieler}, {Stalder},
  {Staniszewski}, {Stark}, {Story}, {Stubbs}, {Suhada}, {van Engelen},
  {Vanderlinde}, {Vieira}, {Vikhlinin}, {Williamson}, {Zahn}, \&
  {Zenteno}}]{benson13}
{Benson}, B.~A., {de Haan}, T., {Dudley}, J.~P., {et~al.} 2013, \apj, 763, 147

\bibitem[{{Bregman} {et~al.}(2006){Bregman}, {Fabian}, {Miller}, \&
  {Irwin}}]{bregman06}
{Bregman}, J.~N., {Fabian}, A.~C., {Miller}, E.~D., \& {Irwin}, J.~A. 2006,
  \apj, 642, 746

\bibitem[{{Bregman} {et~al.}(2001){Bregman}, {Miller}, \& {Irwin}}]{bregman01}
{Bregman}, J.~N., {Miller}, E.~D., \& {Irwin}, J.~A. 2001, \apjl, 553, L125

\bibitem[{{Brodwin} {et~al.}(2010){Brodwin}, {Ruel}, {Ade}, {Aird},
  {Andersson}, {Ashby}, {Bautz}, {Bazin}, {Benson}, {Bleem}, {Carlstrom},
  {Chang}, {Crawford}, {Crites}, {de Haan}, {Desai}, {Dobbs}, {Dudley},
  {Fazio}, {Foley}, {Forman}, {Garmire}, {George}, {Gladders}, {Gonzalez},
  {Halverson}, {High}, {Holder}, {Holzapfel}, {Hrubes}, {Jones}, {Joy},
  {Keisler}, {Knox}, {Lee}, {Leitch}, {Lueker}, {Marrone}, {McMahon}, {Mehl},
  {Meyer}, {Mohr}, {Montroy}, {Murray}, {Padin}, {Plagge}, {Pryke},
  {Reichardt}, {Rest}, {Ruhl}, {Schaffer}, {Shaw}, {Shirokoff}, {Song},
  {Spieler}, {Stalder}, {Stanford}, {Staniszewski}, {Stark}, {Stubbs},
  {Vanderlinde}, {Vieira}, {Vikhlinin}, {Williamson}, {Yang}, {Zahn}, \&
  {Zenteno}}]{brodwin10}
{Brodwin}, M., {Ruel}, J., {Ade}, P.~A.~R., {et~al.} 2010, \apj, 721, 90

\bibitem[{{Carlstrom} {et~al.}(2011){Carlstrom}, {Ade}, {Aird}, {Benson},
  {Bleem}, {Busetti}, {Chang}, {Chauvin}, {Cho}, {Crawford}, {Crites}, {Dobbs},
  {Halverson}, {Heimsath}, {Holzapfel}, {Hrubes}, {Joy}, {Keisler}, {Lanting},
  {Lee}, {Leitch}, {Leong}, {Lu}, {Lueker}, {Luong-van}, {McMahon}, {Mehl},
  {Meyer}, {Mohr}, {Montroy}, {Padin}, {Plagge}, {Pryke}, {Ruhl}, {Schaffer},
  {Schwan}, {Shirokoff}, {Spieler}, {Staniszewski}, {Stark}, {Tucker},
  {Vanderlinde}, {Vieira}, \& {Williamson}}]{carlstrom11}
{Carlstrom}, J.~E., {Ade}, P.~A.~R., {Aird}, K.~A., {et~al.} 2011, \pasp, 123,
  568

\bibitem[{{Cavagnolo} {et~al.}(2009){Cavagnolo}, {Donahue}, {Voit}, \&
  {Sun}}]{cavagnolo09}
{Cavagnolo}, K.~W., {Donahue}, M., {Voit}, G.~M., \& {Sun}, M. 2009, \apjs,
  182, 12

\bibitem[{{Churazov} {et~al.}(2001){Churazov}, {Br{\"u}ggen}, {Kaiser},
  {B{\"o}hringer}, \& {Forman}}]{churazov01}
{Churazov}, E., {Br{\"u}ggen}, M., {Kaiser}, C.~R., {B{\"o}hringer}, H., \&
  {Forman}, W. 2001, \apj, 554, 261

\bibitem[{{Crawford} {et~al.}(1999){Crawford}, {Allen}, {Ebeling}, {Edge}, \&
  {Fabian}}]{crawford99}
{Crawford}, C.~S., {Allen}, S.~W., {Ebeling}, H., {Edge}, A.~C., \& {Fabian},
  A.~C. 1999, \mnras, 306, 857

\bibitem[{{Donahue} {et~al.}(2011){Donahue}, {de Messi{\`e}res}, {O'Connell},
  {Voit}, {Hoffer}, {McNamara}, \& {Nulsen}}]{donahue11}
{Donahue}, M., {de Messi{\`e}res}, G.~E., {O'Connell}, R.~W., {et~al.} 2011,
  \apj, 732, 40

\bibitem[{{Donahue} {et~al.}(1992){Donahue}, {Stocke}, \& {Gioia}}]{donahue92}
{Donahue}, M., {Stocke}, J.~T., \& {Gioia}, I.~M. 1992, \apj, 385, 49

\bibitem[{{Donahue} {et~al.}(2010){Donahue}, {Bruch}, {Wang}, {Voit}, {Hicks},
  {Haarsma}, {Croston}, {Pratt}, {Pierini}, {O'Connell}, \&
  {B{\"o}hringer}}]{donahue10}
{Donahue}, M., {Bruch}, S., {Wang}, E., {et~al.} 2010, \apj, 715, 881

\bibitem[{{Duffy} {et~al.}(2008){Duffy}, {Schaye}, {Kay}, \& {Dalla
  Vecchia}}]{duffy08}
{Duffy}, A.~R., {Schaye}, J., {Kay}, S.~T., \& {Dalla Vecchia}, C. 2008,
  \mnras, 390, L64

\bibitem[{{Edge}(2001)}]{edge01}
{Edge}, A.~C. 2001, \mnras, 328, 762

\bibitem[{{Edge} \& {Frayer}(2003)}]{edge03}
{Edge}, A.~C., \& {Frayer}, D.~T. 2003, \apjl, 594, L13

\bibitem[{{Edge} {et~al.}(2010){Edge}, {Oonk}, {Mittal}, {Allen}, {Baum},
  {B{\"o}hringer}, {Bregman}, {Bremer}, {Combes}, {Crawford}, {Donahue},
  {Egami}, {Fabian}, {Ferland}, {Hamer}, {Hatch}, {Jaffe}, {Johnstone},
  {McNamara}, {O'Dea}, {Popesso}, {Quillen}, {Salom{\'e}}, {Sarazin}, {Voit},
  {Wilman}, \& {Wise}}]{edge10}
{Edge}, A.~C., {Oonk}, J.~B.~R., {Mittal}, R., {et~al.} 2010, \aap, 518, L46

\bibitem[{{Edwards} {et~al.}(2007){Edwards}, {Hudson}, {Balogh}, \&
  {Smith}}]{edwards07}
{Edwards}, L.~O.~V., {Hudson}, M.~J., {Balogh}, M.~L., \& {Smith}, R.~J. 2007,
  \mnras, 379, 100

\bibitem[{{Fabian}(1994)}]{fabian94}
{Fabian}, A.~C. 1994, \araa, 32, 277

\bibitem[{{Fabian}(2012)}]{fabian12}
---. 2012, ArXiv e-prints, arXiv:1204.4114

\bibitem[{{Foley} {et~al.}(2011){Foley}, {Andersson}, {Bazin}, {de Haan},
  {Ruel}, {Ade}, {Aird}, {Armstrong}, {Ashby}, {Bautz}, {Benson}, {Bleem},
  {Bonamente}, {Brodwin}, {Carlstrom}, {Chang}, {Clocchiatti}, {Crawford},
  {Crites}, {Desai}, {Dobbs}, {Dudley}, {Fazio}, {Forman}, {Garmire}, {George},
  {Gladders}, {Gonzalez}, {Halverson}, {High}, {Holder}, {Holzapfel}, {Hoover},
  {Hrubes}, {Jones}, {Joy}, {Keisler}, {Knox}, {Lee}, {Leitch}, {Lueker},
  {Luong-Van}, {Marrone}, {McMahon}, {Mehl}, {Meyer}, {Mohr}, {Montroy},
  {Murray}, {Padin}, {Plagge}, {Pryke}, {Reichardt}, {Rest}, {Ruhl},
  {Saliwanchik}, {Saro}, {Schaffer}, {Shaw}, {Shirokoff}, {Song}, {Spieler},
  {Stalder}, {Stanford}, {Staniszewski}, {Stark}, {Story}, {Stubbs},
  {Vanderlinde}, {Vieira}, {Vikhlinin}, {Williamson}, \& {Zenteno}}]{foley11}
{Foley}, R.~J., {Andersson}, K., {Bazin}, G., {et~al.} 2011, \apj, 731, 86

\bibitem[{{G{\'o}mez} {et~al.}(2002){G{\'o}mez}, {Loken}, {Roettiger}, \&
  {Burns}}]{gomez02}
{G{\'o}mez}, P.~L., {Loken}, C., {Roettiger}, K., \& {Burns}, J.~O. 2002, \apj,
  569, 122

\bibitem[{{Hatch} {et~al.}(2007){Hatch}, {Crawford}, \& {Fabian}}]{hatch07}
{Hatch}, N.~A., {Crawford}, C.~S., \& {Fabian}, A.~C. 2007, \mnras, 380, 33

\bibitem[{{Heckman} {et~al.}(1989){Heckman}, {Baum}, {van Breugel}, \&
  {McCarthy}}]{heckman89}
{Heckman}, T.~M., {Baum}, S.~A., {van Breugel}, W.~J.~M., \& {McCarthy}, P.
  1989, \apj, 338, 48

\bibitem[{{Hicks} {et~al.}(2010){Hicks}, {Mushotzky}, \& {Donahue}}]{hicks10}
{Hicks}, A.~K., {Mushotzky}, R., \& {Donahue}, M. 2010, \apj, 719, 1844

\bibitem[{{Hlavacek-Larrondo} {et~al.}(2012){Hlavacek-Larrondo}, {Fabian},
  {Edge}, {Ebeling}, {Sanders}, {Hogan}, \& {Taylor}}]{hlavacek-larrondo12}
{Hlavacek-Larrondo}, J., {Fabian}, A.~C., {Edge}, A.~C., {et~al.} 2012, \mnras,
  421, 1360

\bibitem[{{Hu} {et~al.}(1985){Hu}, {Cowie}, \& {Wang}}]{Hu85}
{Hu}, E.~M., {Cowie}, L.~L., \& {Wang}, Z. 1985, \apjs, 59, 447

\bibitem[{{Hudson} {et~al.}(2010){Hudson}, {Mittal}, {Reiprich}, {Nulsen},
  {Andernach}, \& {Sarazin}}]{hudson10}
{Hudson}, D.~S., {Mittal}, R., {Reiprich}, T.~H., {et~al.} 2010, \aap, 513, A37

\bibitem[{{Jaffe} {et~al.}(2005){Jaffe}, {Bremer}, \& {Baker}}]{jaffe05}
{Jaffe}, W., {Bremer}, M.~N., \& {Baker}, K. 2005, \mnras, 360, 748

\bibitem[{{Johnstone} {et~al.}(1987){Johnstone}, {Fabian}, \&
  {Nulsen}}]{johnstone87}
{Johnstone}, R.~M., {Fabian}, A.~C., \& {Nulsen}, P.~E.~J. 1987, \mnras, 224,
  75

\bibitem[{{Kalberla} {et~al.}(2005){Kalberla}, {Burton}, {Hartmann}, {Arnal},
  {Bajaja}, {Morras}, \& {P{\"o}ppel}}]{kalberla05}
{Kalberla}, P.~M.~W., {Burton}, W.~B., {Hartmann}, D., {et~al.} 2005, \aap,
  440, 775

\bibitem[{{Koester} {et~al.}(2007){Koester}, {McKay}, {Annis}, {Wechsler},
  {Evrard}, {Bleem}, {Becker}, {Johnston}, {Sheldon}, {Nichol}, {Miller},
  {Scranton}, {Bahcall}, {Barentine}, {Brewington}, {Brinkmann}, {Harvanek},
  {Kleinman}, {Krzesinski}, {Long}, {Nitta}, {Schneider}, {Sneddin}, {Voges},
  \& {York}}]{koester07}
{Koester}, B.~P., {McKay}, T.~A., {Annis}, J., {et~al.} 2007, \apj, 660, 239

\bibitem[{{Lim} {et~al.}(2008){Lim}, {Ao}, \& {Dinh-V-Trung}}]{lim08}
{Lim}, J., {Ao}, Y., \& {Dinh-V-Trung}. 2008, \apj, 672, 252

\bibitem[{{Lim} {et~al.}(2012){Lim}, {Ohyama}, {Chi-Hung}, {Dinh-V-Trung}, \&
  {Shiang-Yu}}]{lim12}
{Lim}, J., {Ohyama}, Y., {Chi-Hung}, Y., {Dinh-V-Trung}, \& {Shiang-Yu}, W.
  2012, \apj, 744, 112

\bibitem[{{Lin} {et~al.}(2009){Lin}, {Partridge}, {Pober}, {Bouchefry},
  {Burke}, {Klein}, {Coish}, \& {Huffenberger}}]{lin09}
{Lin}, Y.-T., {Partridge}, B., {Pober}, J.~C., {et~al.} 2009, \apj, 694, 992

\bibitem[{{Marriage} {et~al.}(2011){Marriage}, {Acquaviva}, {Ade}, {Aguirre},
  {Amiri}, {Appel}, {Barrientos}, {Battistelli}, {Bond}, {Brown}, {Burger},
  {Chervenak}, {Das}, {Devlin}, {Dicker}, {Bertrand Doriese}, {Dunkley},
  {D{\"u}nner}, {Essinger-Hileman}, {Fisher}, {Fowler}, {Hajian}, {Halpern},
  {Hasselfield}, {Hern{\'a}ndez-Monteagudo}, {Hilton}, {Hilton}, {Hincks},
  {Hlozek}, {Huffenberger}, {Handel Hughes}, {Hughes}, {Infante}, {Irwin},
  {Baptiste Juin}, {Kaul}, {Klein}, {Kosowsky}, {Lau}, {Limon}, {Lin},
  {Lupton}, {Marsden}, {Martocci}, {Mauskopf}, {Menanteau}, {Moodley},
  {Moseley}, {Netterfield}, {Niemack}, {Nolta}, {Page}, {Parker}, {Partridge},
  {Quintana}, {Reese}, {Reid}, {Sehgal}, {Sherwin}, {Sievers}, {Spergel},
  {Staggs}, {Swetz}, {Switzer}, {Thornton}, {Trac}, {Tucker}, {Warne},
  {Wilson}, {Wollack}, \& {Zhao}}]{marriage11}
{Marriage}, T.~A., {Acquaviva}, V., {Ade}, P.~A.~R., {et~al.} 2011, \apj, 737,
  61

\bibitem[{{Mathews}(2009)}]{mathews09}
{Mathews}, W.~G. 2009, \apjl, 695, L49

\bibitem[{{McDonald}(2011)}]{mcdonald11c}
{McDonald}, M. 2011, \apjl, 742, L35

\bibitem[{{McDonald} {et~al.}(2013){McDonald}, {Benson}, {Veilleux}, {Bautz},
  \& {Reichardt}}]{mcdonald13}
{McDonald}, M., {Benson}, B., {Veilleux}, S., {Bautz}, M.~W., \& {Reichardt},
  C.~L. 2013, \apjl, 765, L37

\bibitem[{{McDonald} {et~al.}(2011{\natexlab{a}}){McDonald}, {Veilleux}, \&
  {Mushotzky}}]{mcdonald11a}
{McDonald}, M., {Veilleux}, S., \& {Mushotzky}, R. 2011{\natexlab{a}}, \apj,
  731, 33

\bibitem[{{McDonald} {et~al.}(2010){McDonald}, {Veilleux}, {Rupke}, \&
  {Mushotzky}}]{mcdonald10}
{McDonald}, M., {Veilleux}, S., {Rupke}, D.~S.~N., \& {Mushotzky}, R. 2010,
  \apj, 721, 1262

\bibitem[{{McDonald} {et~al.}(2011{\natexlab{b}}){McDonald}, {Veilleux},
  {Rupke}, {Mushotzky}, \& {Reynolds}}]{mcdonald11b}
{McDonald}, M., {Veilleux}, S., {Rupke}, D.~S.~N., {Mushotzky}, R., \&
  {Reynolds}, C. 2011{\natexlab{b}}, \apj, 734, 95

\bibitem[{{McDonald} {et~al.}(2012{\natexlab{a}}){McDonald}, {Wei}, \&
  {Veilleux}}]{mcdonald12b}
{McDonald}, M., {Wei}, L.~H., \& {Veilleux}, S. 2012{\natexlab{a}}, \apjl, 755,
  L24

\bibitem[{{McDonald} {et~al.}(2012{\natexlab{b}}){McDonald}, {Bayliss},
  {Benson}, {Foley}, {Ruel}, {Sullivan}, {Veilleux}, {Aird}, {Ashby}, {Bautz},
  {Bazin}, {Bleem}, {Brodwin}, {Carlstrom}, {Chang}, {Cho}, {Clocchiatti},
  {Crawford}, {Crites}, {de Haan}, {Desai}, {Dobbs}, {Dudley}, {Egami},
  {Forman}, {Garmire}, {George}, {Gladders}, {Gonzalez}, {Halverson},
  {Harrington}, {High}, {Holder}, {Holzapfel}, {Hoover}, {Hrubes}, {Jones},
  {Joy}, {Keisler}, {Knox}, {Lee}, {Leitch}, {Liu}, {Lueker}, {Luong-van},
  {Mantz}, {Marrone}, {McMahon}, {Mehl}, {Meyer}, {Miller}, {Mocanu}, {Mohr},
  {Montroy}, {Murray}, {Natoli}, {Padin}, {Plagge}, {Pryke}, {Rawle},
  {Reichardt}, {Rest}, {Rex}, {Ruhl}, {Saliwanchik}, {Saro}, {Sayre},
  {Schaffer}, {Shaw}, {Shirokoff}, {Simcoe}, {Song}, {Spieler}, {Stalder},
  {Staniszewski}, {Stark}, {Story}, {Stubbs}, {{\v S}uhada}, {van Engelen},
  {Vanderlinde}, {Vieira}, {Vikhlinin}, {Williamson}, {Zahn}, \&
  {Zenteno}}]{mcdonald12c}
{McDonald}, M., {Bayliss}, M., {Benson}, B.~A., {et~al.} 2012{\natexlab{b}},
  \nat, 488, 349

\bibitem[{{McNamara} \& {Nulsen}(2007)}]{mcnamara07}
{McNamara}, B.~R., \& {Nulsen}, P.~E.~J. 2007, \araa, 45, 117

\bibitem[{{McNamara} \& {Nulsen}(2012)}]{mcnamara12}
---. 2012, New Journal of Physics, 14, 055023

\bibitem[{{McNamara} \& {O'Connell}(1989)}]{mcnamara89}
{McNamara}, B.~R., \& {O'Connell}, R.~W. 1989, \aj, 98, 2018

\bibitem[{{Motl} {et~al.}(2005){Motl}, {Hallman}, {Burns}, \&
  {Norman}}]{motl05}
{Motl}, P.~M., {Hallman}, E.~J., {Burns}, J.~O., \& {Norman}, M.~L. 2005,
  \apjl, 623, L63

\bibitem[{{Nagai} {et~al.}(2007){Nagai}, {Kravtsov}, \& {Vikhlinin}}]{nagai07}
{Nagai}, D., {Kravtsov}, A.~V., \& {Vikhlinin}, A. 2007, \apj, 668, 1

\bibitem[{{Navarro} {et~al.}(1997){Navarro}, {Frenk}, \& {White}}]{nfw}
{Navarro}, J.~F., {Frenk}, C.~S., \& {White}, S.~D.~M. 1997, \apj, 490, 493

\bibitem[{{O'Dea} {et~al.}(2008){O'Dea}, {Baum}, {Privon}, {Noel-Storr},
  {Quillen}, {Zufelt}, {Park}, {Edge}, {Russell}, {Fabian}, {Donahue},
  {Sarazin}, {McNamara}, {Bregman}, \& {Egami}}]{odea08}
{O'Dea}, C.~P., {Baum}, S.~A., {Privon}, G., {et~al.} 2008, \apj, 681, 1035

\bibitem[{{Oegerle} {et~al.}(2001){Oegerle}, {Cowie}, {Davidsen}, {Hu},
  {Hutchings}, {Murphy}, {Sembach}, \& {Woodgate}}]{oegerle01}
{Oegerle}, W.~R., {Cowie}, L., {Davidsen}, A., {et~al.} 2001, \apj, 560, 187

\bibitem[{{Peres} {et~al.}(1998){Peres}, {Fabian}, {Edge}, {Allen},
  {Johnstone}, \& {White}}]{peres98}
{Peres}, C.~B., {Fabian}, A.~C., {Edge}, A.~C., {et~al.} 1998, \mnras, 298, 416

\bibitem[{{Peterson} \& {Fabian}(2006)}]{peterson06}
{Peterson}, J.~R., \& {Fabian}, A.~C. 2006, \physrep, 427, 1

\bibitem[{{Pfrommer} {et~al.}(2012){Pfrommer}, {Chang}, \&
  {Broderick}}]{pfrommer12}
{Pfrommer}, C., {Chang}, P., \& {Broderick}, A.~E. 2012, \apj, 752, 24

\bibitem[{{Pipino} \& {Pierpaoli}(2010)}]{pipino10}
{Pipino}, A., \& {Pierpaoli}, E. 2010, \mnras, 404, 1603

\bibitem[{{Planck Collaboration} {et~al.}(2011){Planck Collaboration},
  {Aghanim}, {Arnaud}, {Ashdown}, {Aumont}, {Baccigalupi}, {Balbi}, {Banday},
  {Barreiro}, {Bartelmann}, {Bartlett}, {Battaner}, {Benabed}, {Beno{\^i}t},
  {Bernard}, {Bersanelli}, {Bhatia}, {Bock}, {Bonaldi}, {Bond}, {Borrill},
  {Bouchet}, {Brown}, {Bucher}, {Burigana}, {Cabella}, {Cardoso}, {Catalano},
  {Cay{\'o}n}, {Challinor}, {Chamballu}, {Chary}, {Chiang}, {Chiang}, {Chon},
  {Christensen}, {Churazov}, {Clements}, {Colafrancesco}, {Colombi}, {Couchot},
  {Coulais}, {Crill}, {Cuttaia}, {da Silva}, {Dahle}, {Danese}, {de Bernardis},
  {de Gasperis}, {de Rosa}, {de Zotti}, {Delabrouille}, {Delouis},
  {D{\'e}sert}, {Diego}, {Dolag}, {Donzelli}, {Dor{\'e}}, {D{\"o}rl},
  {Douspis}, {Dupac}, {Efstathiou}, {En{\ss}lin}, {Finelli}, {Flores-Cacho},
  {Forni}, {Frailis}, {Franceschi}, {Fromenteau}, {Galeotta}, {Ganga},
  {G{\'e}nova-Santos}, {Giard}, {Giardino}, {Giraud-H{\'e}raud},
  {Gonz{\'a}lez-Nuevo}, {G{\'o}rski}, {Gratton}, {Gregorio}, {Gruppuso},
  {Harrison}, {Henrot-Versill{\'e}}, {Hern{\'a}ndez-Monteagudo}, {Herranz},
  {Hildebrandt}, {Hivon}, {Hobson}, {Holmes}, {Hovest}, {Hoyland},
  {Huffenberger}, {Jaffe}, {Jones}, {Juvela}, {Keih{\"a}nen}, {Keskitalo},
  {Kisner}, {Kneissl}, {Knox}, {Kurki-Suonio}, {Lagache}, {Lamarre}, {Lasenby},
  {Laureijs}, {Lawrence}, {Leach}, {Leonardi}, {Linden-V{\o}rnle},
  {L{\'o}pez-Caniego}, {Lubin}, {Mac{\'{\i}}as-P{\'e}rez}, {MacTavish},
  {Maffei}, {Maino}, {Mandolesi}, {Mann}, {Maris}, {Marleau},
  {Mart{\'{\i}}nez-Gonz{\'a}lez}, {Masi}, {Matarrese}, {Matthai}, {Mazzotta},
  {Melchiorri}, {Melin}, {Mendes}, {Mennella}, {Mitra},
  {Miville-Desch{\^e}nes}, {Moneti}, {Montier}, {Morgante}, {Mortlock},
  {Munshi}, {Murphy}, {Naselsky}, {Natoli}, {Netterfield},
  {N{\o}rgaard-Nielsen}, {Noviello}, {Novikov}, {Novikov}, {Osborne}, {Pajot},
  {Pasian}, {Patanchon}, {Perdereau}, {Perotto}, {Perrotta}, {Piacentini},
  {Piat}, {Pierpaoli}, {Piffaretti}, {Plaszczynski}, {Pointecouteau},
  {Polenta}, {Ponthieu}, {Poutanen}, {Pratt}, {Pr{\'e}zeau}, {Prunet}, {Puget},
  {Rebolo}, {Reinecke}, {Renault}, {Ricciardi}, {Riller}, {Ristorcelli},
  {Rocha}, {Rosset}, {Rubi{\~n}o-Mart{\'{\i}}n}, {Rusholme}, {Sandri},
  {Santos}, {Schaefer}, {Scott}, {Seiffert}, {Smoot}, {Starck}, {Stivoli},
  {Stolyarov}, {Sunyaev}, {Sygnet}, {Tauber}, {Terenzi}, {Toffolatti},
  {Tomasi}, {Tristram}, {Tuovinen}, {Valenziano}, {Vibert}, {Vielva}, {Villa},
  {Vittorio}, {Wandelt}, {White}, {White}, {Yvon}, {Zacchei}, \&
  {Zonca}}]{planck11}
{Planck Collaboration}, {Aghanim}, N., {Arnaud}, M., {et~al.} 2011, \aap, 536,
  A10

\bibitem[{{Planck Collaboration} {et~al.}(2012){Planck Collaboration}, {Ade},
  {Aghanim}, {Arnaud}, {Ashdown}, {Atrio-Barandela}, {Aumont}, {Baccigalupi},
  {Balbi}, {Banday}, \& et~al.}]{planck12}
{Planck Collaboration}, {Ade}, P.~A.~R., {Aghanim}, N., {et~al.} 2012, ArXiv
  e-prints, arXiv:1207.4061

\bibitem[{{Pratt} {et~al.}(2010){Pratt}, {Arnaud}, {Piffaretti},
  {B{\"o}hringer}, {Ponman}, {Croston}, {Voit}, {Borgani}, \&
  {Bower}}]{pratt10}
{Pratt}, G.~W., {Arnaud}, M., {Piffaretti}, R., {et~al.} 2010, \aap, 511, A85

\bibitem[{{Predehl} {et~al.}(2007){Predehl}, {Andritschke}, {Bornemann},
  {Br{\"a}uninger}, {Briel}, {Brunner}, {Burkert}, {Dennerl}, {Eder},
  {Freyberg}, {Friedrich}, {F{\"u}rmetz}, {Hartmann}, {Hartner}, {Hasinger},
  {Herrmann}, {Holl}, {Huber}, {Kendziorra}, {Kink}, {Meidinger}, {M{\"u}ller},
  {Pavlinsky}, {Pfeffermann}, {Roh{\'e}}, {Santangelo}, {Schmitt}, {Schwope},
  {Steinmetz}, {Str{\"u}der}, {Sunyaev}, {Tiedemann}, {Vongehr}, {Wilms},
  {Erhard}, {Gutruf}, {Jugler}, {Kampf}, {Graue}, {Citterio}, {Valsecci},
  {Vernani}, \& {Zimmerman}}]{erosita}
{Predehl}, P., {Andritschke}, R., {Bornemann}, W., {et~al.} 2007, in Society of
  Photo-Optical Instrumentation Engineers (SPIE) Conference Series, Vol. 6686,
  Society of Photo-Optical Instrumentation Engineers (SPIE) Conference Series

\bibitem[{{Rafferty} {et~al.}(2008){Rafferty}, {McNamara}, \&
  {Nulsen}}]{rafferty08}
{Rafferty}, D.~A., {McNamara}, B.~R., \& {Nulsen}, P.~E.~J. 2008, \apj, 687,
  899

\bibitem[{{Reichardt} {et~al.}(2013){Reichardt}, {Stalder}, {Bleem}, {Montroy},
  {Aird}, {Andersson}, {Armstrong}, {Ashby}, {Bautz}, {Bayliss}, {Bazin},
  {Benson}, {Brodwin}, {Carlstrom}, {Chang}, {Cho}, {Clocchiatti}, {Crawford},
  {Crites}, {de Haan}, {Desai}, {Dobbs}, {Dudley}, {Foley}, {Forman}, {George},
  {Gladders}, {Gonzalez}, {Halverson}, {Harrington}, {High}, {Holder},
  {Holzapfel}, {Hoover}, {Hrubes}, {Jones}, {Joy}, {Keisler}, {Knox}, {Lee},
  {Leitch}, {Liu}, {Lueker}, {Luong-Van}, {Mantz}, {Marrone}, {McDonald},
  {McMahon}, {Mehl}, {Meyer}, {Mocanu}, {Mohr}, {Murray}, {Natoli}, {Padin},
  {Plagge}, {Pryke}, {Rest}, {Ruel}, {Ruhl}, {Saliwanchik}, {Saro}, {Sayre},
  {Schaffer}, {Shaw}, {Shirokoff}, {Song}, {Spieler}, {Staniszewski}, {Stark},
  {Story}, {Stubbs}, {{\v S}uhada}, {van Engelen}, {Vanderlinde}, {Vieira},
  {Vikhlinin}, {Williamson}, {Zahn}, \& {Zenteno}}]{reichardt13}
{Reichardt}, C.~L., {Stalder}, B., {Bleem}, L.~E., {et~al.} 2013, \apj, 763,
  127

\bibitem[{{Russell} {et~al.}(2012){Russell}, {Fabian}, {Taylor}, {Sanders},
  {Blundell}, {Crawford}, {Johnstone}, \& {Belsole}}]{russell12}
{Russell}, H.~R., {Fabian}, A.~C., {Taylor}, G.~B., {et~al.} 2012, \mnras, 422,
  590

\bibitem[{{Salom{\'e}} \& {Combes}(2003)}]{salome03}
{Salom{\'e}}, P., \& {Combes}, F. 2003, \aap, 412, 657

\bibitem[{{Salom{\'e}} {et~al.}(2008){Salom{\'e}}, {Combes}, {Revaz}, {Edge},
  {Hatch}, {Fabian}, \& {Johnstone}}]{salome08}
{Salom{\'e}}, P., {Combes}, F., {Revaz}, Y., {et~al.} 2008, \aap, 484, 317

\bibitem[{{Samuele} {et~al.}(2011){Samuele}, {McNamara}, {Vikhlinin}, \&
  {Mullis}}]{samuele11}
{Samuele}, R., {McNamara}, B.~R., {Vikhlinin}, A., \& {Mullis}, C.~R. 2011,
  \apj, 731, 31

\bibitem[{{Sanders} {et~al.}(2010){Sanders}, {Fabian}, {Smith}, \&
  {Peterson}}]{sanders10}
{Sanders}, J.~S., {Fabian}, A.~C., {Smith}, R.~K., \& {Peterson}, J.~R. 2010,
  \mnras, 402, L11

\bibitem[{{Santos} {et~al.}(2008){Santos}, {Rosati}, {Tozzi}, {B{\"o}hringer},
  {Ettori}, \& {Bignamini}}]{santos08}
{Santos}, J.~S., {Rosati}, P., {Tozzi}, P., {et~al.} 2008, \aap, 483, 35

\bibitem[{{Santos} {et~al.}(2010){Santos}, {Tozzi}, {Rosati}, \&
  {B{\"o}hringer}}]{santos10}
{Santos}, J.~S., {Tozzi}, P., {Rosati}, P., \& {B{\"o}hringer}, H. 2010, \aap,
  521, A64+

\bibitem[{{Santos} {et~al.}(2012){Santos}, {Tozzi}, {Rosati}, {Nonino}, \&
  {Giovannini}}]{santos12}
{Santos}, J.~S., {Tozzi}, P., {Rosati}, P., {Nonino}, M., \& {Giovannini}, G.
  2012, \aap, 539, A105

\bibitem[{{Sayers} {et~al.}(2013){Sayers}, {Mroczkowski}, {Czakon}, {Golwala},
  {Mantz}, {Ameglio}, {Downes}, {Koch}, {Lin}, {Molnar}, {Moustakas},
  {Muchovej}, {Pierpaoli}, {Shitanishi}, {Siegel}, \& {Umetsu}}]{sayers13}
{Sayers}, J., {Mroczkowski}, T., {Czakon}, N.~G., {et~al.} 2013, \apj, 764, 152

\bibitem[{{Sehgal} {et~al.}(2010){Sehgal}, {Bode}, {Das},
  {Hernandez-Monteagudo}, {Huffenberger}, {Lin}, {Ostriker}, \&
  {Trac}}]{sehgal10}
{Sehgal}, N., {Bode}, P., {Das}, S., {et~al.} 2010, \apj, 709, 920

\bibitem[{{Semler} {et~al.}(2012){Semler}, {{\v S}uhada}, {Aird}, {Ashby},
  {Bautz}, {Bayliss}, {Bazin}, {Bocquet}, {Benson}, {Bleem}, {Brodwin},
  {Carlstrom}, {Chang}, {Cho}, {Clocchiatti}, {Crawford}, {Crites}, {de Haan},
  {Desai}, {Dobbs}, {Dudley}, {Foley}, {George}, {Gladders}, {Gonzalez},
  {Halverson}, {Harrington}, {High}, {Holder}, {Holzapfel}, {Hoover}, {Hrubes},
  {Jones}, {Joy}, {Keisler}, {Knox}, {Lee}, {Leitch}, {Liu}, {Lueker},
  {Luong-Van}, {Mantz}, {Marrone}, {McDonald}, {McMahon}, {Mehl}, {Meyer},
  {Mocanu}, {Mohr}, {Montroy}, {Murray}, {Natoli}, {Padin}, {Plagge}, {Pryke},
  {Reichardt}, {Rest}, {Ruel}, {Ruhl}, {Saliwanchik}, {Saro}, {Sayre},
  {Schaffer}, {Shaw}, {Shirokoff}, {Song}, {Spieler}, {Stalder},
  {Staniszewski}, {Stark}, {Story}, {Stubbs}, {van Engelen}, {Vanderlinde},
  {Vieira}, {Vikhlinin}, {Williamson}, {Zahn}, \& {Zenteno}}]{semler12}
{Semler}, D.~R., {{\v S}uhada}, R., {Aird}, K.~A., {et~al.} 2012, ArXiv
  e-prints, arXiv:1208.3368

\bibitem[{{Siemiginowska} {et~al.}(2010){Siemiginowska}, {Burke}, {Aldcroft},
  {Worrall}, {Allen}, {Bechtold}, {Clarke}, \& {Cheung}}]{siemiginowska10}
{Siemiginowska}, A., {Burke}, D.~J., {Aldcroft}, T.~L., {et~al.} 2010, \apj,
  722, 102

\bibitem[{{Song} {et~al.}(2012){Song}, {Zenteno}, {Stalder}, {Desai}, {Bleem},
  {Aird}, {Armstrong}, {Ashby}, {Bayliss}, {Bazin}, {Benson}, {Bertin},
  {Brodwin}, {Carlstrom}, {Chang}, {Cho}, {Clocchiatti}, {Crawford}, {Crites},
  {de Haan}, {Dobbs}, {Dudley}, {Foley}, {George}, {Gettings}, {Gladders},
  {Gonzalez}, {Halverson}, {Harrington}, {High}, {Holder}, {Holzapfel},
  {Hoover}, {Hrubes}, {Joy}, {Keisler}, {Knox}, {Lee}, {Leitch}, {Liu},
  {Lueker}, {Luong-Van}, {Marrone}, {McDonald}, {McMahon}, {Mehl}, {Meyer},
  {Mocanu}, {Mohr}, {Montroy}, {Natoli}, {Nurgaliev}, {Padin}, {Plagge},
  {Pryke}, {Reichardt}, {Rest}, {Ruel}, {Ruhl}, {Saliwanchik}, {Saro}, {Sayre},
  {Schaffer}, {Shaw}, {Shirokoff}, {Suhada}, {Spieler}, {Stanford},
  {Staniszewski}, {Stark}, {Story}, {Stubbs}, {van Engelen}, {Vanderlinde},
  {Vieira}, {Williamson}, \& {Zahn}}]{song12}
{Song}, J., {Zenteno}, A., {Stalder}, B., {et~al.} 2012, ArXiv e-prints,
  arXiv:1207.4369

\bibitem[{{Sparks} {et~al.}(2012){Sparks}, {Pringle}, {Carswell}, {Donahue},
  {Martin}, {Voit}, {Cracraft}, {Manset}, \& {Hough}}]{sparks12}
{Sparks}, W.~B., {Pringle}, J.~E., {Carswell}, R.~F., {et~al.} 2012, \apjl,
  750, L5

\bibitem[{{Staniszewski} {et~al.}(2009){Staniszewski}, {Ade}, {Aird}, {Benson},
  {Bleem}, {Carlstrom}, {Chang}, {Cho}, {Crawford}, {Crites}, {de Haan},
  {Dobbs}, {Halverson}, {Holder}, {Holzapfel}, {Hrubes}, {Joy}, {Keisler},
  {Lanting}, {Lee}, {Leitch}, {Loehr}, {Lueker}, {McMahon}, {Mehl}, {Meyer},
  {Mohr}, {Montroy}, {Ngeow}, {Padin}, {Plagge}, {Pryke}, {Reichardt}, {Ruhl},
  {Schaffer}, {Shaw}, {Shirokoff}, {Spieler}, {Stalder}, {Stark},
  {Vanderlinde}, {Vieira}, {Zahn}, \& {Zenteno}}]{staniszewski09}
{Staniszewski}, Z., {Ade}, P.~A.~R., {Aird}, K.~A., {et~al.} 2009, \apj, 701,
  32

\bibitem[{{Sun}(2009)}]{sun09b}
{Sun}, M. 2009, \apj, 704, 1586

\bibitem[{{Sun} {et~al.}(2009){Sun}, {Voit}, {Donahue}, {Jones}, {Forman}, \&
  {Vikhlinin}}]{sun09a}
{Sun}, M., {Voit}, G.~M., {Donahue}, M., {et~al.} 2009, \apj, 693, 1142

\bibitem[{{Sunyaev} \& {Zeldovich}(1972)}]{sunyaev72}
{Sunyaev}, R.~A., \& {Zeldovich}, Y.~B. 1972, Comments on Astrophysics and
  Space Physics, 4, 173

\bibitem[{{Sutherland} \& {Dopita}(1993)}]{sutherland93}
{Sutherland}, R.~S., \& {Dopita}, M.~A. 1993, \apjs, 88, 253

\bibitem[{{Tyson}(2002)}]{lsst}
{Tyson}, J.~A. 2002, in Society of Photo-Optical Instrumentation Engineers
  (SPIE) Conference Series, Vol. 4836, Society of Photo-Optical Instrumentation
  Engineers (SPIE) Conference Series, ed. J.~A. {Tyson} \& S.~{Wolff}, 10--20

\bibitem[{{Vanderlinde} {et~al.}(2010){Vanderlinde}, {Crawford}, {de Haan},
  {Dudley}, {Shaw}, {Ade}, {Aird}, {Benson}, {Bleem}, {Brodwin}, {Carlstrom},
  {Chang}, {Crites}, {Desai}, {Dobbs}, {Foley}, {George}, {Gladders}, {Hall},
  {Halverson}, {High}, {Holder}, {Holzapfel}, {Hrubes}, {Joy}, {Keisler},
  {Knox}, {Lee}, {Leitch}, {Loehr}, {Lueker}, {Marrone}, {McMahon}, {Mehl},
  {Meyer}, {Mohr}, {Montroy}, {Ngeow}, {Padin}, {Plagge}, {Pryke}, {Reichardt},
  {Rest}, {Ruel}, {Ruhl}, {Schaffer}, {Shirokoff}, {Song}, {Spieler},
  {Stalder}, {Staniszewski}, {Stark}, {Stubbs}, {van Engelen}, {Vieira},
  {Williamson}, {Yang}, {Zahn}, \& {Zenteno}}]{vanderlinde10}
{Vanderlinde}, K., {Crawford}, T.~M., {de Haan}, T., {et~al.} 2010, \apj, 722,
  1180

\bibitem[{{Vikhlinin} {et~al.}(2007){Vikhlinin}, {Burenin}, {Forman}, {Jones},
  {Hornstrup}, {Murray}, \& {Quintana}}]{vikhlinin07}
{Vikhlinin}, A., {Burenin}, R., {Forman}, W.~R., {et~al.} 2007, in Heating
  versus Cooling in Galaxies and Clusters of Galaxies, ed. {H.~B{\"o}hringer,
  G.~W.~Pratt, A.~Finoguenov, \& P.~Schuecker }, 48--+

\bibitem[{{Vikhlinin} {et~al.}(2006){Vikhlinin}, {Kravtsov}, {Forman}, {Jones},
  {Markevitch}, {Murray}, \& {Van Speybroeck}}]{vikhlinin06}
{Vikhlinin}, A., {Kravtsov}, A., {Forman}, W., {et~al.} 2006, \apj, 640, 691

\bibitem[{{Vikhlinin} {et~al.}(2005){Vikhlinin}, {Markevitch}, {Murray},
  {Jones}, {Forman}, \& {Van Speybroeck}}]{vikhlinin05}
{Vikhlinin}, A., {Markevitch}, M., {Murray}, S.~S., {et~al.} 2005, \apj, 628,
  655

\bibitem[{{Vikhlinin} {et~al.}(1998){Vikhlinin}, {McNamara}, {Forman}, {Jones},
  {Quintana}, \& {Hornstrup}}]{vikhlinin98}
{Vikhlinin}, A., {McNamara}, B.~R., {Forman}, W., {et~al.} 1998, \apjl, 498,
  L21

\bibitem[{{Vikhlinin} {et~al.}(2009){Vikhlinin}, {Burenin}, {Ebeling},
  {Forman}, {Hornstrup}, {Jones}, {Kravtsov}, {Murray}, {Nagai}, {Quintana}, \&
  {Voevodkin}}]{vikhlinin09}
{Vikhlinin}, A., {Burenin}, R.~A., {Ebeling}, H., {et~al.} 2009, \apj, 692,
  1033

\bibitem[{{White} {et~al.}(1997){White}, {Jones}, \& {Forman}}]{white97}
{White}, D.~A., {Jones}, C., \& {Forman}, W. 1997, \mnras, 292, 419

\bibitem[{{Williamson} {et~al.}(2011){Williamson}, {Benson}, {High},
  {Vanderlinde}, {Ade}, {Aird}, {Andersson}, {Armstrong}, {Ashby}, {Bautz},
  {Bazin}, {Bertin}, {Bleem}, {Bonamente}, {Brodwin}, {Carlstrom}, {Chang},
  {Chapman}, {Clocchiatti}, {Crawford}, {Crites}, {de Haan}, {Desai}, {Dobbs},
  {Dudley}, {Fazio}, {Foley}, {Forman}, {Garmire}, {George}, {Gladders},
  {Gonzalez}, {Halverson}, {Holder}, {Holzapfel}, {Hoover}, {Hrubes}, {Jones},
  {Joy}, {Keisler}, {Knox}, {Lee}, {Leitch}, {Lueker}, {Luong-Van}, {Marrone},
  {McMahon}, {Mehl}, {Meyer}, {Mohr}, {Montroy}, {Murray}, {Padin}, {Plagge},
  {Pryke}, {Reichardt}, {Rest}, {Ruel}, {Ruhl}, {Saliwanchik}, {Saro},
  {Schaffer}, {Shaw}, {Shirokoff}, {Song}, {Spieler}, {Stalder}, {Stanford},
  {Staniszewski}, {Stark}, {Story}, {Stubbs}, {Vieira}, {Vikhlinin}, \&
  {Zenteno}}]{williamson11}
{Williamson}, R., {Benson}, B.~A., {High}, F.~W., {et~al.} 2011, \apj, 738, 139

\bibitem[{{Wyithe} {et~al.}(2001){Wyithe}, {Turner}, \& {Spergel}}]{wyithe01}
{Wyithe}, J.~S.~B., {Turner}, E.~L., \& {Spergel}, D.~N. 2001, \apj, 555, 504

\bibitem[{{Zhao}(1996)}]{zhao96}
{Zhao}, H. 1996, \mnras, 278, 488

\end{thebibliography}

\clearpage
\appendix
\section{Temperature and Density Profiles -- Data and Models}
Below, we show the X-ray data for all 83 clusters presented in this work. For each cluster, we show a smoothed X-ray image (0.5--5.0 keV), a surface brightness profile (\S2.3), and a 3-bin projected temperature profile (\S2.4). On both the surface brightness and temperature profiles, we overlay (gray curve) the best-fit models from our mass-modeling technique (\S2.4). These images demonstrate the ability of our algorithm to accurately predict the central temperature given only a coarse temperature profile combined with a well-sampled gas density profile and assumptions about the dark matter halo. With few exceptions, this technique provides an excellent fit to the projected temperature profiles, regardless of whether the cluster is relaxed (e.g., SPT-CLJ2344-4243) or disturbed (e.g., SPT-CLJ0542-4100).

\begin{figure*}[h!]
\centering
\begin{tabular}{c c}
\vspace{-0.05in}
\includegraphics[width=0.49\textwidth]{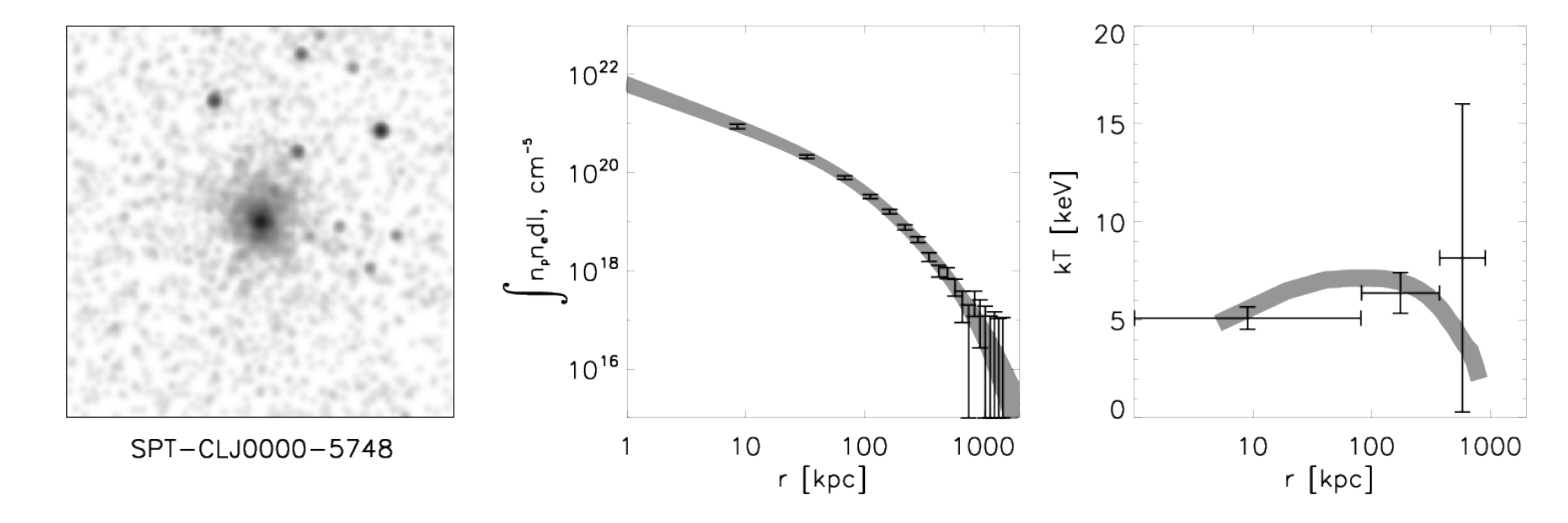} &
\includegraphics[width=0.49\textwidth]{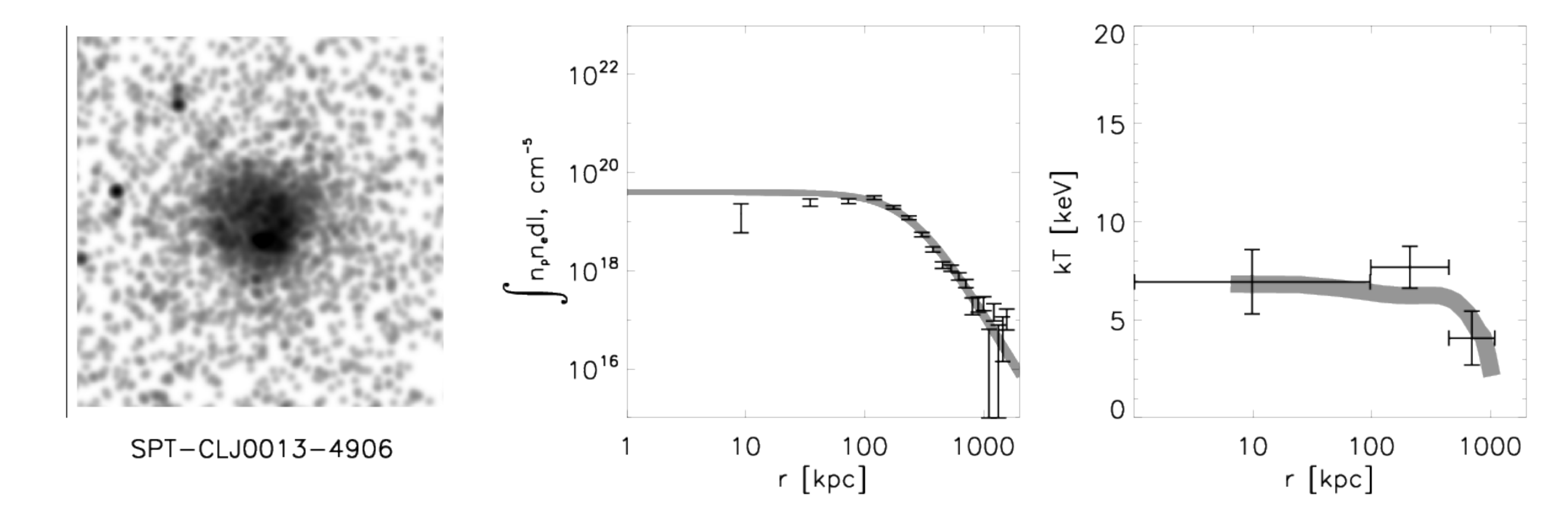} \\
\vspace{-0.05in} 
\includegraphics[width=0.49\textwidth]{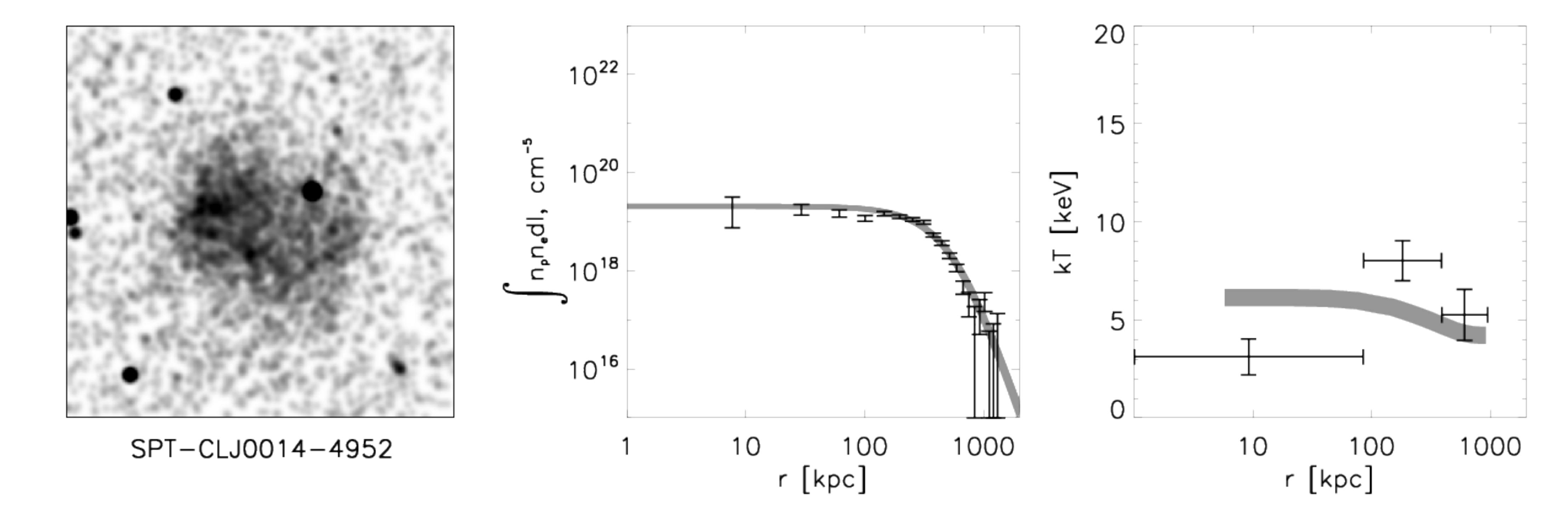} &
\includegraphics[width=0.49\textwidth]{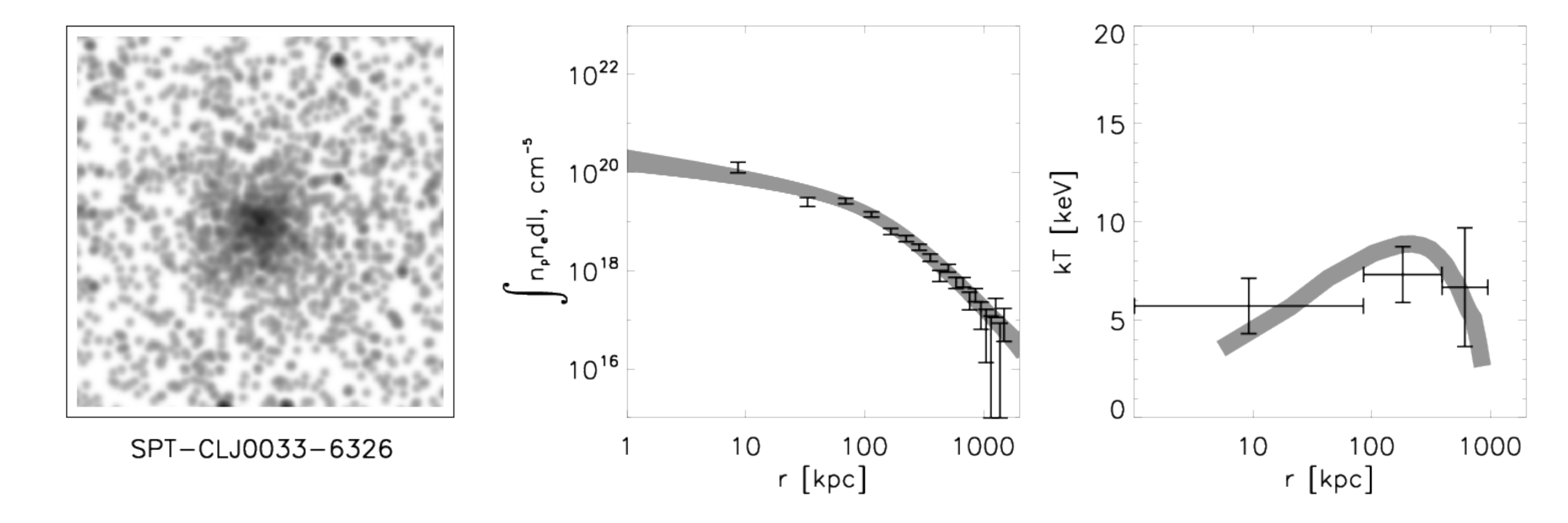} \\
\vspace{-0.05in}
\includegraphics[width=0.49\textwidth]{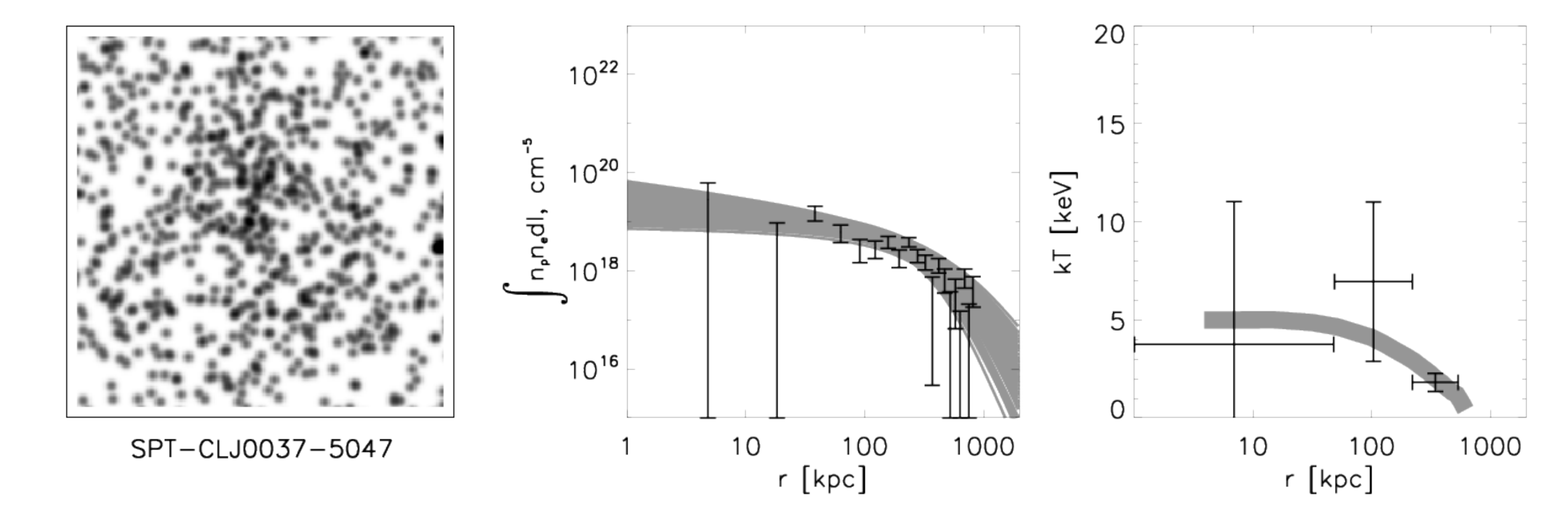} &
\includegraphics[width=0.49\textwidth]{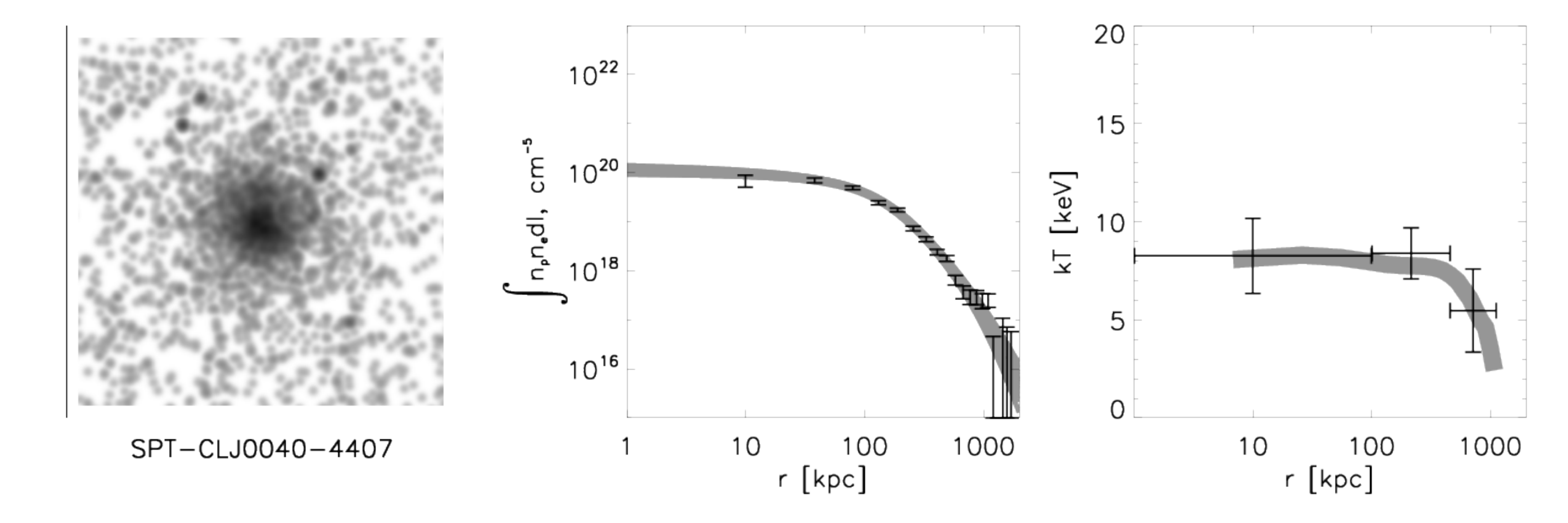} \\
\vspace{-0.05in}
\includegraphics[width=0.49\textwidth]{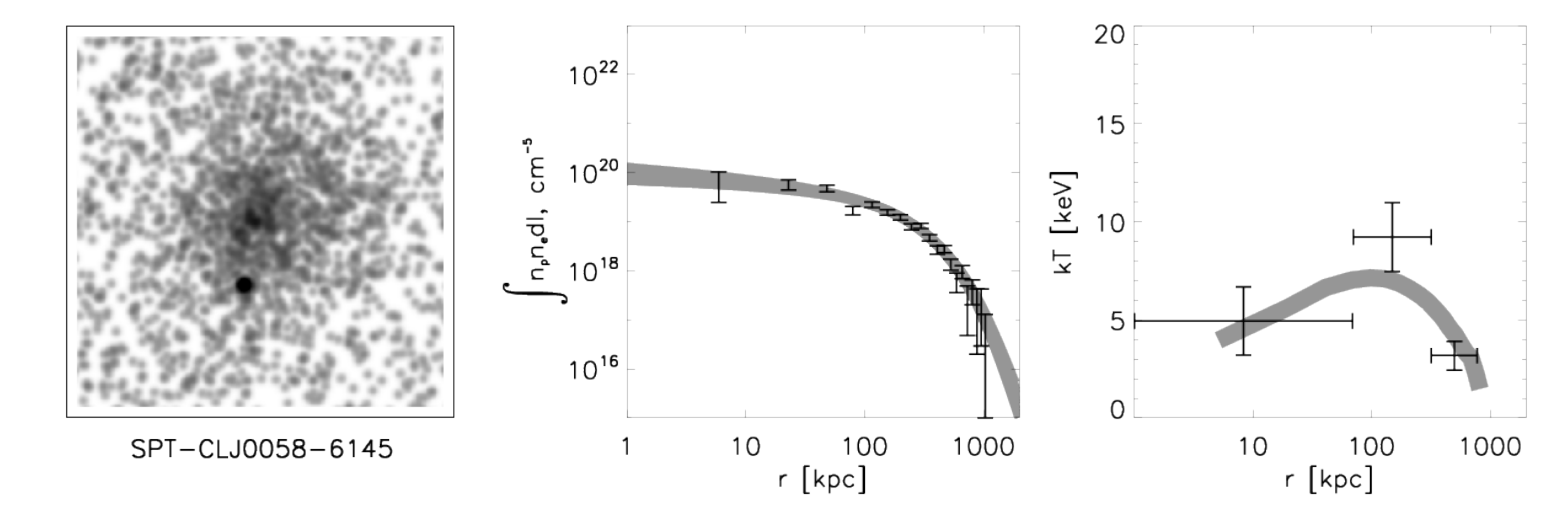} &
\includegraphics[width=0.49\textwidth]{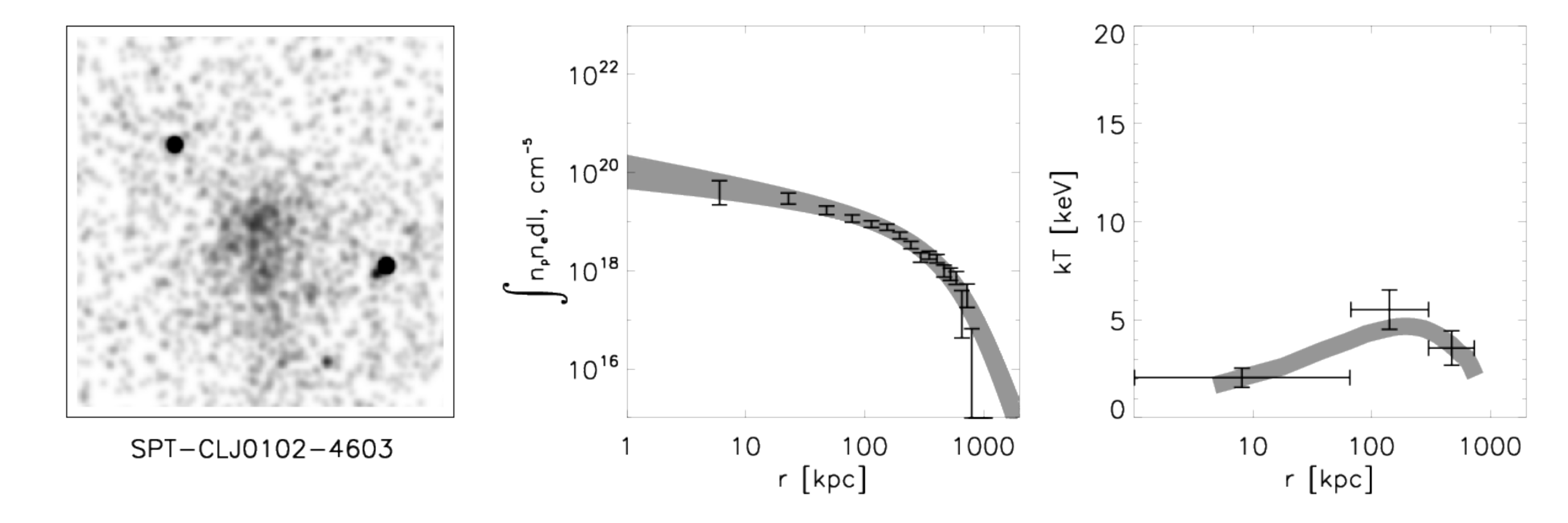} \\
\vspace{-0.05in}
\includegraphics[width=0.49\textwidth]{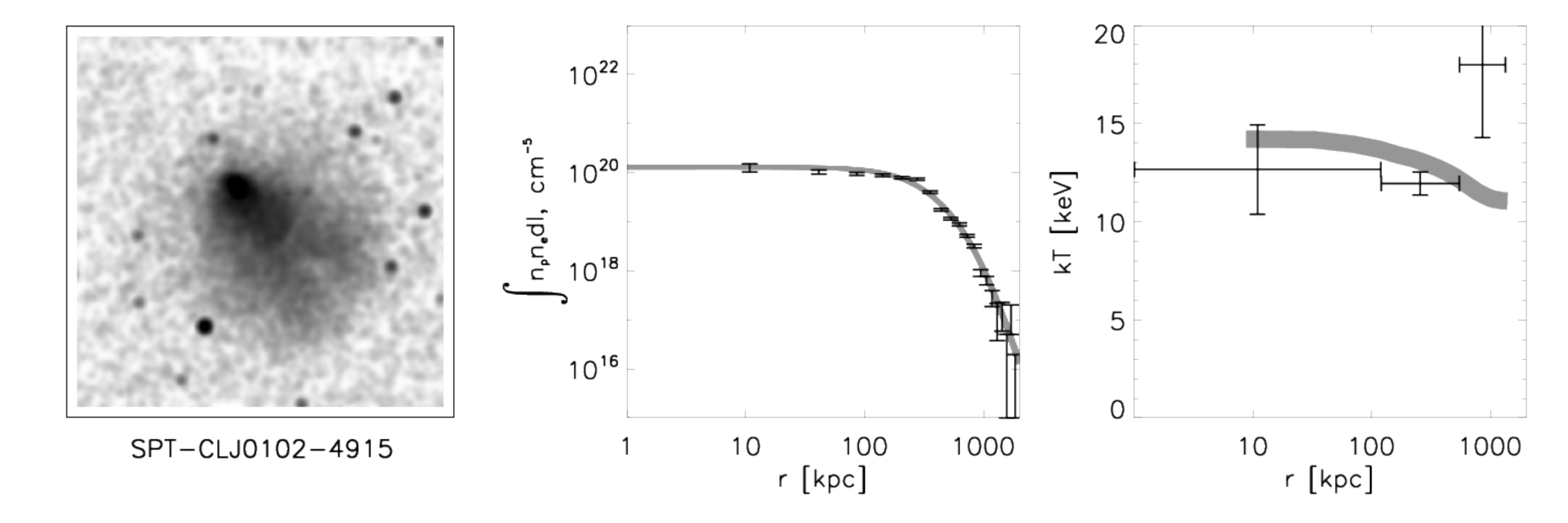} &
\includegraphics[width=0.49\textwidth]{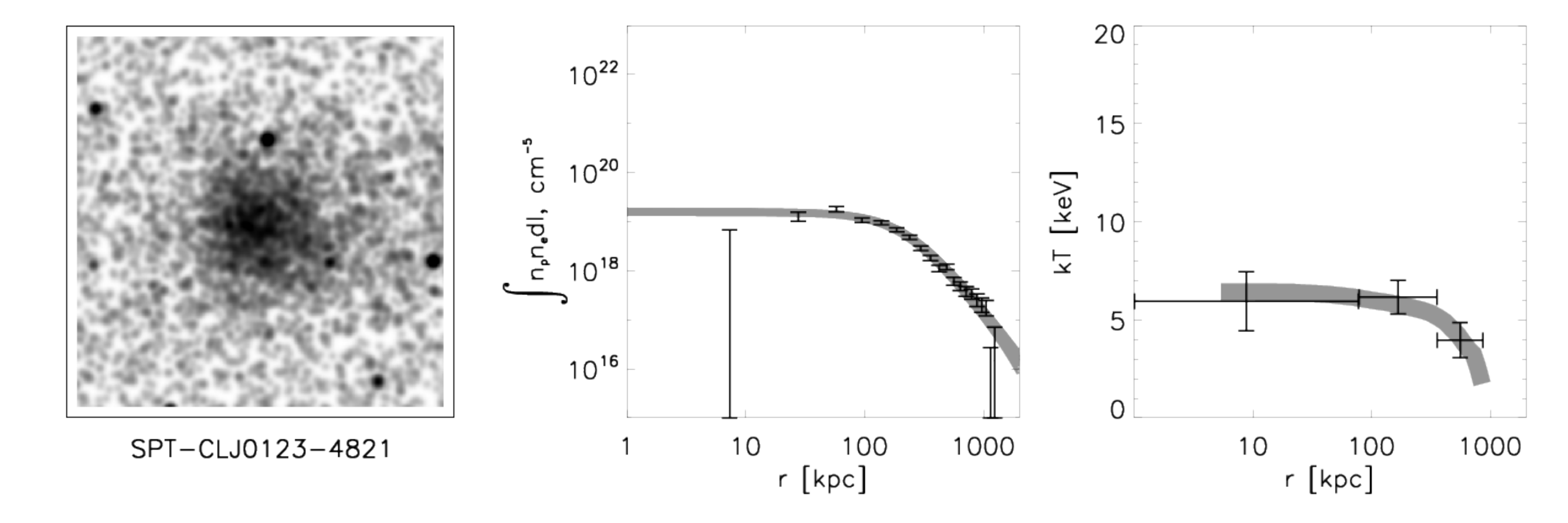} \\
\vspace{-0.05in}
\includegraphics[width=0.49\textwidth]{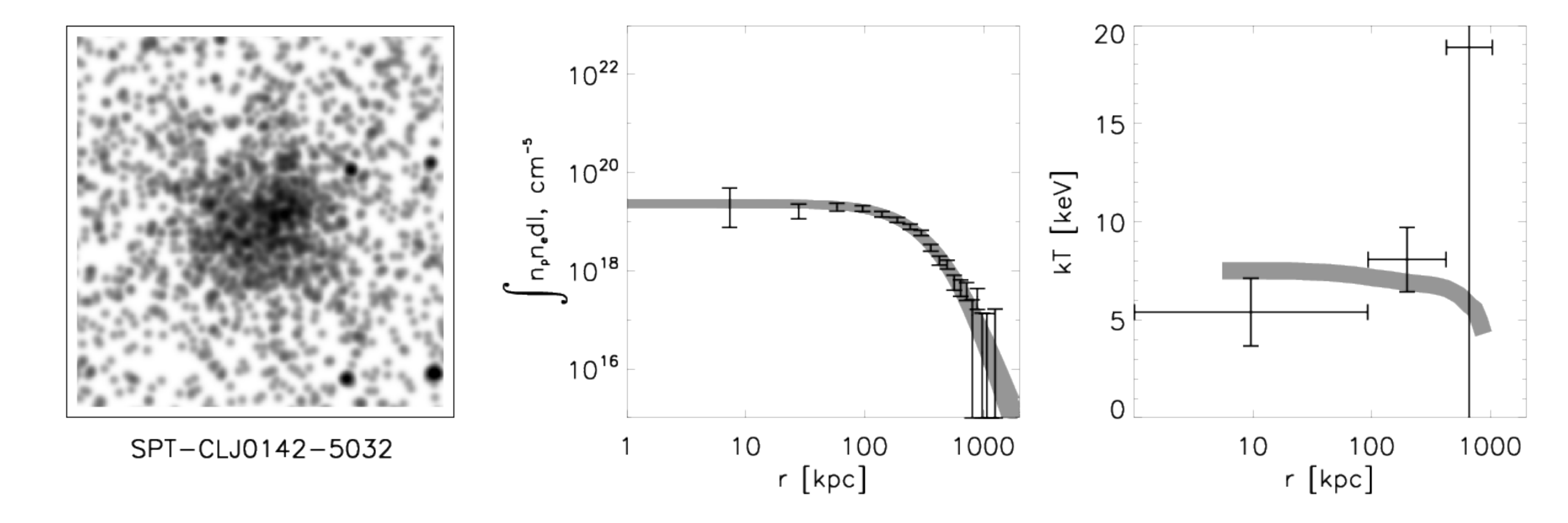} &
\includegraphics[width=0.49\textwidth]{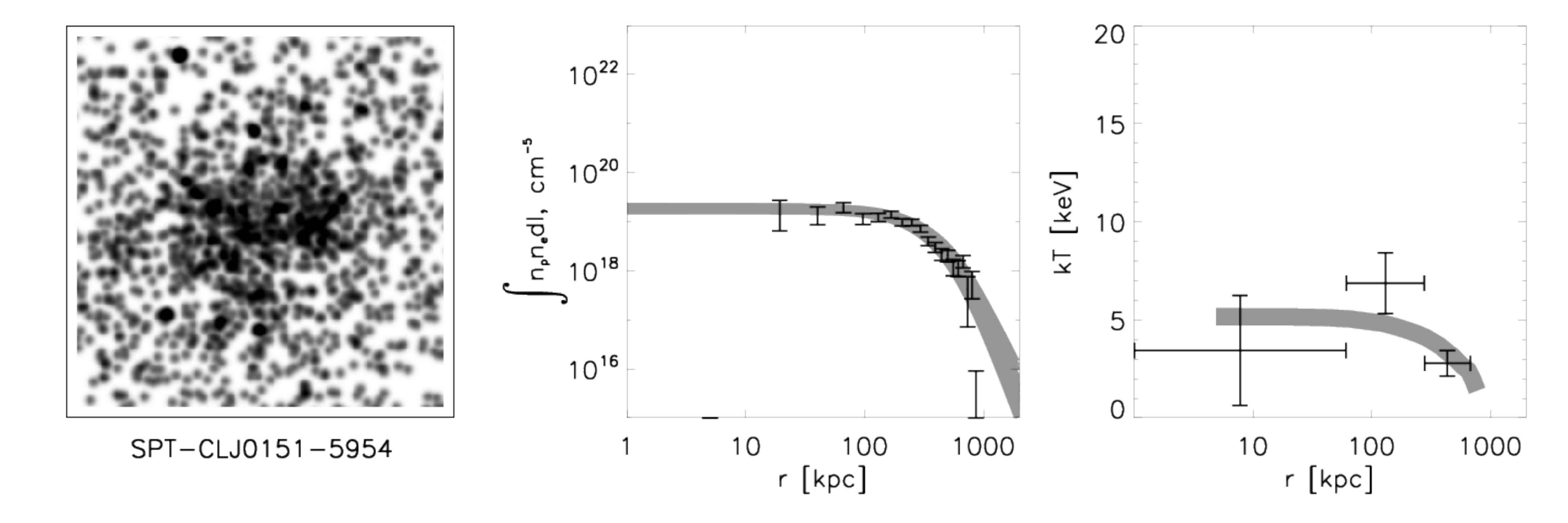} \\
\end{tabular}
\end{figure*}

\begin{figure*}[h!]
\centering
\begin{tabular}{c c}
\vspace{-0.05in}
\includegraphics[width=0.49\textwidth]{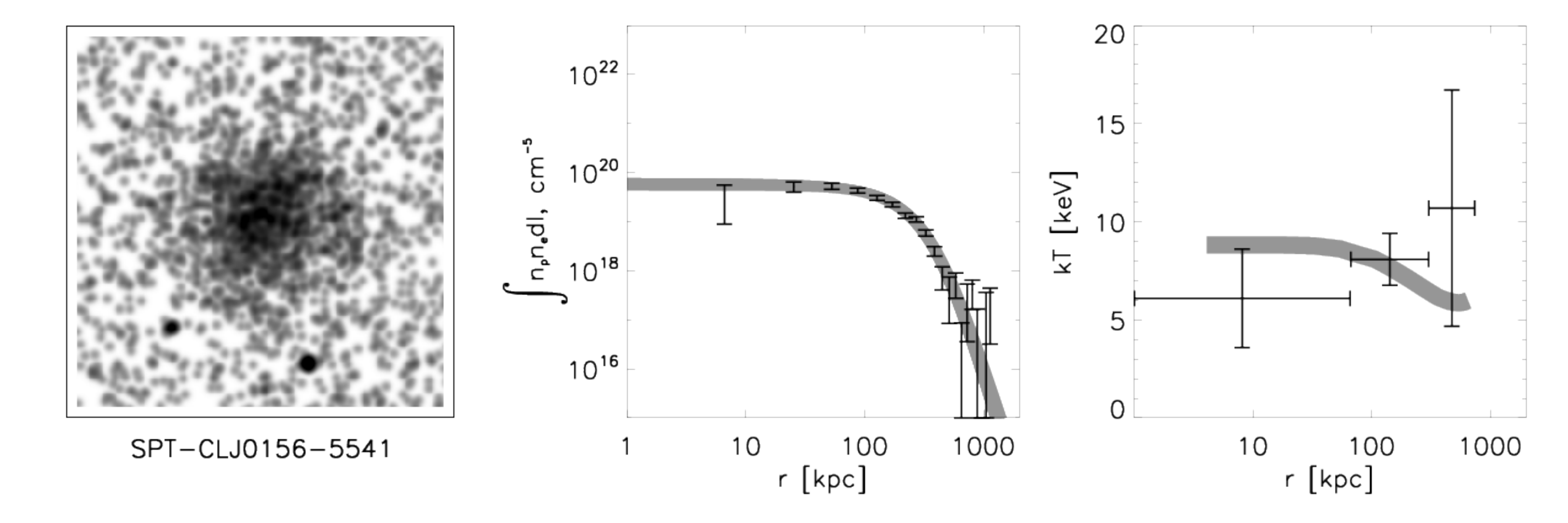} &
\includegraphics[width=0.49\textwidth]{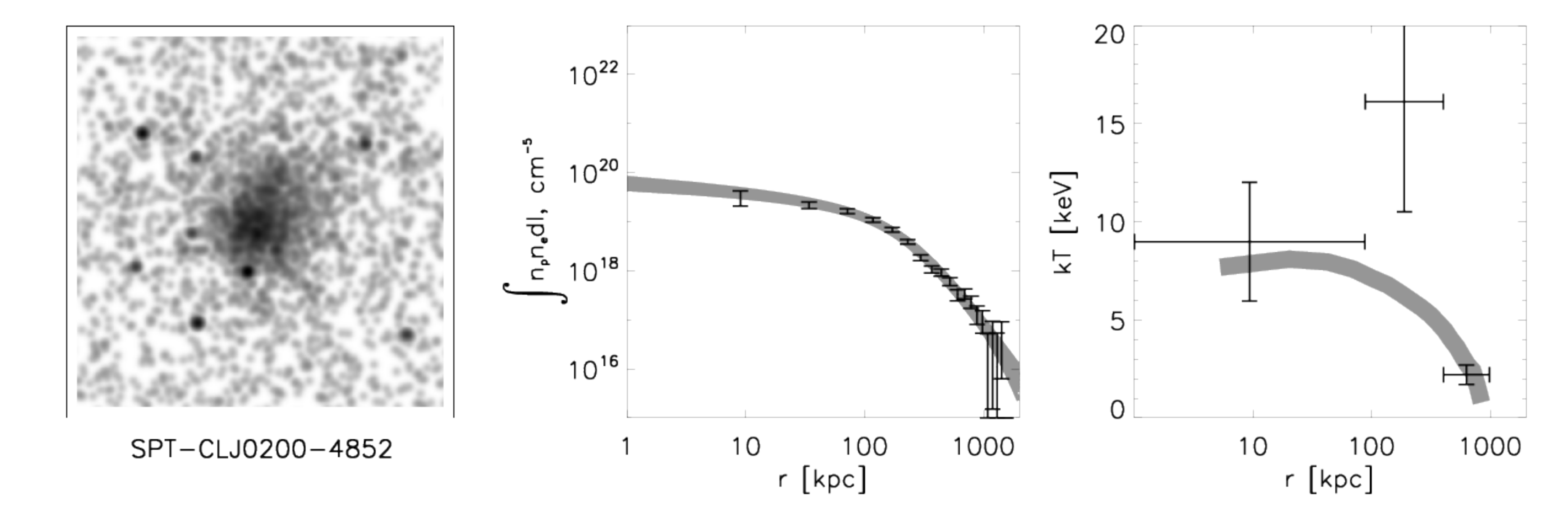} \\
\vspace{-0.05in}
\includegraphics[width=0.49\textwidth]{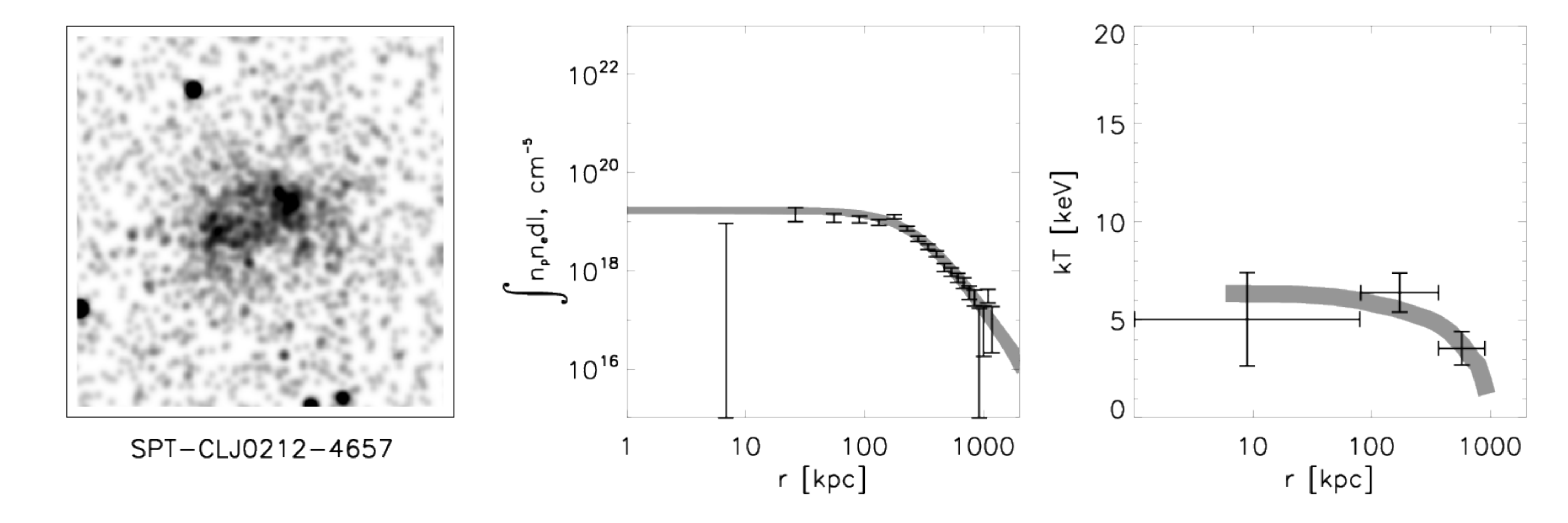} &
\includegraphics[width=0.49\textwidth]{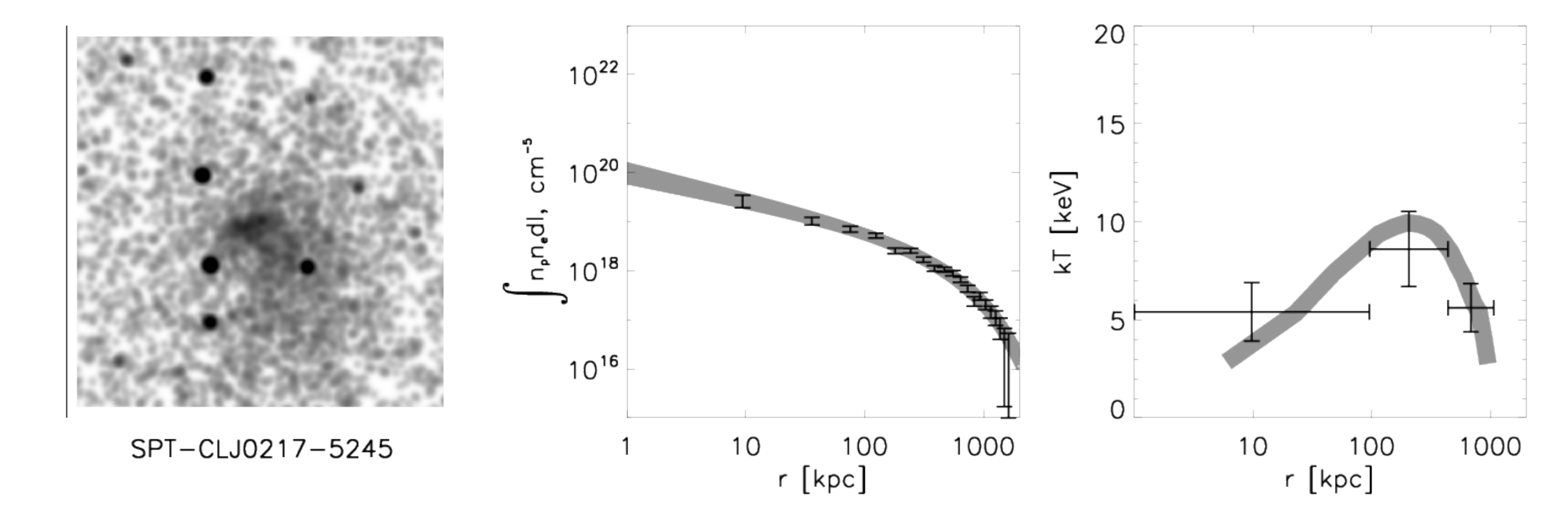} \\
\vspace{-0.05in}
\includegraphics[width=0.49\textwidth]{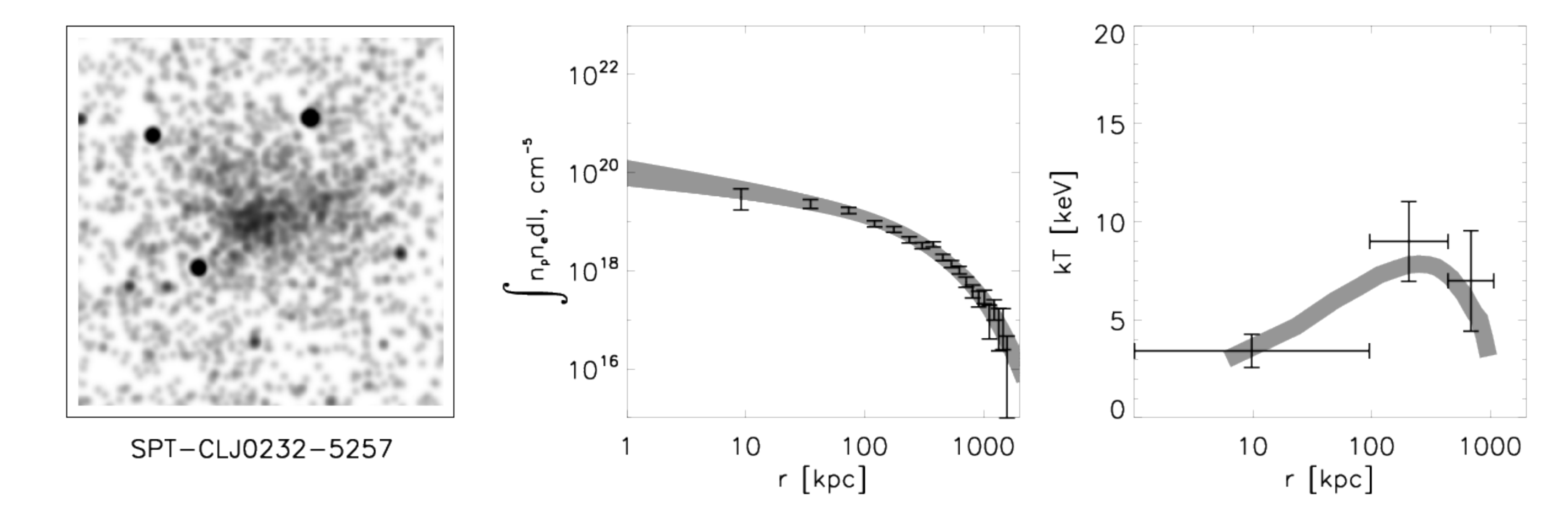} &
\includegraphics[width=0.49\textwidth]{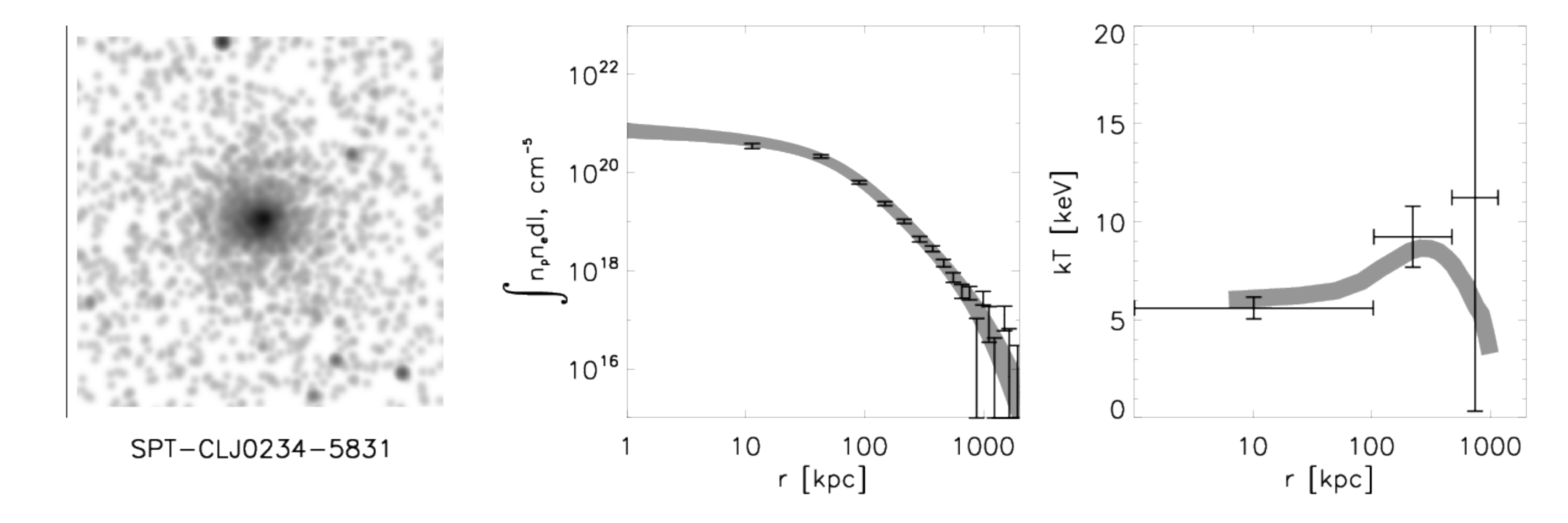} \\
\vspace{-0.05in}
\includegraphics[width=0.49\textwidth]{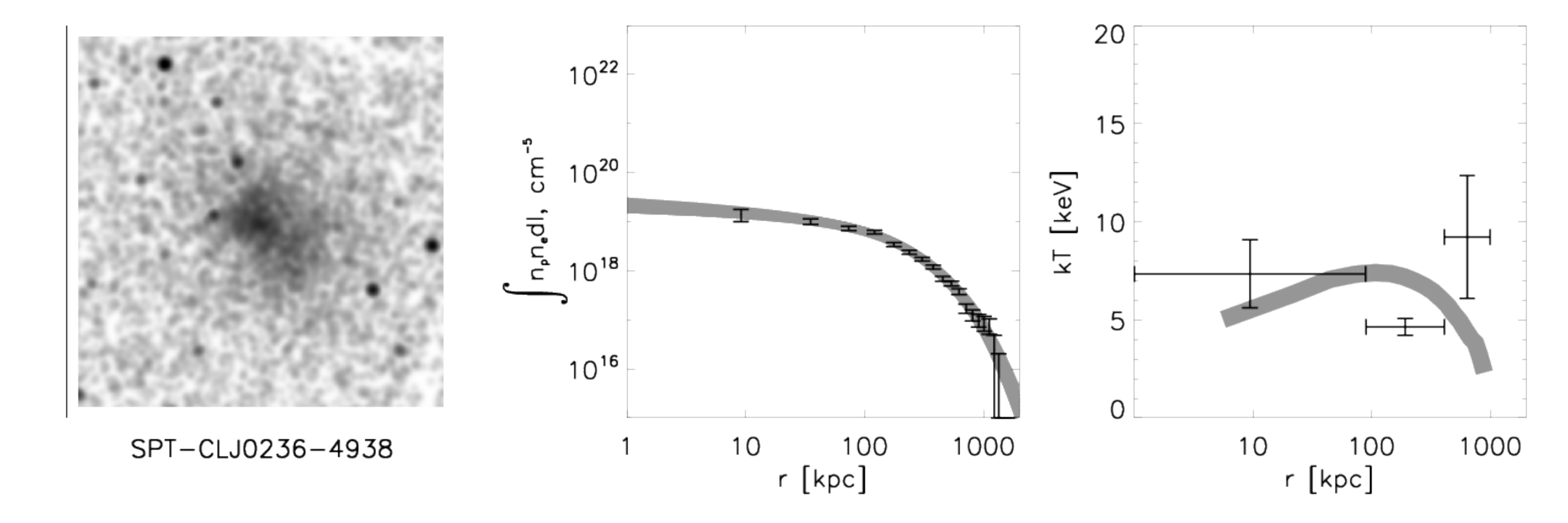} &
\includegraphics[width=0.49\textwidth]{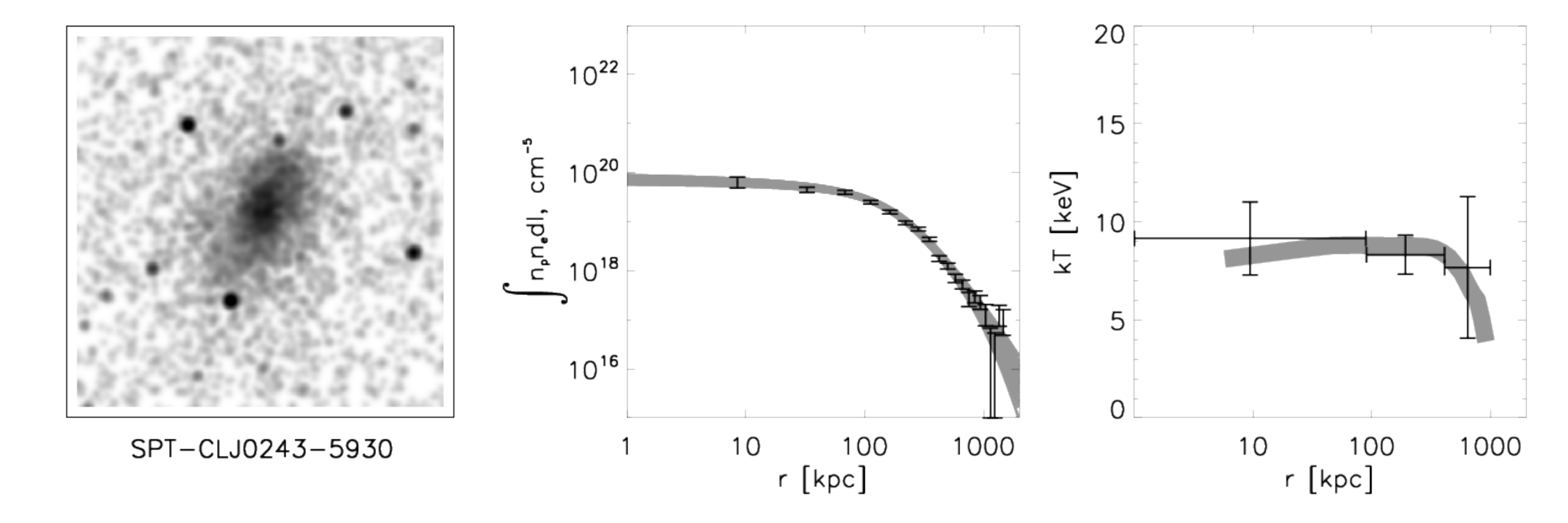} \\
\vspace{-0.05in}
\includegraphics[width=0.49\textwidth]{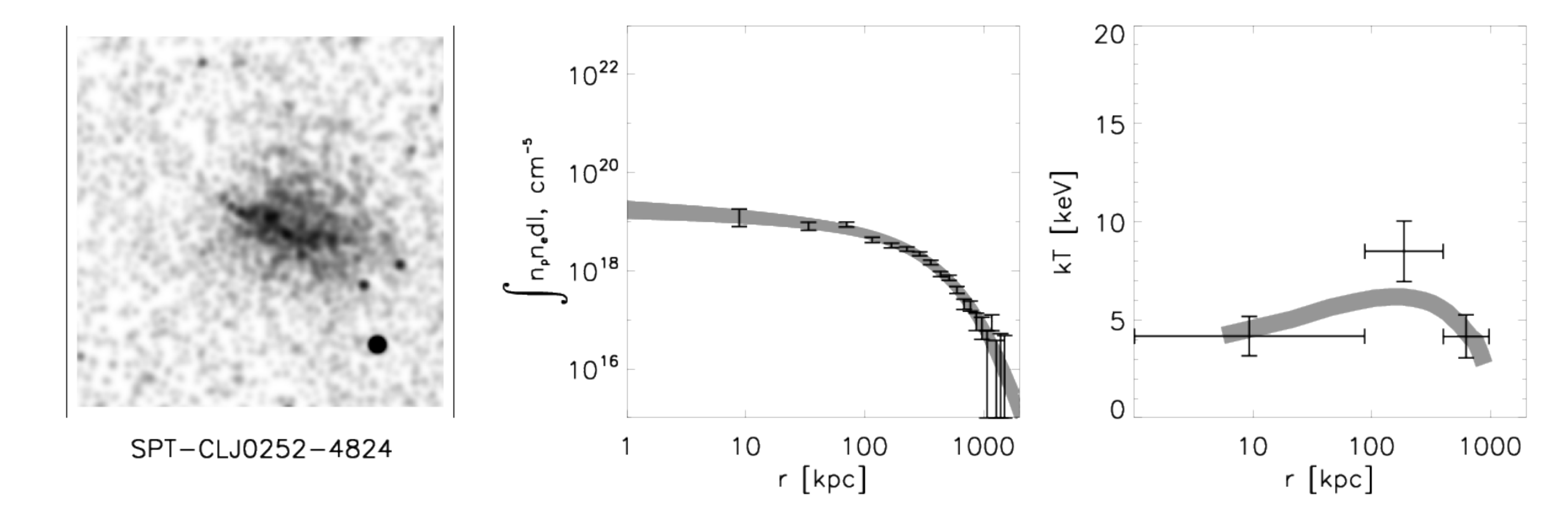} &
\includegraphics[width=0.49\textwidth]{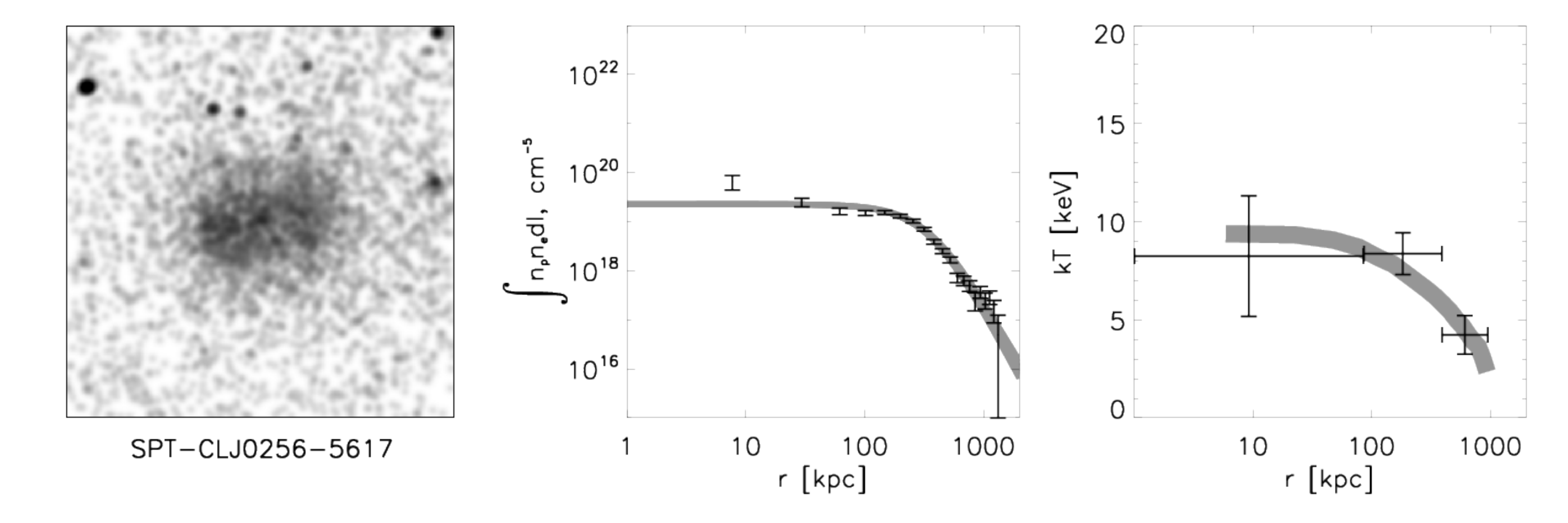} \\
\vspace{-0.05in}
\includegraphics[width=0.49\textwidth]{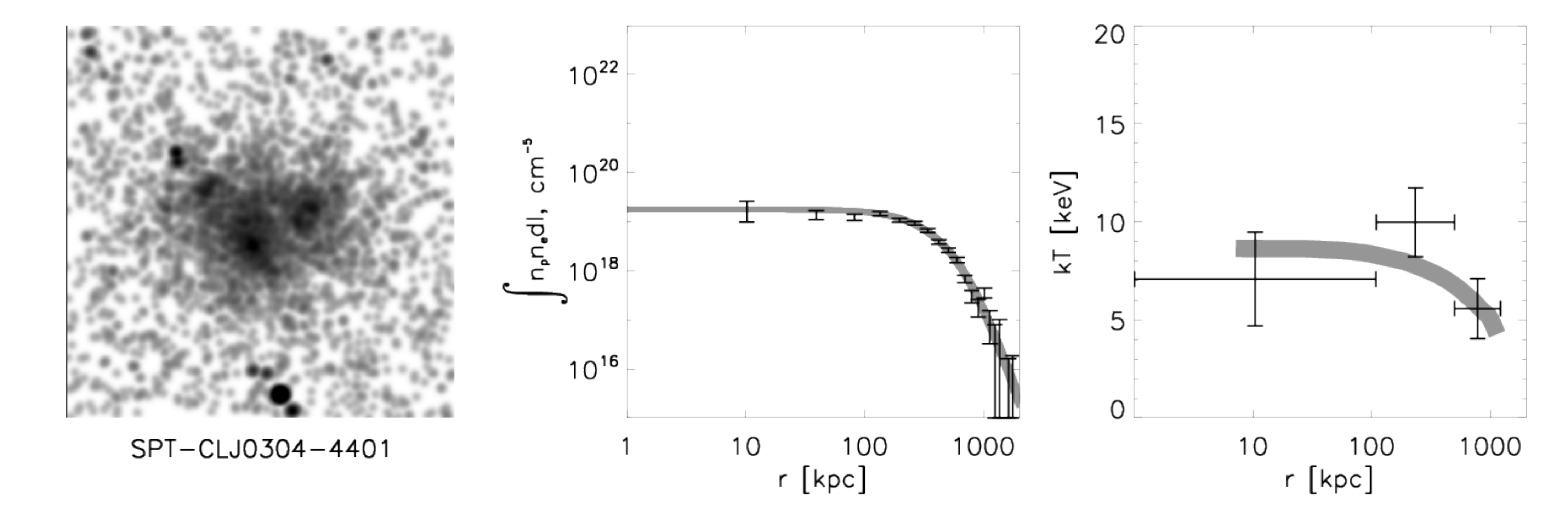} &
\includegraphics[width=0.49\textwidth]{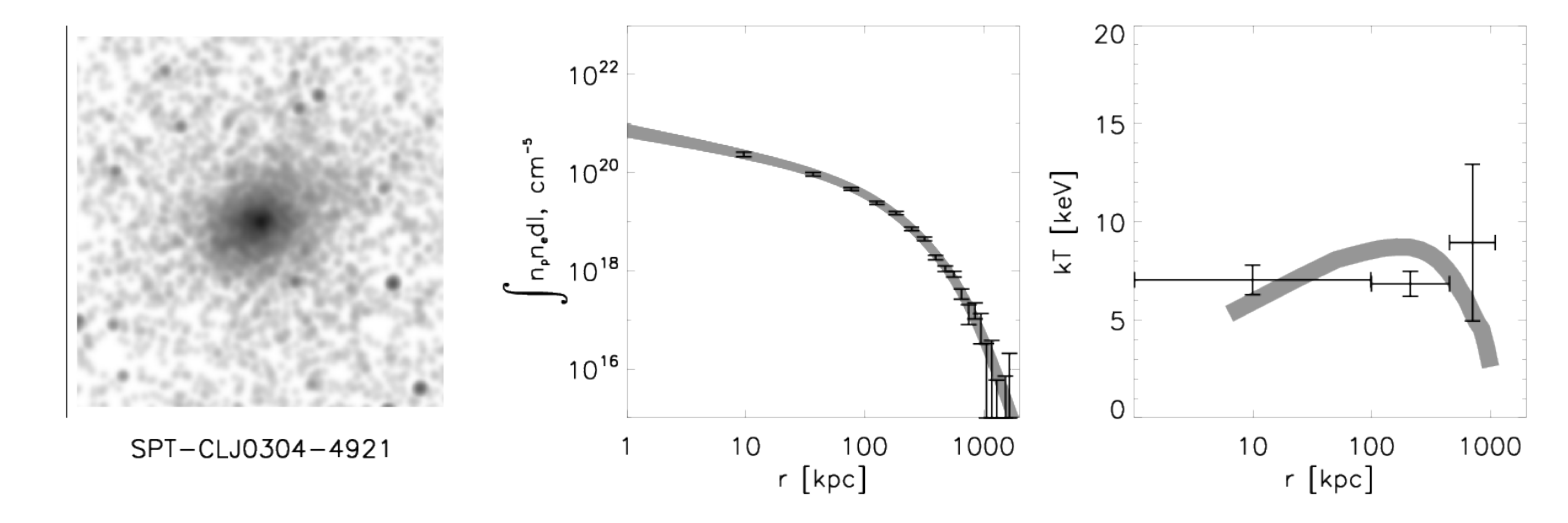} \\
\vspace{-0.05in}
\includegraphics[width=0.49\textwidth]{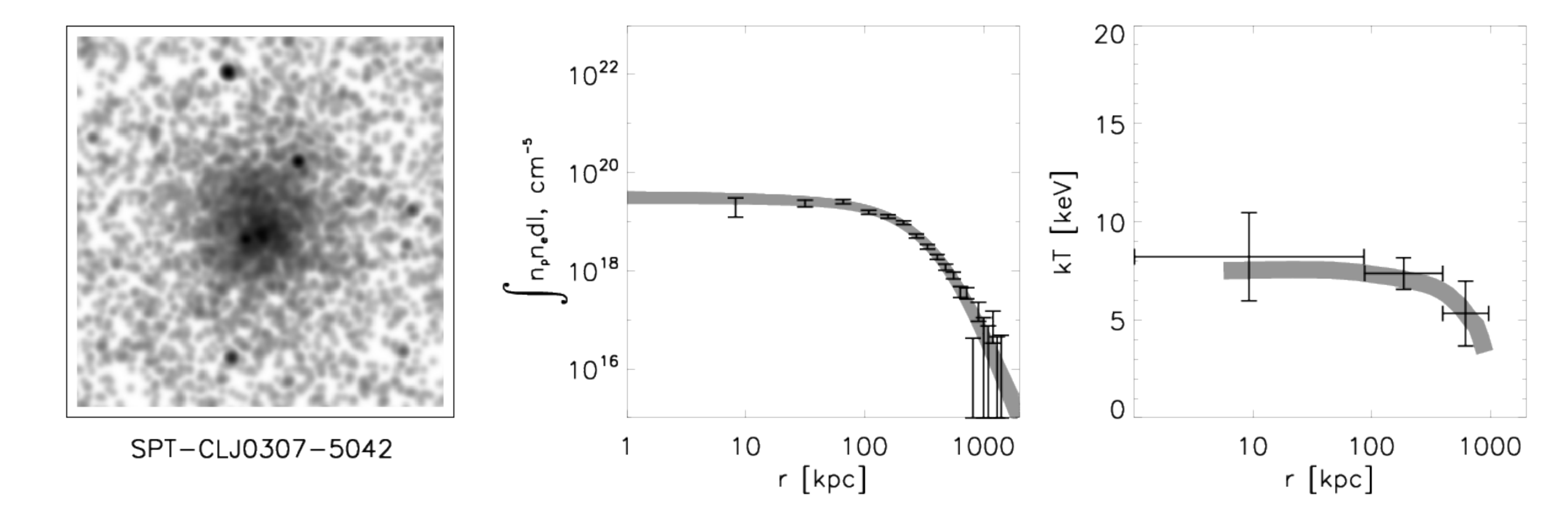} &
\includegraphics[width=0.49\textwidth]{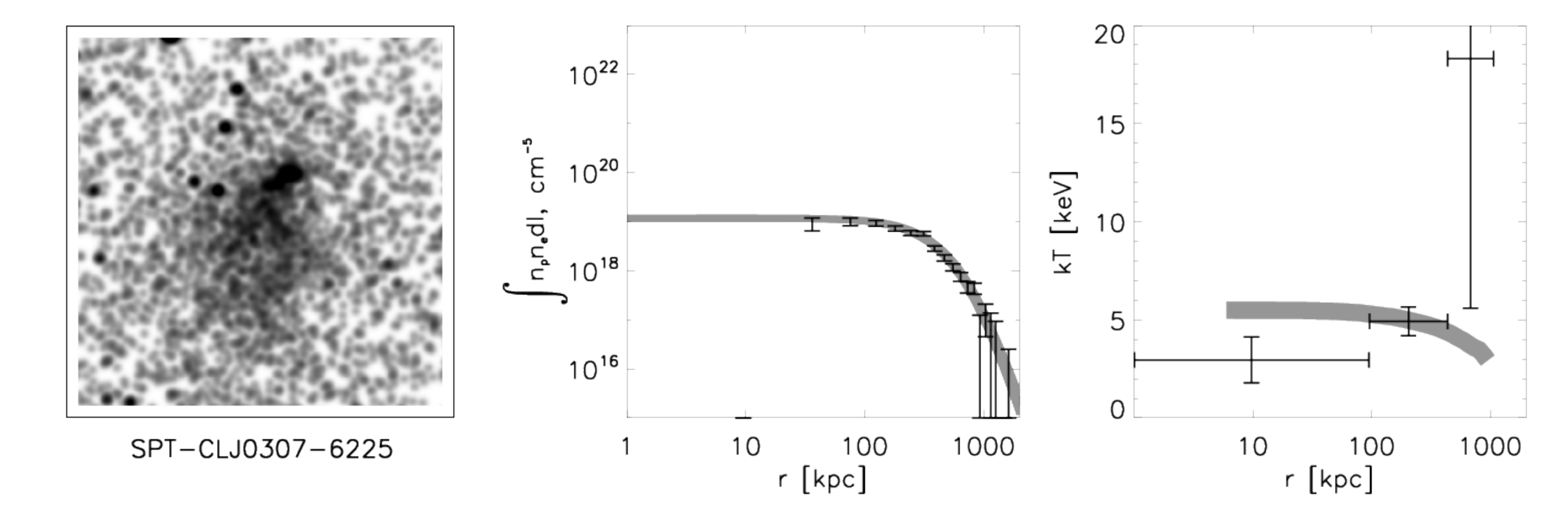} \\
\vspace{-0.05in}
\includegraphics[width=0.49\textwidth]{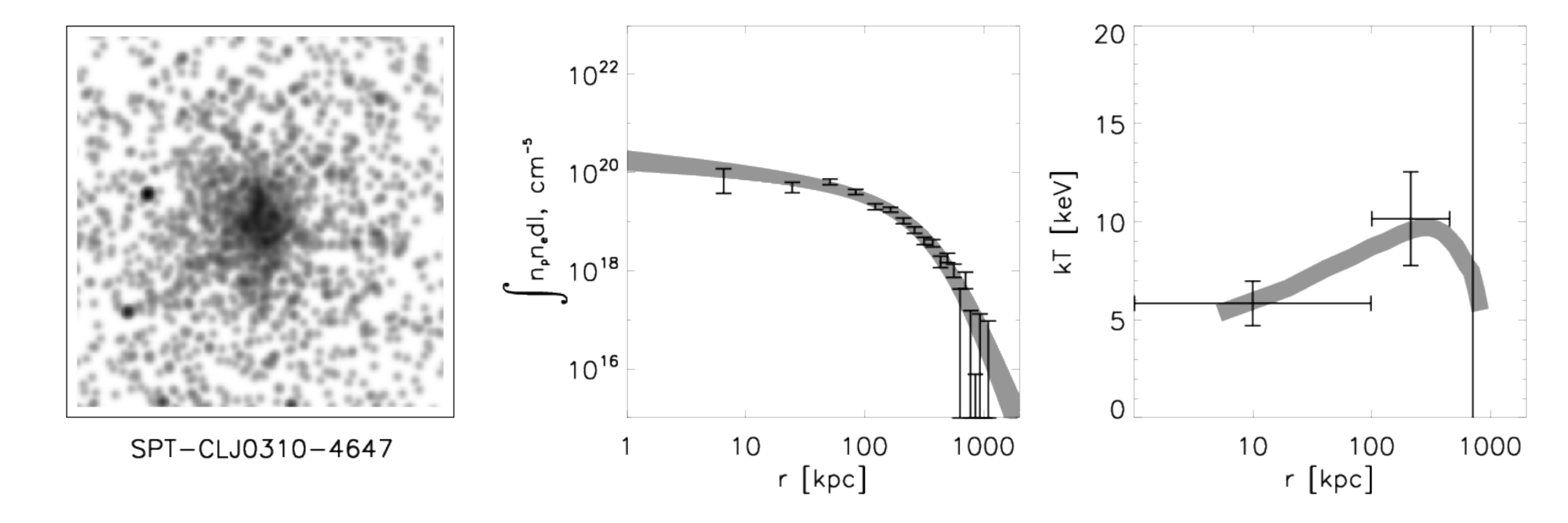} &
\includegraphics[width=0.49\textwidth]{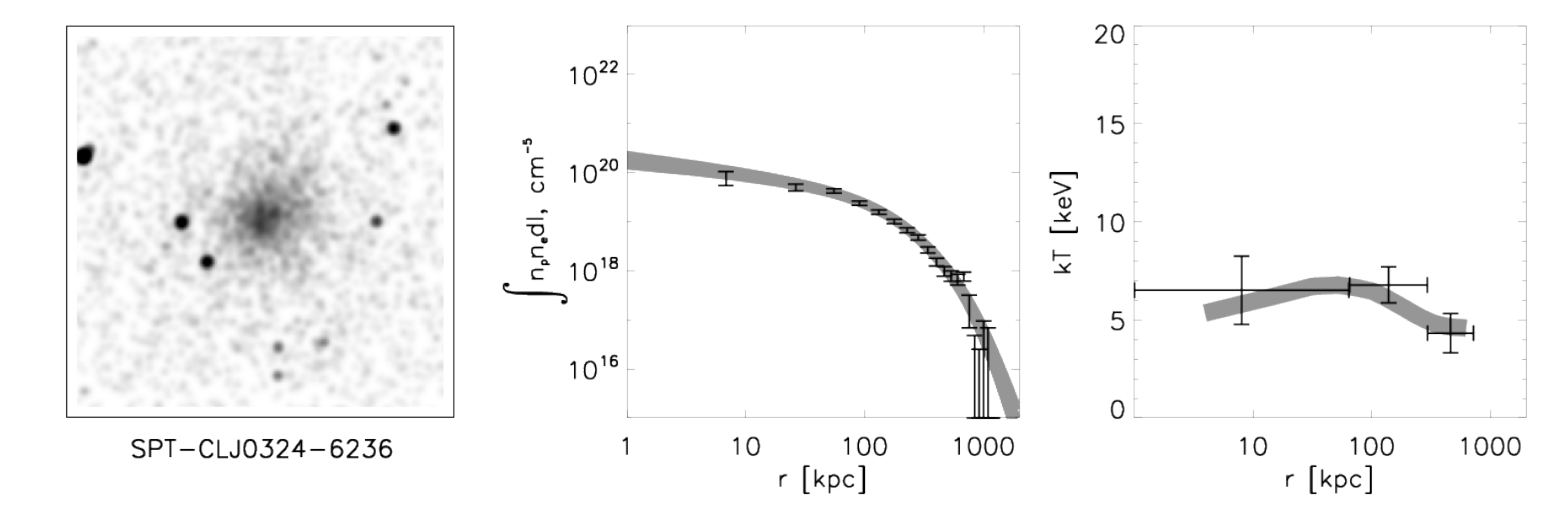} \\
\end{tabular}
\end{figure*}

\begin{figure*}[h!]
\centering
\begin{tabular}{c c}
\vspace{-0.05in}
\includegraphics[width=0.49\textwidth]{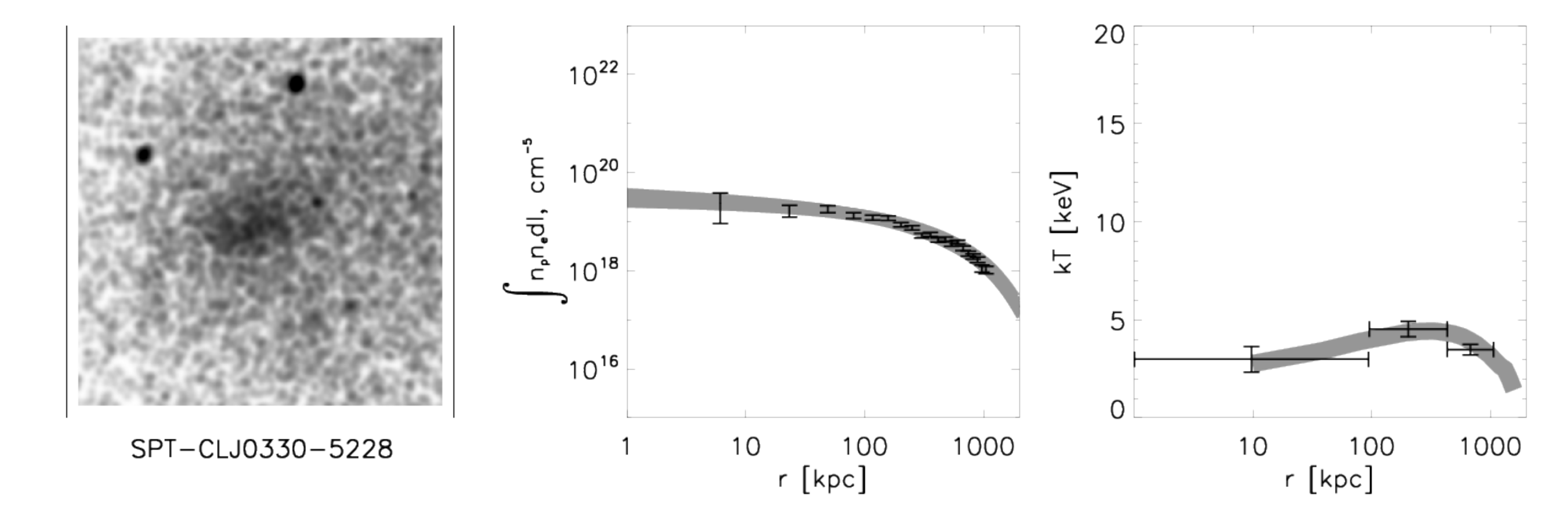} &
\includegraphics[width=0.49\textwidth]{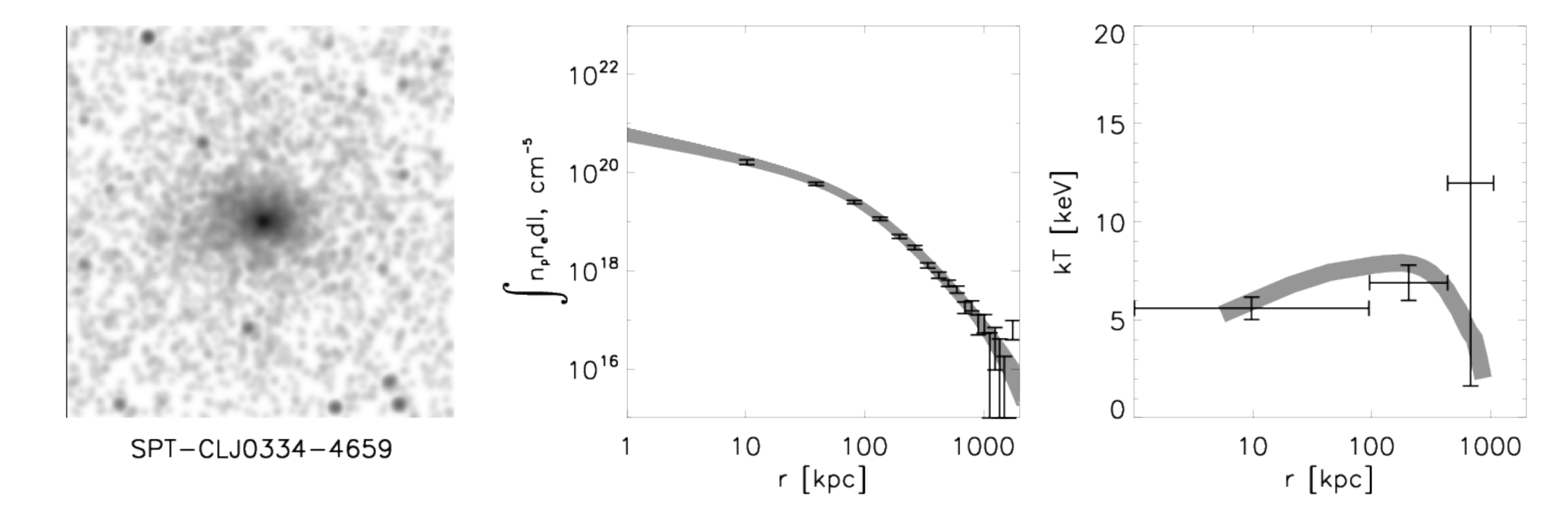} \\
\vspace{-0.05in}
\includegraphics[width=0.49\textwidth]{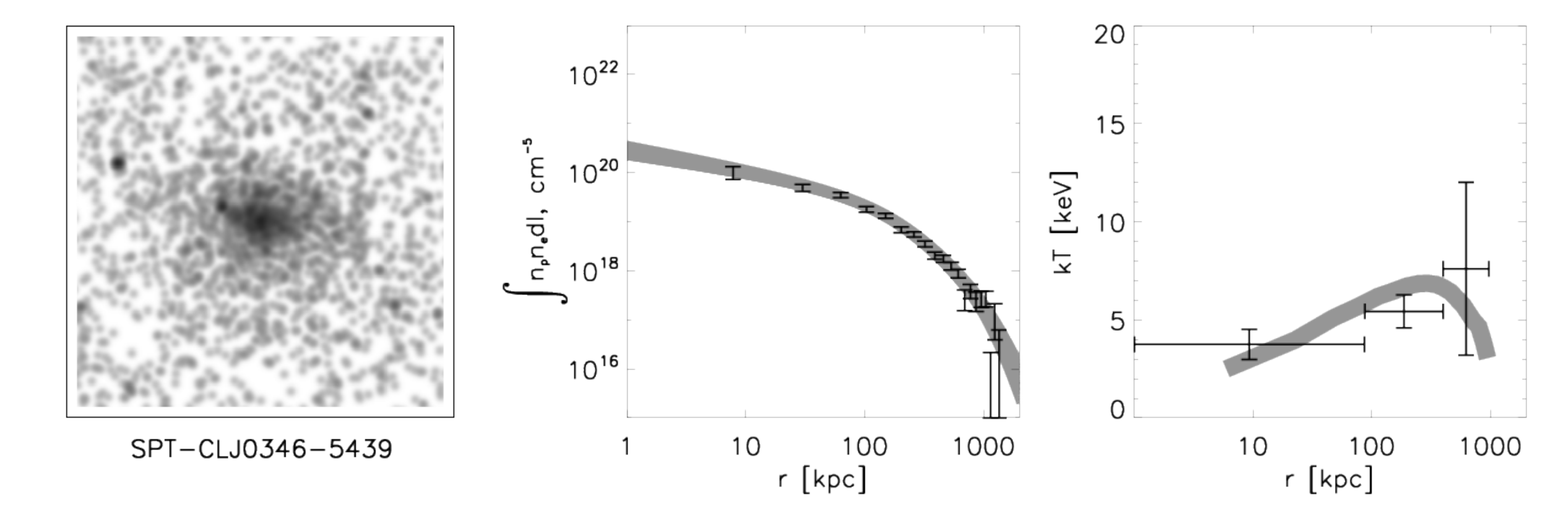} &
\includegraphics[width=0.49\textwidth]{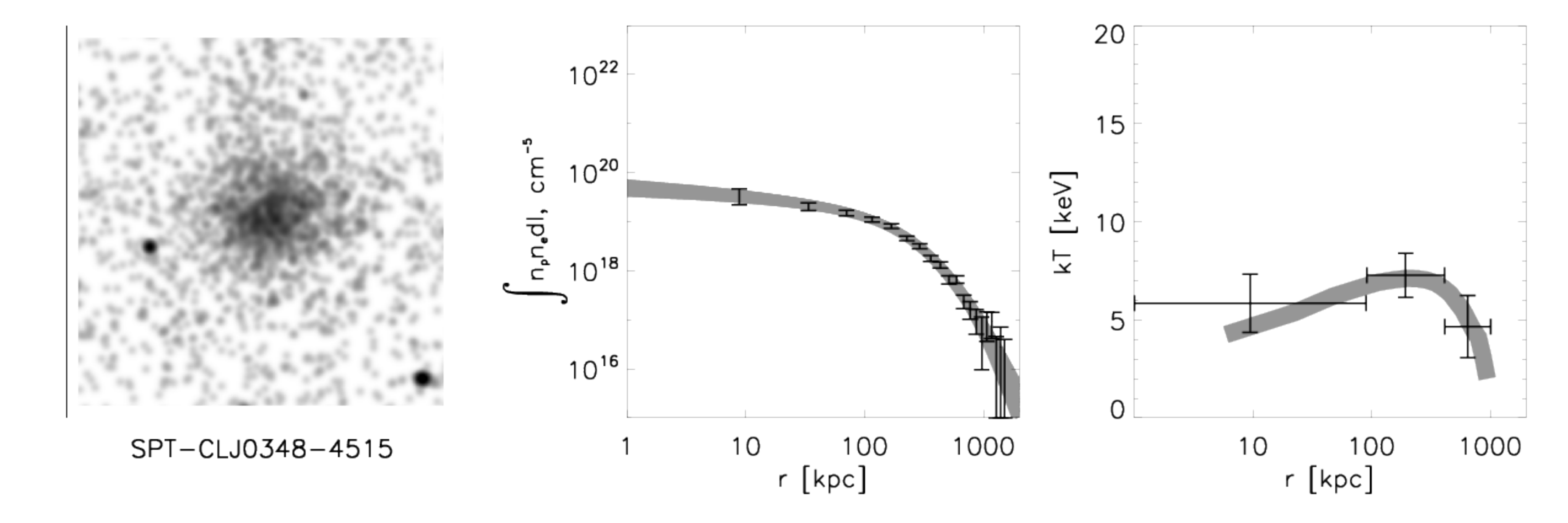} \\
\vspace{-0.05in}
\includegraphics[width=0.49\textwidth]{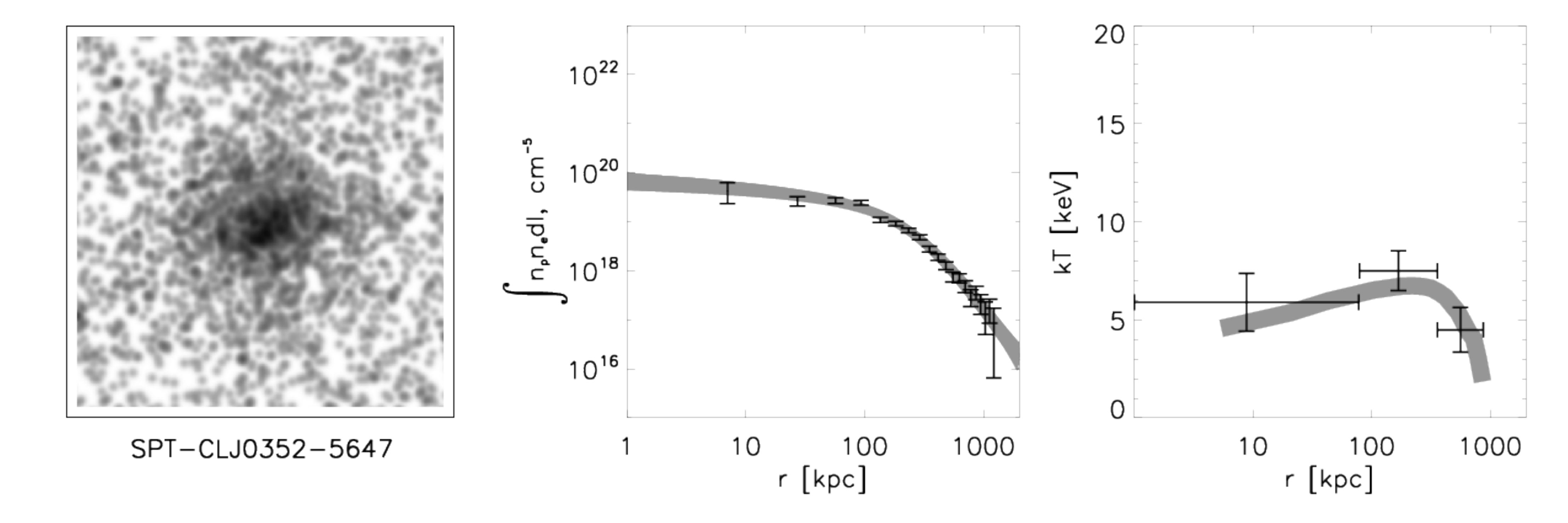} &
\includegraphics[width=0.49\textwidth]{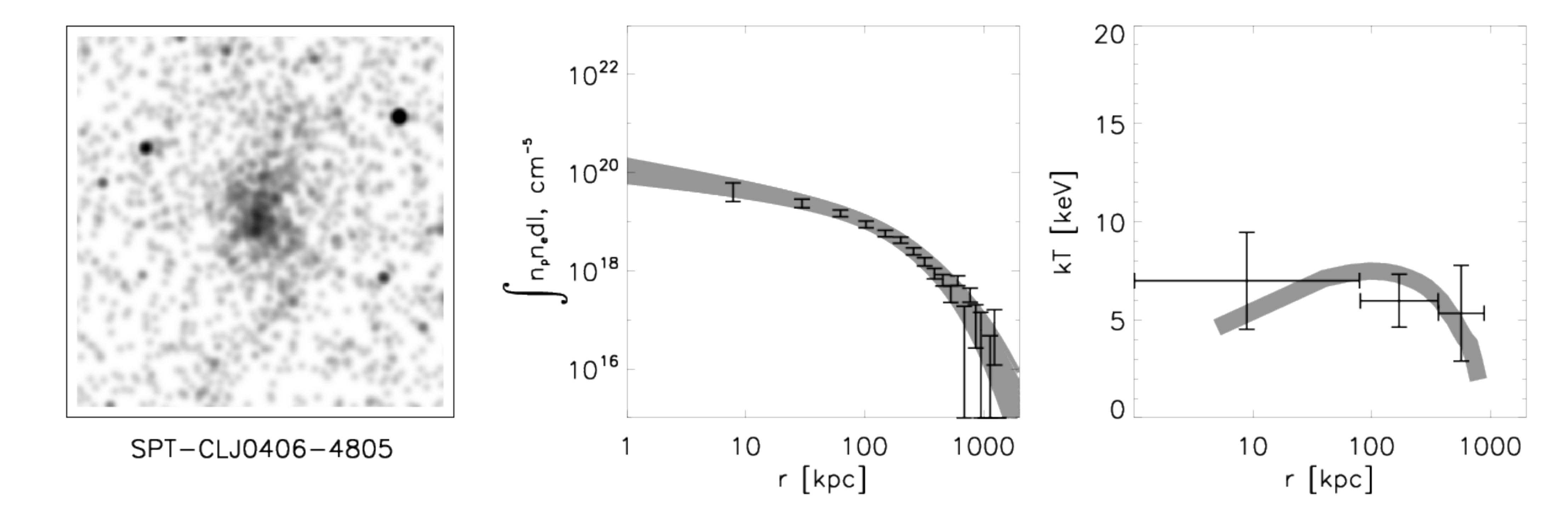} \\
\vspace{-0.05in}
\includegraphics[width=0.49\textwidth]{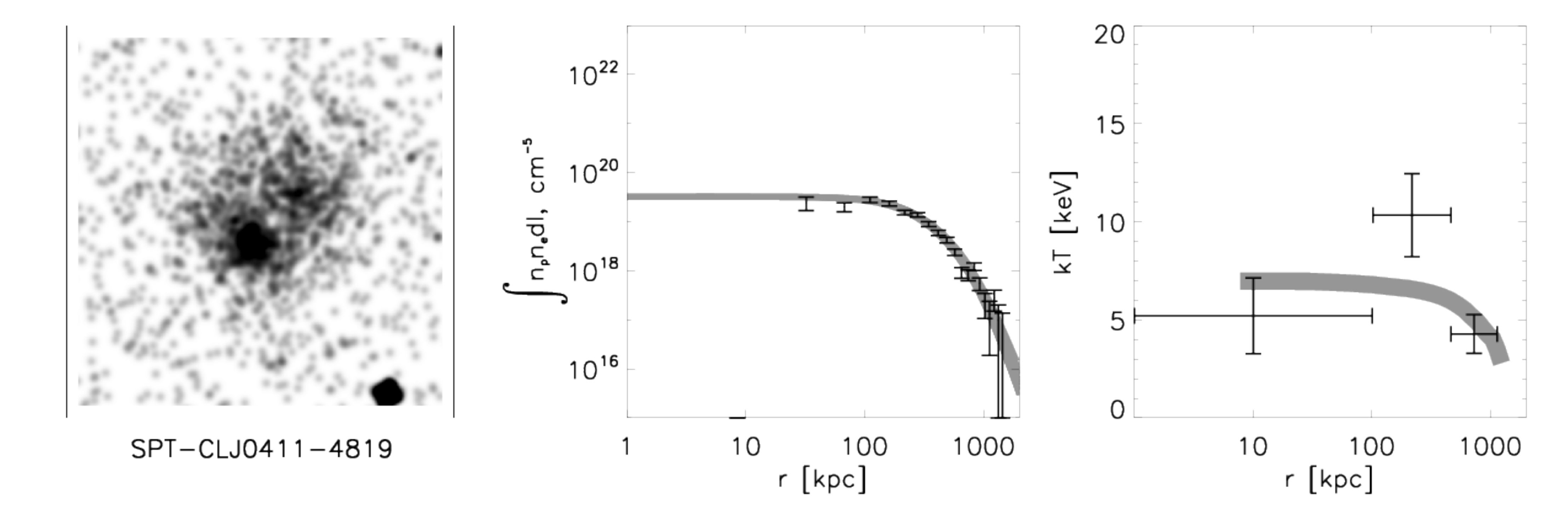} &
\includegraphics[width=0.49\textwidth]{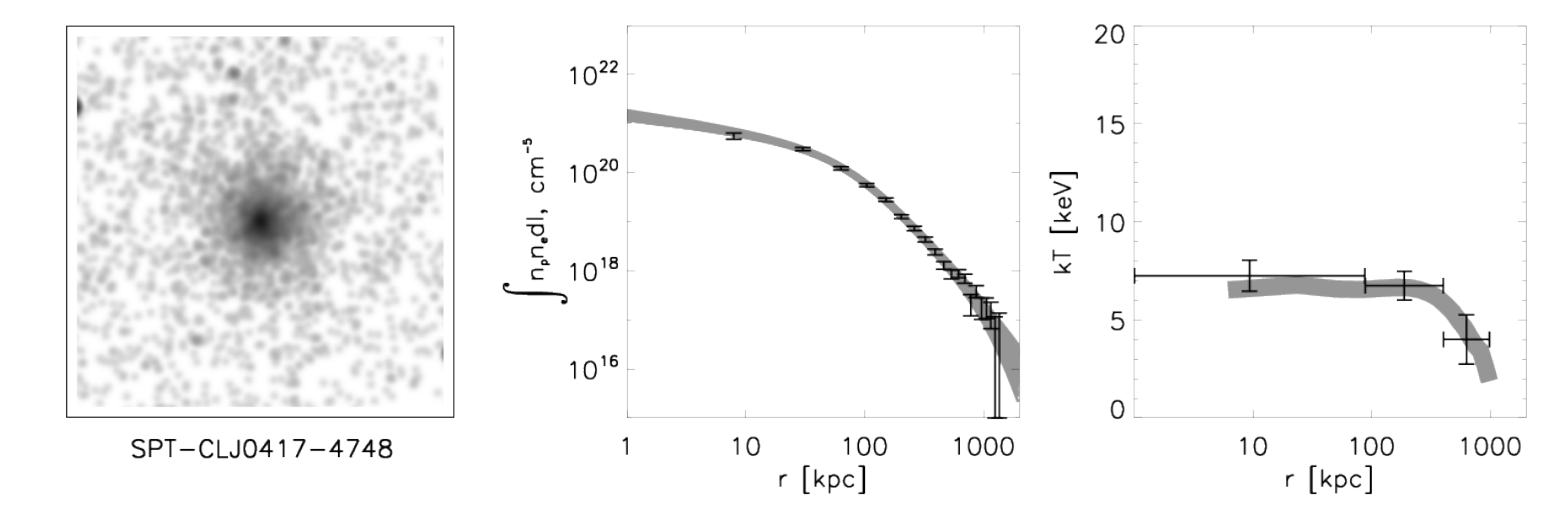} \\
\vspace{-0.05in}
\includegraphics[width=0.49\textwidth]{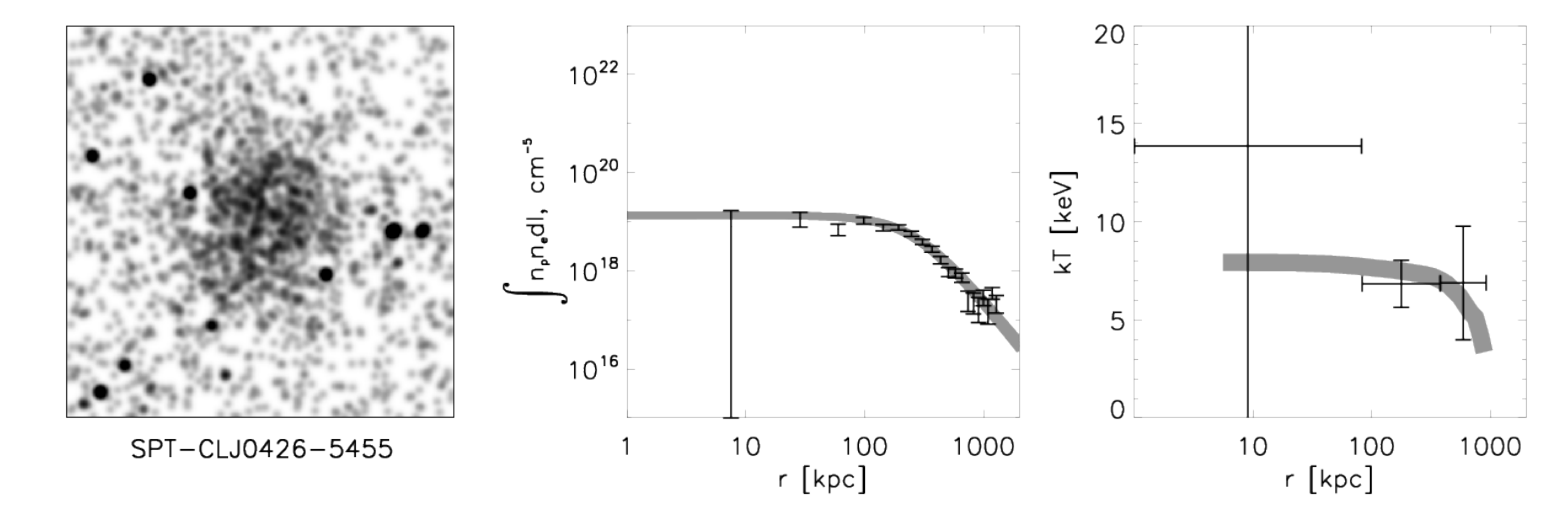} &
\includegraphics[width=0.49\textwidth]{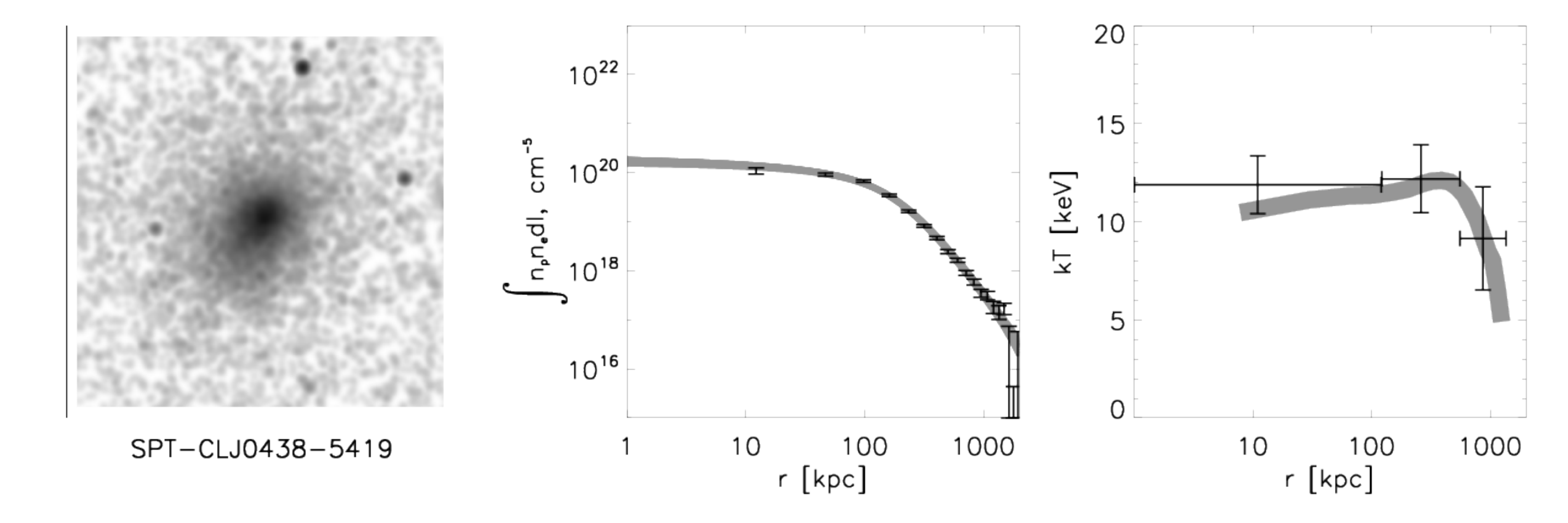} \\
\vspace{-0.05in}
\includegraphics[width=0.49\textwidth]{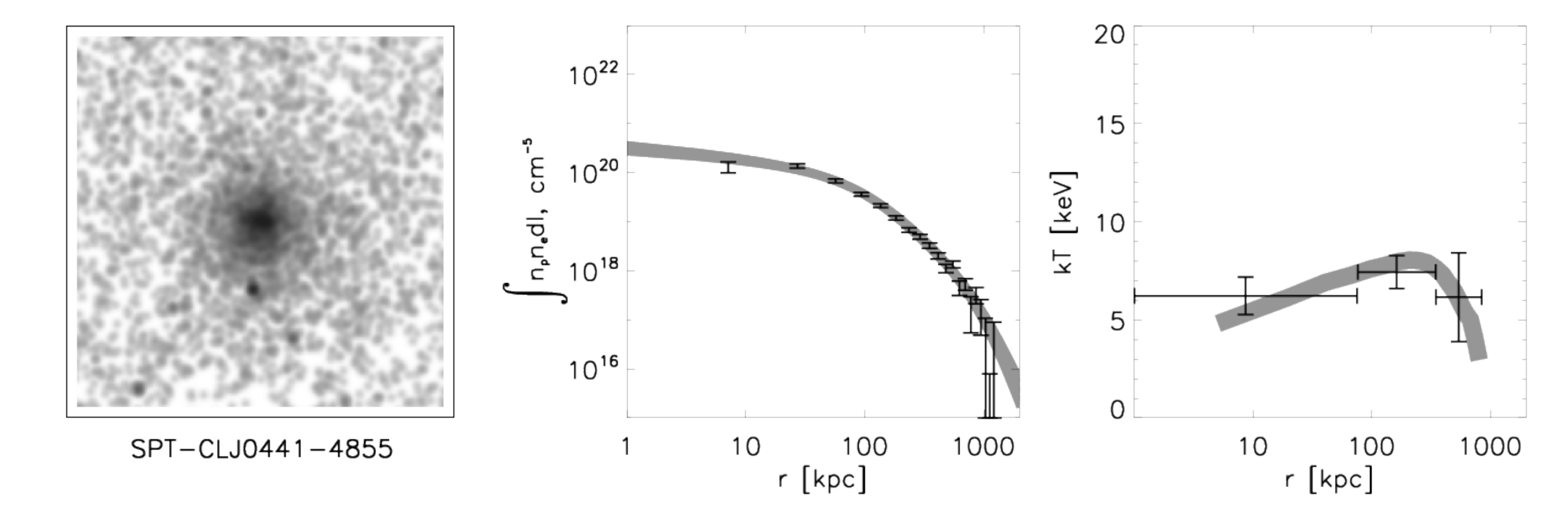} &
\includegraphics[width=0.49\textwidth]{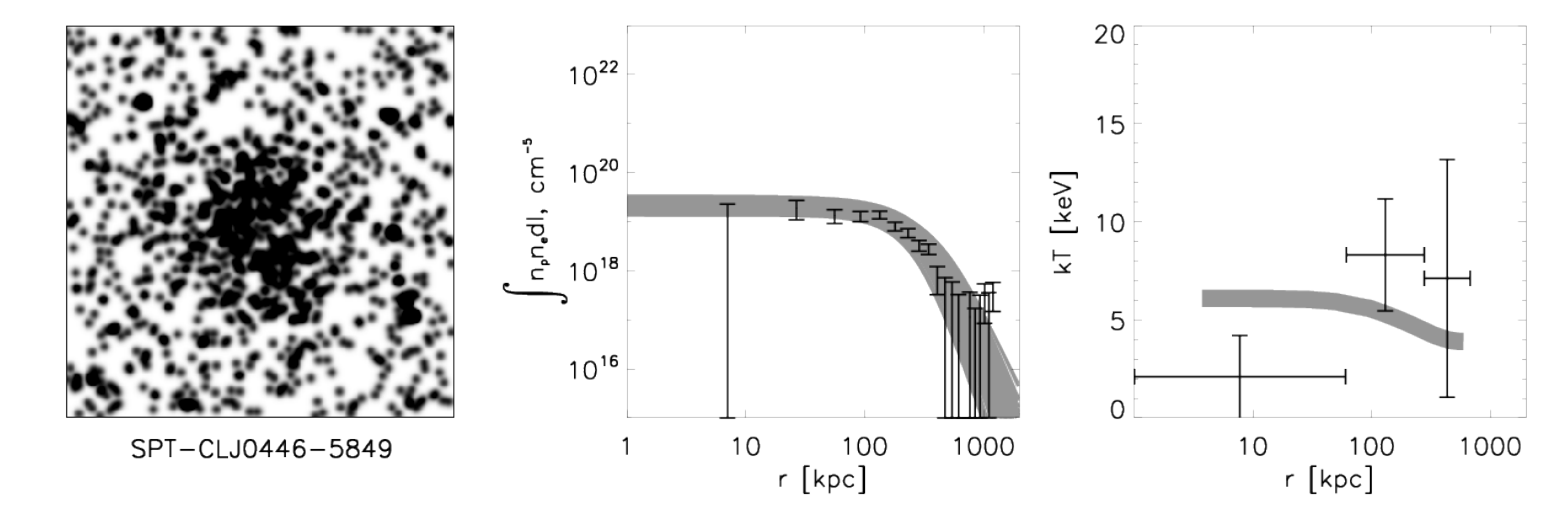} \\
\vspace{-0.05in}
\includegraphics[width=0.49\textwidth]{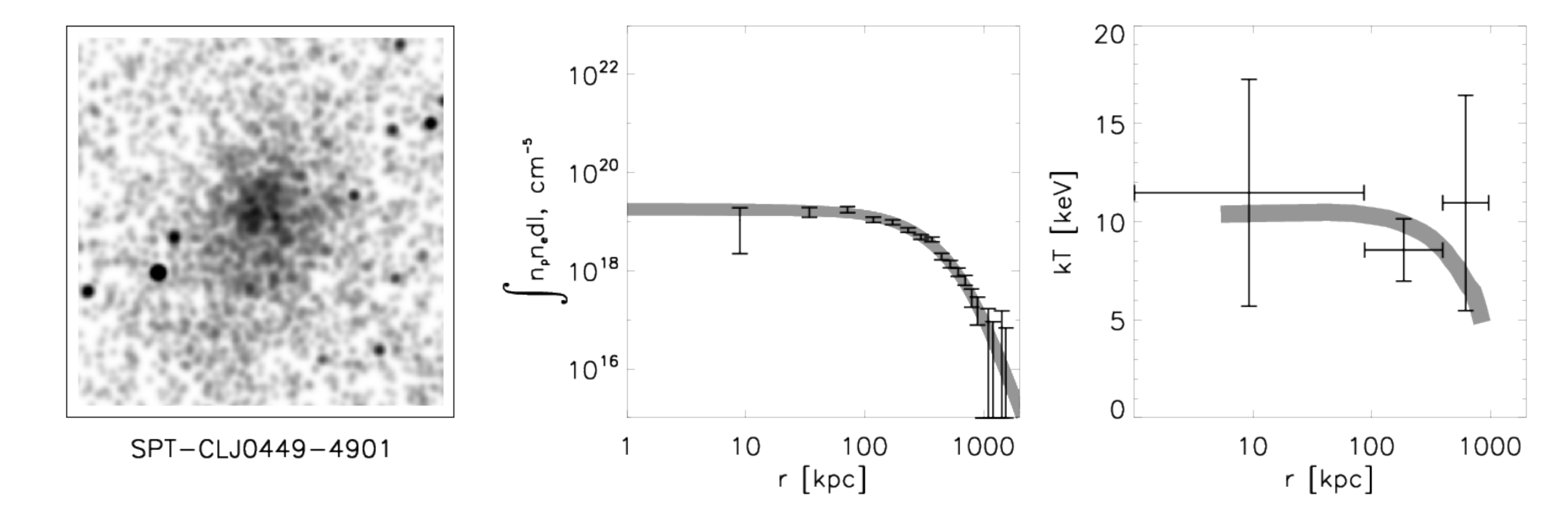} &
\includegraphics[width=0.49\textwidth]{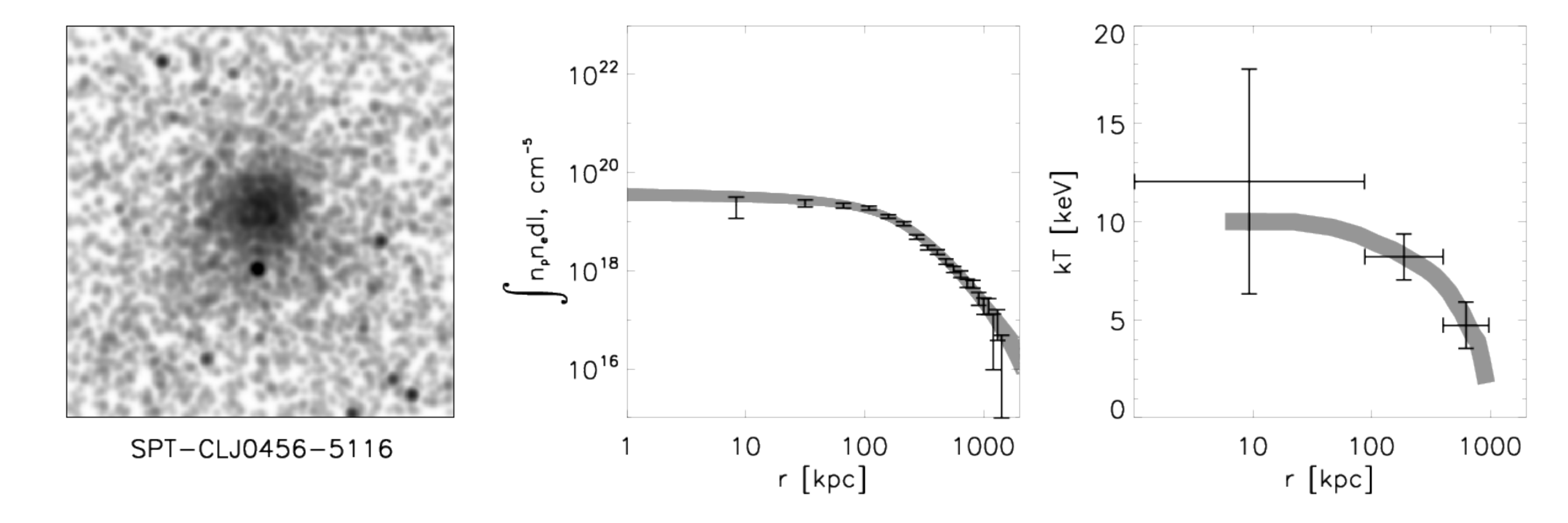} \\
\vspace{-0.05in}
\includegraphics[width=0.49\textwidth]{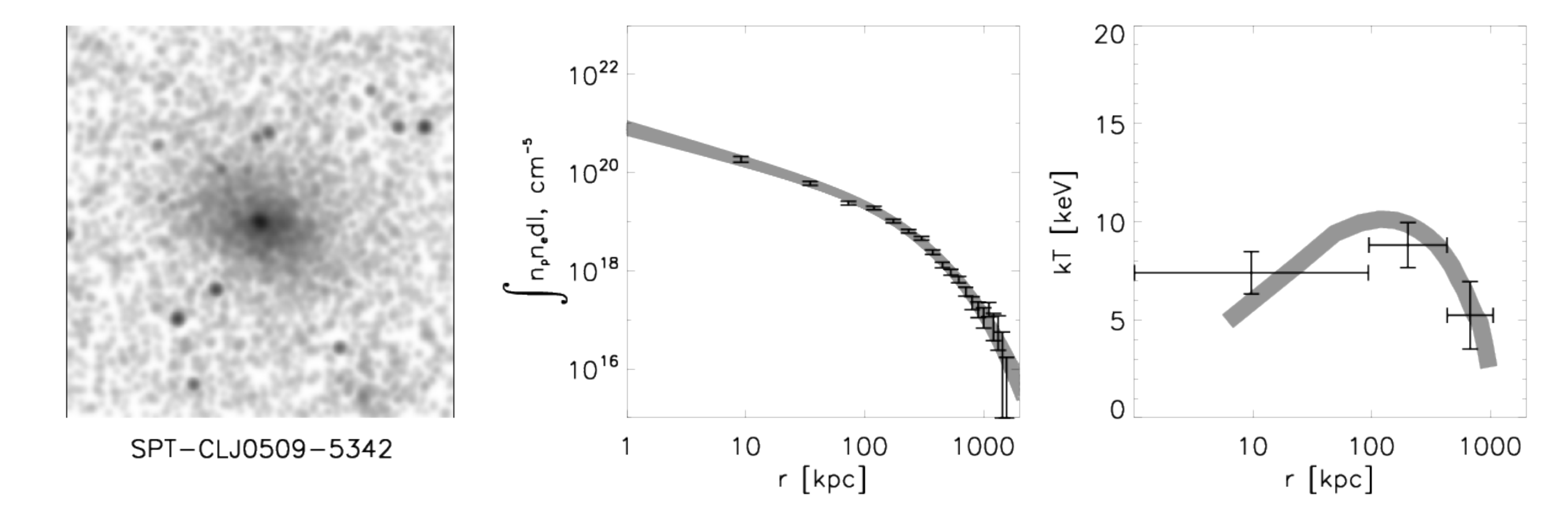} &
\includegraphics[width=0.49\textwidth]{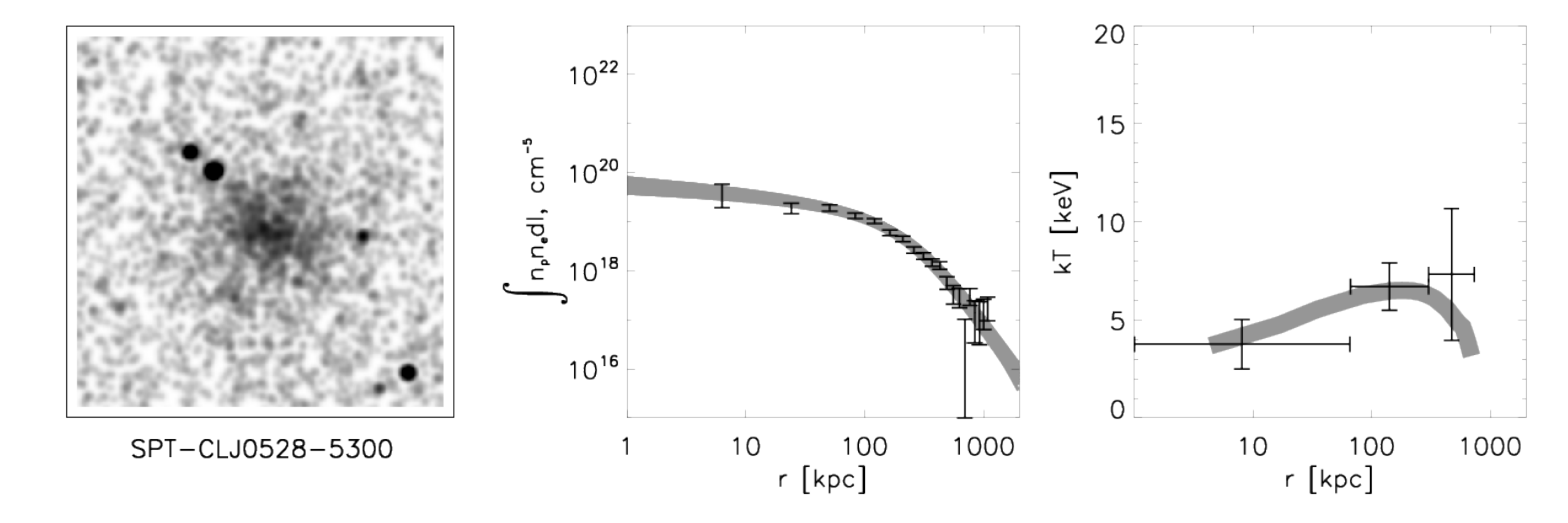} \\
\end{tabular}
\end{figure*}

\begin{figure*}[h!]
\centering
\begin{tabular}{c c}
\vspace{-0.05in}
\includegraphics[width=0.49\textwidth]{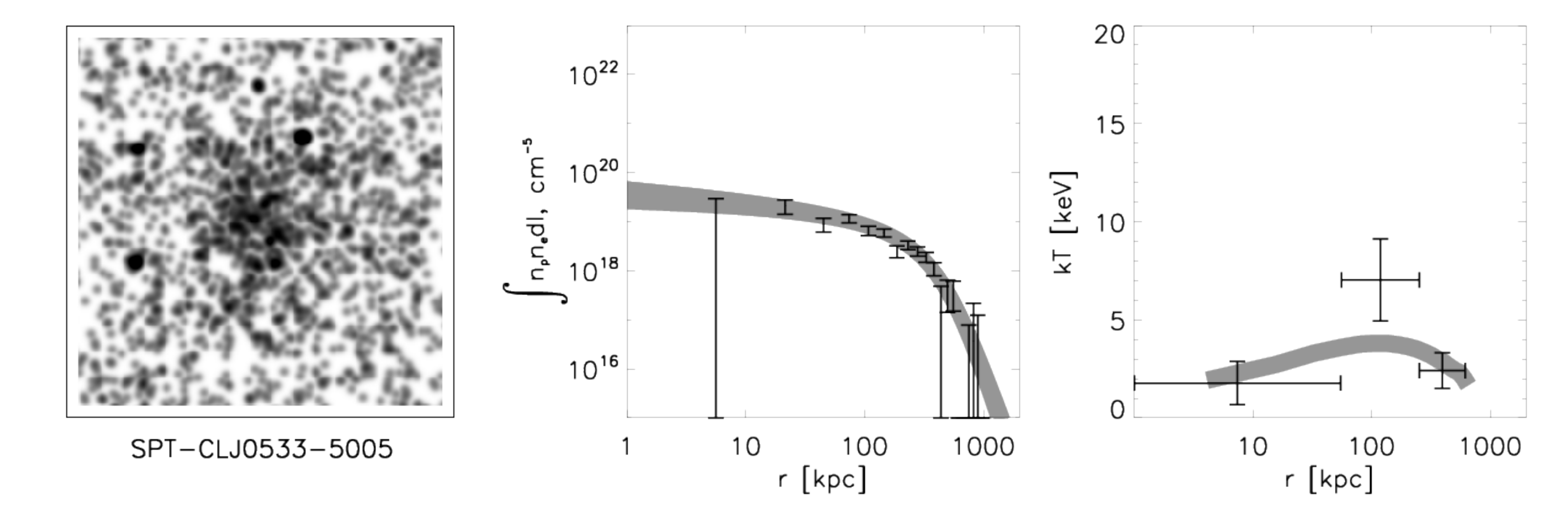} &
\includegraphics[width=0.49\textwidth]{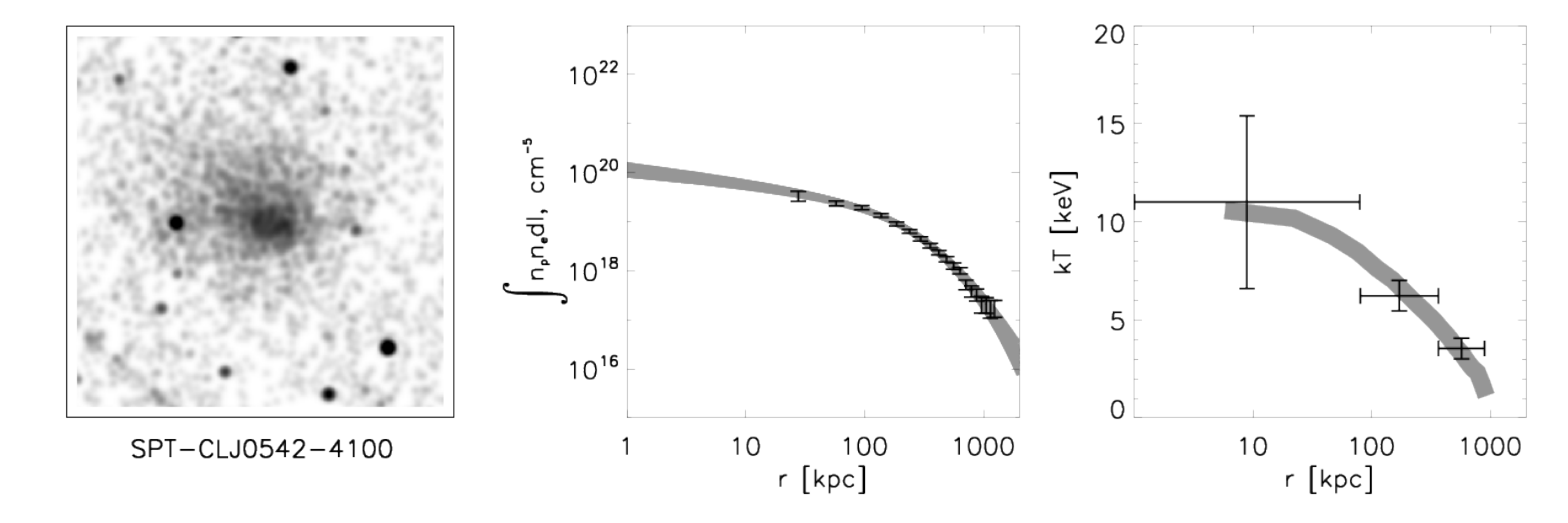} \\
\vspace{-0.05in}
\includegraphics[width=0.49\textwidth]{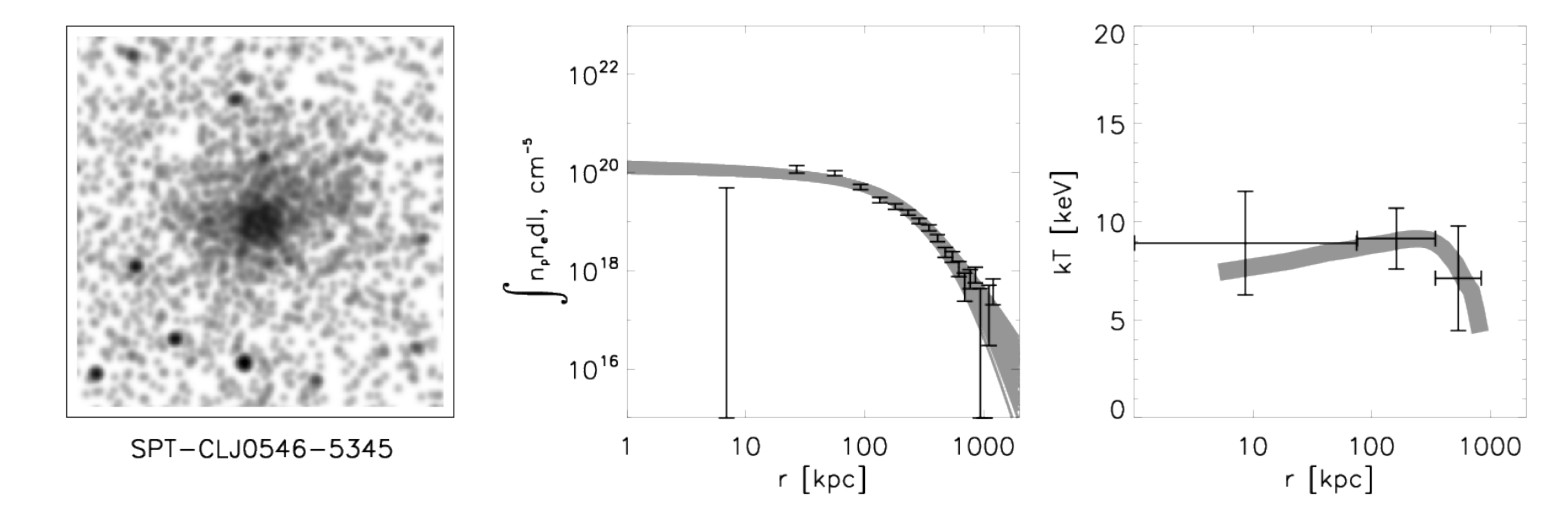} &
\includegraphics[width=0.49\textwidth]{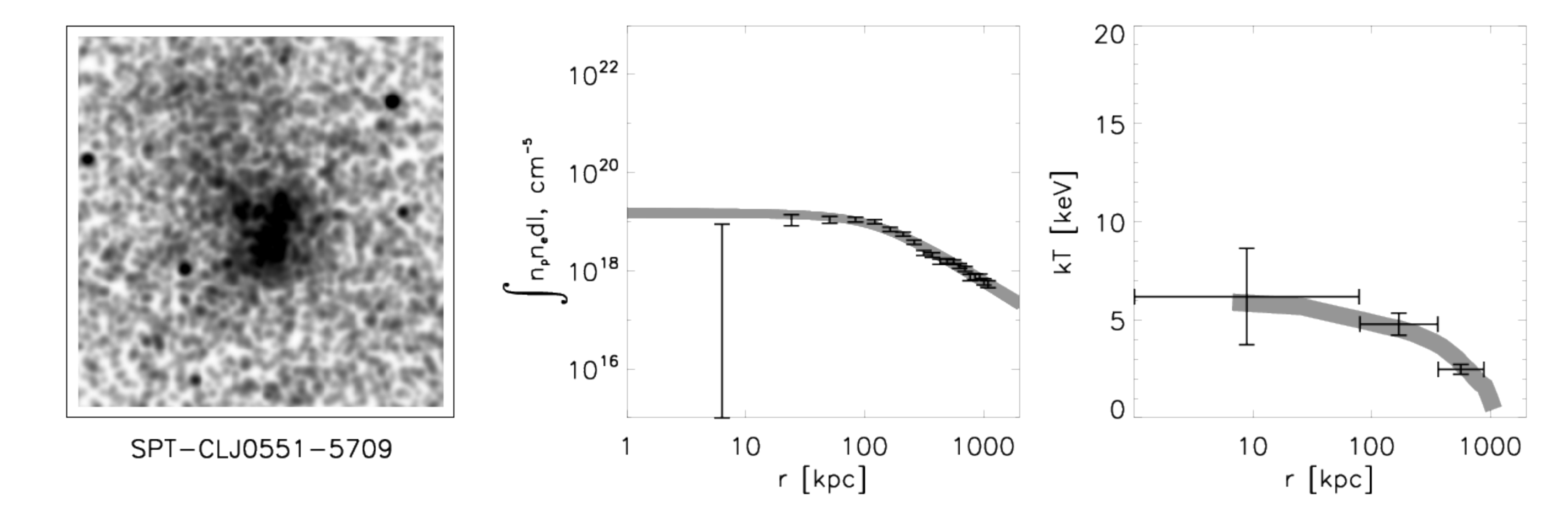} \\
\vspace{-0.05in}
\includegraphics[width=0.49\textwidth]{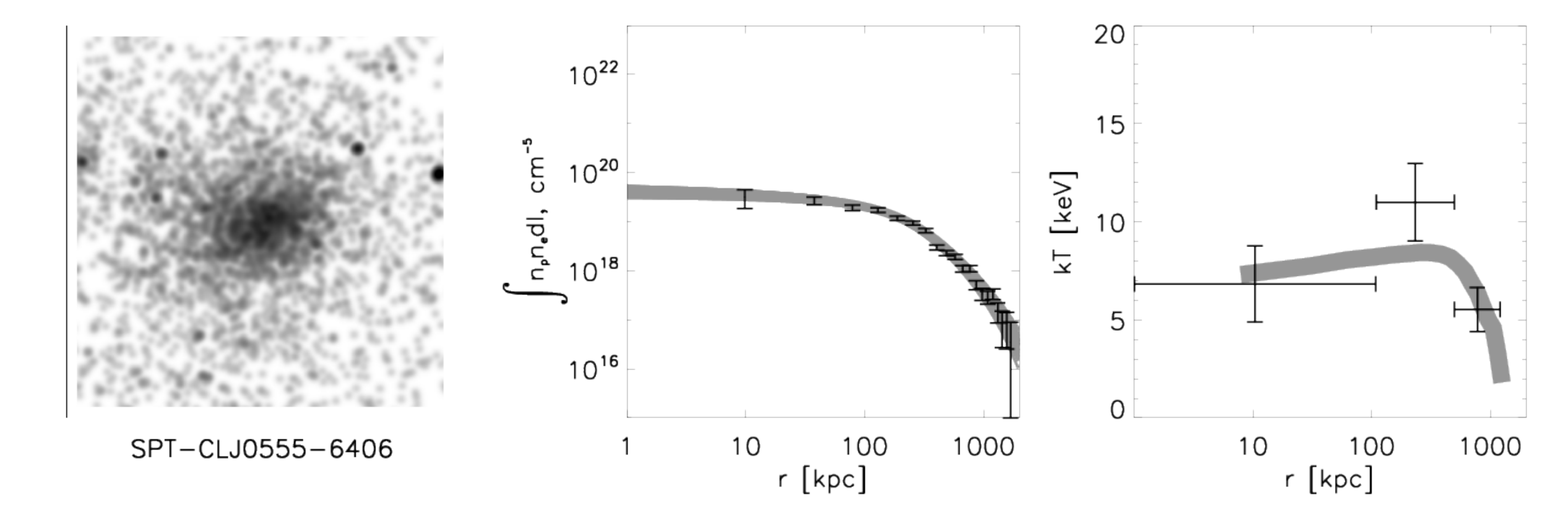} &
\includegraphics[width=0.49\textwidth]{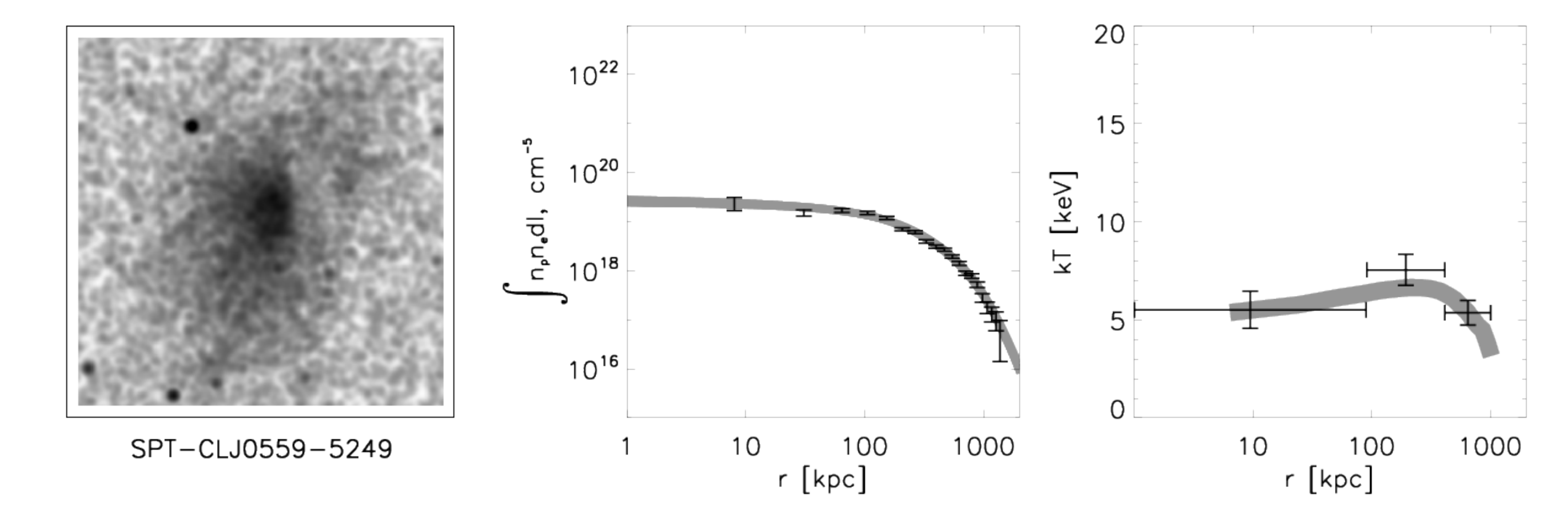} \\
\vspace{-0.05in}
\includegraphics[width=0.49\textwidth]{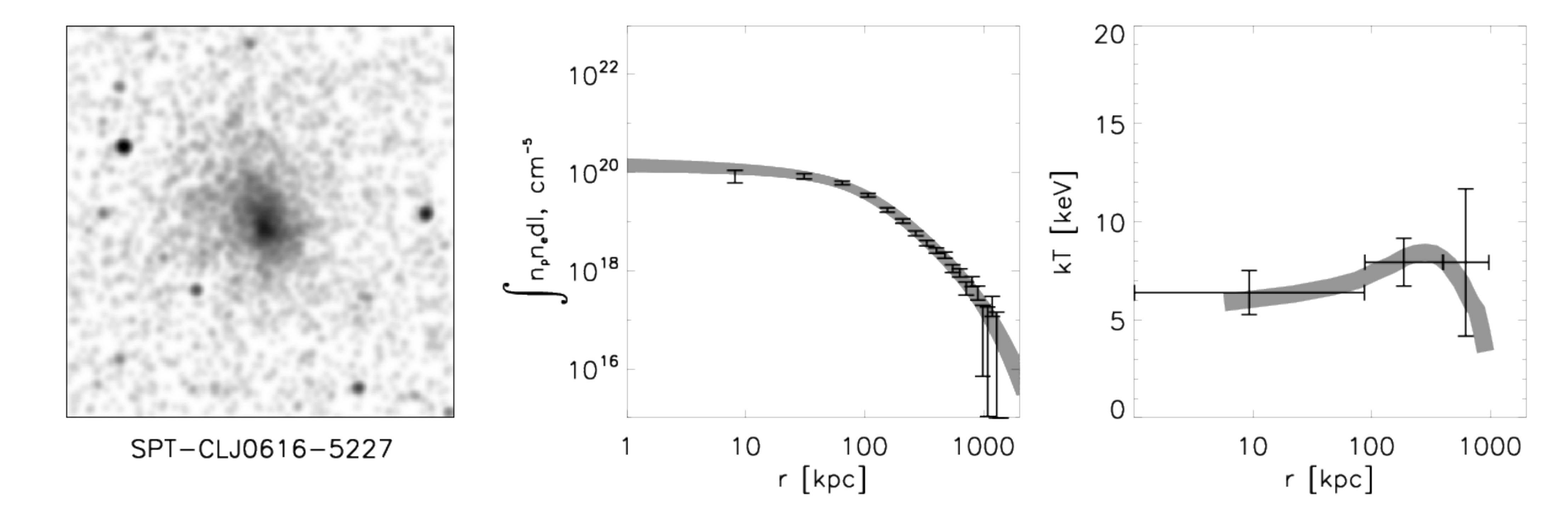} &
\includegraphics[width=0.49\textwidth]{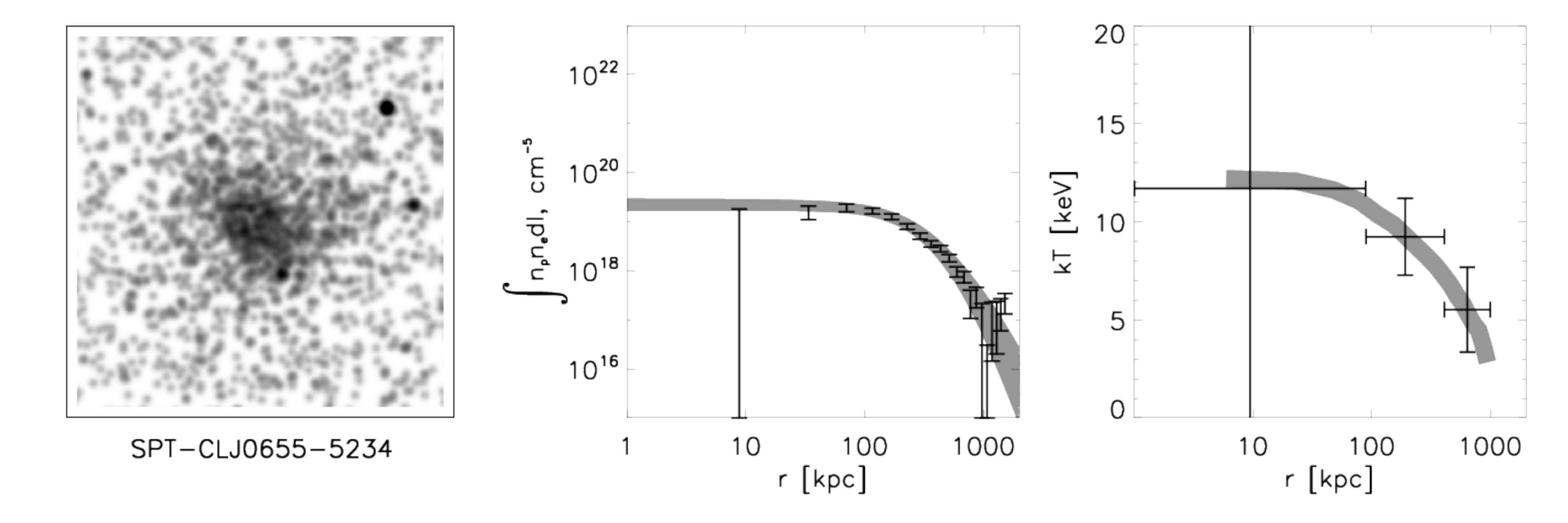} \\
\vspace{-0.05in}
\includegraphics[width=0.49\textwidth]{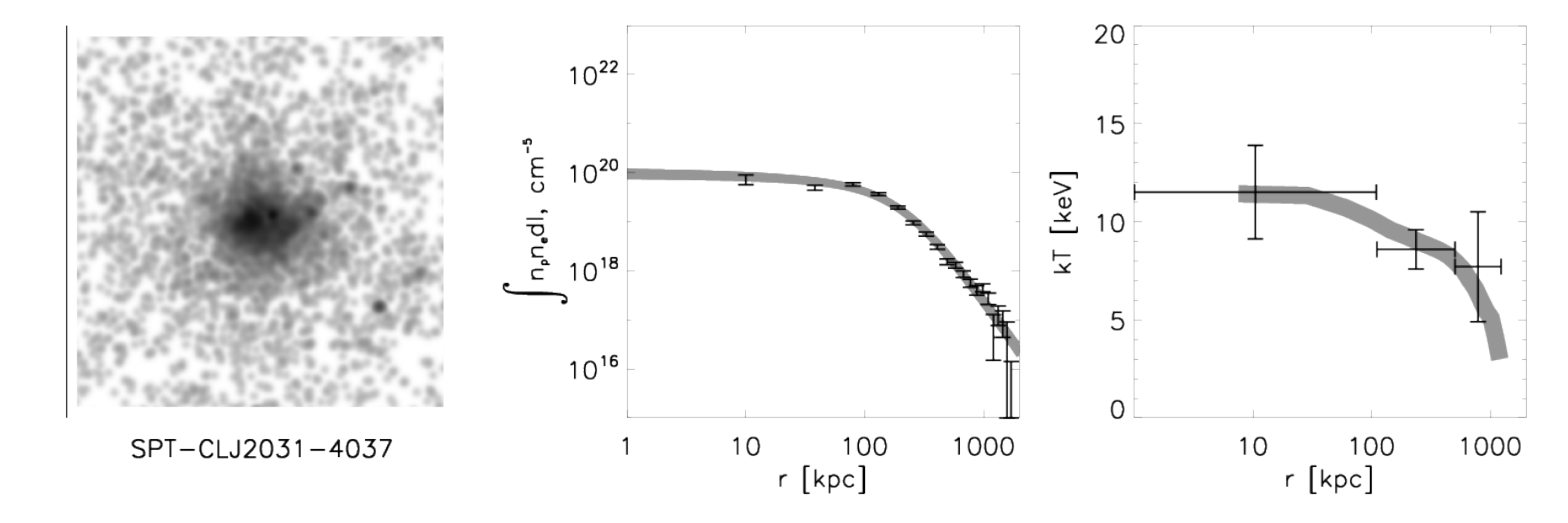} &
\includegraphics[width=0.49\textwidth]{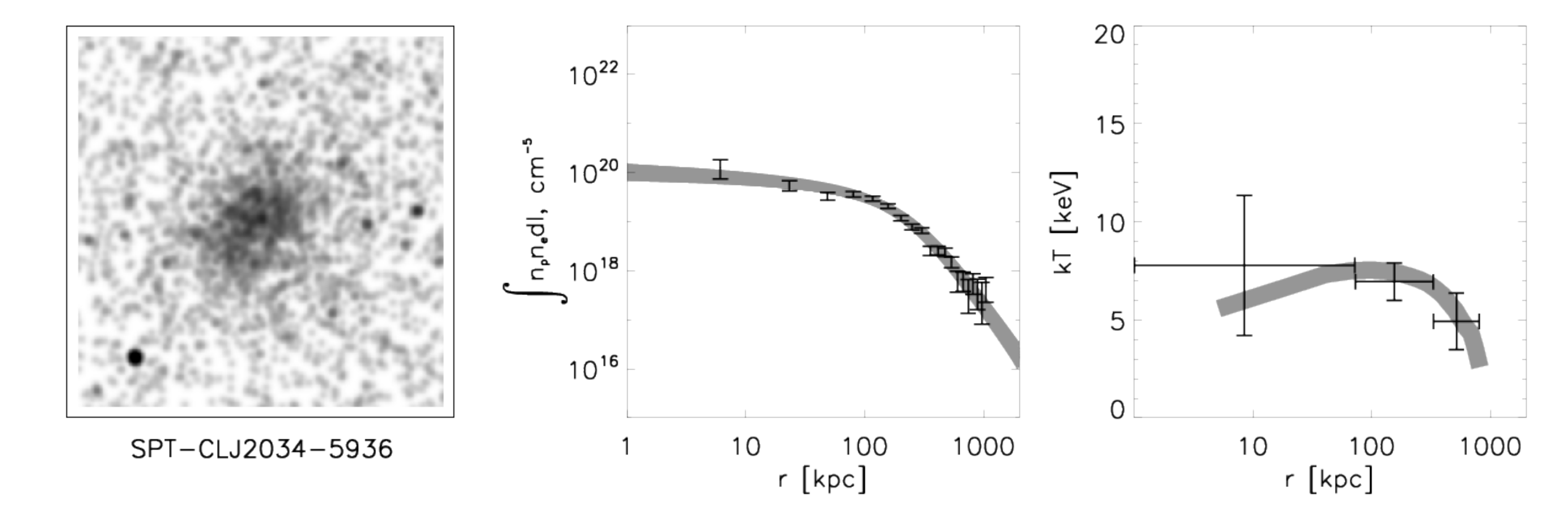} \\
\vspace{-0.05in}
\includegraphics[width=0.49\textwidth]{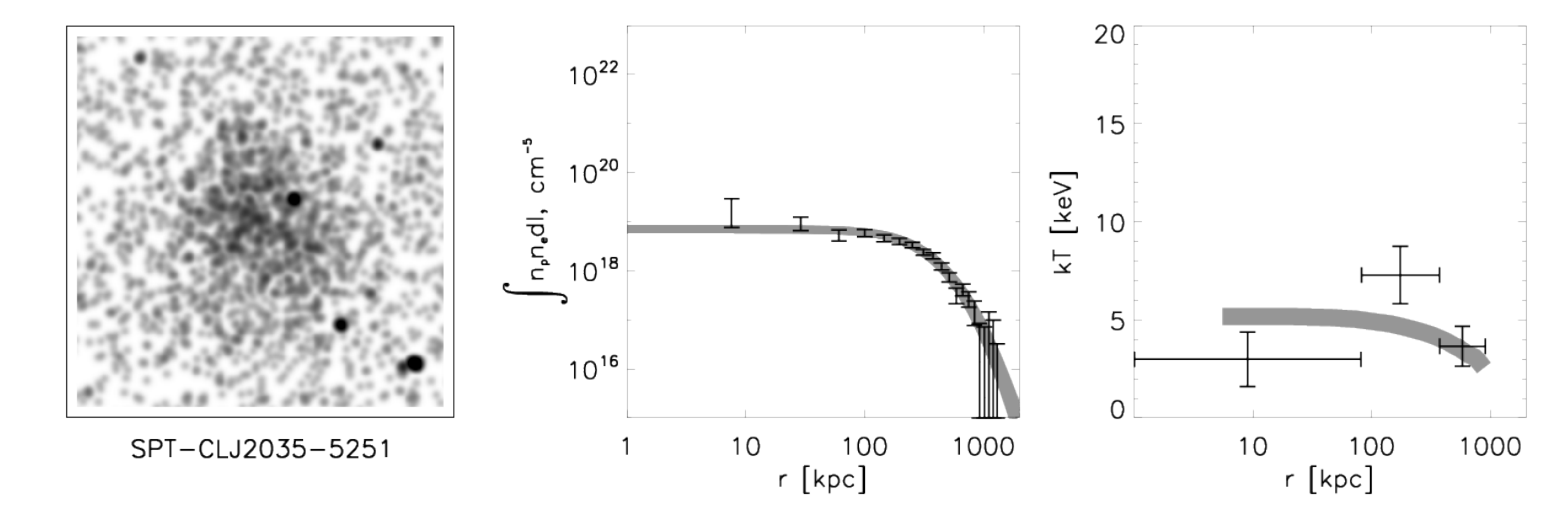} &
\includegraphics[width=0.49\textwidth]{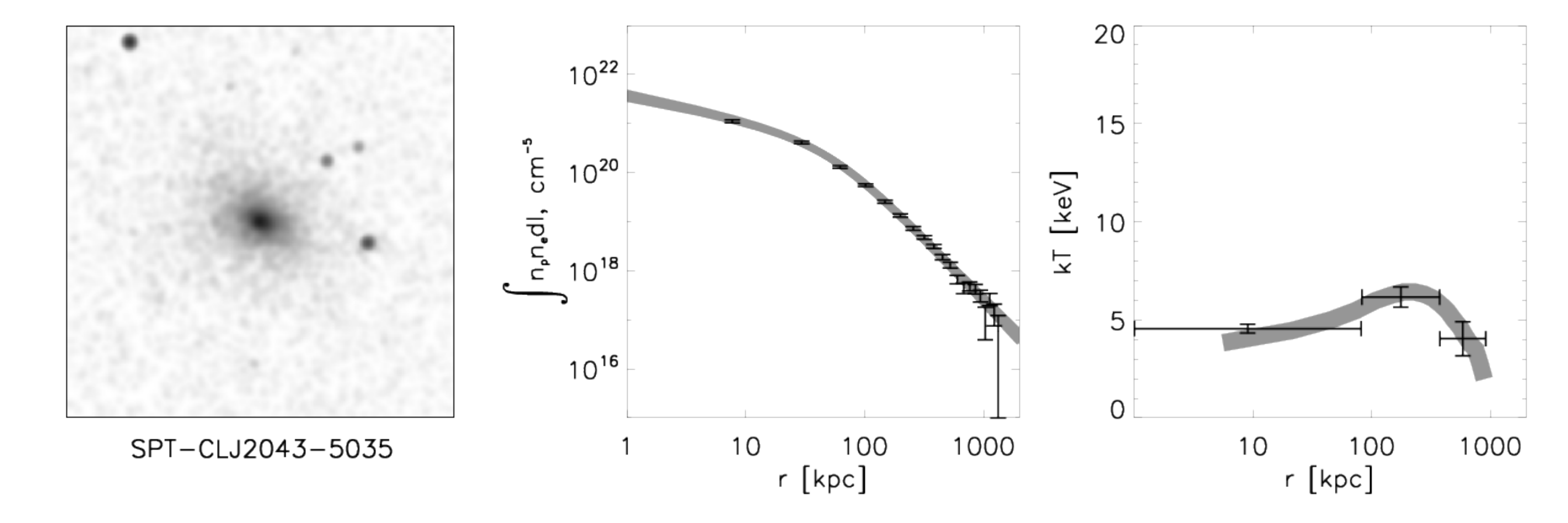} \\
\vspace{-0.05in}
\includegraphics[width=0.49\textwidth]{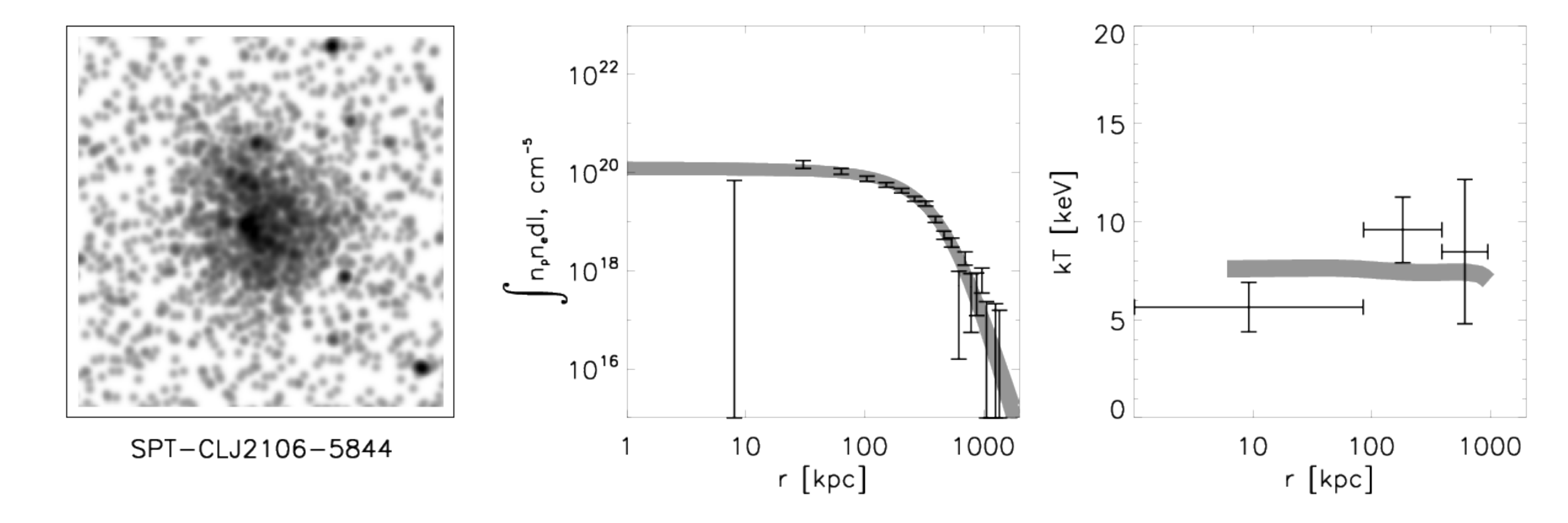} &
\includegraphics[width=0.49\textwidth]{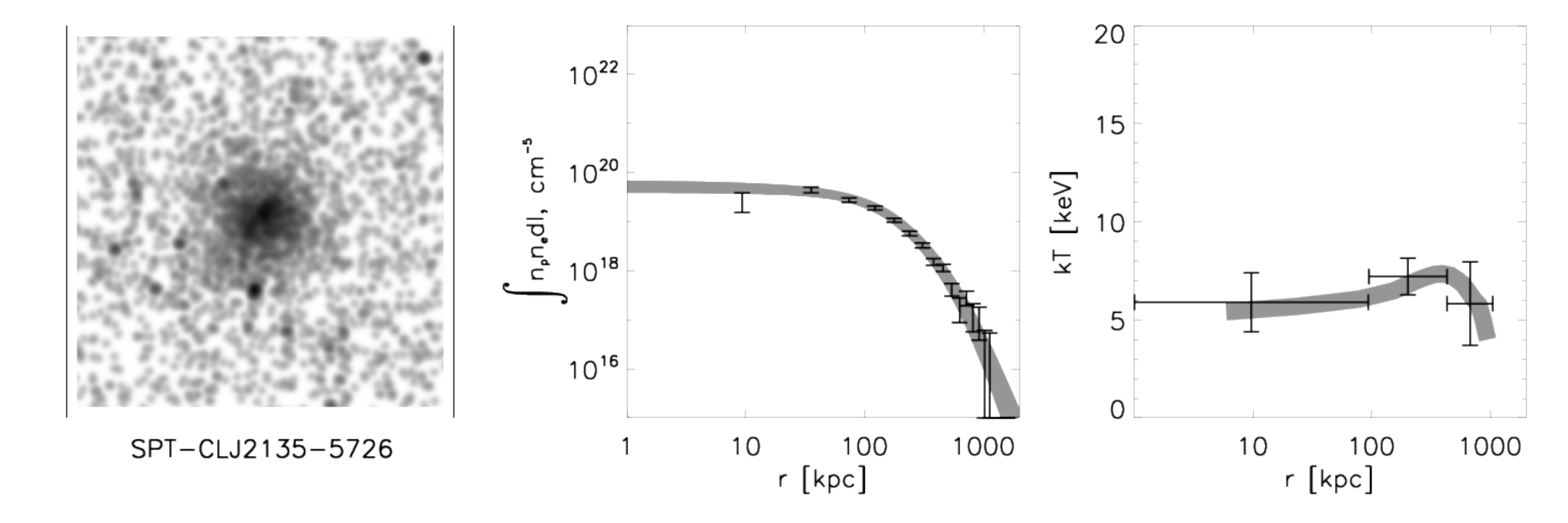} \\
\vspace{-0.05in}
\includegraphics[width=0.49\textwidth]{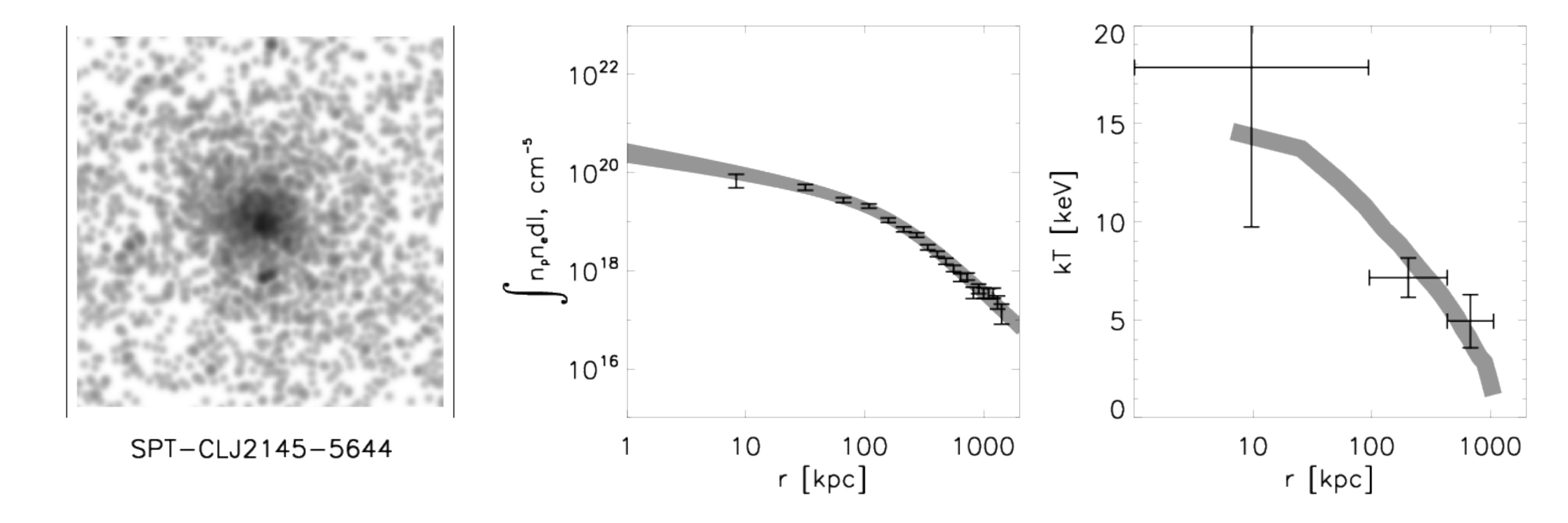} &
\includegraphics[width=0.49\textwidth]{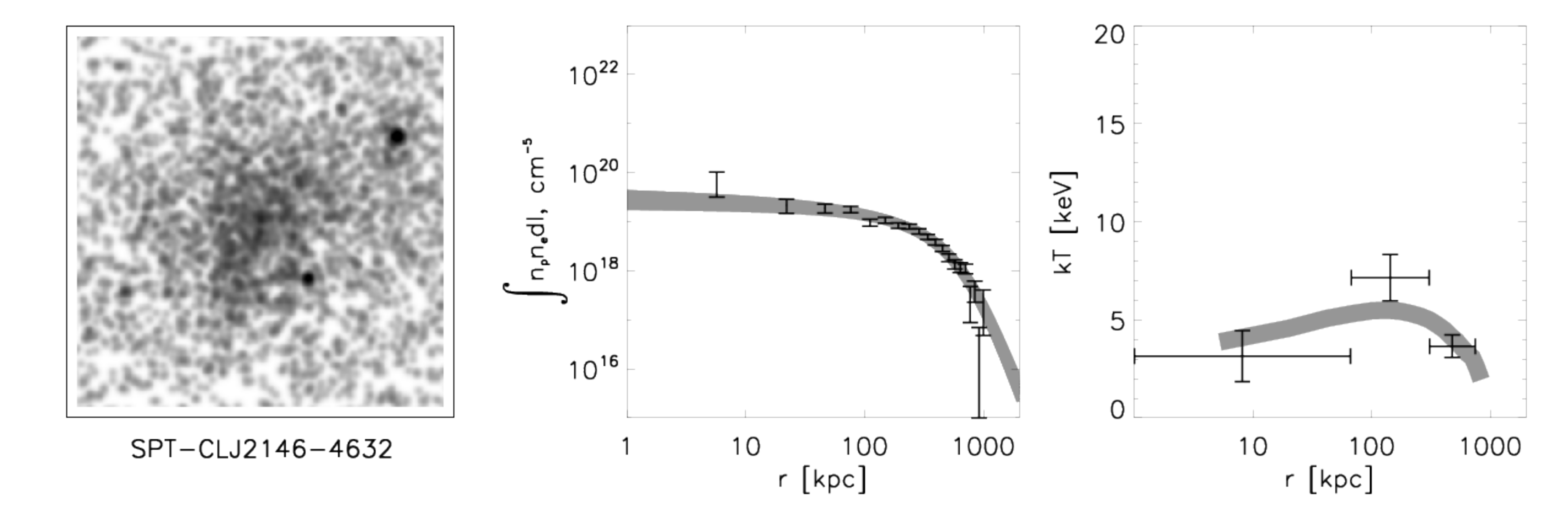} \\
\end{tabular}
\end{figure*}

\begin{figure*}[h!]
\centering
\begin{tabular}{c c}
\vspace{-0.05in}
\includegraphics[width=0.49\textwidth]{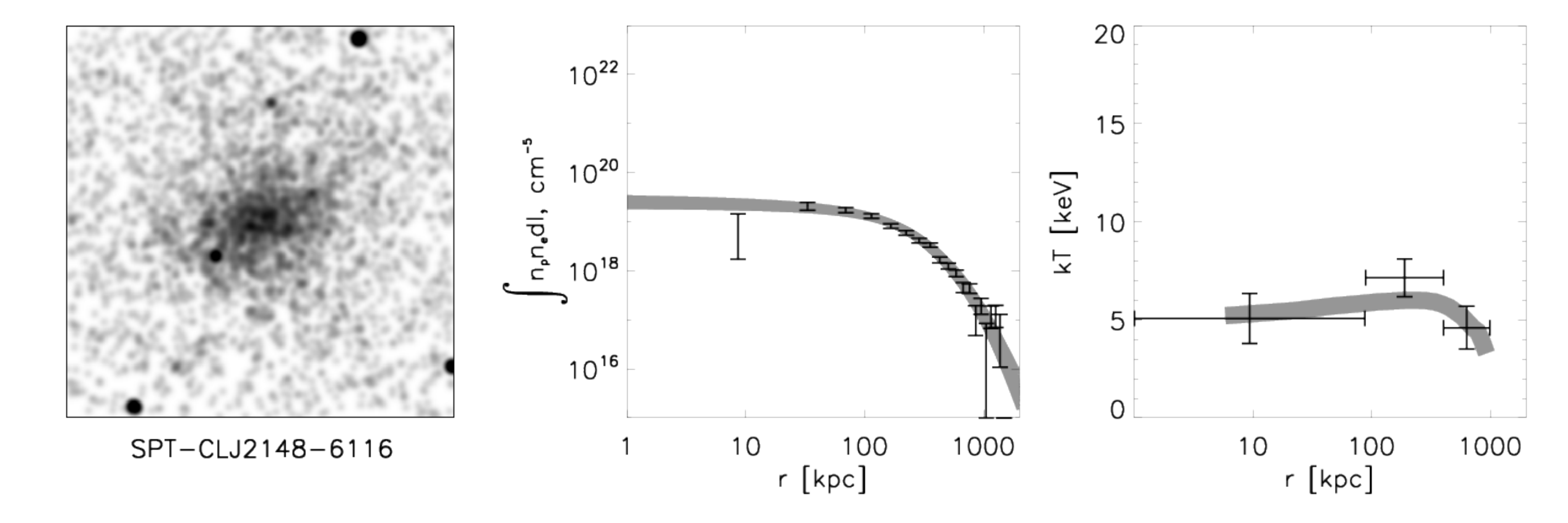} &
\includegraphics[width=0.49\textwidth]{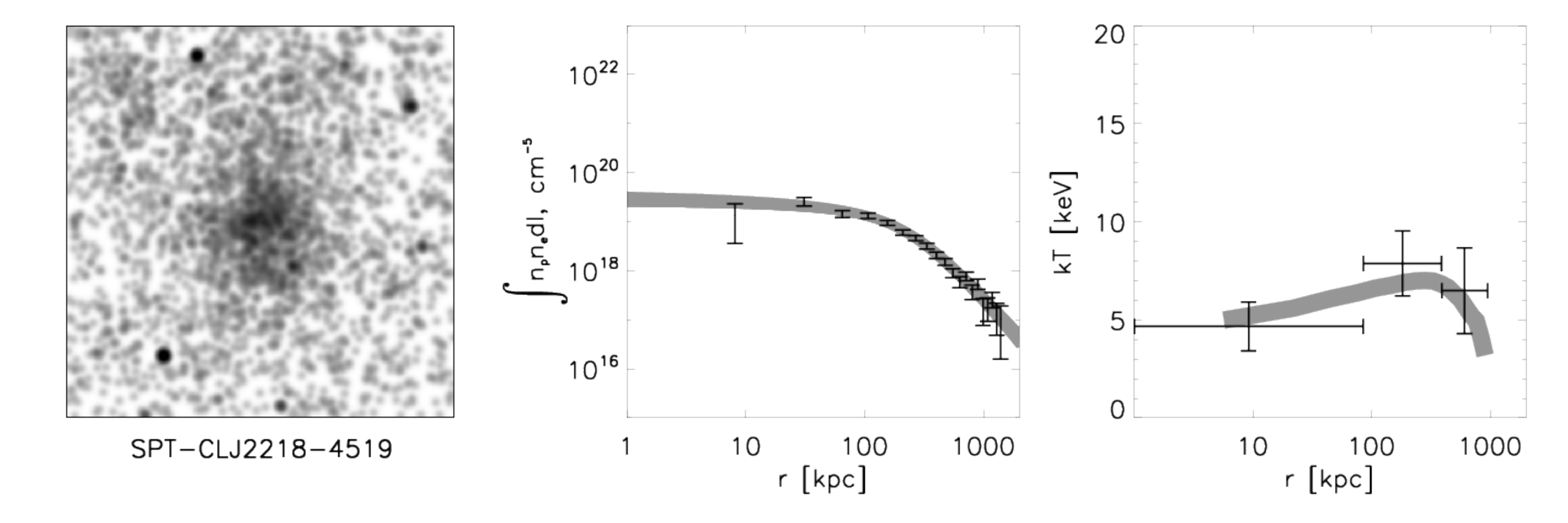} \\
\vspace{-0.05in}
\includegraphics[width=0.49\textwidth]{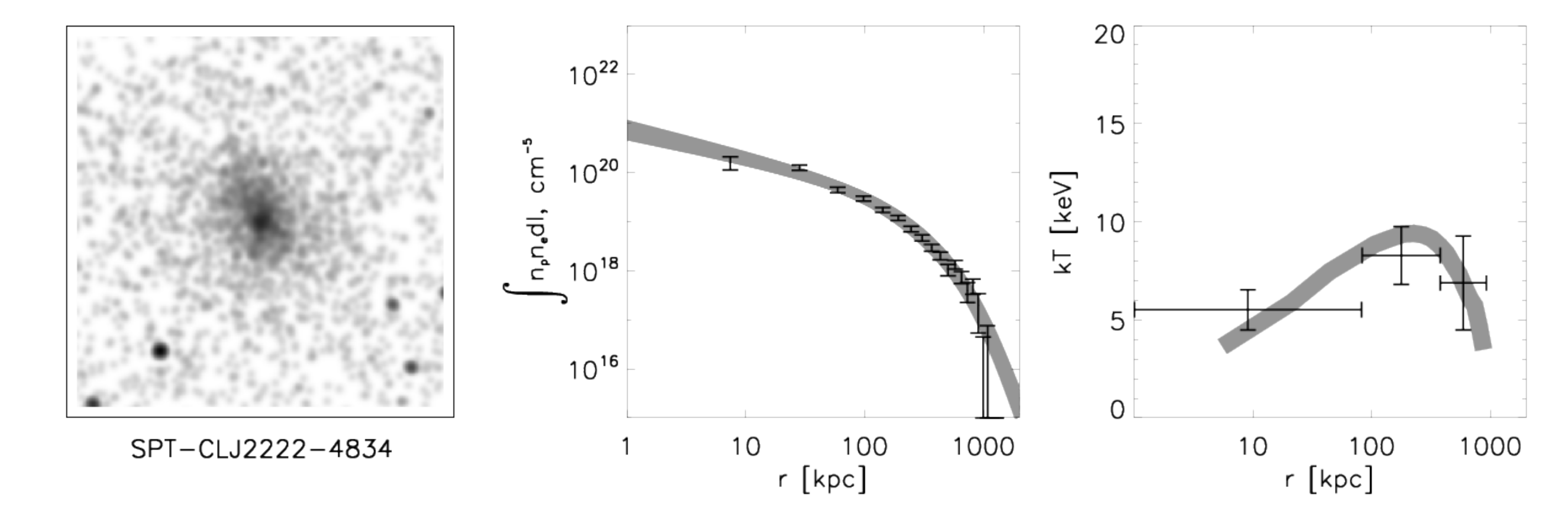} &
\includegraphics[width=0.49\textwidth]{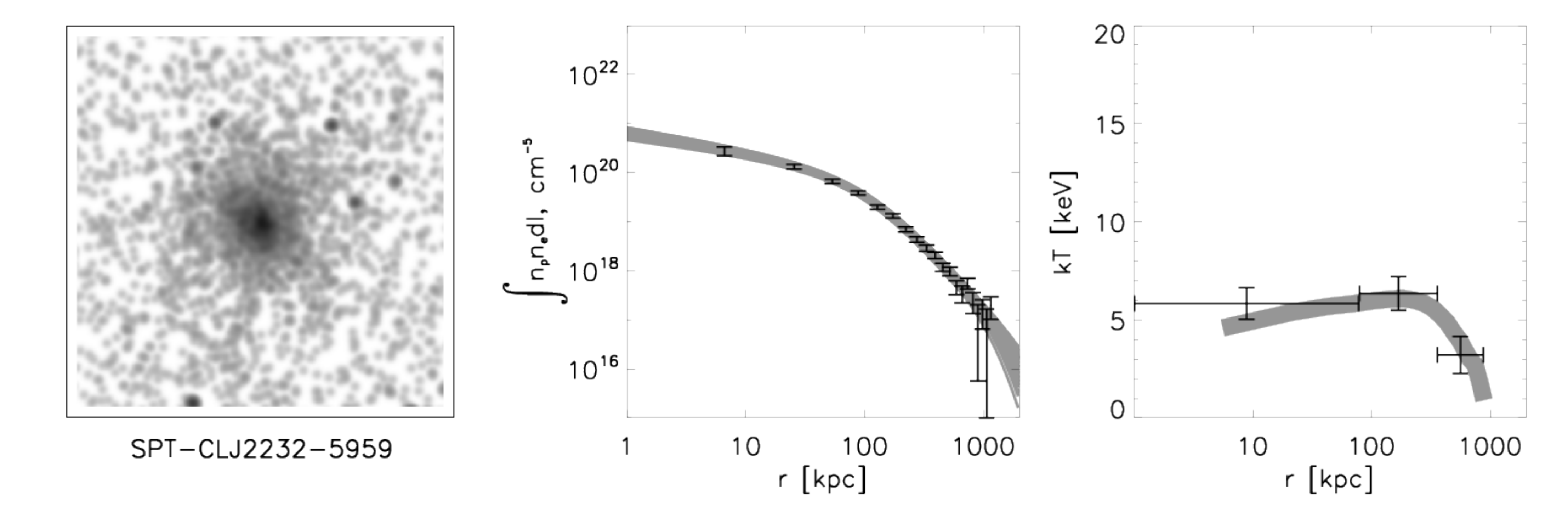} \\
\vspace{-0.05in}
\includegraphics[width=0.49\textwidth]{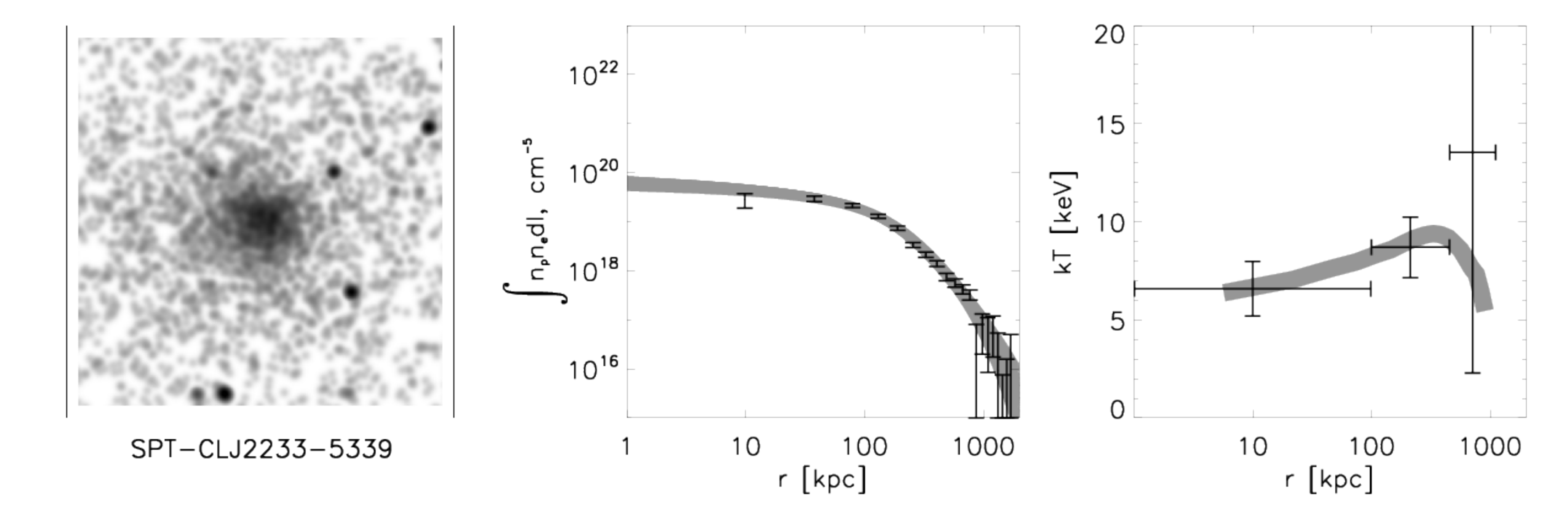} &
\includegraphics[width=0.49\textwidth]{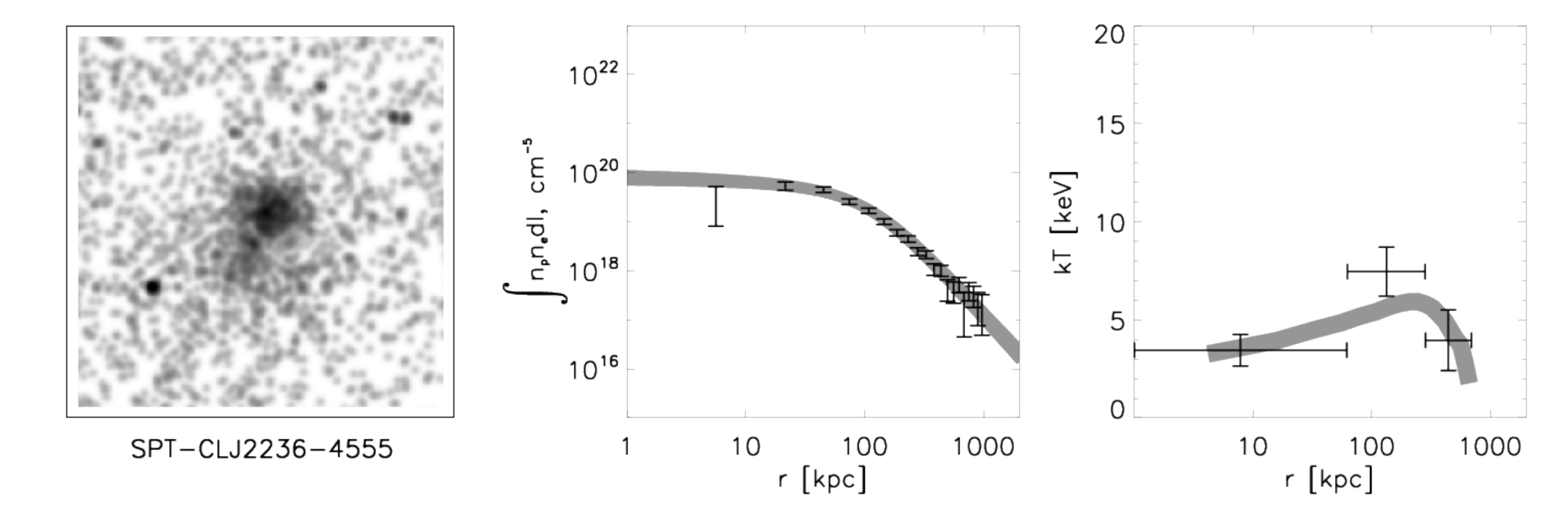} \\
\vspace{-0.05in}
\includegraphics[width=0.49\textwidth]{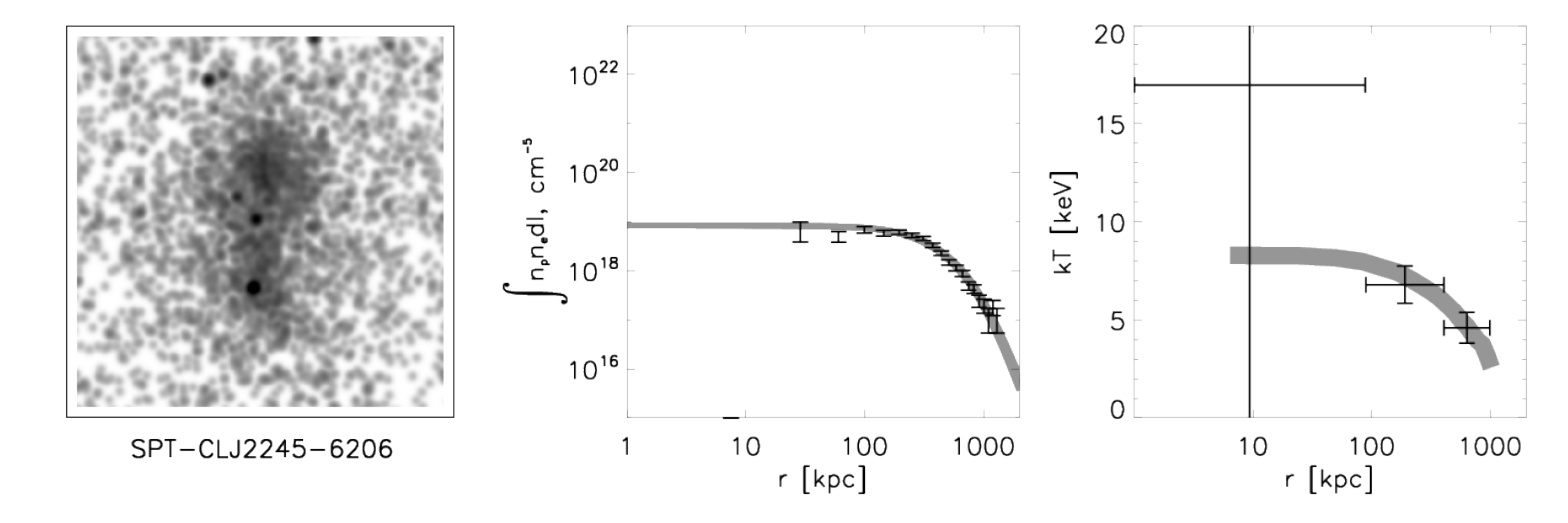} &
\includegraphics[width=0.49\textwidth]{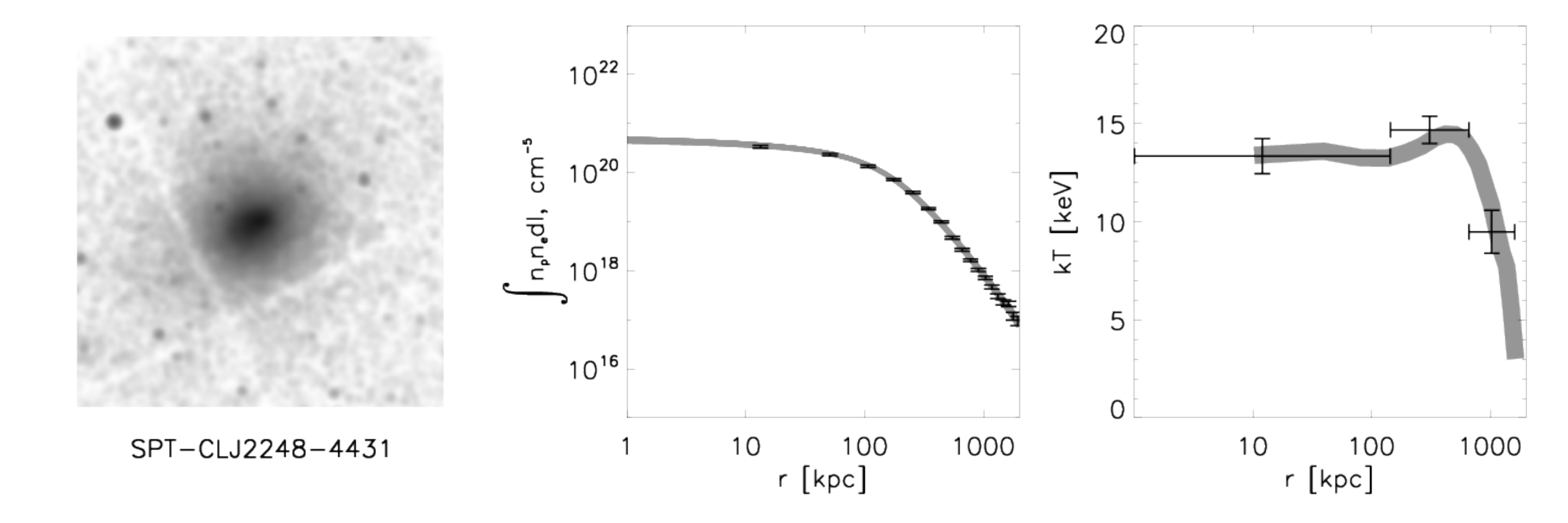} \\
\vspace{-0.05in}
\includegraphics[width=0.49\textwidth]{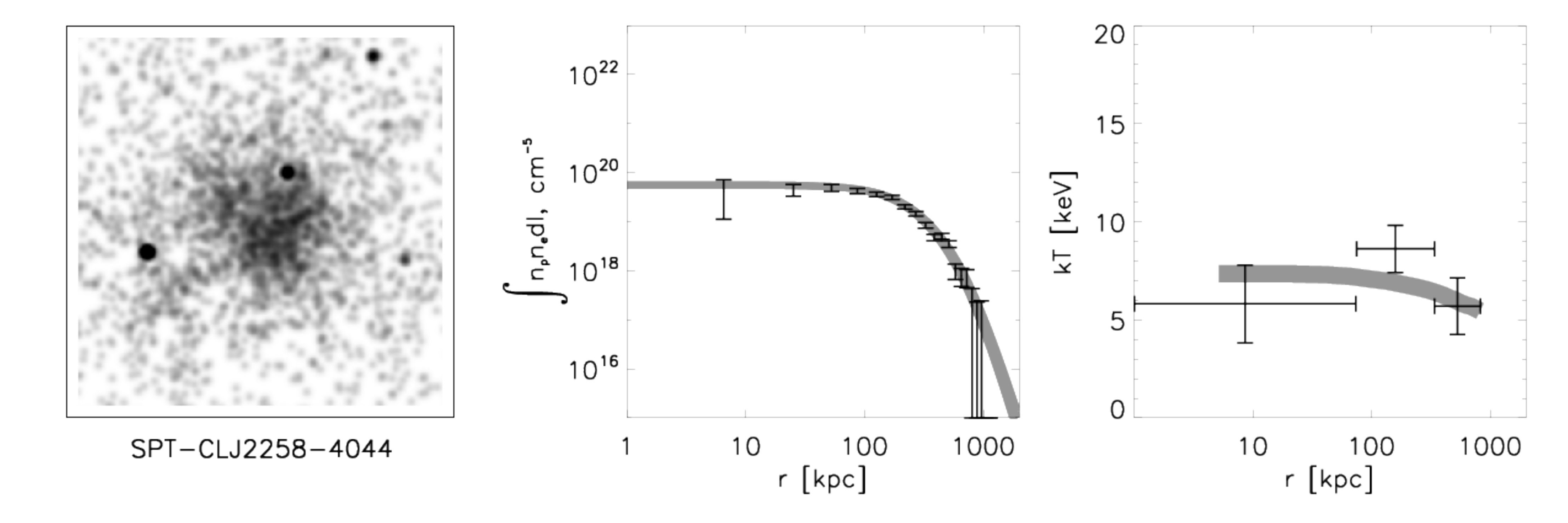} &
\includegraphics[width=0.49\textwidth]{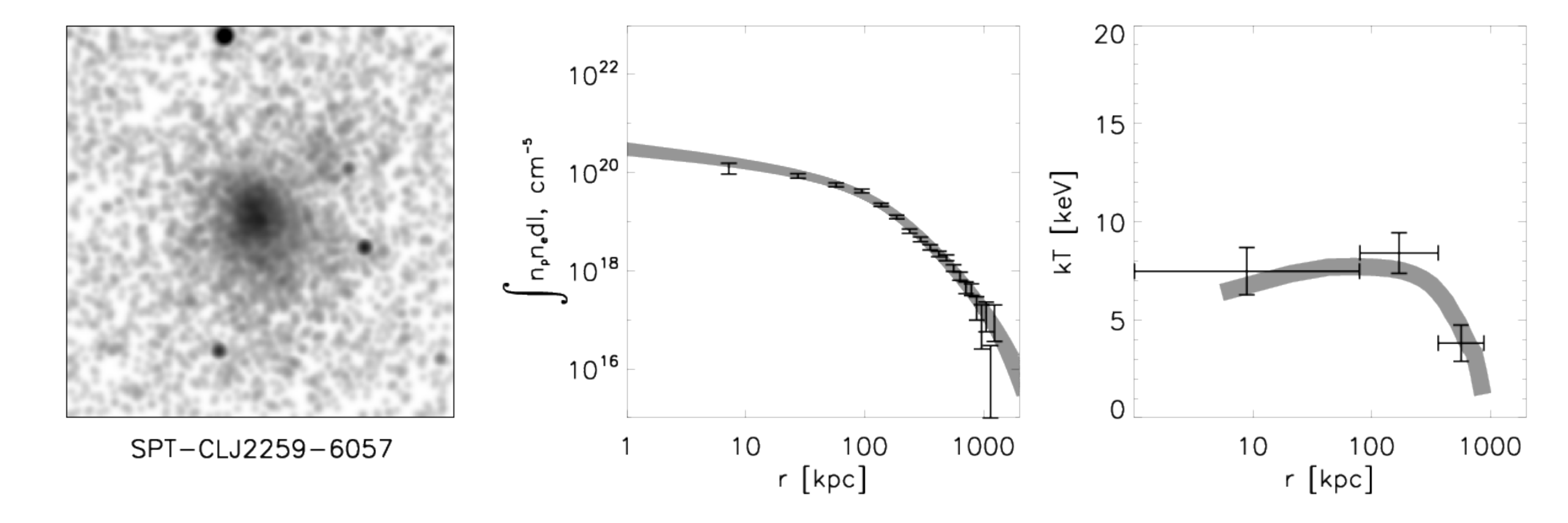} \\
\vspace{-0.05in}
\includegraphics[width=0.49\textwidth]{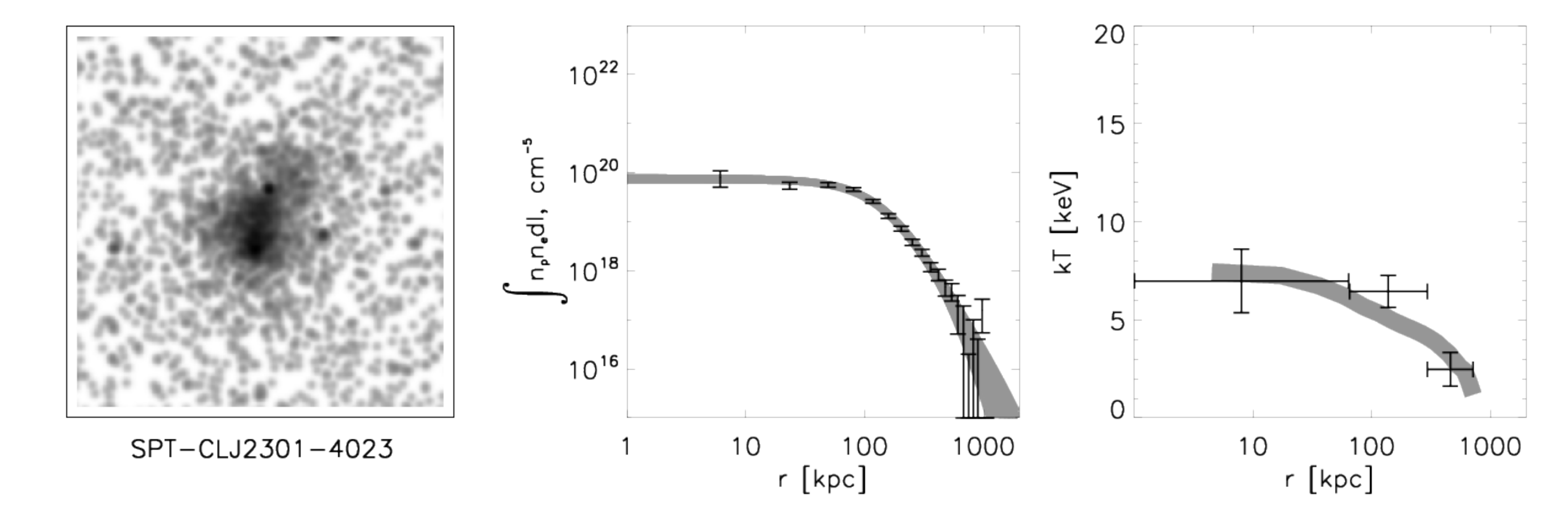} &
\includegraphics[width=0.49\textwidth]{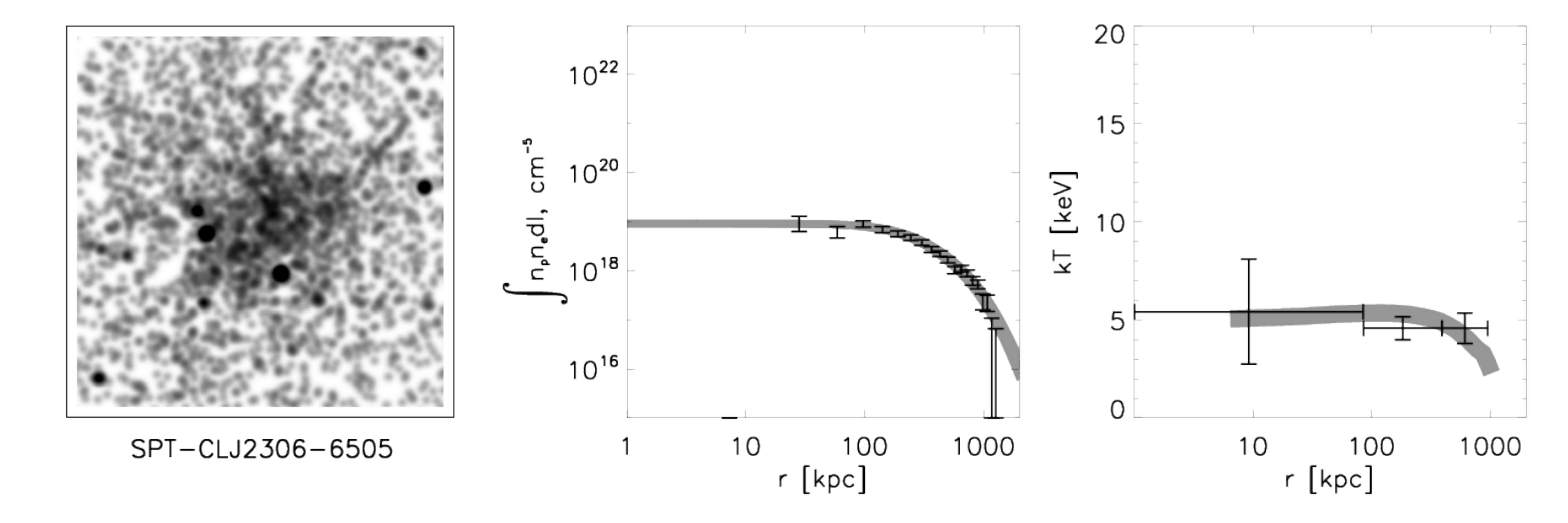} \\
\vspace{-0.05in}
\includegraphics[width=0.49\textwidth]{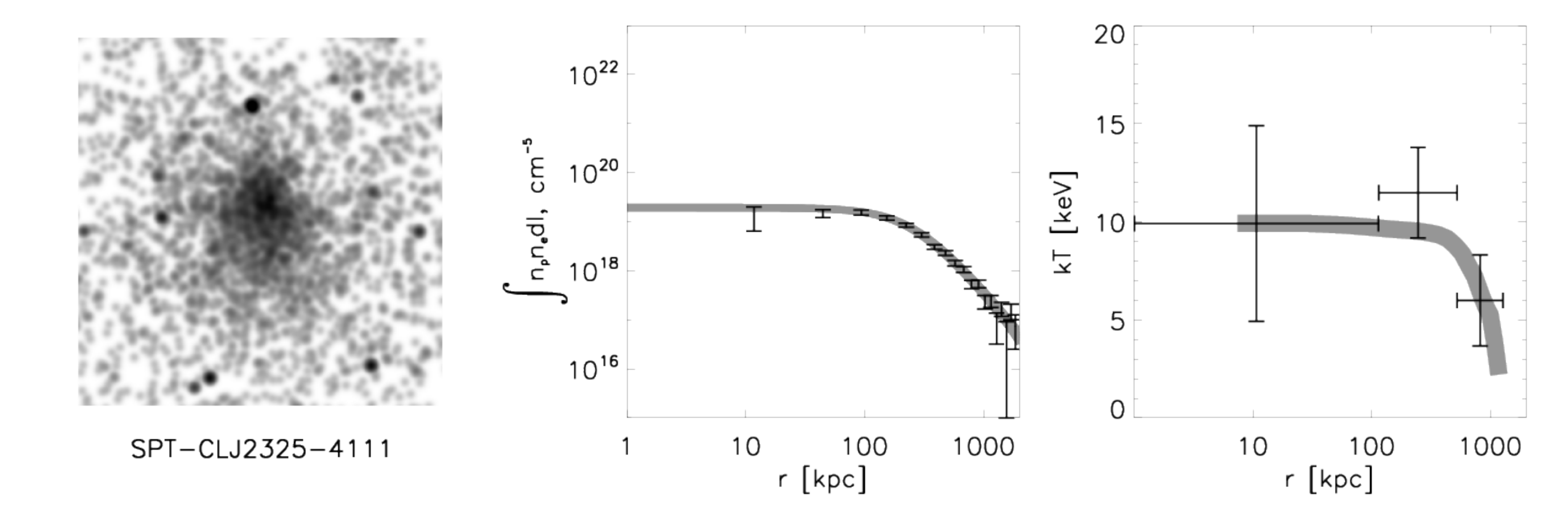} &
\includegraphics[width=0.49\textwidth]{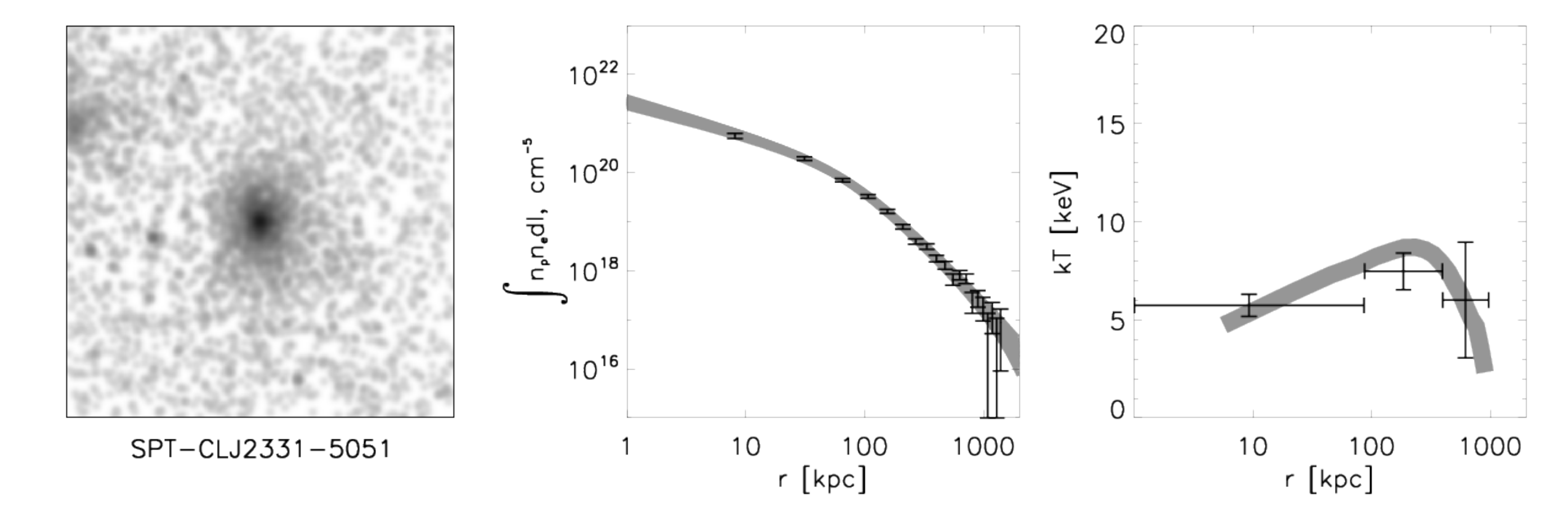} \\
\vspace{-0.05in}
\includegraphics[width=0.49\textwidth]{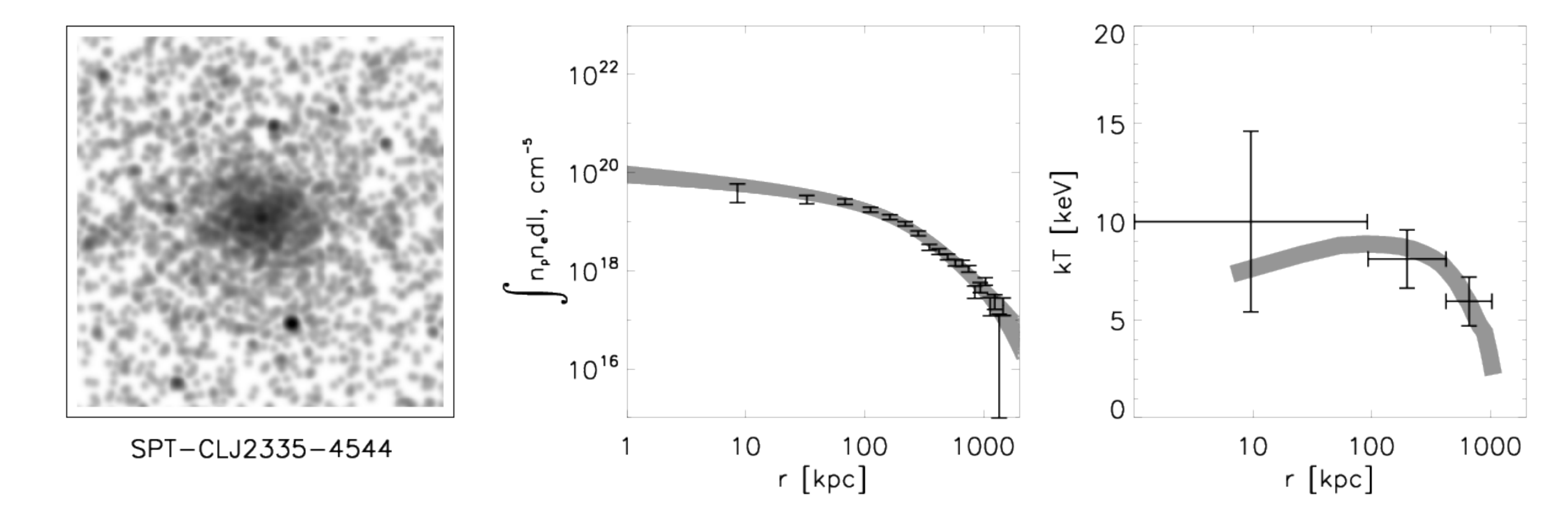} &
\includegraphics[width=0.49\textwidth]{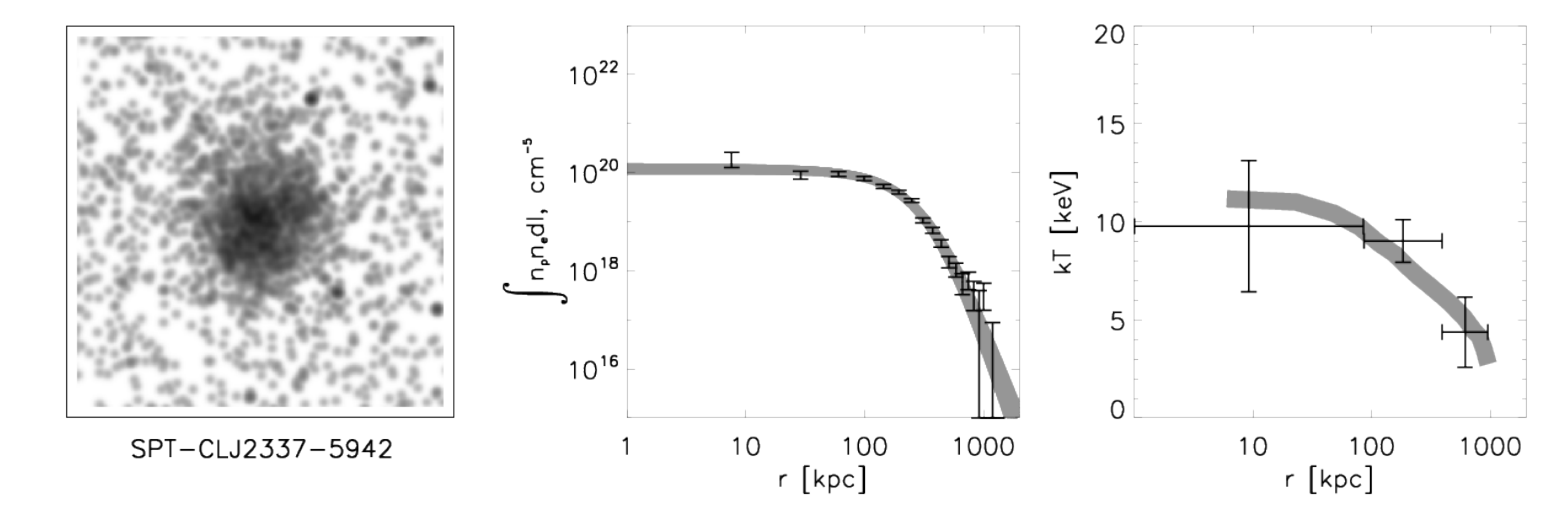} \\
\end{tabular}
\end{figure*}

\begin{figure*}[h!]
\centering
\begin{tabular}{c c}
\vspace{-0.05in}
\includegraphics[width=0.49\textwidth]{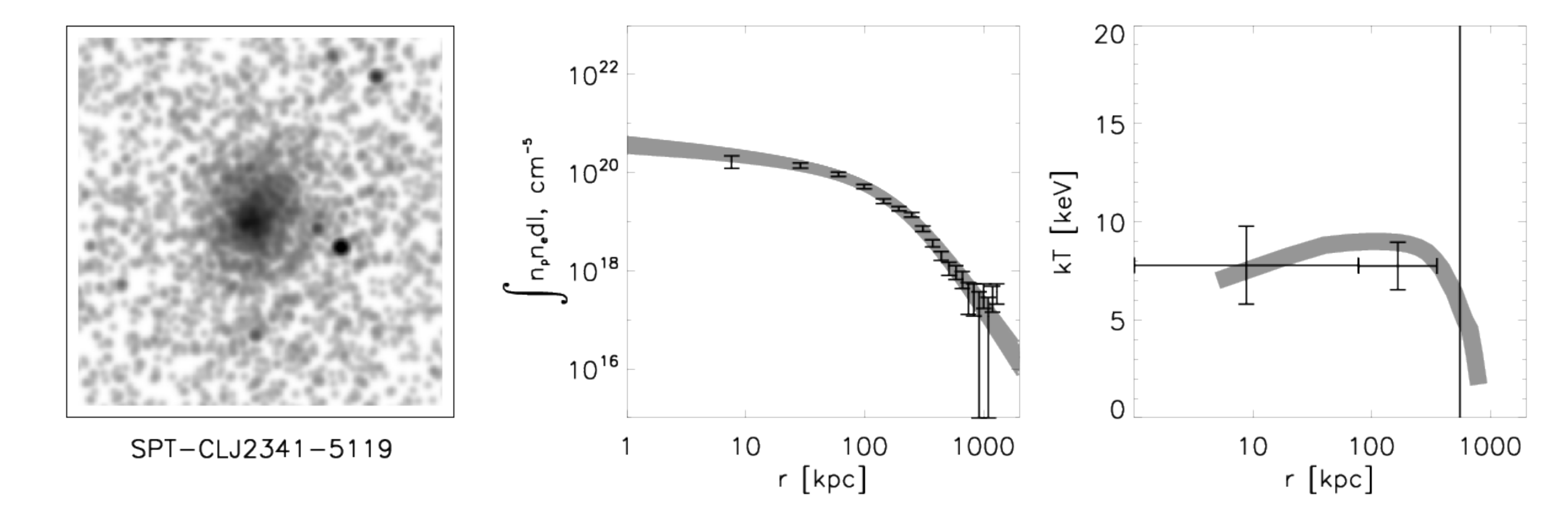} &
\includegraphics[width=0.49\textwidth]{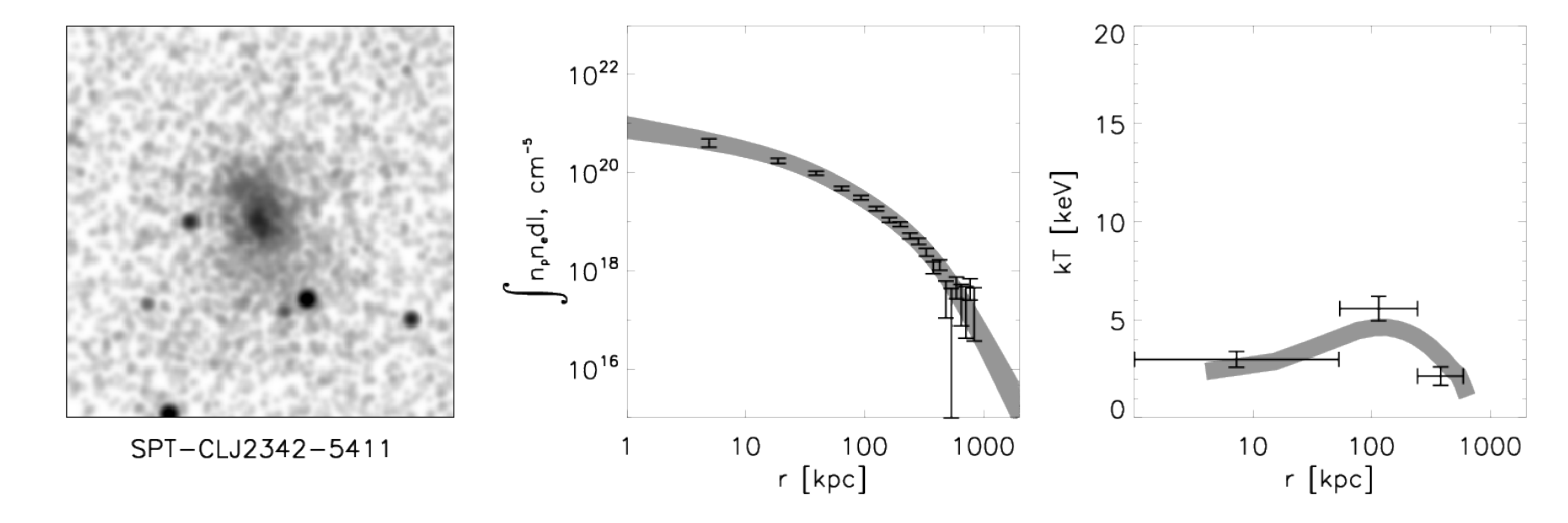} \\
\vspace{-0.05in}
\includegraphics[width=0.49\textwidth]{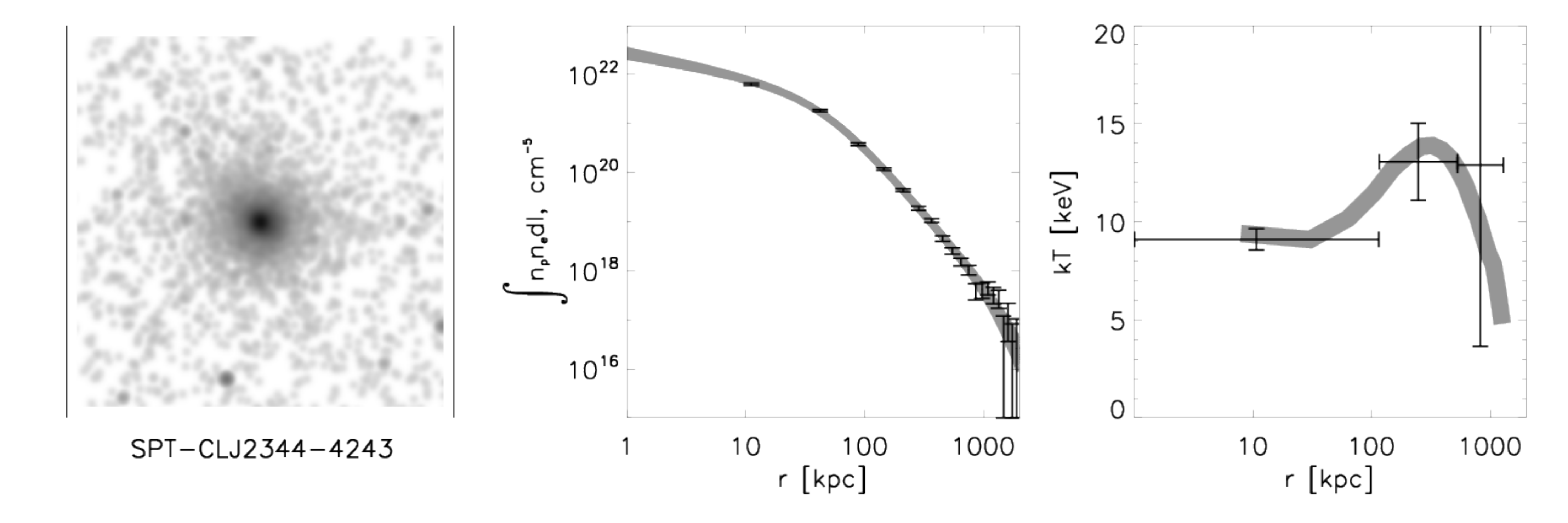} &
\includegraphics[width=0.49\textwidth]{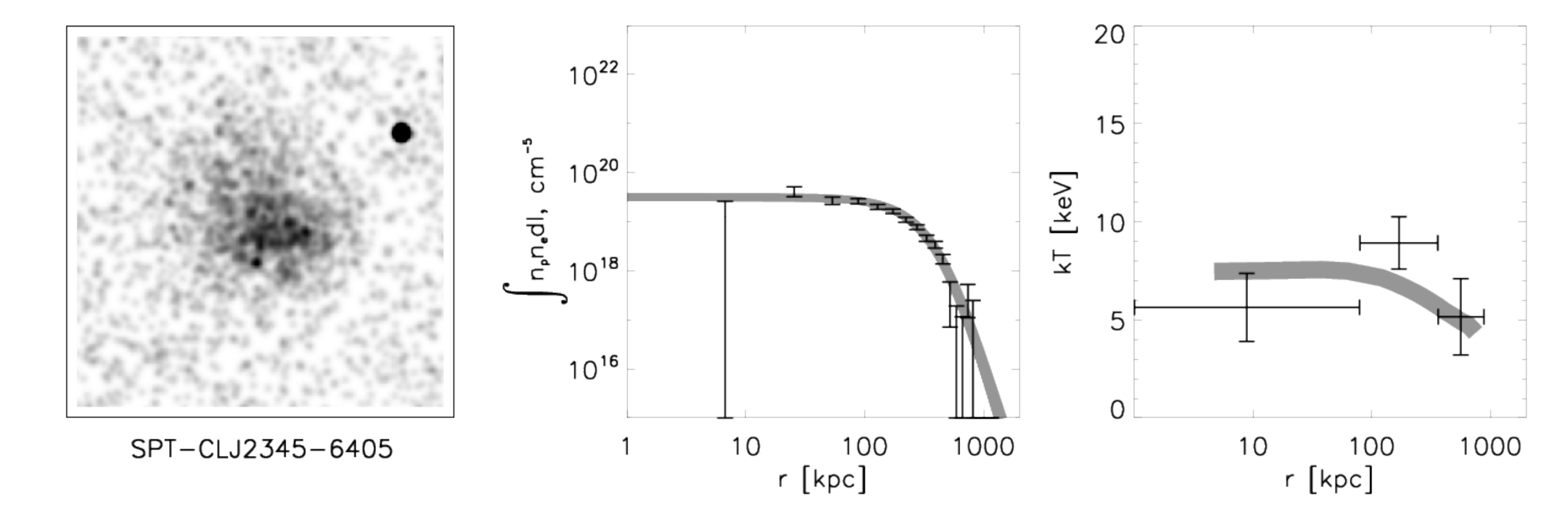} \\
\vspace{-0.05in}
\includegraphics[width=0.49\textwidth]{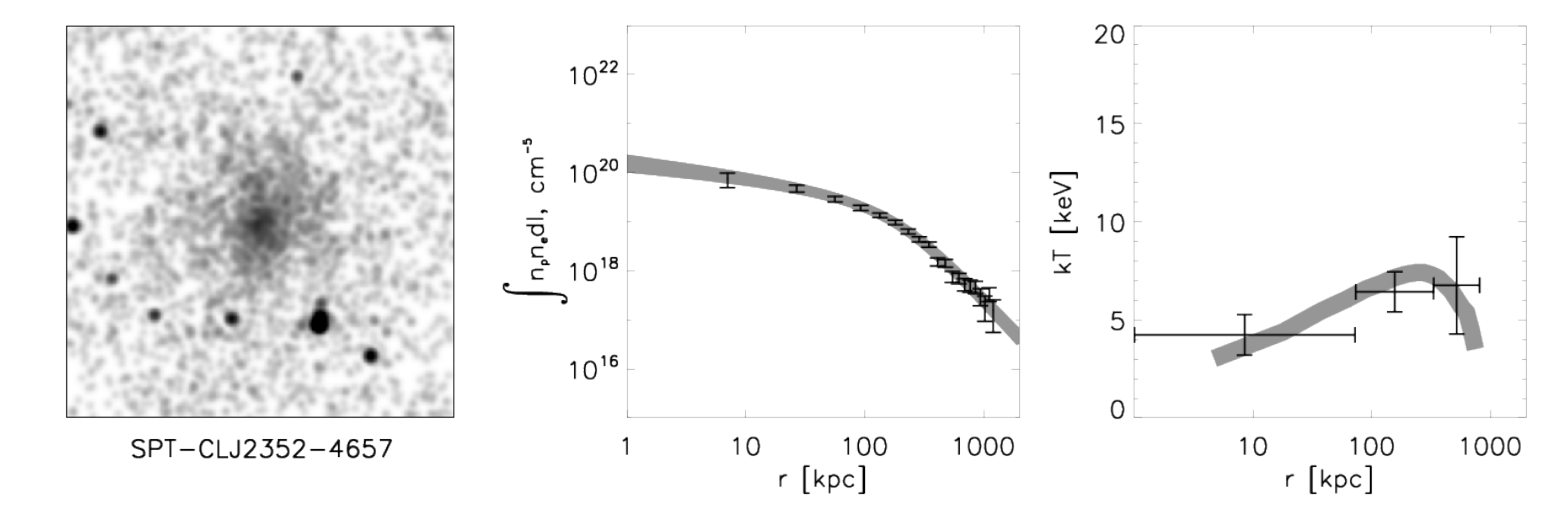} &
\includegraphics[width=0.49\textwidth]{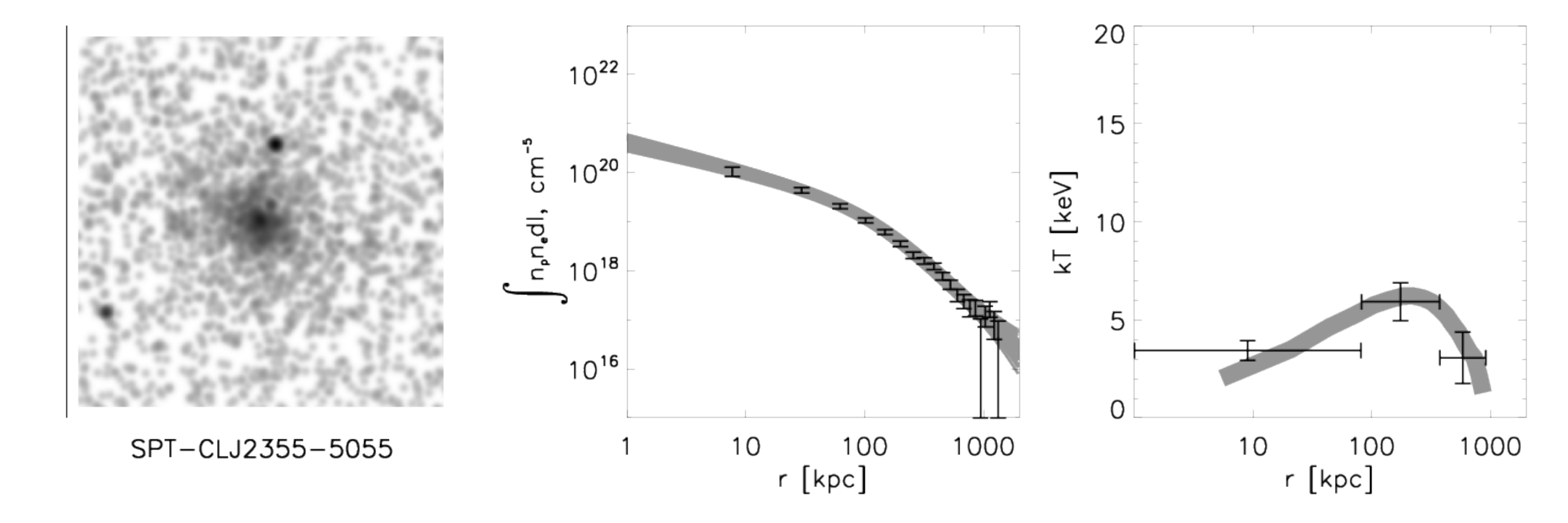} \\
\vspace{-0.05in}
\includegraphics[width=0.49\textwidth]{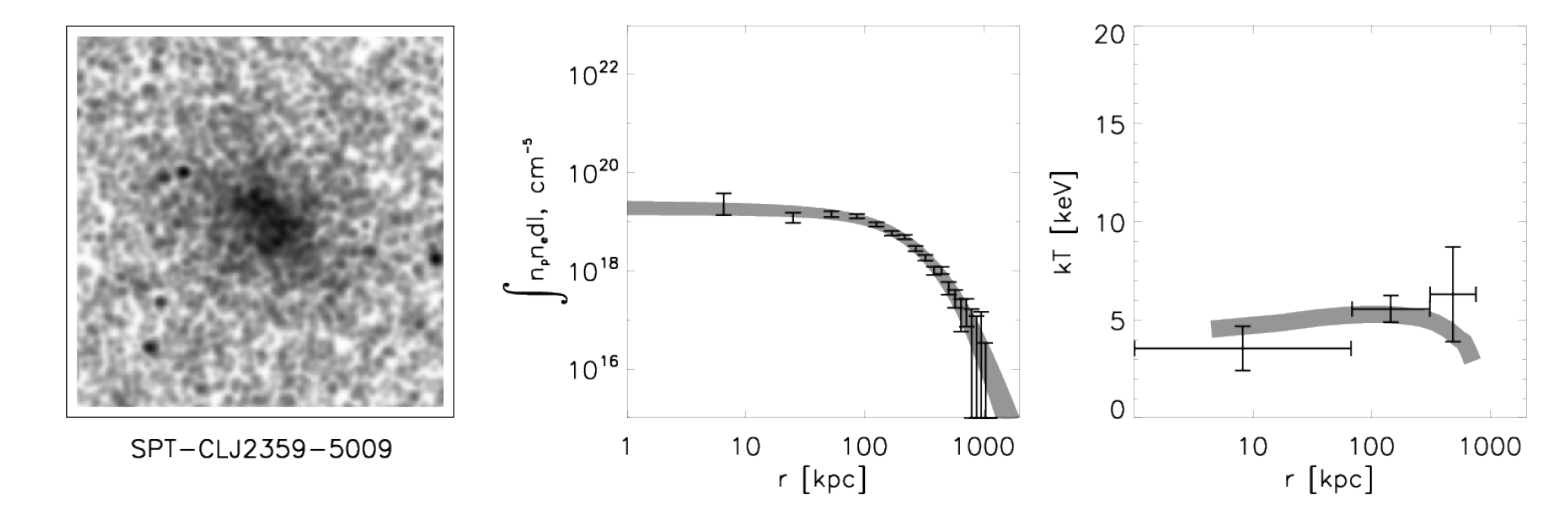} & \\
\end{tabular}
\end{figure*}

\clearpage
\section{Cool Core / Cooling Flow Properties}
Below we provide cool core and cooling flow properties for the full sample of 83 galaxy clusters at $0.3<z<1.2$. A complete description of how these values were measured is presented in \S2. As in \S2--\S4, we use the subscript ``0'' to represent the measured quantity in our smallest annulus: $r<0.012{\rm R}_{500}$. All of these data are deprojected, using our mass-modeling technique described in \S3.
\begin{longtable}{@{\extracolsep{\fill}}c c c c c c c c c c c}
\hline\hline
(1) & (2) & (3) & (4) & (5) & (6) & (7) & (8) & (9) \\
Cluster Name & $n_{e,0}$ & kT$_0$ & c$_{SB}$ & $\alpha$ & K$_0$ & $t_{\textrm{cool},0}$ & $d$M$/dt_{7.7}$ & $d$M$/dt_{\textrm{Univ}}$ \\
 & [cm$^{-3}$] & [keV] & & & [keV cm$^2$] & [Gyr] & [M$_{\odot}$ yr$^{-1}$] & [M$_{\odot}$ yr$^{-1}$]\\
%Name & $\alpha$ [$^{\circ}$] & $\delta$ [$^{\circ}$]& OBSIDs \\
\hline
SPT-CLJ0000-5748 & 0.164$^{+0.004}_{-0.004}$ &  4.2 $\pm$  0.6 & 0.29$^{+0.03}_{-0.04}$ & 1.00 $\pm$ 0.04 &  14.1$^{+   2.2}_{-   2.1}$ &  0.18$^{+ 0.03}_{- 0.03}$ &  437.3$^{+  68.1}_{-  51.9}$ &  437.3$^{+  68.1}_{-  51.9}$ \\
SPT-CLJ0013-4906 & 0.007$^{+0.000}_{-0.000}$ &  8.0 $\pm$  1.6 & 0.03$^{+0.00}_{-0.02}$ & 0.09 $\pm$ 0.01 & 229.8$^{+  50.9}_{-  49.5}$ &  6.83$^{+ 1.57}_{- 1.51}$ &--&   33.3$^{+   8.5}_{-   5.6}$ \\
SPT-CLJ0014-4952 & 0.004$^{+0.000}_{-0.000}$ &  7.4 $\pm$  0.9 & 0.03$^{+0.00}_{-0.01}$ & 0.03 $\pm$ 0.00 & 309.6$^{+  42.1}_{-  41.2}$ & 11.44$^{+ 1.63}_{- 1.58}$ &--&--\\
SPT-CLJ0033-6326 & 0.042$^{+0.002}_{-0.002}$ &  2.3 $\pm$  1.4 & 0.08$^{+0.01}_{-0.03}$ & 0.77 $\pm$ 0.06 &  19.2$^{+  12.4}_{-  11.8}$ &  0.47$^{+ 0.31}_{- 0.29}$ &   57.1$^{+  87.8}_{-  21.5}$ &   83.1$^{+ 127.8}_{-  31.4}$ \\
SPT-CLJ0037-5047 & 0.009$^{+0.001}_{-0.001}$ &  5.1 $\pm$  7.3 & 0.08$^{+0.03}_{-0.04}$ & 0.41 $\pm$ 0.14 & 115.7$^{+ 183.8}_{- 161.9}$ &  3.67$^{+ 6.13}_{- 5.08}$ &    0.0$^{+  -0.0}_{-   0.0}$ &    0.0$^{+  -0.0}_{-   0.0}$ \\
SPT-CLJ0040-4407 & 0.018$^{+0.000}_{-0.000}$ &  7.5 $\pm$  1.9 & 0.07$^{+0.00}_{-0.02}$ & 0.38 $\pm$ 0.04 & 108.0$^{+  29.9}_{-  28.9}$ &  2.34$^{+ 0.67}_{- 0.64}$ &   86.2$^{+  29.6}_{-  17.6}$ &  159.8$^{+  54.9}_{-  32.5}$ \\
SPT-CLJ0058-6145 & 0.031$^{+0.001}_{-0.001}$ &  2.4 $\pm$  1.7 & 0.06$^{+0.00}_{-0.02}$ & 0.63 $\pm$ 0.06 &  24.0$^{+  18.5}_{-  17.8}$ &  0.64$^{+ 0.51}_{- 0.48}$ &   25.4$^{+  70.3}_{-  10.8}$ &   15.2$^{+  42.1}_{-   6.4}$ \\
SPT-CLJ0102-4603 & 0.025$^{+0.001}_{-0.001}$ &  0.9 $\pm$  0.5 & 0.07$^{+0.01}_{-0.02}$ & 0.66 $\pm$ 0.09 &  10.3$^{+   6.1}_{-   5.8}$ &  0.29$^{+ 0.18}_{- 0.16}$ &   58.7$^{+  74.1}_{-  21.0}$ &   58.7$^{+  74.1}_{-  21.0}$ \\
SPT-CLJ0102-4915 & 0.009$^{+0.000}_{-0.000}$ & 16.2 $\pm$  2.3 & 0.03$^{+0.01}_{-0.01}$ & 0.06 $\pm$ 0.00 & 364.6$^{+  54.5}_{-  53.8}$ &  7.10$^{+ 1.09}_{- 1.07}$ &--&--\\
SPT-CLJ0123-4821 & 0.004$^{+0.000}_{-0.000}$ &  7.3 $\pm$  1.5 & 0.04$^{+0.00}_{-0.02}$ & 0.09 $\pm$ 0.04 & 276.3$^{+  61.9}_{-  60.1}$ &  9.77$^{+ 2.28}_{- 2.18}$ &--&--\\
SPT-CLJ0142-5032 & 0.005$^{+0.000}_{-0.000}$ &  8.5 $\pm$  1.7 & 0.03$^{+0.00}_{-0.02}$ & 0.06 $\pm$ 0.01 & 307.7$^{+  70.5}_{-  67.7}$ & 10.03$^{+ 2.43}_{- 2.29}$ &--&--\\
SPT-CLJ0151-5954 & 0.004$^{+0.000}_{-0.000}$ &  6.2 $\pm$  2.8 & 0.04$^{+0.00}_{-0.02}$ & 0.02 $\pm$ 0.02 & 264.9$^{+ 130.5}_{- 123.3}$ & 10.62$^{+ 5.46}_{- 5.02}$ &--&--\\
SPT-CLJ0156-5541 & 0.008$^{+0.000}_{-0.000}$ &  9.7 $\pm$  2.5 & 0.06$^{+0.00}_{-0.02}$ & 0.07 $\pm$ 0.06 & 238.6$^{+  70.9}_{-  66.9}$ &  6.06$^{+ 1.92}_{- 1.76}$ &--&--\\
SPT-CLJ0200-4852 & 0.017$^{+0.000}_{-0.000}$ &  7.0 $\pm$  3.0 & 0.05$^{+0.00}_{-0.02}$ & 0.52 $\pm$ 0.05 & 105.5$^{+  48.6}_{-  47.0}$ &  2.42$^{+ 1.14}_{- 1.09}$ &--&    2.4$^{+   1.9}_{-   0.7}$ \\
SPT-CLJ0212-4657 & 0.004$^{+0.000}_{-0.000}$ &  8.2 $\pm$  2.4 & 0.04$^{+0.00}_{-0.02}$ & 0.04 $\pm$ 0.01 & 340.7$^{+ 108.6}_{- 104.5}$ & 12.14$^{+ 4.03}_{- 3.81}$ &--&--\\
SPT-CLJ0217-5245 & 0.019$^{+0.000}_{-0.000}$ &  1.6 $\pm$  1.5 & 0.06$^{+0.00}_{-0.02}$ & 0.75 $\pm$ 0.04 &  21.8$^{+  21.3}_{-  20.7}$ &  0.73$^{+ 0.72}_{- 0.69}$ &    5.8$^{+ 100.7}_{-   2.8}$ &    5.8$^{+ 100.7}_{-   2.8}$ \\
SPT-CLJ0232-5257 & 0.020$^{+0.001}_{-0.001}$ &  1.6 $\pm$  0.8 & 0.04$^{+0.00}_{-0.02}$ & 0.64 $\pm$ 0.05 &  21.3$^{+  12.2}_{-  11.6}$ &  0.70$^{+ 0.41}_{- 0.38}$ &   40.4$^{+  46.6}_{-  14.1}$ &   40.4$^{+  46.6}_{-  14.1}$ \\
SPT-CLJ0234-5831 & 0.058$^{+0.001}_{-0.001}$ &  5.7 $\pm$  0.6 & 0.18$^{+0.01}_{-0.03}$ & 0.80 $\pm$ 0.05 &  38.1$^{+   4.4}_{-   4.3}$ &  0.63$^{+ 0.08}_{- 0.08}$ &  427.8$^{+  46.7}_{-  38.4}$ &  427.8$^{+  46.7}_{-  38.4}$ \\
SPT-CLJ0236-4938 & 0.011$^{+0.000}_{-0.000}$ &  3.1 $\pm$  1.7 & 0.06$^{+0.00}_{-0.02}$ & 0.54 $\pm$ 0.04 &  61.8$^{+  35.9}_{-  35.0}$ &  2.20$^{+ 1.30}_{- 1.25}$ &    2.2$^{+   2.8}_{-   0.8}$ &    2.2$^{+   2.8}_{-   0.8}$ \\
SPT-CLJ0243-5930 & 0.015$^{+0.000}_{-0.000}$ &  6.4 $\pm$  1.9 & 0.05$^{+0.00}_{-0.02}$ & 0.35 $\pm$ 0.05 & 104.3$^{+  31.9}_{-  31.1}$ &  2.55$^{+ 0.80}_{- 0.77}$ &   15.2$^{+   6.2}_{-   3.4}$ &   15.2$^{+   6.2}_{-   3.4}$ \\
SPT-CLJ0252-4824 & 0.008$^{+0.000}_{-0.000}$ &  2.3 $\pm$  1.0 & 0.03$^{+0.00}_{-0.02}$ & 0.44 $\pm$ 0.06 &  58.6$^{+  26.5}_{-  25.7}$ &  2.47$^{+ 1.15}_{- 1.09}$ &    2.3$^{+   1.7}_{-   0.7}$ &    2.3$^{+   1.7}_{-   0.7}$ \\
SPT-CLJ0256-5617 & 0.004$^{+0.000}_{-0.000}$ & 14.2 $\pm$  3.1 & 0.04$^{+0.00}_{-0.02}$ & 0.04 $\pm$ 0.01 & 547.9$^{+ 130.1}_{- 125.9}$ & 14.71$^{+ 3.64}_{- 3.47}$ &--&--\\
SPT-CLJ0304-4401 & 0.003$^{+0.000}_{-0.000}$ & 10.1 $\pm$  2.4 & 0.02$^{+0.01}_{-0.01}$ & 0.04 $\pm$ 0.01 & 441.8$^{+ 112.6}_{- 109.6}$ & 14.72$^{+ 3.88}_{- 3.73}$ &--&--\\
SPT-CLJ0304-4921 & 0.055$^{+0.001}_{-0.001}$ &  4.0 $\pm$  0.8 & 0.13$^{+0.01}_{-0.02}$ & 0.76 $\pm$ 0.03 &  28.0$^{+   5.6}_{-   5.5}$ &  0.53$^{+ 0.11}_{- 0.11}$ &  139.0$^{+  31.9}_{-  21.9}$ &  194.7$^{+  44.7}_{-  30.7}$ \\
SPT-CLJ0307-5042 & 0.007$^{+0.000}_{-0.000}$ &  7.2 $\pm$  2.2 & 0.05$^{+0.00}_{-0.02}$ & 0.16 $\pm$ 0.05 & 204.9$^{+  67.4}_{-  65.6}$ &  6.31$^{+ 2.13}_{- 2.05}$ &--&--\\
SPT-CLJ0307-6225 & 0.003$^{+0.000}_{-0.000}$ &  6.4 $\pm$  1.2 & 0.03$^{+0.00}_{-0.02}$ & 0.04 $\pm$ 0.01 & 322.7$^{+  66.6}_{-  64.1}$ & 13.91$^{+ 3.04}_{- 2.87}$ &--&--\\
SPT-CLJ0310-4647 & 0.025$^{+0.001}_{-0.001}$ &  3.1 $\pm$  1.1 & 0.06$^{+0.00}_{-0.02}$ & 0.51 $\pm$ 0.05 &  35.9$^{+  14.4}_{-  13.7}$ &  0.97$^{+ 0.40}_{- 0.38}$ &   26.3$^{+  15.3}_{-   7.1}$ &   13.9$^{+   8.1}_{-   3.7}$ \\
SPT-CLJ0324-6236 & 0.025$^{+0.001}_{-0.001}$ &  3.5 $\pm$  1.7 & 0.09$^{+0.02}_{-0.03}$ & 0.44 $\pm$ 0.07 &  40.8$^{+  21.6}_{-  20.7}$ &  1.05$^{+ 0.57}_{- 0.54}$ &   49.3$^{+  48.8}_{-  16.4}$ &   49.3$^{+  48.8}_{-  16.4}$ \\
SPT-CLJ0330-5228 & 0.008$^{+0.000}_{-0.000}$ &  1.5 $\pm$  0.7 & 0.03$^{+0.00}_{-0.02}$ & 0.47 $\pm$ 0.05 &  37.1$^{+  16.2}_{-  15.9}$ &  1.63$^{+ 0.72}_{- 0.71}$ &   20.6$^{+  15.2}_{-   6.1}$ &   20.6$^{+  15.2}_{-   6.1}$ \\
SPT-CLJ0334-4659 & 0.057$^{+0.001}_{-0.001}$ &  4.3 $\pm$  0.6 & 0.18$^{+0.01}_{-0.03}$ & 0.84 $\pm$ 0.03 &  29.4$^{+   4.4}_{-   4.2}$ &  0.53$^{+ 0.08}_{- 0.08}$ &   73.9$^{+  11.4}_{-   8.7}$ &   73.9$^{+  11.4}_{-   8.7}$ \\
SPT-CLJ0346-5439 & 0.037$^{+0.001}_{-0.001}$ &  1.5 $\pm$  0.8 & 0.08$^{+0.01}_{-0.02}$ & 0.71 $\pm$ 0.04 &  13.7$^{+   7.3}_{-   7.0}$ &  0.37$^{+ 0.20}_{- 0.19}$ &  104.0$^{+ 105.5}_{-  34.8}$ &  104.0$^{+ 105.5}_{-  34.8}$ \\
SPT-CLJ0348-4515 & 0.015$^{+0.000}_{-0.000}$ &  2.4 $\pm$  1.5 & 0.06$^{+0.00}_{-0.02}$ & 0.50 $\pm$ 0.05 &  40.7$^{+  26.0}_{-  25.0}$ &  1.39$^{+ 0.91}_{- 0.86}$ &   20.1$^{+  31.2}_{-   7.6}$ &   20.1$^{+  31.2}_{-   7.6}$ \\
SPT-CLJ0352-5647 & 0.015$^{+0.000}_{-0.000}$ &  2.7 $\pm$  1.5 & 0.07$^{+0.02}_{-0.03}$ & 0.43 $\pm$ 0.05 &  44.2$^{+  25.4}_{-  24.5}$ &  1.46$^{+ 0.86}_{- 0.81}$ &   23.8$^{+  28.5}_{-   8.4}$ &   23.8$^{+  28.5}_{-   8.4}$ \\
SPT-CLJ0406-4805 & 0.025$^{+0.001}_{-0.001}$ &  3.1 $\pm$  2.5 & 0.09$^{+0.01}_{-0.03}$ & 0.66 $\pm$ 0.07 &  36.5$^{+  30.4}_{-  28.9}$ &  0.98$^{+ 0.84}_{- 0.78}$ &    9.9$^{+  36.2}_{-   4.3}$ &    9.9$^{+  36.2}_{-   4.3}$ \\
SPT-CLJ0411-4819 & 0.005$^{+0.000}_{-0.000}$ &  7.8 $\pm$  1.9 & 0.03$^{+0.00}_{-0.02}$ & 0.05 $\pm$ 0.01 & 274.3$^{+  74.5}_{-  71.9}$ &  9.18$^{+ 2.60}_{- 2.47}$ &--&--\\
SPT-CLJ0417-4748 & 0.086$^{+0.002}_{-0.002}$ &  6.3 $\pm$  0.8 & 0.21$^{+0.02}_{-0.03}$ & 0.80 $\pm$ 0.03 &  32.5$^{+   4.5}_{-   4.4}$ &  0.45$^{+ 0.07}_{- 0.06}$ &  517.0$^{+  74.2}_{-  57.7}$ &  517.0$^{+  74.2}_{-  57.7}$ \\
SPT-CLJ0426-5455 & 0.003$^{+0.000}_{-0.000}$ &  8.9 $\pm$ 15.1 & 0.03$^{+0.00}_{-0.02}$ & 0.04 $\pm$ 0.01 & 407.6$^{+ 718.4}_{- 688.2}$ & 14.77$^{+26.46}_{-24.83}$ &    0.0$^{+  -0.0}_{-   0.0}$ &    0.0$^{+  -0.0}_{-   0.0}$ \\
SPT-CLJ0438-5419 & 0.023$^{+0.000}_{-0.000}$ &  8.9 $\pm$  1.5 & 0.07$^{+0.00}_{-0.02}$ & 0.44 $\pm$ 0.03 & 110.3$^{+  19.4}_{-  19.1}$ &  2.08$^{+ 0.38}_{- 0.37}$ &   91.7$^{+  18.3}_{-  13.1}$ &  207.4$^{+  41.3}_{-  29.5}$ \\
SPT-CLJ0441-4855 & 0.047$^{+0.001}_{-0.001}$ &  3.4 $\pm$  1.0 & 0.13$^{+0.02}_{-0.03}$ & 0.62 $\pm$ 0.03 &  26.1$^{+   7.7}_{-   7.5}$ &  0.55$^{+ 0.17}_{- 0.16}$ &  230.1$^{+  89.1}_{-  50.2}$ &  154.7$^{+  59.9}_{-  33.7}$ \\
SPT-CLJ0446-5849 & 0.004$^{+0.001}_{-0.001}$ &  7.6 $\pm$  2.1 & 0.02$^{+0.00}_{-0.02}$ & 0.03 $\pm$ 0.02 & 286.9$^{+ 110.5}_{-  95.5}$ &  9.95$^{+ 4.41}_{- 3.58}$ &--&--\\
SPT-CLJ0449-4901 & 0.004$^{+0.000}_{-0.000}$ &  9.8 $\pm$  5.8 & 0.03$^{+0.00}_{-0.02}$ & 0.12 $\pm$ 0.06 & 358.9$^{+ 223.3}_{- 215.3}$ & 11.14$^{+ 7.09}_{- 6.72}$ &--&--\\
SPT-CLJ0456-5116 & 0.008$^{+0.000}_{-0.000}$ & 10.8 $\pm$  5.7 & 0.04$^{+0.00}_{-0.02}$ & 0.24 $\pm$ 0.05 & 264.5$^{+ 146.4}_{- 142.4}$ &  6.32$^{+ 3.57}_{- 3.42}$ &--&--\\
SPT-CLJ0509-5342 & 0.056$^{+0.001}_{-0.001}$ &  3.7 $\pm$  1.1 & 0.13$^{+0.01}_{-0.03}$ & 0.86 $\pm$ 0.02 &  25.0$^{+   7.7}_{-   7.6}$ &  0.48$^{+ 0.15}_{- 0.15}$ &   39.2$^{+  16.3}_{-   8.9}$ &   39.2$^{+  16.3}_{-   8.9}$ \\
SPT-CLJ0528-5300 & 0.018$^{+0.001}_{-0.001}$ &  2.0 $\pm$  1.3 & 0.07$^{+0.02}_{-0.03}$ & 0.50 $\pm$ 0.06 &  29.2$^{+  19.3}_{-  18.5}$ &  0.97$^{+ 0.65}_{- 0.62}$ &    9.6$^{+  16.2}_{-   3.7}$ &    9.6$^{+  16.2}_{-   3.7}$ \\
SPT-CLJ0533-5005 & 0.020$^{+0.001}_{-0.001}$ &  1.0 $\pm$  1.1 & 0.04$^{+0.01}_{-0.03}$ & 0.57 $\pm$ 0.12 &  13.5$^{+  16.2}_{-  15.0}$ &  0.41$^{+ 0.51}_{- 0.46}$ &   15.2$^{+ 146.9}_{-   8.0}$ &    9.1$^{+  87.8}_{-   4.8}$ \\
SPT-CLJ0542-4100 & 0.025$^{+0.000}_{-0.000}$ & 11.3 $\pm$  4.4 & 0.01$^{+0.01}_{-0.01}$ & 0.61 $\pm$ 0.04 & 133.8$^{+  54.3}_{-  52.9}$ &  2.17$^{+ 0.90}_{- 0.87}$ &    2.0$^{+   1.2}_{-   0.5}$ &    2.0$^{+   1.2}_{-   0.5}$ \\
SPT-CLJ0546-5345 & 0.020$^{+0.001}_{-0.001}$ &  5.3 $\pm$  2.6 & 0.07$^{+0.01}_{-0.02}$ & 0.33 $\pm$ 0.05 &  70.4$^{+  37.6}_{-  36.1}$ &  1.69$^{+ 0.93}_{- 0.88}$ &  168.2$^{+ 169.4}_{-  56.2}$ &   37.3$^{+  37.6}_{-  12.5}$ \\
SPT-CLJ0551-5709 & 0.005$^{+0.000}_{-0.000}$ &  9.3 $\pm$  2.5 & 0.03$^{+0.00}_{-0.02}$ & 0.17 $\pm$ 0.05 & 329.4$^{+  91.8}_{-  89.7}$ & 10.26$^{+ 2.93}_{- 2.84}$ &--&--\\
SPT-CLJ0555-6406 & 0.009$^{+0.000}_{-0.000}$ &  5.4 $\pm$  1.9 & 0.04$^{+0.00}_{-0.02}$ & 0.31 $\pm$ 0.05 & 128.3$^{+  48.7}_{-  47.2}$ &  4.06$^{+ 1.58}_{- 1.51}$ &--&    4.8$^{+   2.7}_{-   1.3}$ \\
SPT-CLJ0559-5249 & 0.008$^{+0.000}_{-0.000}$ &  3.5 $\pm$  1.0 & 0.03$^{+0.00}_{-0.01}$ & 0.30 $\pm$ 0.04 &  90.1$^{+  25.7}_{-  25.2}$ &  3.48$^{+ 1.01}_{- 0.99}$ &    4.2$^{+   1.6}_{-   0.9}$ &    4.2$^{+   1.6}_{-   0.9}$ \\
SPT-CLJ0616-5227 & 0.024$^{+0.000}_{-0.000}$ &  4.6 $\pm$  1.1 & 0.09$^{+0.01}_{-0.03}$ & 0.44 $\pm$ 0.05 &  56.6$^{+  14.6}_{-  14.2}$ &  1.35$^{+ 0.36}_{- 0.35}$ &  247.1$^{+  78.3}_{-  47.9}$ &  158.2$^{+  50.2}_{-  30.7}$ \\
SPT-CLJ0655-5234 & 0.005$^{+0.000}_{-0.000}$ & 18.2 $\pm$ 11.9 & 0.04$^{+0.00}_{-0.02}$ & 0.09 $\pm$ 0.06 & 655.7$^{+ 458.4}_{- 434.8}$ & 15.41$^{+11.13}_{-10.29}$ &--&--\\
SPT-CLJ2031-4037 & 0.016$^{+0.000}_{-0.000}$ & 12.2 $\pm$  2.4 & 0.05$^{+0.00}_{-0.02}$ & 0.37 $\pm$ 0.04 & 189.8$^{+  39.9}_{-  38.9}$ &  3.43$^{+ 0.75}_{- 0.72}$ &   35.8$^{+   8.7}_{-   5.9}$ &   97.6$^{+  23.7}_{-  15.9}$ \\
SPT-CLJ2034-5936 & 0.026$^{+0.001}_{-0.001}$ &  3.7 $\pm$  3.6 & 0.06$^{+0.00}_{-0.02}$ & 0.51 $\pm$ 0.05 &  41.9$^{+  41.9}_{-  40.2}$ &  1.04$^{+ 1.06}_{- 1.00}$ &   56.2$^{+1323.3}_{-  27.5}$ &   28.0$^{+ 658.7}_{-  13.7}$ \\
SPT-CLJ2035-5251 & 0.002$^{+0.000}_{-0.000}$ &  6.2 $\pm$  1.4 & 0.05$^{+0.00}_{-0.02}$ & 0.03 $\pm$ 0.02 & 362.8$^{+  93.4}_{-  88.8}$ & 17.13$^{+ 4.70}_{- 4.36}$ &--&--\\
SPT-CLJ2043-5035 & 0.143$^{+0.002}_{-0.002}$ &  3.5 $\pm$  0.2 & 0.31$^{+0.02}_{-0.04}$ & 1.00 $\pm$ 0.03 &  12.7$^{+   1.0}_{-   1.0}$ &  0.18$^{+ 0.02}_{- 0.01}$ &  749.9$^{+  53.6}_{-  46.9}$ &  749.9$^{+  53.6}_{-  46.9}$ \\
SPT-CLJ2106-5844 & 0.012$^{+0.000}_{-0.000}$ &  7.2 $\pm$  1.3 & 0.04$^{+0.00}_{-0.02}$ & 0.12 $\pm$ 0.05 & 135.8$^{+  26.8}_{-  25.9}$ &  3.42$^{+ 0.71}_{- 0.68}$ &  381.5$^{+  81.5}_{-  57.1}$ &   28.2$^{+   6.0}_{-   4.2}$ \\
SPT-CLJ2135-5726 & 0.012$^{+0.000}_{-0.000}$ &  4.2 $\pm$  1.5 & 0.06$^{+0.00}_{-0.02}$ & 0.31 $\pm$ 0.05 &  79.4$^{+  30.2}_{-  29.3}$ &  2.43$^{+ 0.95}_{- 0.91}$ &   27.5$^{+  15.4}_{-   7.3}$ &   50.1$^{+  28.1}_{-  13.2}$ \\
SPT-CLJ2145-5644 & 0.032$^{+0.001}_{-0.001}$ & 15.7 $\pm$  8.1 & 0.10$^{+0.01}_{-0.02}$ & 0.71 $\pm$ 0.04 & 155.6$^{+  84.1}_{-  81.4}$ &  2.03$^{+ 1.12}_{- 1.07}$ &    5.9$^{+   6.3}_{-   2.0}$ &    5.9$^{+   6.3}_{-   2.0}$ \\
SPT-CLJ2146-4632 & 0.012$^{+0.000}_{-0.000}$ &  2.1 $\pm$  1.3 & 0.05$^{+0.00}_{-0.02}$ & 0.42 $\pm$ 0.07 &  40.3$^{+  26.6}_{-  25.6}$ &  1.54$^{+ 1.04}_{- 0.98}$ &   11.5$^{+  19.5}_{-   4.4}$ &    4.4$^{+   7.5}_{-   1.7}$ \\
SPT-CLJ2148-6116 & 0.007$^{+0.000}_{-0.000}$ &  3.7 $\pm$  1.3 & 0.04$^{+0.00}_{-0.02}$ & 0.26 $\pm$ 0.06 &  98.5$^{+  36.2}_{-  35.1}$ &  3.78$^{+ 1.43}_{- 1.36}$ &    1.1$^{+   0.6}_{-   0.3}$ &    1.1$^{+   0.6}_{-   0.3}$ \\
SPT-CLJ2218-4519 & 0.010$^{+0.000}_{-0.000}$ &  3.0 $\pm$  1.2 & 0.05$^{+0.00}_{-0.02}$ & 0.37 $\pm$ 0.06 &  65.0$^{+  28.6}_{-  27.5}$ &  2.41$^{+ 1.09}_{- 1.03}$ &    2.3$^{+   1.6}_{-   0.7}$ &    2.3$^{+   1.6}_{-   0.7}$ \\
SPT-CLJ2222-4834 & 0.060$^{+0.002}_{-0.002}$ &  2.4 $\pm$  1.0 & 0.13$^{+0.01}_{-0.03}$ & 0.77 $\pm$ 0.04 &  15.8$^{+   7.2}_{-   6.9}$ &  0.34$^{+ 0.16}_{- 0.15}$ &  165.4$^{+ 122.4}_{-  49.4}$ &  165.4$^{+ 122.4}_{-  49.4}$ \\
SPT-CLJ2232-5959 & 0.054$^{+0.001}_{-0.001}$ &  3.8 $\pm$  0.8 & 0.15$^{+0.01}_{-0.03}$ & 0.70 $\pm$ 0.05 &  26.8$^{+   6.0}_{-   5.8}$ &  0.51$^{+ 0.12}_{- 0.12}$ &  228.3$^{+  59.3}_{-  39.0}$ &  228.3$^{+  59.3}_{-  39.0}$ \\
SPT-CLJ2233-5339 & 0.013$^{+0.000}_{-0.000}$ &  4.3 $\pm$  1.4 & 0.06$^{+0.00}_{-0.02}$ & 0.40 $\pm$ 0.05 &  77.0$^{+  26.8}_{-  25.9}$ &  2.26$^{+ 0.82}_{- 0.78}$ &   21.7$^{+  10.5}_{-   5.3}$ &   48.2$^{+  23.2}_{-  11.8}$ \\
SPT-CLJ2236-4555 & 0.026$^{+0.001}_{-0.001}$ &  2.0 $\pm$  0.8 & 0.08$^{+0.01}_{-0.03}$ & 0.47 $\pm$ 0.06 &  22.7$^{+   9.9}_{-   9.5}$ &  0.67$^{+ 0.30}_{- 0.28}$ &   68.7$^{+  46.6}_{-  19.8}$ &   68.7$^{+  46.6}_{-  19.8}$ \\
SPT-CLJ2245-6206 & 0.002$^{+0.000}_{-0.000}$ & 11.3 $\pm$ 22.5 & 0.02$^{+0.01}_{-0.01}$ & 0.03 $\pm$ 0.01 & 680.8$^{+1385.5}_{-1345.6}$ & 25.04$^{+51.52}_{-49.32}$ &    0.0$^{+  -0.0}_{-   0.0}$ &    0.0$^{+  -0.0}_{-   0.0}$ \\
SPT-CLJ2248-4431 & 0.033$^{+0.000}_{-0.000}$ & 13.0 $\pm$  0.9 & 0.07$^{+0.00}_{-0.02}$ & 0.48 $\pm$ 0.02 & 128.0$^{+   9.5}_{-   9.4}$ &  1.79$^{+ 0.14}_{- 0.14}$ &  652.4$^{+  48.2}_{-  42.0}$ &  997.1$^{+  73.6}_{-  64.1}$ \\
SPT-CLJ2258-4044 & 0.007$^{+0.000}_{-0.000}$ &  7.9 $\pm$  2.0 & 0.04$^{+0.00}_{-0.02}$ & 0.06 $\pm$ 0.05 & 207.9$^{+  56.4}_{-  54.7}$ &  5.98$^{+ 1.68}_{- 1.61}$ &   10.9$^{+   3.7}_{-   2.2}$ &--\\
SPT-CLJ2259-6057 & 0.037$^{+0.001}_{-0.001}$ &  5.1 $\pm$  1.2 & 0.09$^{+0.00}_{-0.02}$ & 0.58 $\pm$ 0.03 &  45.8$^{+  11.4}_{-  11.2}$ &  0.91$^{+ 0.23}_{- 0.23}$ &   68.6$^{+  21.2}_{-  13.1}$ &   68.6$^{+  21.2}_{-  13.1}$ \\
SPT-CLJ2301-4023 & 0.017$^{+0.000}_{-0.000}$ &  9.6 $\pm$  1.6 & 0.10$^{+0.01}_{-0.02}$ & 0.25 $\pm$ 0.05 & 147.9$^{+  27.8}_{-  27.0}$ &  3.00$^{+ 0.59}_{- 0.57}$ &   96.2$^{+  19.6}_{-  13.9}$ &   52.3$^{+  10.6}_{-   7.6}$ \\
SPT-CLJ2306-6505 & 0.004$^{+0.000}_{-0.000}$ &  3.8 $\pm$  2.7 & 0.03$^{+0.00}_{-0.02}$ & 0.19 $\pm$ 0.08 & 158.8$^{+ 115.7}_{- 111.6}$ &  7.45$^{+ 5.55}_{- 5.26}$ &--&--\\
SPT-CLJ2325-4111 & 0.004$^{+0.000}_{-0.000}$ & 10.4 $\pm$  5.0 & 0.03$^{+0.00}_{-0.02}$ & 0.10 $\pm$ 0.04 & 417.4$^{+ 210.9}_{- 204.0}$ & 13.06$^{+ 6.76}_{- 6.44}$ &--&--\\
SPT-CLJ2331-5051 & 0.102$^{+0.002}_{-0.002}$ &  3.9 $\pm$  0.6 & 0.22$^{+0.02}_{-0.04}$ & 0.91 $\pm$ 0.03 &  18.0$^{+   2.9}_{-   2.8}$ &  0.28$^{+ 0.05}_{- 0.04}$ &  236.0$^{+  40.0}_{-  29.9}$ &  236.0$^{+  40.0}_{-  29.9}$ \\
SPT-CLJ2335-4544 & 0.017$^{+0.000}_{-0.000}$ &  5.5 $\pm$  4.6 & 0.06$^{+0.00}_{-0.02}$ & 0.50 $\pm$ 0.05 &  83.6$^{+  72.4}_{-  70.1}$ &  2.10$^{+ 1.85}_{- 1.77}$ &    3.5$^{+  17.6}_{-   1.6}$ &    3.5$^{+  17.6}_{-   1.6}$ \\
SPT-CLJ2337-5942 & 0.013$^{+0.000}_{-0.000}$ & 15.6 $\pm$  3.3 & 0.05$^{+0.00}_{-0.02}$ & 0.14 $\pm$ 0.05 & 281.7$^{+  65.8}_{-  63.9}$ &  4.98$^{+ 1.21}_{- 1.16}$ &   84.6$^{+  23.2}_{-  15.0}$ &   32.9$^{+   9.0}_{-   5.8}$ \\
SPT-CLJ2341-5119 & 0.046$^{+0.001}_{-0.001}$ &  5.2 $\pm$  2.0 & 0.08$^{+0.00}_{-0.02}$ & 0.55 $\pm$ 0.04 &  40.4$^{+  16.3}_{-  15.8}$ &  0.74$^{+ 0.31}_{- 0.29}$ &  253.8$^{+ 156.1}_{-  70.0}$ &  152.8$^{+  94.0}_{-  42.1}$ \\
SPT-CLJ2342-5411 & 0.077$^{+0.002}_{-0.002}$ &  1.9 $\pm$  0.4 & 0.17$^{+0.03}_{-0.04}$ & 0.81 $\pm$ 0.08 &  10.3$^{+   2.4}_{-   2.4}$ &  0.21$^{+ 0.05}_{- 0.05}$ &  235.7$^{+  65.2}_{-  42.0}$ &  198.4$^{+  54.9}_{-  35.3}$ \\
SPT-CLJ2344-4243 & 0.286$^{+0.004}_{-0.004}$ &  9.5 $\pm$  0.5 & 0.38$^{+0.03}_{-0.05}$ & 1.29 $\pm$ 0.03 &  22.0$^{+   1.4}_{-   1.4}$ &  0.17$^{+ 0.01}_{- 0.01}$ & 1881.1$^{+ 113.5}_{- 101.3}$ & 1881.1$^{+ 113.5}_{- 101.3}$ \\
SPT-CLJ2345-6405 & 0.007$^{+0.000}_{-0.000}$ &  6.9 $\pm$  1.7 & 0.04$^{+0.00}_{-0.02}$ & 0.14 $\pm$ 0.07 & 186.0$^{+  51.6}_{-  49.5}$ &  5.67$^{+ 1.65}_{- 1.55}$ &--&--\\
SPT-CLJ2352-4657 & 0.030$^{+0.001}_{-0.001}$ &  1.6 $\pm$  1.0 & 0.07$^{+0.00}_{-0.02}$ & 0.60 $\pm$ 0.05 &  17.1$^{+  11.3}_{-  10.9}$ &  0.49$^{+ 0.33}_{- 0.32}$ &   49.6$^{+  84.8}_{-  19.2}$ &   49.6$^{+  84.8}_{-  19.2}$ \\
SPT-CLJ2355-5055 & 0.046$^{+0.002}_{-0.002}$ &  1.4 $\pm$  0.5 & 0.13$^{+0.01}_{-0.03}$ & 0.84 $\pm$ 0.04 &  10.7$^{+   4.2}_{-   4.0}$ &  0.27$^{+ 0.11}_{- 0.10}$ &  105.4$^{+  60.1}_{-  28.1}$ &  105.4$^{+  60.1}_{-  28.1}$ \\
SPT-CLJ2359-5009 & 0.009$^{+0.000}_{-0.000}$ &  3.0 $\pm$  1.1 & 0.06$^{+0.00}_{-0.02}$ & 0.30 $\pm$ 0.05 &  71.0$^{+  28.6}_{-  27.6}$ &  2.79$^{+ 1.15}_{- 1.10}$ &--&--\\

\hline
\\
\multicolumn{9}{l}{(1):  Cluster name}\\
\multicolumn{9}{l}{(2):  Central ($r<0.012\textrm{R}_{500}$) electron density}\\
\multicolumn{9}{l}{(3):  Central ($r<0.012\textrm{R}_{500}$) temperature}\\
\multicolumn{9}{l}{(4):  Surface brightness concentration (Santos \etal 2008)}\\
\multicolumn{9}{l}{(5):  Electron density slope at 0.04R$_{500}$}\\
\multicolumn{9}{l}{(6):  Central ($r<0.012\textrm{R}_{500}$) entropy}\\
\multicolumn{9}{l}{(7):  Central ($r<0.012\textrm{R}_{500}$) cooling time}\\
\multicolumn{9}{l}{(8):  Mass deposition rate within $r(t_{cool} < 7.7 \textrm{Gyr}$)}\\
\multicolumn{9}{l}{(9):  Mass deposition rate within $r(t_{cool} < t_{\rm{Univ}}$)}\\
%\end{tabular}
\label{table:sample}
%}
%\end{table*}
\end{longtable}

\end{document}